\documentclass{article}

\usepackage[preprint]{neurips_2026}

\usepackage[utf8]{inputenc} % allow utf-8 input
\usepackage[T1]{fontenc}    % use 8-bit T1 fonts
\usepackage{hyperref}       % hyperlinks
\usepackage{url}            % simple URL typesetting
\usepackage{booktabs}       % professional-quality tables
\usepackage{amsfonts}       % blackboard math symbols
\usepackage{nicefrac}       % compact symbols for 1/2, etc.
\usepackage{microtype}      % microtypography
\usepackage{xcolor}         % colors
\usepackage{multirow}
\usepackage{graphicx}
\usepackage{todonotes}
\usepackage{pifont}
\usepackage{tcolorbox}
\usepackage[normalem]{ulem}
\usepackage{amsmath}
\usepackage{amssymb}
\usepackage{mathtools}
\usepackage{amsthm}
\usepackage{enumitem}
\usepackage{pifont}
\usepackage{placeins}

\usepackage{enumitem}

\title{Task-conditioned probing of instruction-tuned multimodal LLMs: Region-specific brain alignment patterns under naturalistic stimuli}

\author{}

\author{
Subba Reddy Oota$^{1}$, Khushbu Pahwa$^{2,3}$\thanks{Work was done while at Rice University, prior to role at Amazon.},
 Prachi Jindal$^{4}$, Satya Sai Srinath Namburi$^{5}$ \\
\textbf{Maneesh Singh$^6$, Tanmoy Chakraborty$^4$, Bapi S. Raju$^7$, Manish Gupta$^8$} \\
\normalsize $^1$Technische Universität Berlin, Germany, 
$^2$Rice University, USA, 
$^3$AWS AI Labs, Amazon \\
$^4$IIT Delhi, India, 
$^5$University of Wisconsin - Madison, USA, 
$^6$Spector Inc, USA \\ 
$^7$IIIT-Hyderabad, India, 
$^8$Microsoft, India \\
\small \texttt{subba.reddy.oota@tu-berlin.de, gmanish@microsoft.com}
}

\begin{document}

\maketitle

\begin{abstract}
Recent brain encoding studies have shown that multimodal large language models (MLLMs) exhibit a higher degree of brain alignment compared to unimodal models.
%Recent brain-encoding studies have shown that multimodal large language models (MLLMs) align more closely with brain activity than unimodal models.
More recently, instruction-tuned (IT) multimodal models 
have been shown to generate task-specific representations that align strongly with brain activity, yet most prior evaluations focus on unimodal stimuli or non-instruction-tuned models under multimodal stimuli.
%However, prior work evaluating the brain alignment of MLLMs has primarily focused on unimodal settings or relied on non-instruction-tuned models for multimodal stimuli. 
We still lack a clear understanding of whether instruction-tuning is associated with IT-MLLMs organizing their representations around functional task demands or if they simply reflect surface semantics. 
To address this, we estimate brain alignment by predicting fMRI responses recorded during naturalistic movie watching (video with audio) from MLLM representations. Using instruction-specific embeddings from six video and two audio IT-MLLMs, 
across 13 video task instructions, we find that instruction-tuned video MLLMs show higher brain alignment than in-context learning (ICL) multimodal models ($\sim$9\%), non-instruction-tuned multimodal models ($\sim$15\%), and unimodal baselines ($\sim$20\%).
Our evaluation of MLLMs across video and audio tasks, and language-guided probing produces distinct task-specific MLLM representations that vary across brain regions.
%Experiments on 13 video task-specific instructions show that instruction-tuned video MLLMs significantly outperform in-context learning multimodal models (by $\sim$9\%), non-instruction-tuned multimodal models (by $\sim$15\%) and unimodal models (by $\sim$20\%). 
%Our evaluation of MLLMs for both video and audio tasks using language-guided instructions shows clear disentanglement in task-specific representations from MLLMs, consistent with differentiation of multimodal functional processing in the brain. 
We also find that ICL models show strong semantic organization ($r$=0.78), while IT models show weak coupling to instruction-text semantics ($r$=0.14), consistent with task-conditioned subspaces associated with higher brain alignment.
%MLLM layers align hierarchically with the brain, with early sensory areas showing strong alignment with early layers, while higher-level visual and language regions align more with middle-to-late layers. 
These findings are consistent with an association between task-specific instructions and stronger brain-MLLM alignment, and open new avenues for mapping joint information processing in both systems.
\end{abstract}

\section{Introduction}
%\sri{Subba - We can change citations to numbers if you feel that will save space. I've shared screenshot to your WhatsApp related to that. Delete this comment later!!}

\begin{figure}[t]
    \centering
    \includegraphics[width=\linewidth]{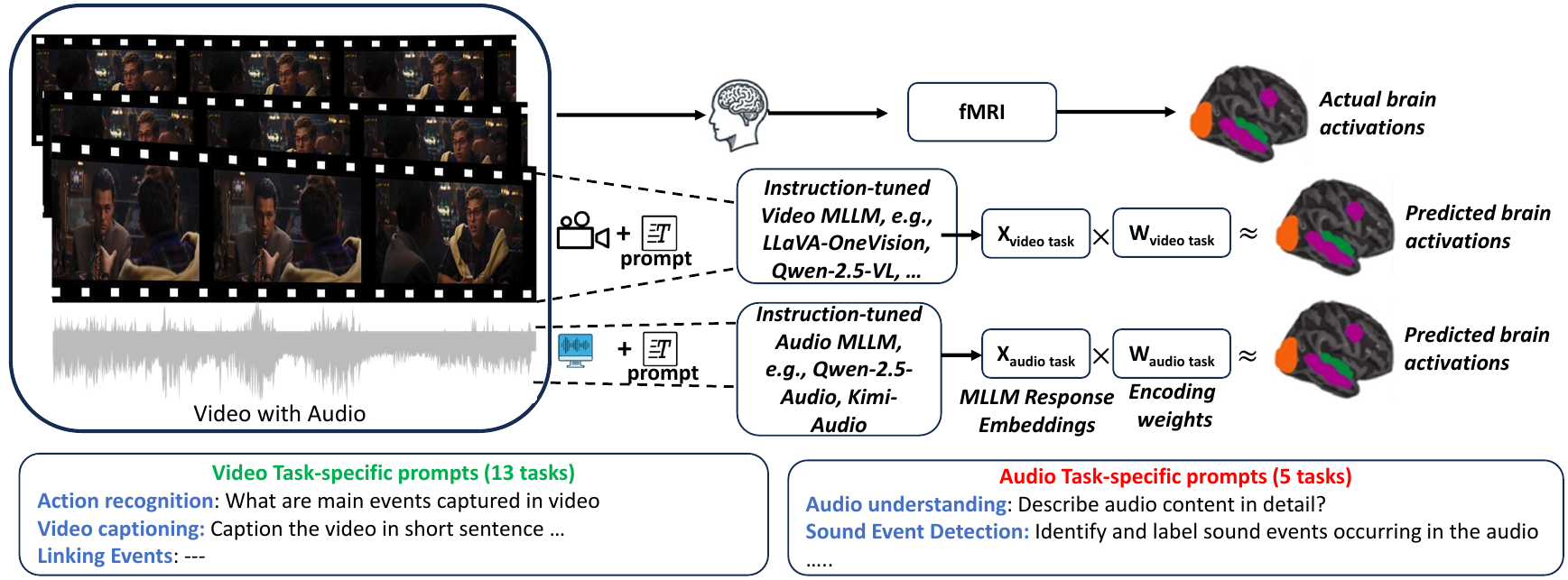}
    %\vspace{-0.1cm}
    \caption{Leveraging instruction-tuned multimodal video and audio models for brain encoding with a diverse set of instructions. For the given movie clip, we can obtain different multimodal representations using instructions that ask the model to (i) generate the caption of the video, (ii) identify whether temporal events are present, (iii) determine the primary colors dominant in the video, etc. Using instruction-specific representations (X), we estimate the alignment using a simple linear function $f$ (ridge regression), which maps MLLM representations to brain recordings. Here, W denotes voxelwise encoding model weights.}
    \label{fig:approach}
\end{figure}

The alignment between internal representations of multimodal Transformer models and cortical activation patterns obtained from naturalistic stimuli has emerged as a key focus in the study of brain-model correspondence. Recent studies on multimodal models in brain encoding commonly consider two settings: (i) multimodal models evaluated with unimodal stimuli~\citep{doerig2022semantic,wang2022natural,oota2022visio,tang2023brain,oota2025correlating,srijith2025aligning}, and (ii) multimodal models evaluated with multimodal stimuli~\citep{subramaniam2024revealing,dong2023interpreting,oota2024multi,sartzetaki2024one}. See App.~\ref{app:multimodal_task_settings}  for details.

In the former setting, 
%brain recordings are obtained from unimodal image stimuli, but representations from multimodal models, which integrate modalities such as vision and language, 
multimodal models achieve a higher degree of brain alignment compared to vision-only models~\citep{doerig2022semantic,wang2022natural,oota2022visio,popham2021visual}.
This observation holds true to instruction-tuned (IT) multimodal large language models (MLLMs), especially when prompted with natural instructions~\citep{oota2025correlating}. In the latter setting, multimodal models exhibit higher degree of brain alignment over unimodal models~\citep{dong2023interpreting,oota2024multi}.

While prior studies have examined brain alignment with IT-MLLMs, they have largely been limited to unimodal stimuli, or have used non-instruction-tuned models in the context of multimodal stimuli. In this work, we address this gap by systematically probing IT-MLLMs under naturalistic multimodal stimuli, measuring how task-specific, language-guided representations align with brain activity across cortical regions from early sensory to higher-order cognitive areas.

Several unimodal studies report that task-specific fine-tuned Transformer models better align with brain activity during text processing~\citep{oota2022neural,aw2022training,oota2022joint}, speech~\citep{oota2023speech,tuckute2022many,oota2024speech}, and vision~\citep{wang2019neural,conwell2022can}, outperforming pretrained models in brain predictivity. However, these models are task-specific, limiting generalization, requiring separate data and training per task. Instruction-tuning~\citep{xu2023multiinstruct,dai2023instructblip,liu2024visual} offers a scalable alternative, fine-tuning a single LLM on various NLP tasks and surpassing task-specific models~\citep{taori2023stanford,touvron2023llama,dubey2024llama}, while showing stronger brain alignment~\citep{sun2023tuning,sun2023fine,loong2023instruction}. See App.~\ref{app:relatedWork} for details. 

Based on these task-specific model analyses, recent work aligns IT-MLLMs with brain data for text~\citep{benara2024crafting} and images~\citep{oota2025correlating}, although limited to unimodal stimuli. Motivated by advances in MLLMs for video and audio tasks and availability of brain data for multimodal stimuli, we study alignment of video-audio IT-MLLMs for multimodal stimuli, and ask the following research questions:
%\sn{Consider removing this enumerate RQ and A. Something seems to be off with formatting}
% \begin{enumerate}[noitemsep, topsep=0pt]
    (1) How do language-guided task instructions shape task-conditioned representations in IT-MLLMs vs. pretrained in-context learning (ICL)-MLLMs, and to what extent are these representations explained by surface semantics?
    (2) How does the degree of brain alignment vary across task-specific instructions for IT-MLLMs, and how do these patterns compare to matched ICL and non-instruction-tuned baselines?
    %How do different task-specific instructions influence the degree of brain alignment in instruction-tuned video and audio MLLMs?
    %\item[RQ2] Do instruction-tuned video MLLMs exhibit better brain alignment than their audio counterparts when exposed to multimodal stimuli? 
    (3) Do IT-MLLM representations show task-dependent differentiation that aligns with distinct brain regions, supporting data-driven task probing beyond traditional stimuli?
(4) How are semantic-categories of task instructions (e.g., narrative understanding, cognitive reasoning) associated with differences in brain alignment across language, auditory, and visual regions?
To our knowledge, this is the first study to use such MLLMs to model fMRI responses across video and audio tasks (workflow in Fig.~\ref{fig:approach}).
Using brain recordings from participants watching several popular movies with audio~\citep{Ecole50613}, we investigate the brain alignment of IT-MLLMs. Specifically, we evaluate six video IT-MLLMs, two audio IT-MLLMs, two pretrained video MLLMs with ICL, two non-instruction-tuned multimodal models (video+audio),  two unimodal models for video, and one unimodal model for audio as shown in Table~\ref{neural_models}. These models are probed with 13 video task-specific instructions and 5 audio task-specific instructions.
%\sri{Subba: Please see if the key results are correct! We need to use instrucion-tuned MLLMs clearly; I might be wrong }

Our analysis of IT-MLLMs and brain alignment with multimodal stimuli reveals several key findings corresponding to the 4 RQs as follows.
%Probing task-specific representational structure suggests that 
(1) IT-MLLM task-conditioned representations are weakly coupled ($r$=0.14) to surface-semantics and are more consistent with task-specific structure, whereas ICL-MLLMs show a much stronger correlation ($r$=0.78) with surface semantics. 
(2a) Among IT-MLLMs, video-based models show higher brain alignment than audio-based models on multimodal stimuli, a modality contrast within a fixed training paradigm.
(2b) As descriptive context, IT Video MLLMs as a class show higher alignment than ICL, non-IT multimodal, and unimodal baselines. We do not attribute these cross-family differences to instruction-tuning given heterogeneity in scale, training data, and architecture. However, two controlled within-family comparisons (InternVL-8B IT vs. ICL and Qwen-2.5-VL IT vs. ICL) show consistent positive $\Delta$ alignment after instruction tuning across all layers, holding architecture, scale, and pretraining corpus fixed.
%This holds across the whole brain, as well as within language, visual, and auditory regions. (iii) On the other hand, Audio IT-MLLMs outperform both non-instruction-tuned multimodal and unimodal models mainly in the auditory cortex (AC) and middle frontal gyrus (MFG) language regions, with comparable alignment across other language ROIs. 
(3) Consistent with weak coupling to surface-semantics, both video and audio IT-MLLMs generate task-conditioned representations that selectively align with different functional brain regions. For example, audio understanding and captioning tasks show stronger alignment with language areas, while sound event detection aligns with the auditory cortex and temporal lobe. %\textit{Further, probing IT-MLLMs reveals that instruction-tuning organizes representations primarily by functional task demands and less by semantic representations.} 
(4) Task-specific representations from IT-MLLMs support finer-grained mapping of multimodal processing across cortical hierarchies. For example, language and narrative understanding category tasks show stronger alignment in angular gyrus and lateral temporal cortex, consistent with event-structure findings in naturalistic stimuli~\citep{baldassano2017discovering}.
%We also group 13 video tasks into five semantic categories and analyze category-specific brain-alignment patterns across cortical sub-regions.
%Grouping instruction-tuned video MLLMs by task type reveals distinct alignment profiles. 
%Tasks involving language and narrative understanding exhibit stronger alignment in language-related sub-regions such as angular gyrus and lateral temporal regions, consistent with prior findings on event structure representation in naturalistic stimuli~\citep{baldassano2017discovering}. 
%Similarly, spatial understanding tasks engage regions of the dorsal visual pathway, particularly the intraparietal sulcus and surrounding parietal cortex. 
Overall, IT-MLLMs exhibit both layer- and task-dependent alignment patterns, making them useful probes for studying functional specialization under naturalistic stimuli and for connecting cognitive modeling with neuroscience. 
\begin{table}[t]
\scriptsize
\centering
\begin{minipage}{0.6\textwidth}
\begin{center} 
\caption{MLLMs for video, audio vs. multimodal, unimodal models (IT: Instruction-tuned) (IC: In-context learning).} 
\label{neural_models} 
%\vspace{-0.1cm}
\begin{tabular}{|l|l|c|l|} 
\hline
\textbf{Model Name}& \textbf{IT}& \textbf{\#Layers} & \textbf{Modality} \\
\hline
InstructBLIPVideo&\ding{51}& 33  & Video+Text\\
Video-LLaVA&\ding{51} & 33 & Video+Text\\ 
LLaVa-NeXT-Video &\ding{51}& 33 & Video+Text\\ 
Qwen-2.5-VL&\ding{51}&  29 & Video+Text\\
Videochat-R1&\ding{51}& 29 & Video+Text\\ 
LLaVA-OneVision&\ding{51}& 28 & Video+Text\\ 
\hline
Qwen-2.5-Audio &\ding{51}&  29 & Audio+Text \\
Kimi-Audio&\ding{51}&  29 & Audio+Text \\ \hline
Qwen-2.5-Omni (ICL) & \ding{53}&  29 & Video+Audio+Text \\ 
InternVL (ICL) & \ding{53}&  29 & Video+Text \\ \hline
VILA & \ding{53}&  29 & Video+Audio \\
TVLT&\ding{53} &  12 & Video+Audio \\ \hline
VideoMAE&\ding{53} &  24 & Video \\
TimeSFormer&\ding{53} &  12 & Video \\
AST&\ding{53}  &24 & Audio \\
\hline
\end{tabular} 
\end{center} 
\end{minipage}
%\hspace{3pt}
\begin{minipage}{0.38\textwidth}
\centering
\caption{Instructions for various MLLM audio tasks. %\sri{Suggestion: We can move these 2 tables to Appendix most likely!}
}
\label{prompt_instructions_audio}
\begin{center}
\scriptsize
%\vspace{-0.2cm}
\begin{tabular}{|p{0.6in}|p{1.2in}|}
\hline
\bf Task&\bf Description\\ 
\hline
Audio Understanding   & Can you describe the audio content in detail? \\ \hline
Audio Comprehension  & What are people doing in the audio? \\                                 \hline
Audio Captioning& Caption the audio in a short sentence. \\ \hline
Sound Event Detection & Identify and label the sound events occurring in the audio. \\ \hline
Speaker Identification & Who is speaking in the audio? \\ \hline
\end{tabular}
\end{center}
\end{minipage}
\end{table}

\section{Dataset and Models}
%\vspace{-3mm}

\subsection{Brain Imaging Dataset} 
\label{sec:dataset}
%\vspace{-2mm}
We experiment with Movie10~\citep{Ecole50613}, a multimodal naturalistic fMRI dataset, obtained from the Courtois NeuroMod databank. This dataset was collected while four human subjects (s1, s2, s3, s5; data for s4 and s6 is not public) passively watched four different movies:  \emph{The Bourne supremacy ($\sim$100 mins)}, \emph{The wolf of wall street ($\sim$170 mins)}, \emph{Hidden figures ($\sim$120 mins)} and \emph{Life ($\sim$50 mins)}. Among these, \emph{Hidden figures} and \emph{Life} are repeated twice, with the repeats used for testing and the remaining movies for training. We use \emph{Life} movie for testing where we average the two repetitions to reduce noise. 
The dataset includes 11,017 TRs (Repetition Time) for training and 2013 TRs for testing. We build encoding models where the train and test sets are totally disjoint. 
%This is among the largest publicly available multimodal fMRI datasets by samples per participant, with 4024 TRs (Time Repetitions) 
%of \emph{The Bourne supremacy} and 6993 TRs of \emph{The wolf of wall street} for training and 2013 TRs of \emph{Life} for test. Train and test sets are totally disjoint. 
%Thus there is no possibility of any information leakage during inference on the test set. 
The fMRI data is collected every 1.49 seconds (= 1 TR).
%More details about dataset and preprocessing is in App.~\ref{app:detailedsubrois}.

The dataset is already preprocessed and projected onto the surface space (``fsaverage6'').
We use the multimodal parcellation of the human cerebral cortex based on the Glasser Atlas (which consists of 180 regions of interest in each hemisphere) to report the ROI (region of interest) analysis for brain maps~\citep{glasser2016multi}. This includes four visual processing regions (early visual cortex (EVC), object-related areas (LOC), face-related areas (OFA) and scene-related areas (PPA)), one early auditory area (AC), and eight language-relevant regions, encompassing broader language regions: angular gyrus (AG), anterior temporal lobe (ATL), posterior temporal lobe (PTL), inferior frontal gyrus (IFG), inferior frontal gyrus orbital (IFGOrb), middle frontal gyrus (MFG), posterior cingulate cortex (PCC) and dorsal medium prefrontal cortex (dmPFC), based on the Fedorenko lab's language parcels~\citep{milton2021parcellation,desai2022proper}.
We show the flatmap with these labeled ROIs, in Appendix Fig.~\ref{fig:language_flatmap} and list the detailed sub-ROIs of these ROIs in App.~\ref{app:detailedsubrois}.

\noindent\textbf{Estimating cross-subject prediction accuracy.}
To account for the intrinsic noise in biological measurements, we adapt 
 \citet{schrimpf2021neural}'s method to estimate the cross-subject prediction accuracy for a model's performance for the Movie10 fMRI dataset. 
Each subject $s$ $\in$ ([1,4]) is chosen as the prediction target and the other three are used to predict this target. 
We use a voxel-wise encoding model (see Section \ref{sec:modelArch}) to predict one participant's response from others.  The detailed approach is described in App.~\ref{app:cross_subject_flatmaps}.
%By subsampling fMRI dataset from four participants, we generate all possible combinations of $s$ participants ($s$ $\in$ [2,4]) for watching movies, and use a voxel-wise encoding model (see Sec. \ref{sec:modelArch}) to predict one participant's response from others.
Note that the estimated cross-subject prediction accuracy is based on the assumption of a perfect model, which might differ from real-world scenarios, yet offers valuable insights into model's performance.
%We estimate cross-subject prediction accuracy by training on the combined brain data from \textit{The Bourne supremacy} and \textit{The wolf of wall street} and testing on the brain data from the movie \textit{Life}.
We present the cross-subject prediction accuracy across voxels for the {Movie10 fMRI} dataset for each of the four participants in App.~\ref{app:cross_subject_flatmaps}. 
%The plots show that across all participants higher activity is observed in the language and visual regions with a max correlation up to 0.4 implying that data has low noise and low cross-subject variability.

%\vspace{-2pt}

%\vspace{-4pt}

\subsection{Instruction-tuned MLLMs for Video and Audio}
To investigate whether IT-MLLMs models, when prompted using natural language-guided instructions, show stronger alignment with fMRI responses to multimodal stimuli, we consider six popular modern IT video MLLMs (InstructBLIPVideo~\citep{dai2023instructblip}, 
Video-LLaVA~\citep{lin2023video}, 
LLaVA-Next-Video~\citep{zhang2024llavanext-video}, 
Qwen-2.5-VL~\citep{Qwen2VL}, 
%Video-LLaMA3~\citep{damonlpsg2025videollama3}, 
Videochat-R1~\citep{li2025videochatr1}, LLaVA-OneVision~\citep{li2025videochatr1}) and two IT audio MLLMs (Qwen-2.5-Audio~\citep{Qwen2-Audio}, Kimi-Audio~\citep{kimi_audio_2024}). We also experiment with two pretrained video ICL-MLLMs (Qwen-2.5-Omni~\citep{xu2025qwen2} and InternVL~\citep{chen2024internvl}), two non-instruction-tuned multimodal (VILA~\citep{lin2023vila} and TVLT~\citep{tang2022tvlt}), two video unimodal models (VideoMAE~\citep{tong2022videomae} and TimeSFormer~\citep{gberta_2021_ICML}), and one audio unimodal (AST~\citep{baade2022mae}) model. Details for these models are reported in Table~\ref{neural_models}.

Several prior brain encoding studies have compared a wide range of language, speech and multimodal models differing in size, architecture, training data, and objectives, with the primary goal of understanding how closely model representations align with brain-relevant semantics~\citep{schrimpf2021neural, toneva2019interpreting, antonello2021low, oota2024multi}. We follow this approach while additionally providing a controlled within-family comparison for Qwen-2.5-VL, InternVL that isolates instruction tuning.

\setlength{\tabcolsep}{2pt}
\begin{table*}[!t]
\caption{Instructions for various MLLM video tasks.
}
\label{prompt_instructions}
%\vspace{-3mm}
\begin{center}
\scriptsize
\begin{tabular}{|l|l|}
\hline
\bf Task&\bf Description\\ 
\hline
\multirow{1}{*}{Action Recognition} & What are the main events captured in the video? \\  
\multirow{1}{*}{Video Understanding}   & Can you describe the video content in detail? \\  
\multirow{1}{*}{Visual Question Answering}  & How many people are in the video, and what are they doing? \\
                                             
\multirow{1}{*}{Video Captioning}& Caption the video in a short sentence. \\ 
Object and Scene Recognition & What are the main objects and people visible in the video? Describe each one briefly. \\ 
Commonsense Reasoning & Why did the character take this action? What could have motivated them to do this? \\
Spatial Understanding & Where is this video taken from? What place/landmark is shown in the video? \\ 
Temporal Ordering & Step-by-step describe the activity shown in the video. \\ 
\multirow{1}{*}{Video reasoning} &  What is unusual about this video? \\ 
Narrative Understanding & Summarize the main storyline of the movie. What is the central conflict, and how is it resolved? \\ 
Emotion and Sentiment Analysis & What emotions do the characters express during the video? How does the video make you feel overall? \\ 
%Cultural Context Understanding & What cultural references or norms are represented in the video? How does this shape the characters' actions? \\ \hline
Global Appearance & Describe changes in characters' appearances throughout the video, including any noticeable outfit changes.  \\ 
Linking Events & Explain how an early event in the video influences later developments. \\ \hline
\end{tabular}
\end{center}
\end{table*}

% \setlength{\tabcolsep}{2pt}
% \begin{table}[t]
% \caption{Instructions for various multimodal audio tasks. %\sri{Suggestion: We can move these 2 tables to Appendix most likely!}
% }
% \label{prompt_instructions_audio}
% \begin{center}
% \scriptsize
% \begin{tabular}{|l|l|}
% \hline
% \bf Task&\bf Description\\ 
% \hline
% \multirow{1}{*}{Audio Understanding}   & Can you describe the audio content in detail? \\ \hline
% Audio Comprehension  & What are people doing in the audio? \\                                 \hline
% Audio Captioning& Caption the audio in a short sentence. \\ \hline
% Sound Event Detection & Identify and label the sound events occurring in the audio. \\ \hline
% Speaker Identification & Who is speaking in the audio? \\ \hline
% \end{tabular}
% \end{center}
% \end{table}
%\vspace{-0.1cm}
\subsection{Natural Language Instructions and Feature Extraction from IT-MLLMs}
\label{feature_extraction}
%\vspace{-0.1cm}
\noindent\textbf{Video-specific tasks.}
To ensure the diversity of task-specific instructions while considering videos as input, we consider 13 instructions, as shown in Table~\ref{prompt_instructions}, and extract the language-guided representations from IT video MLLMs. 
%As shown in Table~\ref{prompt_instructions}, 13 natural instructions cover 13 task categories, including \emph{action recognition, video understanding, visual question answering, video captioning, object and scene recognition, commonsense reasoning, spatial understanding, temporal ordering, video reasoning, narrative understanding, emotion and sentiment analysis, global appearance and linking events}. 
This set of 13 tasks is inspired from  VideoInstruct100K~\citep{Maaz2023VideoChatGPT} and are generally applicable to any video regardless of the contents in the image frames. 
We provide a sample of generated outputs for all the six video MLLMs in Tables~\ref{instruct_model_outputs_prompts} to~\ref{videochat_r1_model_outputs_prompts} in App.~\ref{app:ModelGeneratedOutputs}.
%\sri{SUBBA: I think the tables 4,5,6,7,8,9 has all the 6 models!! If so, please rephrase the above sentence!}.

To extract instruction-specific representations from IT video MLLMs for the brain encoding task, we input a video and task instruction to obtain the embeddings for the language-guided instruction. For ICL models, a video is paired with a natural language prompt without instruction tuning. For TVLT and VILA, we input video and audio. For TimesFormer and VideoMAE we input video only. We perform zero-shot inference on these models. For all IT video MLLMs, we use the pretrained Transformer weights, which generate hidden state representations at each layer. We then average these hidden state representations at layer level of output generated tokens to obtain  final embedding at each layer for each video with respect to task instruction.

\noindent\textbf{Audio-specific tasks.}
We consider five natural instructions while considering audio as input, as shown in Table~\ref{prompt_instructions_audio}, and extract the language-guided representations from audio IT-MLLMs. 
%As shown in Table~\ref{prompt_instructions_audio}, five natural instructions cover five task categories, including {audio understanding, audio comprehension, audio captioning, sound event detection, and speaker identification}. 
We provide generated outputs for a sample using the IT audio models across the five tasks in Tables~\ref{audio_instruct_model_outputs_prompts_qwen2.5} and~\ref{audio_instruct_model_outputs_prompts_kimi} in App.~\ref{app:ModelGeneratedOutputs}.
Similar to IT video models, to extract instruction-specific representations from the IT audio MLLMs for the brain encoding task, we input a audio and task instruction to obtain the embeddings for language-guided instruction. For AST we input audio only. We follow similar feature extraction method as video-tasks to extract audio task representations.

\section{Methodology}
%\vspace{-0.2cm}
\noindent\textbf{Voxel-wise encoding model.}
\label{sec:modelArch}
We train banded ridge regression based voxel-wise encoding models~\citep{la2022feature} to predict the fMRI brain activity associated with the stimulus representations obtained from task-specific instructions from instruction-tuned and other baseline models. Banded ridge regression optimizes a different regularization hyperparameter per feature space, and decomposes the explained variance over feature spaces. This decomposition helps in identifying which task-specific instruction contributes most to the explainable variance in different brain regions. %Overall, banded ridge regression helps to accurately identify the contribution of each task-specific instruction, leading to better prediction accuracy and better interpretability. 
We employ z-score thresholding separately for both input stimulus representations and brain recordings for training and test datasets. For each subject, we account for the delay in the hemodynamic response by modeling hemodynamic response function using a finite response filter (FIR) per voxel with 5 temporal delays corresponding to $\sim$7.5 seconds~\citep{huth2022gallant}. Formally, at each time step $t$, we encode the stimuli as $X_{t}\in \mathbb{R}^{D}$ and brain region voxels $Y_{t}\in \mathbb{R}^{V}$, where $D$ denotes the dimension of the concatenation of delayed 5 TRs, and $V$ denotes the number of voxels. Overall, with $N$ such TRs, we obtain $N$ training examples. Detailed hyper-parameter settings are in App.~\ref{app:hyperparameters_details}. 
%\sri{Appendix C says we used bootstraped ridge regression and this para says we used banded ridge regression!}.

% \noindent\textbf{Train-test setup.}
%  To be completely clear: independent encoding models are trained for each subject using data concatenated from two movies (\textit{The Bourne supremacy}: 4024 TRs and The \textit{wolf of wall street}: 6993 TRs). The test set consisted of data from the \textit{``Life'' movie} (2013 TRs).  

\noindent\textbf{Evaluation metrics.}
We evaluate our models using Pearson Correlation (PC), which is a standard metric for evaluating brain alignment \citep{jain2018incorporating,schrimpf2021neural,goldstein2022shared}. Let TR be \#time repetitions in the test set. Let $Y=\{Y_i\}_{i=1}^{TR}$ and $\hat{Y}=\{\hat{Y}_i\}_{i=1}^{TR}$ denote actual and predicted value vectors for a single voxel. Thus, $Y$ and $\hat{Y}~\in \mathbb{R}^{TR}$. 
%and also $\hat{Y}\in \mathbb{R}^{TR}$. 
PC is computed as correlation between the model's predictions $\hat{Y}$ and neural recordings $Y$. 
To quantify the model predictions, the resulting model prediction correlations are divided by the estimated cross-subject prediction accuracy; resulting in a standardized measure of performance referred to as normalized brain alignment. To calculate \emph{normalized alignment}, we select voxels with cross-subject prediction accuracy $\ge$ 0.05, in line with previous works~\citep{popham2021visual,la2022feature}.

% \sri{After reading the paper; we didn't discuss anything related to PC, cross-subject prediction accuracy in the main paper! I think we need to bring some part of appendix and restructure the paper accordingly}
%\vspace{-0.2cm}
\section{Results}
\label{sec:results}
%\vspace{-0.2cm}
%\sn{If removing RQs above, reformat this. The square brackets seems off with formatting in my opinion}
\subsection*{\textbf{[RQ1]:} ICL-MLLM representations strongly correlate with instruction-text semantics; IT-MLLMs representations show weak semantic coupling and are more task-conditioned.}
%\vspace{-0.2cm}

To understand the difference in the working of IT and ICL MLLMs, we perform additional analysis to identify fundamental differences in the representational organization. For the 13 video tasks, we first compute a 13×13 semantic-similarity matrix using two independent text embedding models (all-MiniLM-L6-v2~\citep{reimers-2019-sentence-bert} and MPNet~\citep{reimers-2019-sentence-bert}). App.~\ref{app:instruction_semantics_representaions} Fig.~\ref{fig:instruction_similarity_sbert} shows pairwise semantic similarity between 13 instruction prompts computed using MiniLM embeddings (left) and MPNet (right). From Fig.~\ref{fig:instruction_similarity_sbert} (left), we observe that the semantic similarity ranges from 0.15 to 0.85 (mean: 0.42 $\pm$ 0.18), with multiple pairs of high-similarity ($>$0.60) identified: (Action Recognition vs. Video Understanding (0.68), Video Understanding vs. Visual Question Answering (0.65), Object \& Scene Recognition vs. Action Recognition (0.68)). We also observe low semantic similarity pairs: (Commonsense Reasoning vs. Most others, Commonsense Reasoning vs. other tasks, Spatial Understanding vs. Emotional/Narrative). This confirms our task-instruction set contains a broad range of semantic relatedness across task instructions.

\begin{table}[t]
\centering
\scriptsize
\caption{IT vs. ICL models: Representational similarity and clustering metrics. *** p$<$0.001, FDR-significant across 29 layers; ns = not significant after FDR correction.}
\label{incontext_it_semantics}
%\vspace{-0.2cm}
\scriptsize
% \resizebox{\textwidth}{!}{
\begin{tabular}{|p{1in} |l |c |c |c |l |l|}
\hline
\textbf{Metric} & \textbf{Model} & \textbf{Layer 5} & \textbf{Layer 14} & \textbf{Layer 25} & \textbf{Winner} & \textbf{Magnitude} \\
\hline

\multirow{2}{1in}{\textbf{Semantic correlation}} 
 & IT (Qwen-2.5-VL-Instruct)    & Pearson $r = 0.077$ (ns) & 0.130 (ns) & 0.143 (ns) & \multirow{2}{*}{ICL} & Weak \\
 & ICL (Qwen-2.5-VL) & 0.690*** & 0.680*** & 0.777*** &  & $\sim$4.2$\times$ stronger \\
\hline
\multirow{2}{1in}{\textbf{Adjusted Rand Index (label alignment)}}
 & IT $\Delta$ (Func $-$ Sem)  & +0.129 & +0.142 & +0.174 & Task-aligned & Weak \\
 & ICL $\Delta$ (Func $-$ Sem) & $-0.442$ & $-0.291$ & $-0.349$ & Semantic-aligned & Very Strong \\
\hline
\end{tabular}
% }
\end{table}

\noindent\textbf{Qwen-2.5-VL: ICL vs. IT MLLM representations.} For the 13 video tasks, we extracted language hidden states from Qwen-2.5-VL-7B across all 29 layers for a pretrained (ICL) checkpoint and its IT variant, using identical video inputs with different instructions to isolate instruction effects. We show detailed results for representative early, middle, and late layers (L5, L14, L25) in App.~\ref{app:instruction_semantics_representaions} Fig.~\ref{fig:instruction_similarity_representations}. While we present detailed results for three representative layers, we computed correlations and clustering metrics across all 29 layers to ensure robustness.
% that the findings are not artifacts of the selection of a specific layer. 

\noindent\textbf{Semantic coupling differs sharply between ICL and IT.}
Table~\ref{incontext_it_semantics} summarizes semantic coupling and functional-vs-semantic clustering differences for representative layers.
For the 13 tasks, we perform representational similarity analysis (RSA) by computing correlations between the upper triangles of the 13×13 semantic-similarity matrix (from App.~\ref{app:instruction_semantics_representaions} Fig.~\ref{fig:instruction_similarity_sbert}) and the corresponding representation-similarity matrix derived from MLLM hidden states (same videos, same pipeline), per layer for one ICL and IT-MLLM. We find that for the IT model, the correlation between instruction semantic similarity and internal representation similarity is weak across layers (e.g., best layer L28: Pearson r=0.183, p=0.109; Spearman $\rho$=0.266, p=0.018); no layer remains significant after FDR correction across 29 layers (see App.~\ref{app:instruction_semantics_representaions} Table~\ref{tab:correlation_depth}). In contrast, ICL-MLLM shows strong RSA coupling (e.g. r$\approx$0.78), indicating that internal states closely track prompt wording, consistent with shallow text-proximal matching.

\noindent\textbf{IT representations decouple from surface semantics with depth.} 
We find that ICL models show stronger coupling to prompt semantics: r=0.78 vs 0.14 (IT), a 4.2x advantage (p<0.001). Table~\ref{incontext_it_semantics} further notes that IT model representations show progressively stronger task-conditioned clustering with depth, as measured by ARI\footnote{\url{https://scikit-learn.org/stable/modules/generated/sklearn.metrics.adjusted_rand_score.html}} $\Delta$=+0.13 to +0.17 across layers, progressive strengthening: L5: +0.129 $\rightarrow$ L25: +0.174 (increasing with depth). Overall, these results suggest that ICL representations are more wording-sensitive, whereas IT representations are less coupled to surface semantics and stronger task-conditioned representational structure, consistent with the IT gains in brain alignment reported in subsequent sections. This is consistent with~\citep{zhu2025internvl3} which showed that instruction tuning leads to task-specific representations supporting superior downstream performance. We extend this finding by showing that these representations also yield improved brain alignment, as reported in subsequent sections. 
We report a more detailed analysis in Appendices~\ref{app:instruction_semantics_representaions} and~\ref{app:instrcution_tuning_icl}.

\begin{figure*}[!t]
    \centering
     \includegraphics[width=\linewidth]{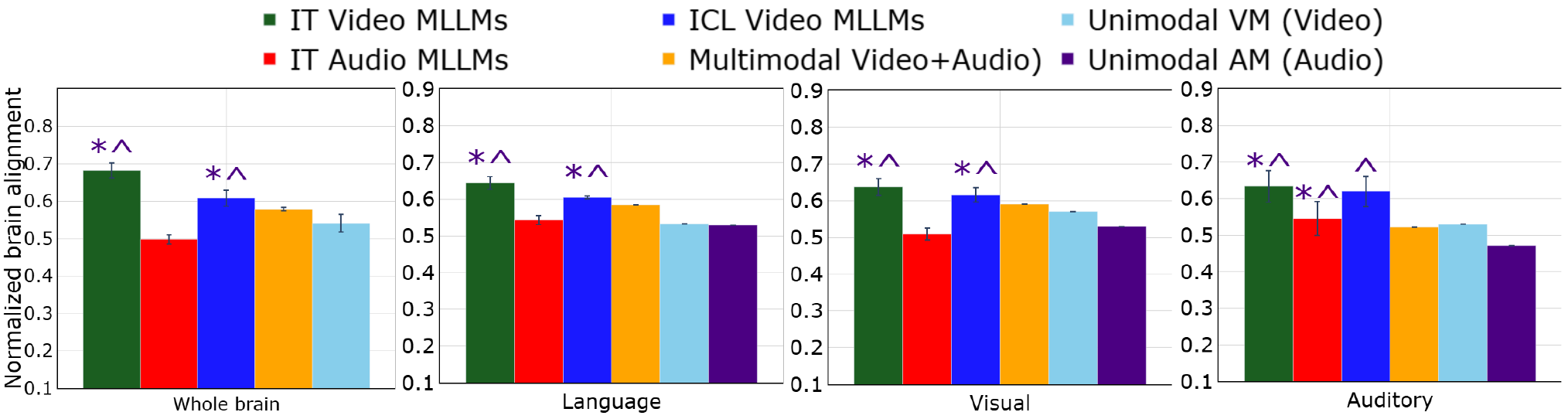}\\
     %\vspace{-0.1cm}
    %   \includegraphics[width=0.22\linewidth]{images/whole_brain_normalized_alignment_video_audio_barplot.pdf}
    % % % % \hspace{-10pt}
    % \includegraphics[width=0.22\linewidth]{images/Language_brain_normalized_alignment_barplot.pdf}
    %  \includegraphics[width=0.22\linewidth]{images/Visual_brain_normalized_alignment_barplot.pdf}
    %  \includegraphics[width=0.22\linewidth]{images/Auditory_brain_normalized_alignment_barplot.pdf}
    %\includegraphics[width=\linewidth]{images/barplots2.pdf}\
   % \vspace{-0.4cm}
    \caption{Average normalized brain alignment of IT video MLLMs vs IT audio MLLMs vs ICL video MLLMs vs multimodal and unimodal models across whole brain, language, visual and auditory regions. Error bars indicate the standard error of the mean across participants. $*$ implies that IT MLLM embeddings are significantly better than multimodal models and $\wedge$ means that IT MLLM embeddings are significantly better unimodal models with p$\leq 0.05$. 
    %IT: Instruction-tuned, ICL: In-context learning
    % \sri{Can we mention about this p-value somewhere in Section 4 or Appendix or near Wilcox-signed test?}
    % \textcolor{green}{\textbf{$\times$}}
    % $*$ and $\wedge$ imply that instruction-tuned MLLM embeddings are significantly better than multimodal and unimodal models with p$\leq 0.05$.  
    }
    \label{fig:normalized_brain_alignment_grouped}
\end{figure*}

%\vspace{-0.2cm}
\subsection*{\textbf{[RQ2]:} IT Video MLLMs representations align well with brain activity across wholebrain, language, visual and auditory regions.}
%\vspace{-0.2cm}

We measure brain alignment from representations of IT video and audio MLLMs at the whole-brain level and within language, visual, and auditory ROIs. For each IT MLLM, we calculate the average normalized brain alignment across 13 tasks for video and 5 tasks for audio, multiple subjects, and best MLLM layer, using the Movie10 fMRI dataset. 
%Similarly, for instruction-tuned Audio MLLMs, we calculate the average normalized brain alignment across five tasks, multiple subjects, and best MLLM layer. 
Additionally, we report the brain alignment performance of ICL video MLLMs, non-instruction-tuned multimodal models, unimodal video models, and unimodal audio model (AST). We treat the non-instruction-tuned multimodal models and unimodal models (audio and video) as the baselines when comparing against the IT-MLLMs.

\noindent\textbf{Whole brain analysis.} Fig.~\ref{fig:normalized_brain_alignment_grouped} (a) provides these insights: (i) At the whole-brain level, the Wilcoxon signed-rank test reveals that the differences in brain alignment between IT video MLLMs and ICL models, the non-instruction-tuned multimodal and unimodal models are statistically significant. In particular, all IT video MLLMs achieve over $\sim$9\% improvement in brain alignment compared to ICL models, and $\sim$15\% improvement compared to other baselines. This contrasts with prior findings on IT image-based MLLMs, which demonstrated comparable performance to multimodal models when evaluated on unimodal image stimuli~\citep{oota2025correlating}, suggesting that IT video MLLMs are more effective at capturing brain-relevant representations. (ii) IT audio MLLM embeddings show less alignment compared to non instruction-tuned multimodal and unimodal video models. 
These findings indicate that IT video MLLM models capture brain-relevant representations and contain additional information beyond the ICL, non-instruction-tuned multimodal and unimodal models.

\begin{figure*}[t]
    \begin{minipage}{0.49\textwidth}
\centering
   \includegraphics[width=\linewidth]{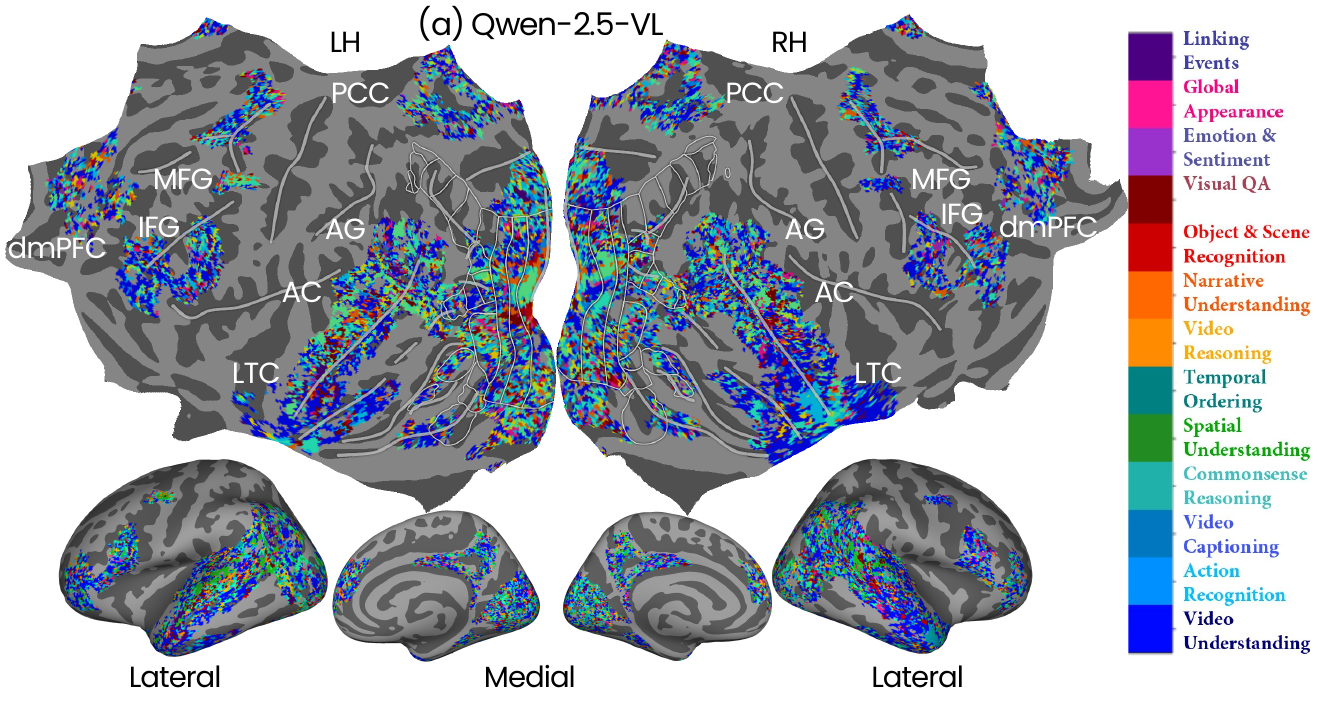}
    \end{minipage}
    \begin{minipage}{0.49\textwidth}
\centering
\includegraphics[width=\linewidth]{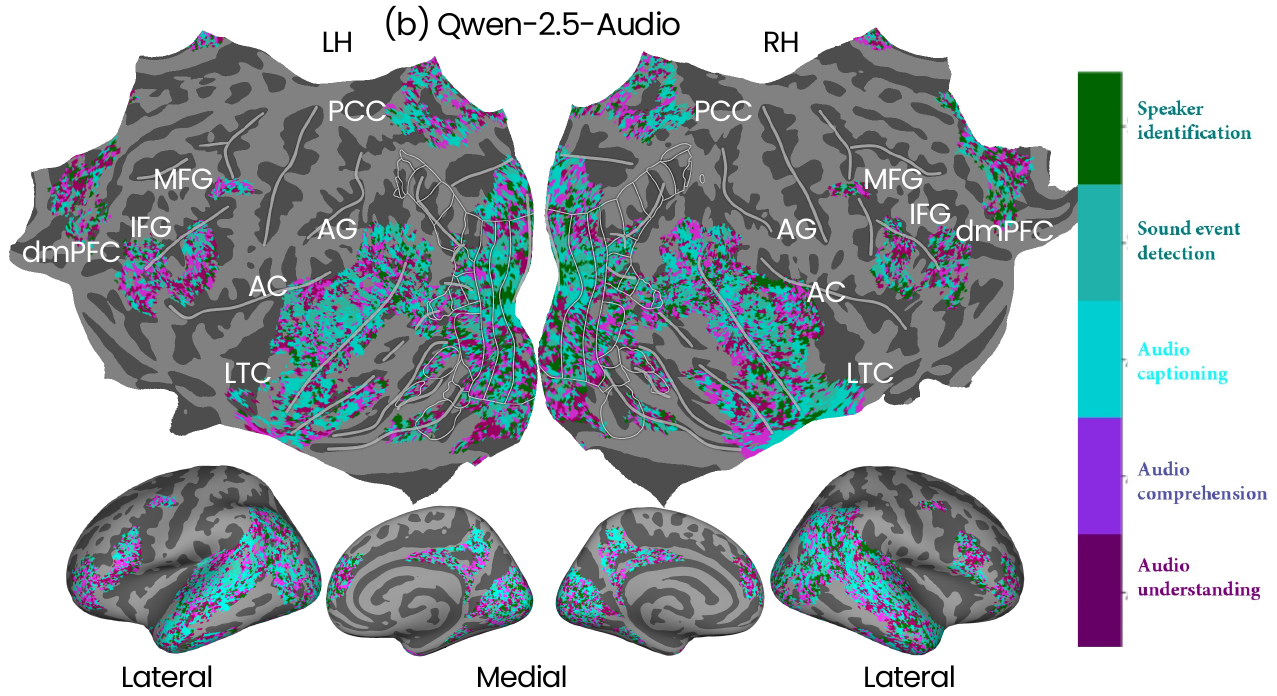}
\end{minipage}
%\vspace{-0.2cm}
    \caption{Each voxel is color-coded with the instruction that led to the highest normalized brain alignment. 
    %The color bar highlights color codes for each instruction. 
    The voxels are projected onto the flattened cortical surface of the `fsaverage' subject.  
    % , with applied hex color codes for the 10 task instructions, 
     (Left): video MLLM (Qwen-2.5-VL). (Right): audio MLLM (Qwen-2.5-Audio).}
    \label{fig:video_instruction_brainmap}
\end{figure*}

\begin{figure*}[t]
    \centering
     \includegraphics[width=0.95\linewidth]{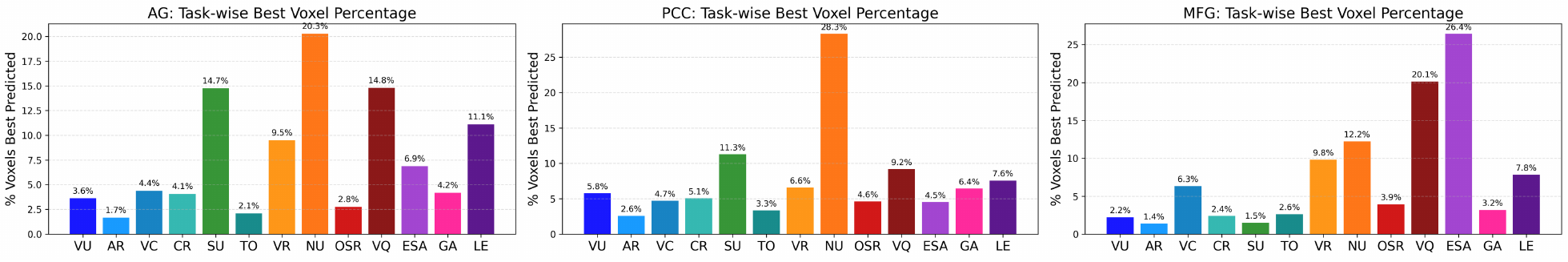}\\
     %\vspace{-0.1cm}
    %   \includegraphics[width=0.22\linewidth]{images/whole_brain_normalized_alignment_video_audio_barplot.pdf}
    % % % % \hspace{-10pt}
    % \includegraphics[width=0.22\linewidth]{images/Language_brain_normalized_alignment_barplot.pdf}
    %  \includegraphics[width=0.22\linewidth]{images/Visual_brain_normalized_alignment_barplot.pdf}
    %  \includegraphics[width=0.22\linewidth]{images/Auditory_brain_normalized_alignment_barplot.pdf}
    %\includegraphics[width=\linewidth]{images/barplots2.pdf}\
    %\vspace{-0.4cm}
    \caption{\% of best predicted voxels per task for an IT video MLLM across 3 lang ROIs.
    }
    \label{fig:qwen_percentage_voxels}
\end{figure*}

\noindent\textbf{Language, visual and auditory region analysis.}
We also present the average normalized brain alignment across language, visual and auditory regions in Fig.~\ref{fig:normalized_brain_alignment_grouped} 
 (b, c \& d). The results from Wilcoxon signed-rank test are consistent with whole-brain performance both in the language and visual regions, i.e., IT video MLLMs embeddings exhibit significantly higher alignment in both language and visual regions compared to ICL video MLLMs, and all baselines. On the other hand, IT audio MLLM embeddings show significant alignment primarily in the auditory cortex and the middle frontal gyrus; when compared to non-instruction-tuned multimodal and unimodal models. Results for individual models are shown in Fig.~\ref{fig:normalized_brain_alignment_allmodels}, detailed language, visual and auditory sub-regions are shown in Fig.~\ref{fig:language_results_subrois} and~\ref{fig:visual_speech_results_subrois} in App.~\ref{app:MultimodalityResults}. 
These results suggest that IT video MLLMs more effectively capture brain-relevant multimodal representations, particularly when processing multimodal stimuli.

Additionally, we present contrast of brainmaps to display the average normalized brain alignment across voxels. Figs.~\ref{fig:qwenvl_internvl_2dmap} and~\ref{fig:qwenvl_omni_2dmap} in App.~\ref{app:instruction_video_incontext_flatmaps} compare IT video MLLMs with ICL video MLLMs. Figs.~\ref{fig:instruction_tuned_video_multimodal_brainmaps_vila} to~\ref{fig:llavanext_tvlt_2dmap} in App.~\ref{app:instruction_video_noninstruction_flatmaps} compare IT video MLLMs with the non-instruction-tuned multimodal models. The results show that IT video MLLMs consistently achieve significantly higher alignment across all brain voxels. However, Figs.~\ref{fig:instruction_tuned_audio_multimodal_brainmaps_qwenaudio} \&~\ref{fig:instruction_tuned_audio_multimodal_brainmaps_Kimiaudio} in App.~\ref{app:instruction_video_noninstruction_flatmaps} reveal clear differences between audio MLLMs and multimodal models: the prediction performance of audio MLLMs lacks brain-relevant semantic information compared to multimodal models. 
% Manish: I don't think it makes sense to suddenly bring in another new model here.
% We have added additional video MLLM, Kimi-VL~\citep{team2025kimi}, to our evaluation and ran it through the same brain-encoding pipeline (identical preprocessing, instruction prompts, voxel-wise mapping, and normalization) (see App.~\ref{app:other_videomllm_kimi} Fig.~\ref{fig:normalized_brain_alignment_allmodels_kimivideo}).

% \sri{SUBBA: I don't have neuroscience background so apologies; but I didn't understand this sentence from the referred figure. Maybe rephrase needed.}.

%\vspace{-0.2cm}
\subsection*{\textbf{[RQ3]:} Video and audio IT-MLLMs successfully differentiate task-specific instructions.}
%\vspace{-0.2cm}
To investigate which instructions are more effective in predicting brain activity and whether IT-MLLMs differentiate task-specific representations and provide clear separation in brain regions, for each voxel, we assign the instruction with the highest normalized brain alignment.

\noindent\textbf{No-prompt baseline.} Task-specific  prompting boosts alignment over a no-prompt baseline (0.746 vs 0.639, +16.7\% for Qwen-2.5-VL), showing gains beyond generic video representations from models.
%To verify that the observed brain alignment is driven by task-specific instructions rather than the model's generic video representations alone, we evaluated a no-prompt baseline using Qwen-2.5-VL with identical video input but without any task instruction. Relative to this baseline (normalized brain alignment = 0.639), task-specific prompting improves alignment to 0.746, corresponding to a relative gain of 16.7\%. This indicates that task-specific prompting contributes substantively to brain alignment beyond the model's generic video representations.

\noindent\textbf{Instruction-tuned video MLLMs.}
Fig.~\ref{fig:video_instruction_brainmap} (left) shows brain maps for Qwen-2.5-VL for video tasks for average normalized brain predictivity across subjects.
%where the voxel color codes are projected onto the flattened cortical surface of the `fsaverage' subject. 
The color-scheme corresponding to each instruction is also reported. We make the following observations: (i) Video understanding exhibits the strongest alignment across the whole brain.
(ii) Tasks such as spatial understanding, narrative understanding, and visual question answering show higher alignment in language-related regions, including the angular gyrus, posterior temporal lobe, and visual regions.
(iii) Higher-order language regions in the frontal cortex are predominantly identified by the video understanding task, with a smaller proportion of voxels also activated by video reasoning and temporal ordering tasks. 

\begin{table}[t]
    \centering
    \caption{Modality control: Mean $\pm$ SD normalized alignment across subjects using Qwen-2.5-Omni. % (backbone and scale held fixed). Normalized brain alignment across whole brain and three ROI groups for video+audio, video-only, and audio-only inputs. Mean $\pm$ SD across subjects.
    }
   % \vspace{-0.2cm}
    \label{tab:modality_control}
    \scriptsize
    \begin{tabular}{|l|c|c|c|c|}
\hline
Modality & Whole Brain & Language & Visual & Auditory \\
\hline
Video + Audio & \textbf{0.670 $\pm$ 0.021} & \textbf{0.685 $\pm$ 0.013} & \textbf{0.689 $\pm$ 0.024} & \textbf{0.681 $\pm$ 0.019} \\
Video only & 0.601 $\pm$ 0.011 & 0.615 $\pm$ 0.027 & 0.629 $\pm$ 0.027 & 0.629 $\pm$ 0.027 \\
Audio only & 0.534 $\pm$ 0.030 & 0.545 $\pm$ 0.019 & 0.521 $\pm$ 0.025 & 0.560 $\pm$ 0.011 \\
\hline
\end{tabular}
\end{table}

\begin{figure*}[t]
\centering
\scriptsize
\begin{minipage}{0.325\textwidth}
\centering
    \includegraphics[width=\linewidth]{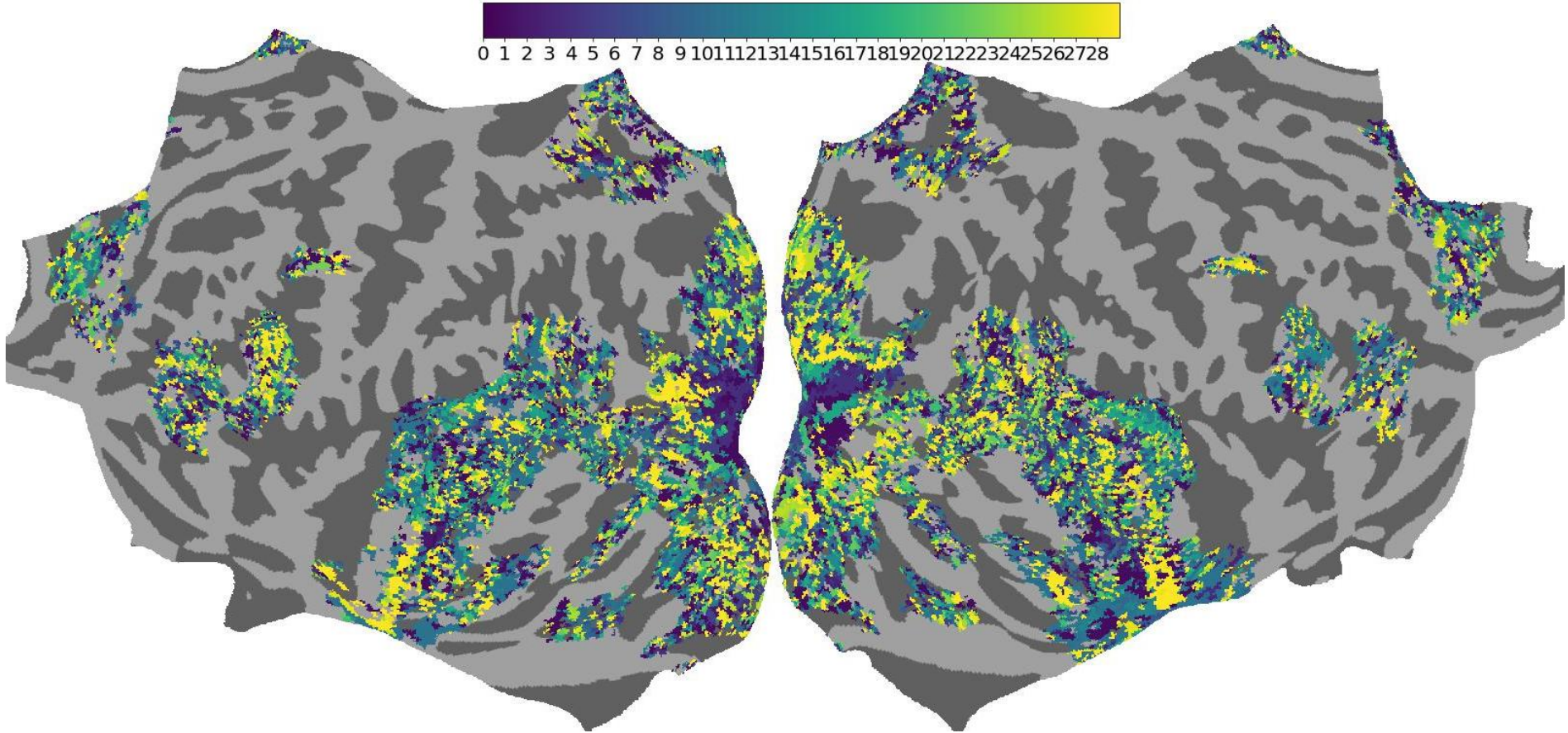}
    \\(a) Qwen-2.5-VL-7B Instruct\\
\end{minipage}
\hfill
\begin{minipage}{0.325\textwidth}
\centering
    \includegraphics[width=\linewidth]{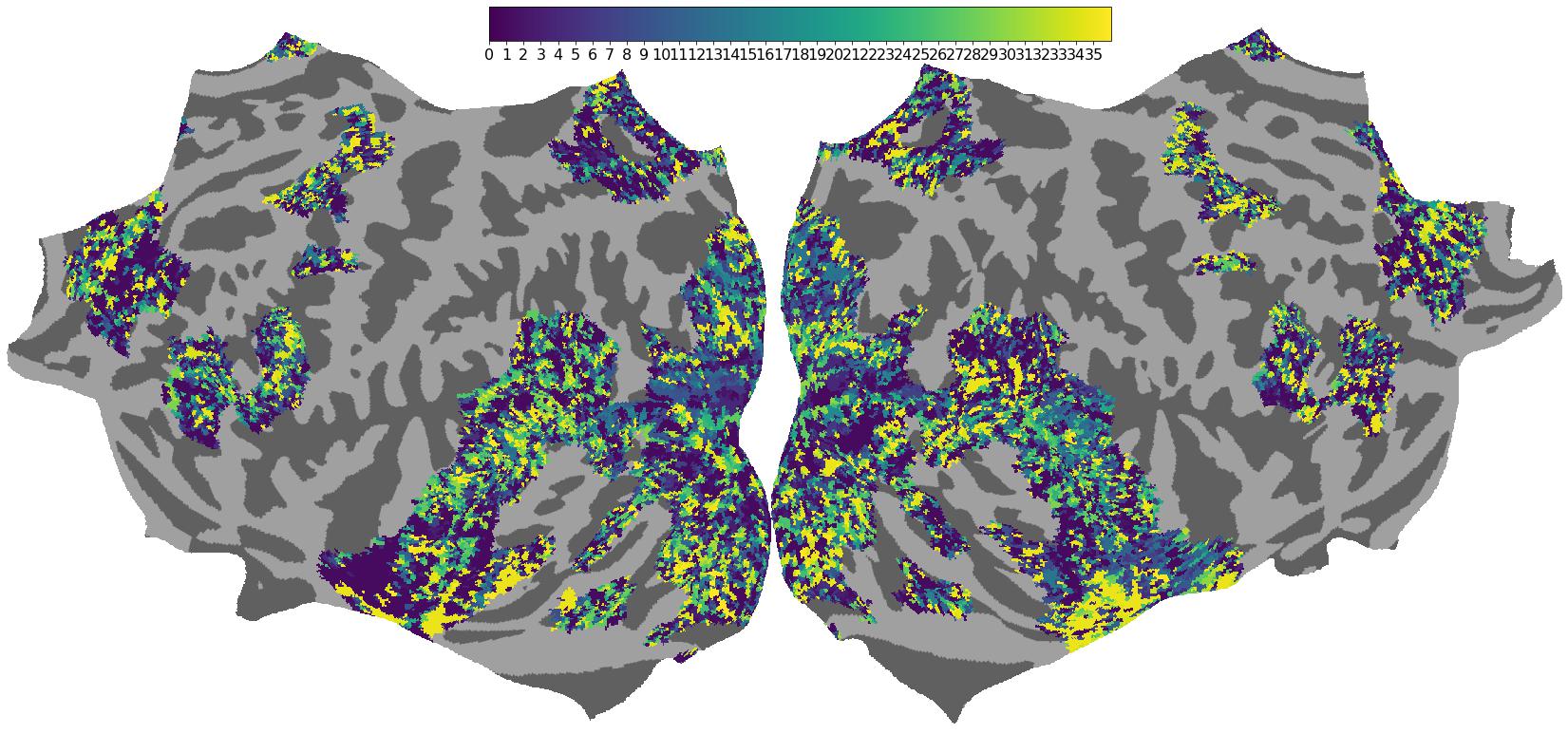}
    \\(b) Qwen-2.5-VL-3B Instruct\\
\end{minipage}
\hfill
\begin{minipage}{0.325\textwidth}
\centering
    \includegraphics[width=\linewidth]{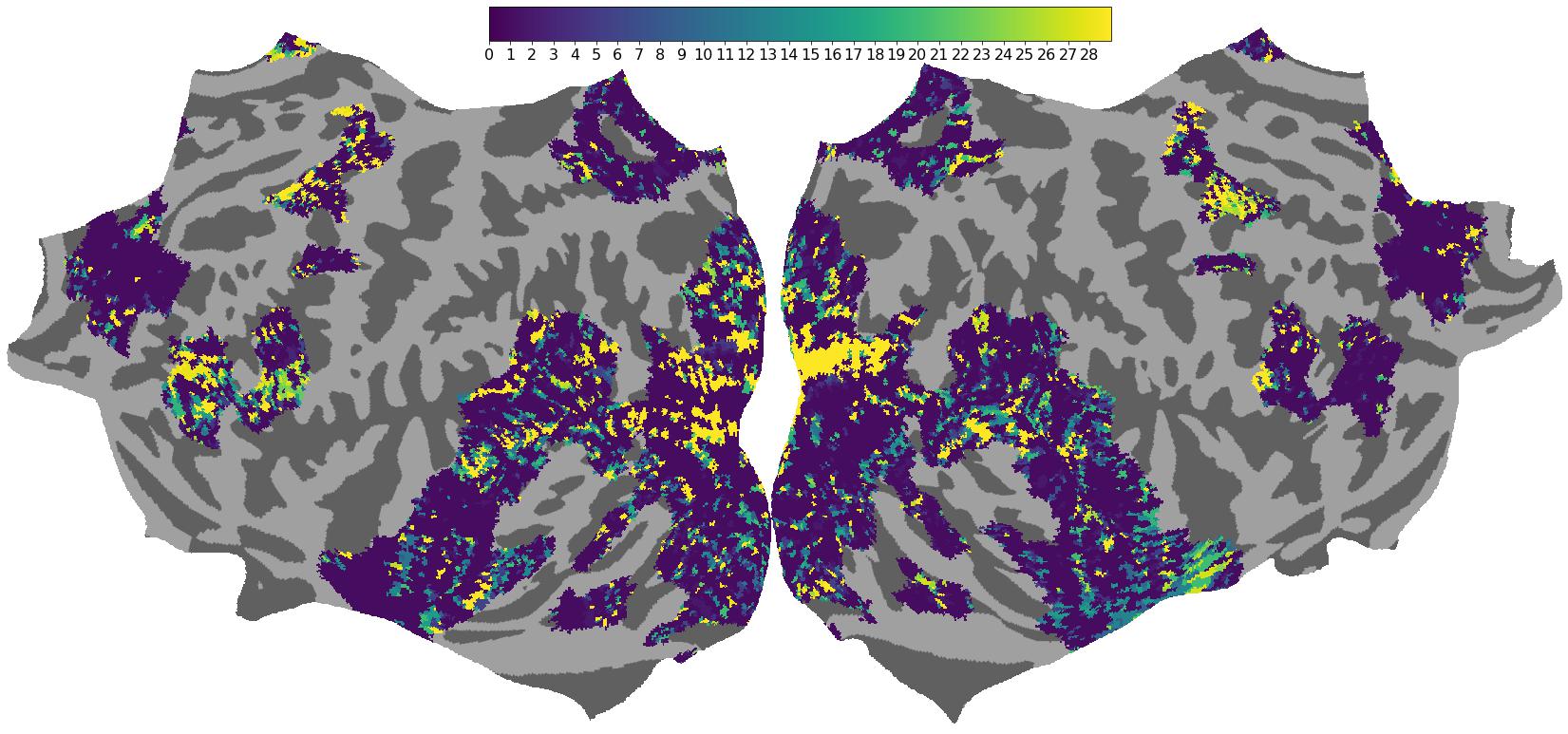}
    \\(c) Qwen-2.5-VL-7B Instruct Random Prompt\\
\end{minipage}
% \begin{minipage}{0.49\textwidth}
% \centering
%     \includegraphics[width=0.85\linewidth]{images/qwen_layers_audio.pdf}
%      \\(b) Qwen-2.5-Audio \\
% \end{minipage}
%\vspace{-0.1cm}
    \caption{
    % (a) Qwen-2.5-Audio: Each voxel is color-coded with the instruction (out of 5) that led to the highest normalized brain alignment. The color bar highlights color codes for each instruction. (b) 
    % (a) Qwen-2.5-VL-7B Instruct, (b) Qwen-2.5-VL-3B Instruct and (c) Qwen-2.5-VL-7B Instruct non-natural language prompt  
    Layer-wise alignment: Each voxel is color coded with the MLLM layer number (out of 29) that led to the highest normalized brain alignment. 
    %The color bar highlights color codes for each layer. 
    The voxels are projected onto the flattened cortical surface of average across subjects on `fsaverage' surface.}
    \label{fig:audio_instruction_brainmap}
\end{figure*}

\noindent\textbf{Higher-order ROI task preferences.}
In Fig.~\ref{fig:qwen_percentage_voxels}, we examine task-wise best-voxel fractions in key higher-order language ROIs to assess how instructions differentially engage them. We observe distinct task-region preferences:
(i) AG is dominated by Narrative Understanding (NU, 20.3\%), Visual QA (VQ, 14.8\%), and Spatial Understanding (SU, 14.7\%), consistent with its role in narrative integration and cross-modal semantic reasoning.
(ii) PCC shows the strongest dominance for Narrative Understanding (NU, 28.3\%) followed by Spatial Understanding (SU, 11.3\%) and Visual QA (VQ, 9.2\%), consistent with its role in episodic memory and discourse-level processing.
(iii) MFG is dominated by Emotion/Sentiment Analysis (ESA, 26.4\%), Visual QA (VQ, 20.1\%), and Narrative Understanding (NU, 12.2\%), consistent with its role in higher-order reasoning, emotional integration, and cognitive control. 
These results suggest that IT video MLLMs more effectively capture brain-relevant multimodal representations, and that different task instructions preferentially engage distinct higher-order ROIs in ways consistent with known functional neuroanatomy. Plots for remaining ROIs are shown in Appendix Fig.~\ref{fig:qwen_percentage_voxels_remaining_rois}. 
% \sri{this para has good info which we can explain more clearly rather than 3 bullet points.}
These findings suggest that IT video MLLMs not only capture modality-specific representations (e.g., visual, linguistic), but also encode task-specific instructions involving semantic integration and event structure (like video understanding). 
%This highlights that these models can encode complex neural patterns. 
We observe similar performance gains in other IT video MLLMs; flatmaps showing task-specific encoding performance for average of subjects are shown in  Figs.~\ref{fig:video_instruction_brainmap_instructblip} and~\ref{fig:video_instruction_brainmap_llavaonevision} in App.~\ref{app:taskSpecificDetailedBrainMaps}.
% ,~\ref{fig:video_instruction_brainmap_videollava} and~\ref{fig:video_instruction_brainmap_llavanext}. 
% \sri{Legends in Fig 13, 14, 15 and Fig 3 are bit different. eg: Video Understanding vs Video Under. basically every legend is cut-off in the Appendix figures}.

\noindent\textbf{Instruction-tuned audio MLLMs.}
Fig.~\ref{fig:video_instruction_brainmap} (right) shows %brainmap for IT audio MLLM (Qwen-2.5-Audio) where the predictions are average across subjects. The voxel color codes are projected onto the flattened cortical surface of the `fsaverage' subject. There is 
a clear distinction between different audio tasks. Audio captioning and sound detection are aligned with the auditory cortex (AC), while audio understanding activates higher-level regions like the inferior temporal (IT) cortex and inferior frontal gyrus (IFG). In contrast, speaker identification  shows very sparse and scattered alignment, with some unexpected activation in the primary visual cortex (V1), suggesting it does not strongly reflect brain-relevant semantic processing. Fig.~\ref{fig:audio_instruction_brainmap_kimi} in App.~\ref{app:taskSpecificDetailedBrainMaps} shows similar brainmap for Kimi-Audio. Plots for ROIs associated with task are shown in Appendix Fig.~\ref{fig:qwen_percentage_voxels_remaining_rois_audio}.

\noindent\textbf{Modality control.}
Using Qwen‑2.5‑Omni for a controlled comparison, video+audio consistently outperforms unimodal inputs, video exceeds audio in most regions, and audio performs best in auditory cortex. Overall this shows that modality drives alignment beyond architecture and scale.
% We provide a controlled within-family modality comparison using Qwen-2.5-Omni, which supports video-only, audio-only, and video+audio inputs while holding backbone architecture and parameter count fixed.
% From Table~\ref{tab:modality_control}, three patterns emerge consistently across regions: (i) Video+Audio outperforms both unimodal settings across all regions, showing that audio adds complementary information beyond what video alone captures; (ii) Video-only substantially exceeds Audio-only across whole-brain, language, and visual regions, consistent with our main finding that video MLLMs achieve higher brain alignment than audio MLLMs in naturalistic movie viewing; (iii) Audio-only performs relatively best in auditory cortex (0.560) compared to visual regions (0.521), consistent with known functional specialization. This provides controlled within-family evidence that input modality contributes to alignment differences beyond backbone architecture and model scale, while broader matched-modality comparisons remain an important direction for future work.

\noindent\textbf{IT-MLLMs capture layer-wise cortical hierarchy.} 
In Fig.~\ref{fig:audio_instruction_brainmap}, for each voxel, we assign the layer with highest normalized brain alignment and visualize this on cortical maps. For Qwen-2.5-VL-7B (Fig.~\ref{fig:audio_instruction_brainmap}, left), we observe a clear gradient: (i) early sensory areas (both visual and auditory) align with lower layers, capturing low-level features; (ii) high-level visual (LOC and PPA) and lang regions (superior temporal sulcus (STS) and AG) align with middle-deep layers, reflecting increasing abstraction and semantically rich representations; (iii) lang regions (IFG, ATL, AG) align most with deepest layers.
Overall, IT‑MLLMs show a layered structure aligned with brain’s hierarchy. See Fig.~\ref{fig:layers_instruction_brainmap} in App.~\ref{app:layerwiseDetails} for other models.

Further, we validate this hierarchy under two settings: (i) scaling the model (Qwen-2.5-VL 3B vs. 7B Instruct) and (ii) a non-language prompt control. Fig.~\ref{fig:audio_instruction_brainmap} (middle) shows the 3B-Instruct model, and (right) the 7B model with a random prompt (like \#\#\#\#\# \#\#\#\# \#\#\#\#). We observe: (i) Non-language control: no hierarchical gradient, with early layers dominating across cortex; (ii) 3B-Instruct: the same early-to-mid/late cortical gradient as the larger model.

% \noindent\textbf{Validation of ``hierarchical correspondence'' between model layers and brain regions.}
% We validate hierarchical correspondence under two settings: (i) Scaling (Qwen-2.5-VL-3B Instruct vs. 7B Instruct) with correct prompt (ii) a non-language prompt control, e.g.: \#\#\#\#\# \#\#\#\# \#\#\#\# this does not provide any meaning while passing video as input. Similarly to the Qwen-2.5-VL-7B Instruct model (Fig.~\ref{fig:audio_instruction_brainmap} (left)), we perform a qualitative analysis by considering that each voxel is color coded with the MLLM layer number (out of 29) that led to the highest normalized brain alignment. Fig.~\ref{fig:audio_instruction_brainmap} (middle) shows the resulting brain maps for the 3B-Instruct model, and Fig.~\ref{fig:audio_instruction_brainmap} (right) displays the 7B-Instruct model with the non-language prompt. We make the following observations: (i) 
% Non-language control (random prompt): There is no hierarchy of information processing observed across layers, where the brain prediction is dominated by early layers across cortex
% (ii) Instruction condition (3B model): The hierarchical pattern remains the same i.e., a systematic early to mid/late gradient across cortex, i.e., the same pattern we observed with the larger model.

\begin{figure}[t] 
\centering
\includegraphics[width=0.49\linewidth]{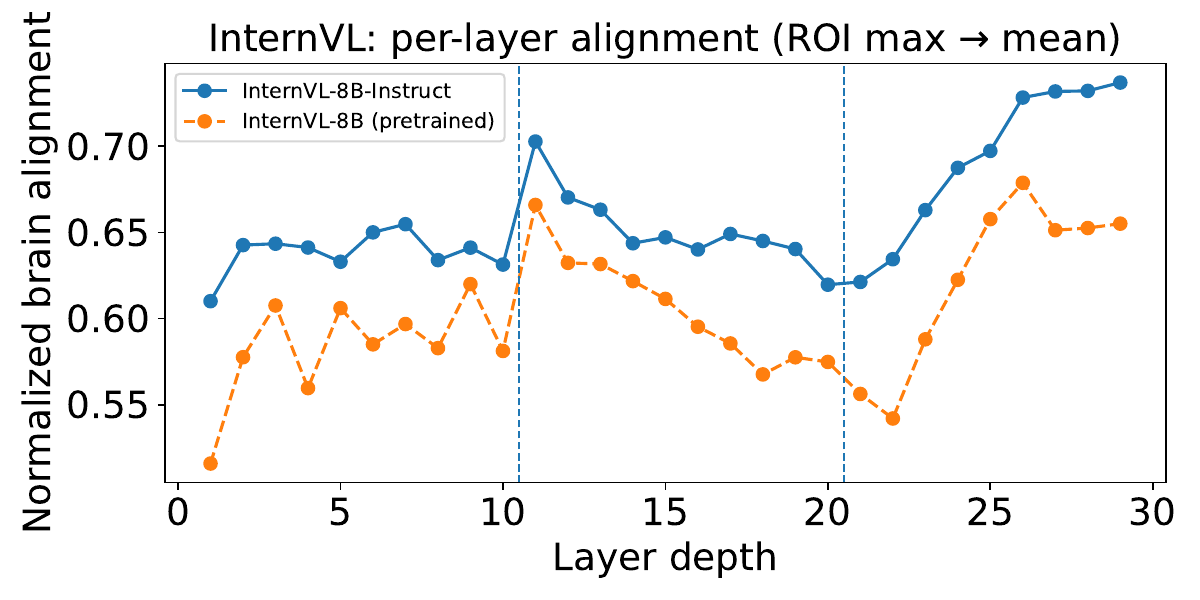}
    \includegraphics[width=0.49\textwidth]{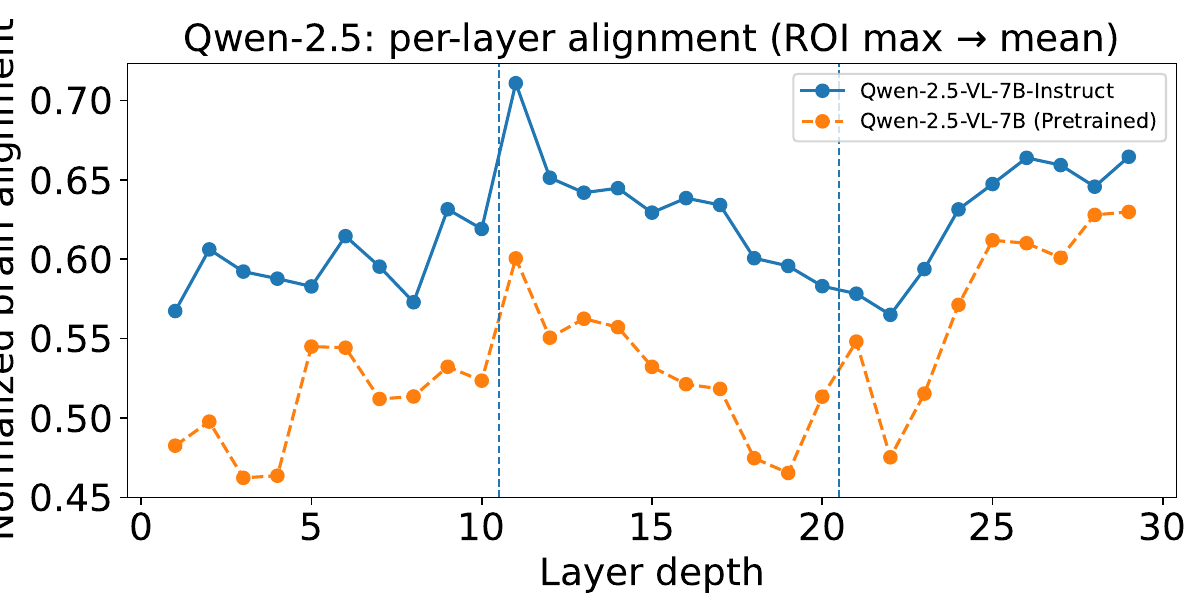}
%\vspace{-0.2cm}
\caption{Normalized brain alignment before vs. after instruction tuning: using brain predictivity across layers for (left) InternVL-8B-Instruct and InternVL-8B models, (right) Qwen2.5-VL-7B-Instruct and Qwen2.5-VL-7B models.
}
\label{fig:internvl_before_after_main}
\end{figure}

\noindent\textbf{Effect of instruction tuning}: We compare same-family models: InternVL-8B pretrained (ICL) vs. InternVL-8B-Instruct and Qwen-2.5-VL-7B pretrained (ICL) vs. Qwen-2.5-VL-7B-Instruct, using the same prompt (``Describe the video'').
For InternVL (Fig.~\ref{fig:internvl_before_after_main}, left), per-layer $\Delta$ alignment (Instruct-pretrained) is positive across all layers (min 0.0557 L5, max 0.1396 L1), with larger gains in mid/late layers (early 0.0934, mid 0.0961, late 0.1100; $\rho\approx0.34$, $p\approx$ 0.07. The improvement centers around layer 15/29, indicating stronger mid/late-layer gains.
Qwen-2.5-VL shows a similar trend (Fig.~\ref{fig:internvl_before_after_main}, right), with consistent positive gains across layers.

%\vspace{-0.1cm}
\subsection*{\textbf{[RQ4]:} Semantic task groups show distinct brain-alignment patterns in IT video MLLMs.}
\label{sec:semTaskGroupResults}
\begin{figure}[t]
    \centering
    \includegraphics[width=0.49\linewidth]{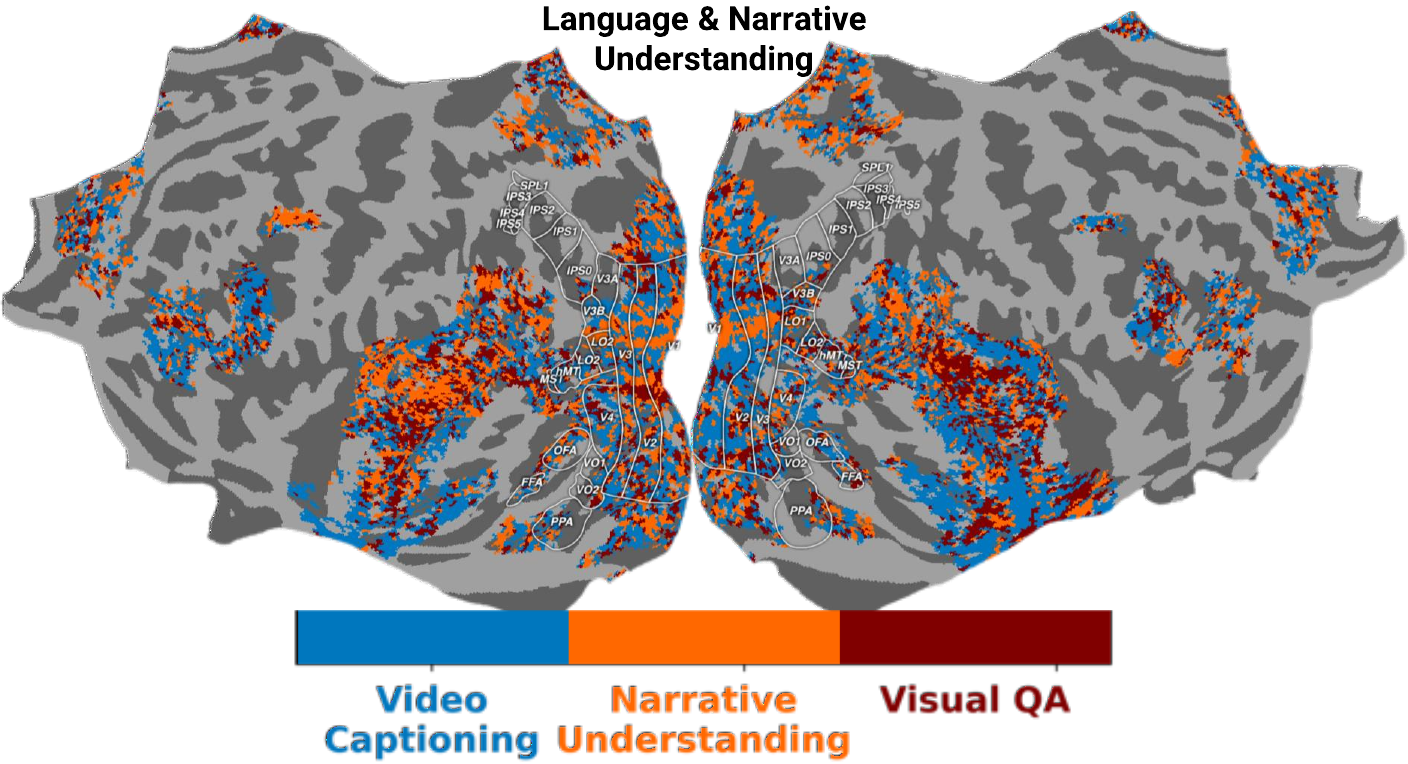}
    \includegraphics[width=0.49\linewidth]{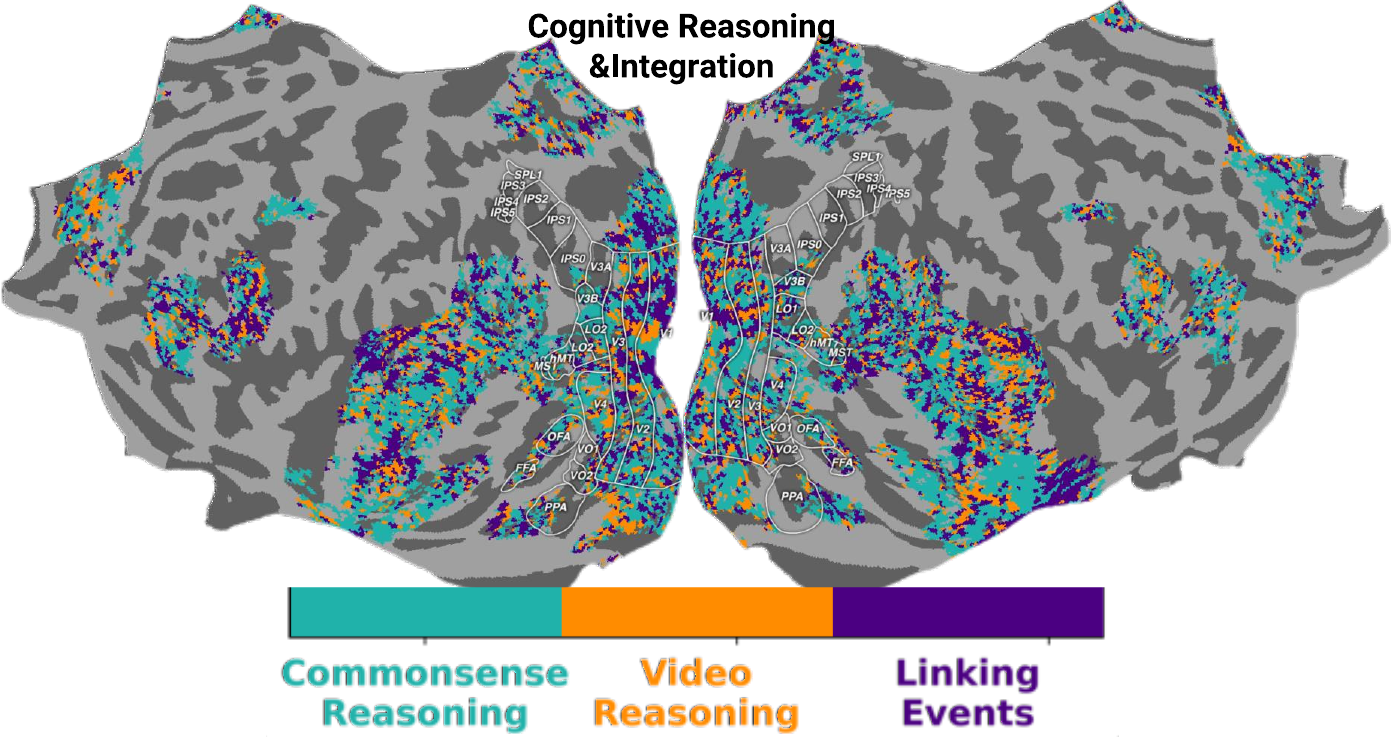}
    %\includegraphics[width=0.24\linewidth]{images/perceptual_qwen.pdf}
    %\vspace{-0.1cm}
    \caption{Semantic Task Group Analysis: Each voxel is colored by the task instruction yielding the highest normalized brain alignment. Voxels are projected onto the flattened cortical surface (averaged across subjects) for Qwen-2.5-VL. Brain maps for remaining 3 task groups are in Fig.~\ref{fig:qwen_instruction_brainmap_disspciation2} in App.~\ref{app:semTaskGroupAnalysis}.} \label{fig:qwen_instruction_brainmap_disspciation}
\end{figure}

%\vspace{-0.1cm}
To examine how task semantics relate to regional alignment patterns, we group 13 video tasks into 5 cognitive categories (perceptual, cognitive reasoning and integration, spatiotemporal understanding, language/narrative understanding, social/affective understanding) and observe clear regional distinctions (Fig.~\ref{fig:qwen_instruction_brainmap_disspciation}).

% \textbf{Language/narrative} tasks show broad activation in temporal–parietal regions (e.g., bilateral temporal lobes, AG, posterior STS, PCC), consistent with multimodal integration needed for narrative comprehension. This aligns with~\citep{mar2011neural,baldassano2017discovering}.
% \textbf{Cognitive reasoning} splits by type: commonsense engages temporal cortex, AG, and higher-order visual areas for semantic processing and world knowledge; video reasoning aligns with early visual regions (V1, V2, V3) for visual perception and motion processing; linking-events tasks involve early auditory cortex, ATL, IFG, and AG for integrating temporal, linguistic, and episodic information. Overall, these category-level patterns indicate that different reasoning tasks are associated with distinct region-wise alignment profiles across modalities and timescales.

% To examine how task semantics relate to regional alignment patterns, we group the 13 video tasks into 5 cognitively grounded categories: Perceptual visual processing, Cognitive reasoning and integration, Spatiotemporal understanding, Language and narrative understanding, and Social and affective understanding. 
% %This categorization allows us to disentangle the functional specificity of brain regions engaged by different task types. 
% Fig.~\ref{fig:qwen_instruction_brainmap_disspciation} illustrates that this grouping captures meaningful distinctions. 

Tasks in the \textbf{Language and narrative understanding} group show broader and denser cortical engagement, particularly across the temporal and parietal cortices, compared to visual and frontal regions. In particular, we observe strong activity in the bilateral temporal lobes for narrative understanding, as well as in the angular gyrus, posterior superior temporal sulcus (pSTS), and posterior cingulate cortex (PCC) regions known to support multimodal integration, which is critical for narrative comprehension. This is aligned with previous work~\citep{mar2011neural,baldassano2017discovering}.
Tasks such as narrative understanding, video captioning, and visual question answering—categorized under high-level language and narrative understanding show broader and denser cortical engagement, particularly across the temporal and parietal cortices, compared to visual and frontal regions. In particular, we observe strong activity in the bilateral temporal lobes for narrative understanding, as well as in the angular gyrus, posterior superior temporal sulcus (pSTS), and posterior cingulate cortex (PCC) regions known to support multimodal integration, which is critical for narrative comprehension. Overall, this result aligns with previous work~\citep{mar2011neural,baldassano2017discovering} that discovered that the movie narrative understanding elicits stronger activity in the aforementioned regions.

\noindent\textbf{Cognitive Reasoning.} Commonsense reasoning shows widespread activation in the temporal cortex, angular gyrus, and higher-order visual regions, reflecting its reliance on semantic processing and world knowledge. In contrast, video reasoning shows strong alignment with early visual areas (V1, V2, V3), indicating a greater dependence on visual perception and motion processing. Linking-events tasks show higher alignment in early auditory cortex and a broader pattern across anterior temporal lobe, inferior frontal gyrus, and angular gyrus, consistent with integrating temporal, linguistic, and episodic information. Overall, these category-level patterns indicate that different reasoning tasks are associated with distinct region-wise alignment profiles across modalities and timescales.

Similar category-specific patterns appear for other groups (App.~\ref{app:semTaskGroupAnalysis}), indicating task-dependent IT-MLLM representations with distinct regional alignment profiles. We also quantify overlap and dissociation in variance explained by different task instructions via variance partitioning (App.~\ref{app:sharedVarianceDetails}).

\section{Conclusion}
%\vspace{-0.2cm}

% Using task-conditioned video/audio IT-MLLM representations, we predict fMRI activity during naturalistic viewing and compare alignment across modalities and ROIs. IT video MLLMs outperform audio, ICL, unimodal, and non-instruction-tuned baselines. Key findings:
Using IT video and audio MLLM representations across task-specific instructions, we evaluate their ability to predict fMRI brain activity during naturalistic movie viewing. We also compare alignment across video and audio MLLMs for whole-brain, language, visual, and auditory regions. IT video MLLMs show significantly stronger brain alignment than audio, ICL, unimodal, and non-instruction-tuned multimodal baselines. Our study on IT-MLLMs and their alignment with multimodal stimuli yields several key findings:
% (1) ICL representations strongly couple to instruction wording, while IT representations are less sensitive to surface semantics.
% (2) IT video MLLMs show strong whole-brain alignment; IT audio MLLMs gains are limited to auditory cortex and a language region (MFG), with performance near non-instruction baselines. This modality asymmetry is consistent with prior studies of audio–brain alignment patterns~\citep{oota2024speech,oota2024multi}.
% (3) Task-specific instructions yield distinct task-conditioned representations in both video and audio MLLMs, producing region-differentiated alignment patterns not observed under unimodal stimuli in prior work~\citep{oota2025correlating}. 
% This supports using instructions as a controlled lever to vary task goals while holding the stimulus fixed. 
% (4) Layer-wise alignment mirrors brain hierarchy, with shallow layers matching sensory areas and deeper layers matching higher-order regions. See App.~\ref{app:extendedDiscussion} for extended discussion, and App.~\ref{app:limitations} for limitations.

(1) ICL-MLLM representations are strongly coupled to instruction-text semantics, whereas IT-MLLM representations show weak semantic coupling to surface wording, suggesting that IT-MLLM representations are less coupled to surface instruction wording than ICL representations.
(2) IT video MLLMs show strong brain alignment across whole brain and major ROIs, whereas IT audio MLLMs show their gains mainly in auditory cortex and a language region (MFG). This modality asymmetry is consistent with prior studies of audio–brain alignment patterns~\citep{oota2024speech,oota2024multi}, but audio IT-MLLM performance remains close to non-instruction-tuned multimodal baselines, indicating substantial room for improvement, consistent with recent efforts to introduce brain-relevant inductive biases in model design~\citep{moussa2025improving,vattikonda2025brainwavlm}.
(3) Task-specific instructions yield distinct task-conditioned representations in both video and audio MLLMs, producing region-differentiated alignment patterns not observed under unimodal stimuli in prior work~\citep{oota2025correlating}. 
This supports using instructions as a controlled lever to vary task goals while holding the stimulus fixed. 
(4) IT-MLLM layers show alignment patterns consistent with the brain's processing hierarchy, with shallow layers aligning with sensory areas and deeper layers aligning with higher-order language regions. This layered correspondence improves interpretability and underscores their potential for studying how the brain organizes task-driven information. 
We provide an extended discussion in App.~\ref{app:extendedDiscussion}. Lastly, we discuss limitations in App.~\ref{app:limitations}.

%These findings suggest that MLLMs with task-specific instructions are a great tool to effectively differentiate task representations and explore functional specialization in the brain.

% \section*{Impact Statement}
% This work aims to advance computational cognitive neuroscience by characterizing how instruction-tuned video–audio MLLMs relate to brain responses under naturalistic multimodal stimuli. We did not create any new neural recordings data as part of this work. We used the Movie10 dataset which is publicly available without any restrictions. Movie10 dataset can be downloaded from \url{https://github.com/courtois-neuromod/movie10/tree/33a97c01503315e5e09b3ac07c6ccadb8b887dcf}. Please read their terms of use\footnote{\url{https://docs.cneuromod.ca/en/latest/ACCESS.html}} for more details. We do not foresee any harmful uses of this technology. 

% \noindent\textbf{Potential positive impacts.} Our findings support the use of task instructions as a scalable and controlled probe for studying task-dependent processing with fixed stimuli, which may accelerate hypothesis generation and evaluation in neuroAI and brain-inspired AI. The analysis may also help guide the design of multimodal models and benchmarks that better reflect neural and cognitive constraints.

%This paper presents work whose goal is to advance the field of cognitive computational neuroscience. 


\bibliography{neurips_2025}
\bibliographystyle{neurips_2026}

\medskip

%%%%%%%%%%%%%%%%%%%%%%%%%%%%%%%%%%%%%%%%%%%%%%%%%%%%%%%%%%%%

\newpage

\appendix
%\onecolumn

\noindent{\Large\textbf{Overview of Appendix Sections}}

\begin{itemize}
\item App.~\ref{app:multimodal_task_settings}: Overview of multimodal model evaluation settings in brain encoding studies
\item App.~\ref{app:relatedWork}: Related work
\item App.~\ref{app:detailedsubrois}: Detailed sub-ROIs of language, visual and auditory regions
\item App.~\ref{app:cross_subject_flatmaps}: Cross-subject prediction accuracy
\item App.~\ref{app:hyperparameters_details}: Implementation details for reproducibility
\item App.~\ref{app:statistical_significance}: Statistical Significance
\item App.~\ref{app:ModelGeneratedOutputs}: Model generated outputs across instructions
\item App.~\ref{app:instruction_semantics_representaions}:  Correlation Between Instruction Semantics and Model Representations
\item App.~\ref{app:instrcution_tuning_icl}: Representational Differences Between IT and ICL Models
\item App.~\ref{app:MultimodalityResults}: Effectiveness of instruction-tuned video MLLMs vs audio MLLMs vs multimodal vs unimodal representations for various brain regions
\item App.~\ref{app:other_videomllm_kimi}: Inclusion of other Video MLLMs
\item App.~\ref{app:instruction_video_incontext_flatmaps}: Contrasting Instruction-tuned Video MLLMs with In-context Learning video MLLMs
\item App.~\ref{app:instruction_video_noninstruction_flatmaps}: Contrasting Instruction-tuned video MLLMs with Non-instruction-tuned Multimodal
\item App.~\ref{app:taskSpecificDetailedBrainMaps}: Brain Maps for Task-specific instructions
\item App.~\ref{app:layerwiseDetails}: Brain Maps showing Layer-wise Details for Video Instruction-based MLLMs
\item App.~\ref{app:layer_hierarchy_qwen}: Validation of ``hierarchical correspondence'' between Model Layers and Brain Regions
\item App.~\ref{app:before_after_tuning}: Self-Controlled Experiments: Comparing Models Before and After Instruction Tuning
\item App.~\ref{app:semTaskGroupAnalysis}: Details of Semantic Task Group Analysis
\item App.~\ref{app:sharedVarianceDetails}: Details of Explained Variance Partitioning
\item App.~\ref{app:extendedDiscussion}: Extended Discussion
%\item App.~\ref{app:llm_usage}: LLM Usage
\item App.~\ref{app:architectural_differences_mllms}: Discussion on Controlling Architectural and Pretraining Differences Across MLLMs
\item App.~\ref{app:normalized_alignment_crosssubject_ev}: Normalized brain alignment: cross-subject vs. repeat-based EV ceiling.
\item App.~\ref{app:cross_subject}: Cross Subject Generalization
\item App.~\ref{app:limitations}: Limitations 
\end{itemize}

\section{Overview of multimodal model evaluation settings in brain encoding studies}
\label{app:multimodal_task_settings}

Recent studies on multimodal models in brain encoding commonly consider two settings, as shown in Table~\ref{tab:multimodal_settings}: (i) multimodal models evaluated with unimodal stimuli, and (ii) multimodal models evaluated with multimodal stimuli. In the former setting, brain recordings are obtained from unimodal image stimuli, but representations from multimodal models, which integrate modalities such as vision and language, achieve a higher degree of brain alignment compared to vision-only models. In the latter setting, where brain recordings are obtained from multimodal stimuli (video with audio), studies show that multimodal models exhibit higher degree of brain alignment over unimodal models

\setlength{\tabcolsep}{1.5pt}
\begin{table*}[!ht]
\centering
\caption{Overview of multimodal model evaluation settings in brain encoding studies.}
\label{tab:multimodal_settings}
\scriptsize
\begin{tabular}{|p{2.5cm}|p{2.8cm}|p{2.8cm}|p{1cm}|p{2cm}|c|}
\hline
\textbf{Study} & \textbf{Model Type} & \textbf{Stimulus Modality} & \textbf{Brain Data} & \textbf{Dataset} &\textbf{Instruction-Tuned}\\
\hline
\citet{doerig2022semantic} & Vision-Language (CLIP) & Unimodal (Images) & fMRI & NSD & \ding{55}\\
\hline
\citet{wang2022natural} & Vision-Language (CLIP) & Unimodal (Images) & fMRI & NSD & \ding{55}\\
\hline
\citet{oota2022visio} & Vision-Language (CLIP, VisualBERT, LXMERT) & Unimodal (Images) & fMRI & BOLD5000 &\ding{55}\\
\hline
\citet{popham2021visual} & Vision-Only CNNs vs. Vision-Language & Unimodal (Silent Videos) & fMRI &  Gallant lab short video clips & \ding{55}\\ \hline
\citet{tang2022tvlt} & non-instruction-tuned multimodal model (BridgeTower) & Unimodal (Silent Videos), Unimodal (listening stories) & fMRI &  Gallant lab short video clips & \ding{55}\\
\hline
\citet{oota2025correlating} & Instruction-tuned Image+Text MLLMs & Unimodal (Images) &fMRI & NSD & \ding{51}\\
\hline
\citet{sartzetaki2024one} & Image Recognition models, Action recognition models & Unimodal (Visual) & fMRI & Bold Moments Dataset & \ding{55}\\
\hline
\citet{nakagi2024brain} & Language models (BERT, GPT-2, Lllama2, OPT) & Multimodal (Videos with audio) &fMRI & 8.3 hours
of video dataset & \ding{55}\\
\hline
\citet{subramaniam2024revealing} & non-instruction-tuned multimodal models (SLIP-CLIP, SimCLR, BLIP, BEIT) & Image frame-text pairs (Movies) & SEEG & AMMT & \ding{55}\\
\hline
\citet{dong2023interpreting} & non-instruction-tuned multimodal models (Merloreserve) & Multimodal (Movies: Videos with audio) & fMRI & Neuromod Friends dataset & \ding{55}\\
\hline
\citet{oota2024multi} & non-instruction-tuned multimodal models (TVLT and ImageBind) & Multimodal (Movies: Videos with audio) & fMRI & Neuromod Movie10 & \ding{55}\\
\hline
Our study & instruction-tuned video and audio MLLMs, in-context learning video and audio MLLMs & Multimodal (Movies: Videos with audio) & fMRI & Neuromod Movie10 & \ding{51}\\
\hline
\end{tabular}
\end{table*}

\section{Related work}
\label{app:relatedWork}

\noindent\textbf{Brain encoding using multimodal models.}
%\sri{Subba: There's multi-modal and multimodal both used in paper! I think we used multimodal in our ICLR paper!}
Our work is closely related to that of~\citet{conwell2022can, wang2022natural,doerig2022semantic,tang2023brain,nakagi2024brain,dong2023vision,oota2024multi}, who proposed using multimodal model representations to study the contribution of  brain alignment in unimodal and multimodal stimuli. The majority of brain encoding studies in using multimodal models focused on a single modality of input – vision alone~\citep{conwell2022can,wang2022natural,doerig2022semantic,wang2022natural,tang2023brain,nakagi2024brain}.  
Recently,~\citet{dong2023vision,oota2022visio} interpreted the effectiveness of multimodal Transformer language models in multimodal naturalistic stimuli. However, these studies focus on pretrained multimodal models which are not generic to tasks and lack the investigation of recent instruction-tuned models. %\khushbu{removed the line from here based on the comment. Will put it in experiments section.}
% Our work is complementary to these studies, \todo{flow is missing here! Also this point need not be in related works; it comes in experiments section}we experiment with instruction-tuned video and audio MLLMs and we perform a comprehensive study using 6 separate instruction-tuned video MLLMs and 2 instruction-tuned audio MLLMs. 

% \mg{I also have a worry that this paper may get into self plagiarism issues. We should use GPT to rephrase sentences rather than copy pasting them as is here.}

\noindent\textbf{Task-based brain alignment.}
Our work is also closely related to that of~\citet{wang2019neural, oota2022neural,aw2022training,sun2023tuning} and~\citet{loong2023instruction}, who propose using task-specific model representations to study the contribution of individual tasks to brain alignment.~\citet{wang2019neural} investigated 21 computer vision tasks to explore which vision tasks are more aligned with the brain while subjects engaged in viewing passive images. Similarly,~\citet{oota2022neural} and~\citet{sun2023tuning} explored 10 GLUE NLP tasks to study which NLP tasks are more brain-aligned during reading and listening to stories.
More recent work by~\citet{loong2023instruction} uses instruction-tuned LLMs to investigate the effect of natural language instruction model representations on brain alignment across layers for language comprehension. Further,~\citet{oota2025correlating} use IT-MLLMs (image+text), using natural language instructions across diverse vision tasks to analyze their alignment with brain activity across layers during visual processing. However, these studies primarily focused on unimodal stimuli and thus do not fully capture the capabilities of multimodal instruction-tuned models under multimodal conditions. 
%Moreover, they often limited their analysis to either language regions for text-based stimuli or visual cortices for visual stimuli, without investigating how task-specific instructions affect both early sensory and high-level language processing. 
We complement these works by examining the impact of a wide range of IT-MLLMs—spanning video and audio-based models with text-based prompts—on their alignment with brain activity from multimodal stimuli. 
%We specifically analyze how task-specific, language-guided instructions from multimodal large language models (MLLMs) influence brain alignment across both early sensory and higher-level language regions.

\section{Detailed sub-ROIs of language, visual and auditory regions}
\label{app:detailedsubrois}

\begin{figure}[!ht]
    \centering
    \includegraphics[width=\linewidth]{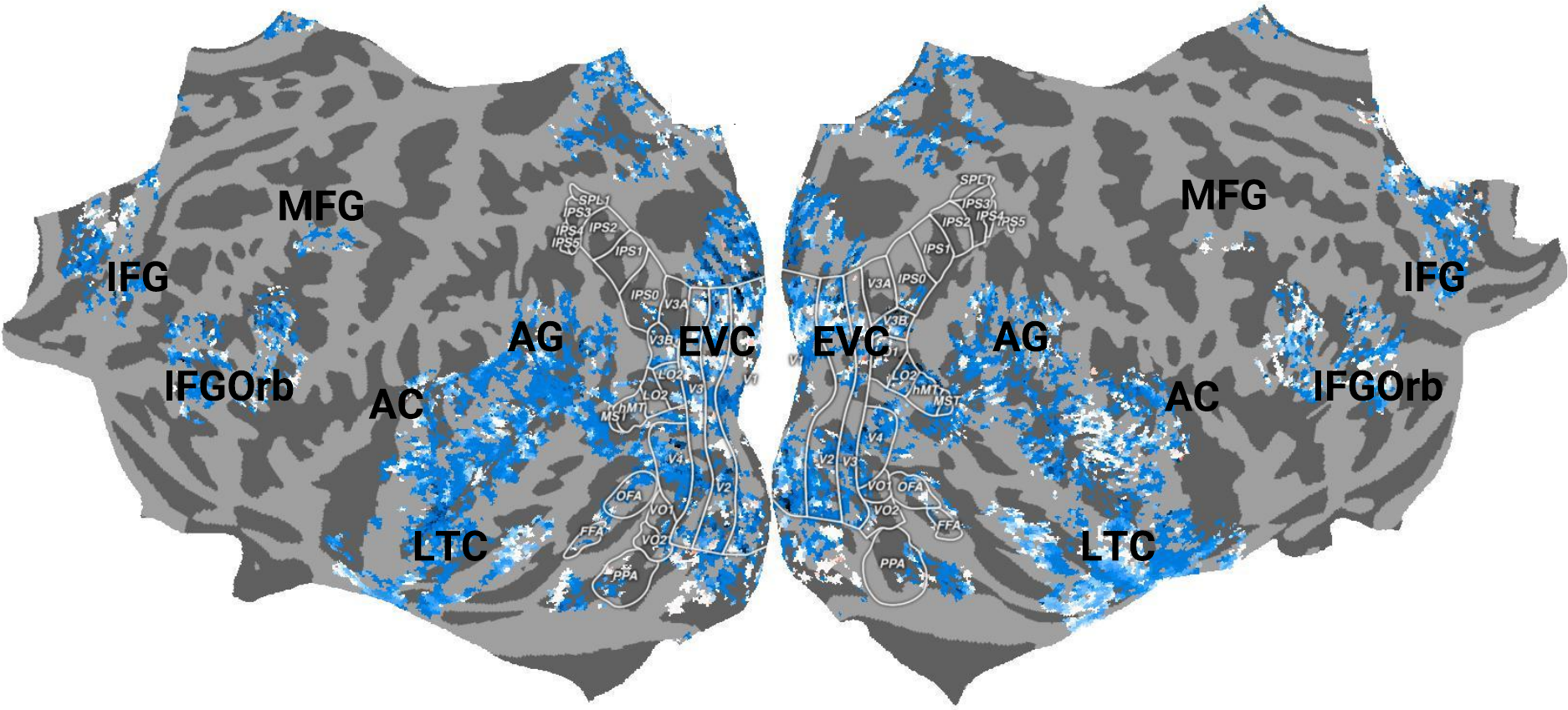}
    \caption{Flattened cortical surfaces for
language-, visual- and auditory-selective regions displayed on the `fsaverage' surface, used as the mask for all participants.}
    \label{fig:language_flatmap}
\end{figure}

The data covers seven brain regions of interest (ROIs) in the human brain with the following sub-divisions: (i) early visual (EV: V1, V2, V3, V3B, and V4); (ii) object-related areas (LO1 and LO2); (iii) face-related areas (OFA), (iv) scene-related areas (PPA), (v) middle temporal (MT: MT, MST, LO3, FST and V3CD), (vi) late language regions, encompassing broader language regions: angular gyrus (AG: PFm, PGs, PGi, TPOJ2, TPOJ3), lateral temporal cortex (LTC: STSda, STSva, STGa, TE1a, TE2a, TGv, TGd, A5, STSdp, STSvp, PSL, STV, TPOJ1), inferior frontal gyrus (IFG: 44, 45, IFJa, IFSp) and middle frontal gyrus (MFG: 55b)~\citep{baker2018connectomic,milton2021parcellation,desai2022proper}. Fig.~\ref{fig:language_flatmap} shows flattened cortical surfaces for
language-, visual- and auditory-selective regions displayed on the `fsaverage' surface, used as the mask for all participants.

\begin{figure*}[!ht] 
\centering
\begin{minipage}{\textwidth}
\centering
    \includegraphics[width=0.6\linewidth]{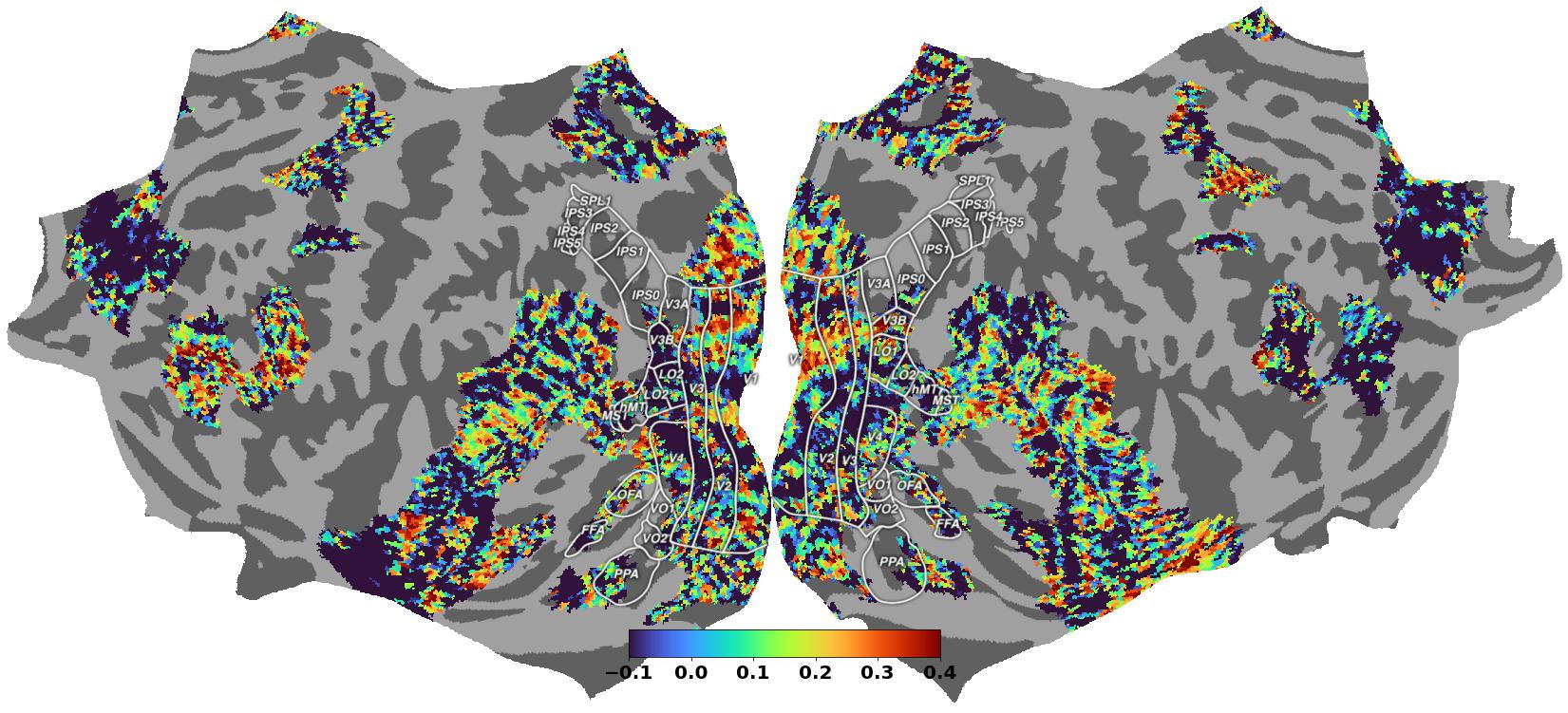}
    \\(a) Subject-01 \\
\end{minipage}
\begin{minipage}{\textwidth}
\centering
    \includegraphics[width=0.6\linewidth]{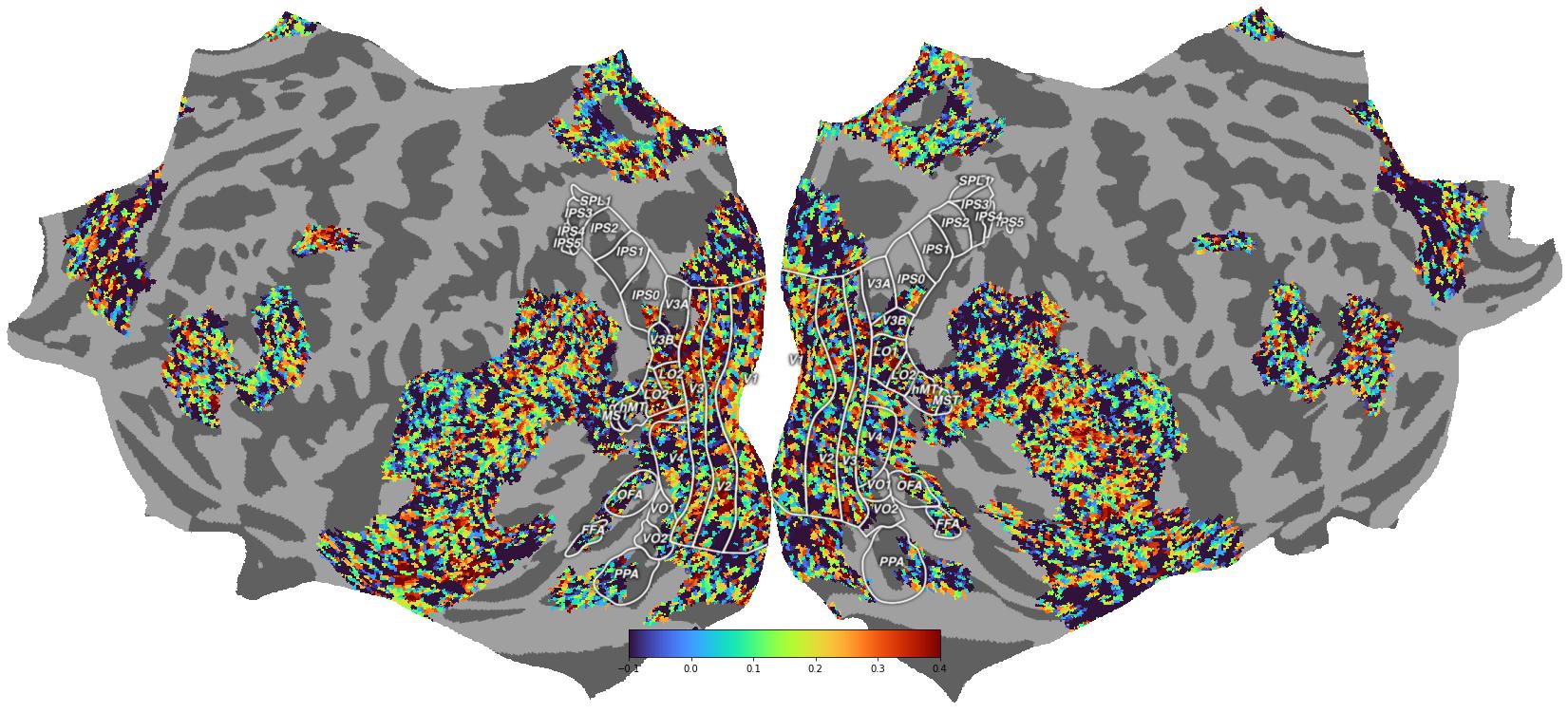}
    \\(a) Subject-02 \\
\end{minipage}
\begin{minipage}{\textwidth}
\centering
    \includegraphics[width=0.6\linewidth]{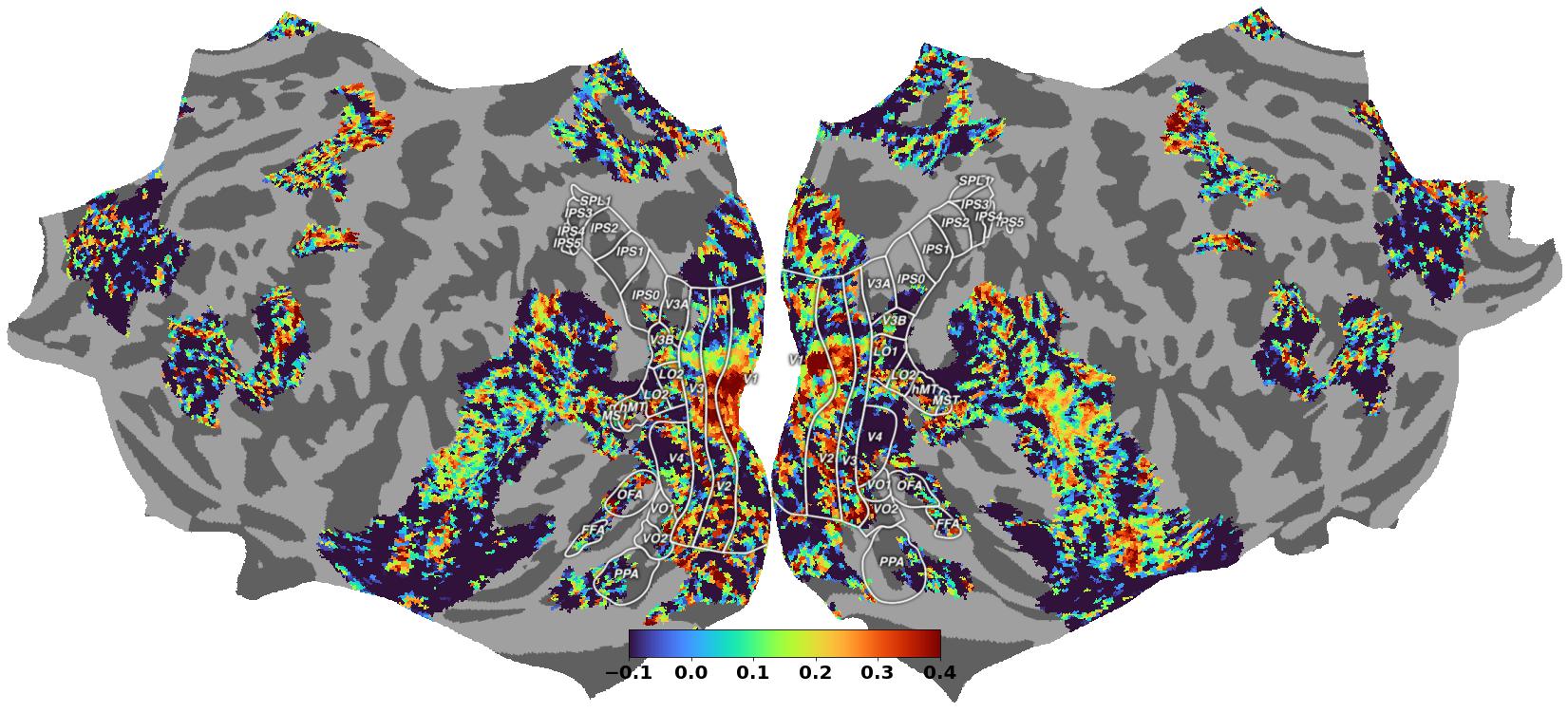}
    \\(b) Subject-03 \\
\end{minipage}
\begin{minipage}{\textwidth}
\centering
    \includegraphics[width=0.6\linewidth]{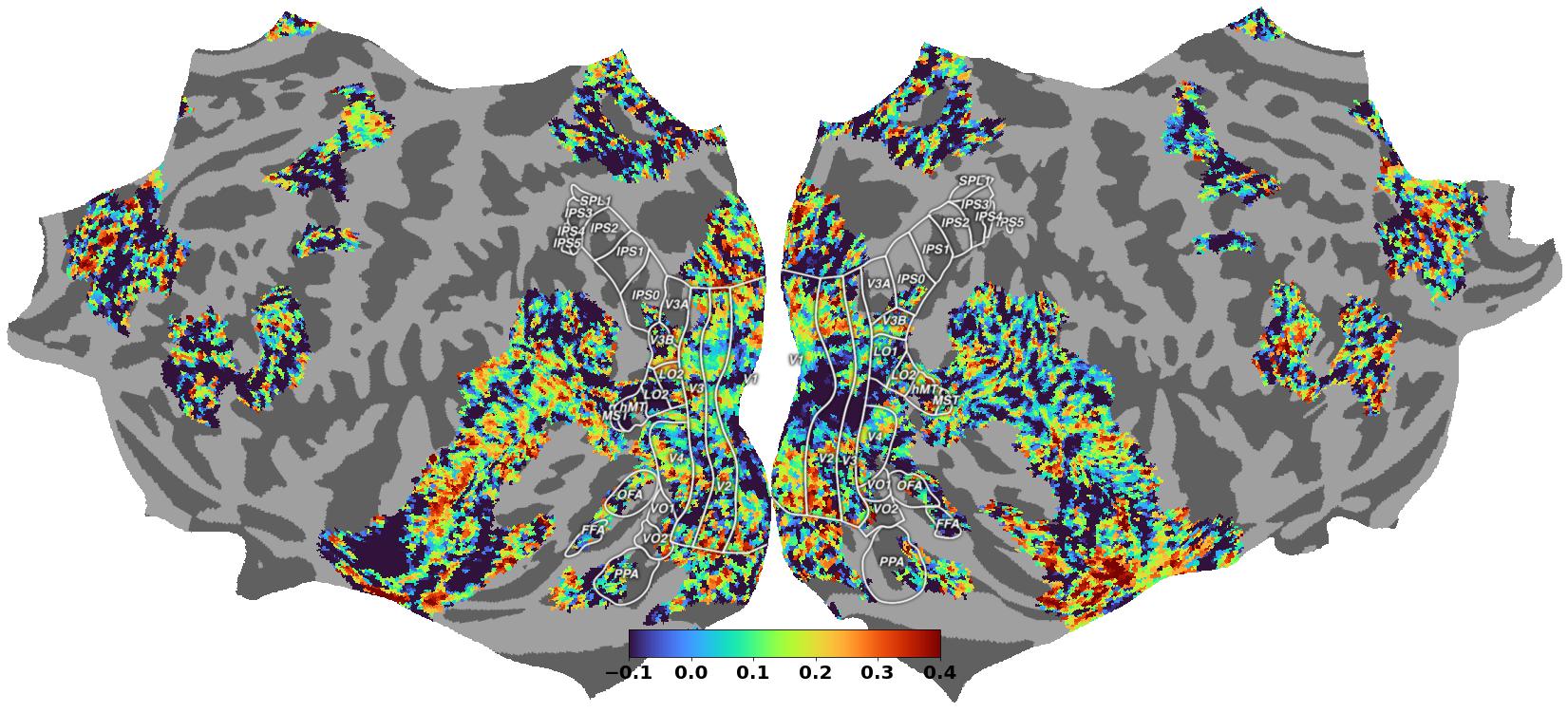}
    \\(c) Subject-05 \\
\end{minipage}
\caption{Estimated cross-subject prediction accuracy for all four participants for the naturalistic movie watching. Pearson correlation scores for each voxel in each subject are projected onto the subject’s flattened cortical surface.}
\label{fig:noise_ceiling_subjects}
\end{figure*}

\section{Cross-subject prediction accuracy}
\label{app:cross_subject_flatmaps}

We follow the method introduced by~\citet{schrimpf2021neural} to estimate how well brain activity in one individual can be predicted from others, using the Movie10 fMRI dataset. Starting with data from $n$ participants (e.g., $n=4$), for each subject $s$ $\in$ ([1,4]) is chosen as the prediction target and the other three are used to predict this target, we use a voxel-wise encoding model (see Sec. \ref{sec:modelArch}) to predict one participant's response from others. For every combination, one participant was randomly chosen as the target, and the model was trained to predict their brain responses using data from the remaining $s-1$ participants. This gave us an average prediction score (correlation) for each voxel at each participant.
To extrapolate to infinitely many humans and thus to obtain the highest possible (most conservative) estimate, as suggested by~\citet{schrimpf2021neural}, we fit the equation $v=v_0\times \left(1-e^{-\frac{x}{\tau_0}}\right)$ where $x$ is each subsample's number of participants, $v$ is each subsample's correlation score and $v_0$ and $\tau_0$ are the fitted parameters. 
This fitting was performed for each sensor independently with 100 bootstraps each to estimate the variance where each bootstrap draws $x$ and $v$ with replacement. The final ceiling value was the median of the per-voxel ceilings $v_0$.

Fig.~\ref{fig:noise_ceiling_subjects} shows the estimated cross-subject prediction accuracy for all four participants for the naturalistic movie watching. Pearson correlation scores for each voxel in each subject are projected onto the subject’s flattened cortical surface. The plots show that across all subjects higher activity is observed in the language and visual regions with a max correlation up to 0.4 implying that data has low noise and low cross-subject variability.

\section{Implementation details for reproducibility}
\label{app:hyperparameters_details}
%\noindent\textbf{Implementation details for reproducibility.}
All feature extraction experiments were conducted on a machine equipped with an NVIDIA A100 GPU with 80 GB of GPU RAM, partitioned into two devices of 40 GB each. The voxelwise encoding models were trained on NVIDIA GeForce RTX 3050 GPU with 4GB of GPU RAM. We used banded ridge-regression with the following parameters: MSE loss function; L2-decay ($\lambda$) varied from  10$^{-1}$ to 10$^{3}$; the best $\lambda$ was chosen by tuning on validation data that comprised a randomly chosen 10\% subset from the train set used only for hyper-parameter tuning. 

\section{Statistical Significance}
\label{app:statistical_significance}

To determine if normalized predictivity scores are  significantly higher than chance, we run a permutation test using blocks of 10 contiguous fMRI TRs (considering the slowness of hemodynamic response) rather than individual TRs. By permuting predictions 5000 times, we create an empirical distribution for chance performance, from which we estimate p-value of the actual performance. The choice of these specific permutation test configurations is based on established methodologies in previous research~\citep{deniz2019representation,reddy2021can,oota2024speech}.
To estimate the statistical significance of performance differences, such as between the model's predictions and chance or residual predictions and chance, we utilized the Wilcoxon signed-rank test~\citep{conover1999practical}, applying it to the mean normalized predictivity for the participants. 
Finally, the Benjamini-Hochberg False Discovery Rate (FDR) correction for multiple comparisons~\citep{benjamini1995controlling} is used for all the tests (appropriate because fMRI data is considered to have positive
dependence~\citep{genovese2000bayesian}).
%In all cases, we denote significant differences (p$\leq 0.05$) with a \textcolor{Indigo}{$\ast$} or \textcolor{Indigo}{$\wedge$}.

\section{Model generated outputs across instructions}
\label{app:ModelGeneratedOutputs}

Tables~\ref{instruct_model_outputs_prompts},~\ref{qwen_model_outputs_prompts},~\ref{video_llava_model_outputs_prompts},~\ref{llava_next_model_outputs_prompts},~\ref{llava_onevision_model_outputs_prompts} and~\ref{videochat_r1_model_outputs_prompts} show model generated outputs for a sample video from the Movie10 dataset using InstructBLIPVideo, Qwen-2.5-VL, Video-LLaVA, LLaVa-NeXT-Video, LLaVA-OneVision and VideoChat-R1 models, respectively. Similarly, Tables~\ref{audio_instruct_model_outputs_prompts_qwen2.5} and~\ref{audio_instruct_model_outputs_prompts_kimi} show model generated outputs for a sample video from the Movie10 dataset using Qwen-2.5 Audio and Kimi-Audio models.

\begin{table*}[!ht]
\caption{Outputs from InstructBLIPVideo for a sample clip from the Movie10 dataset.}
\label{instruct_model_outputs_prompts}
\begin{center}
\scriptsize
\begin{tabular}{|l|p{10cm}|}
\hline
\multicolumn{2}{|c|}{
  \begin{tabular}{ccccc}
    \includegraphics[width=0.19\textwidth]{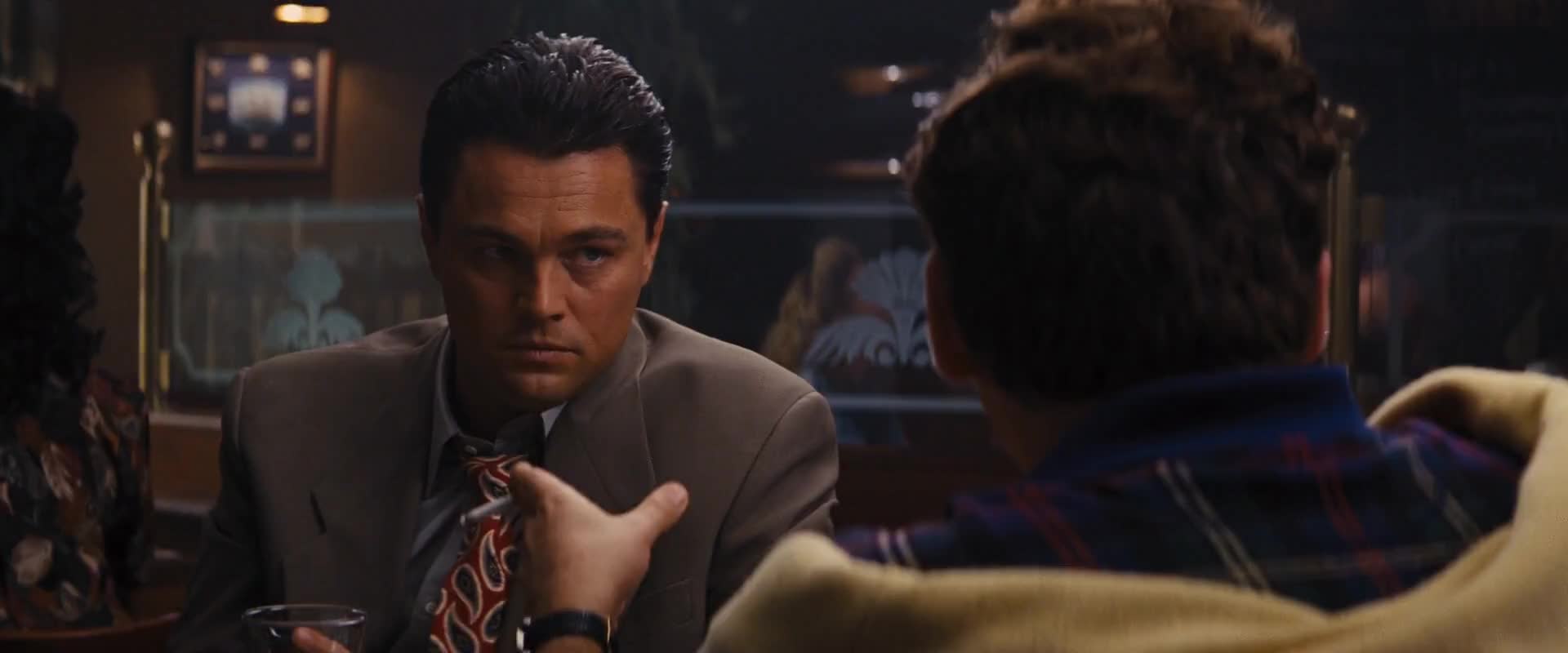} &
    \includegraphics[width=0.19\textwidth]{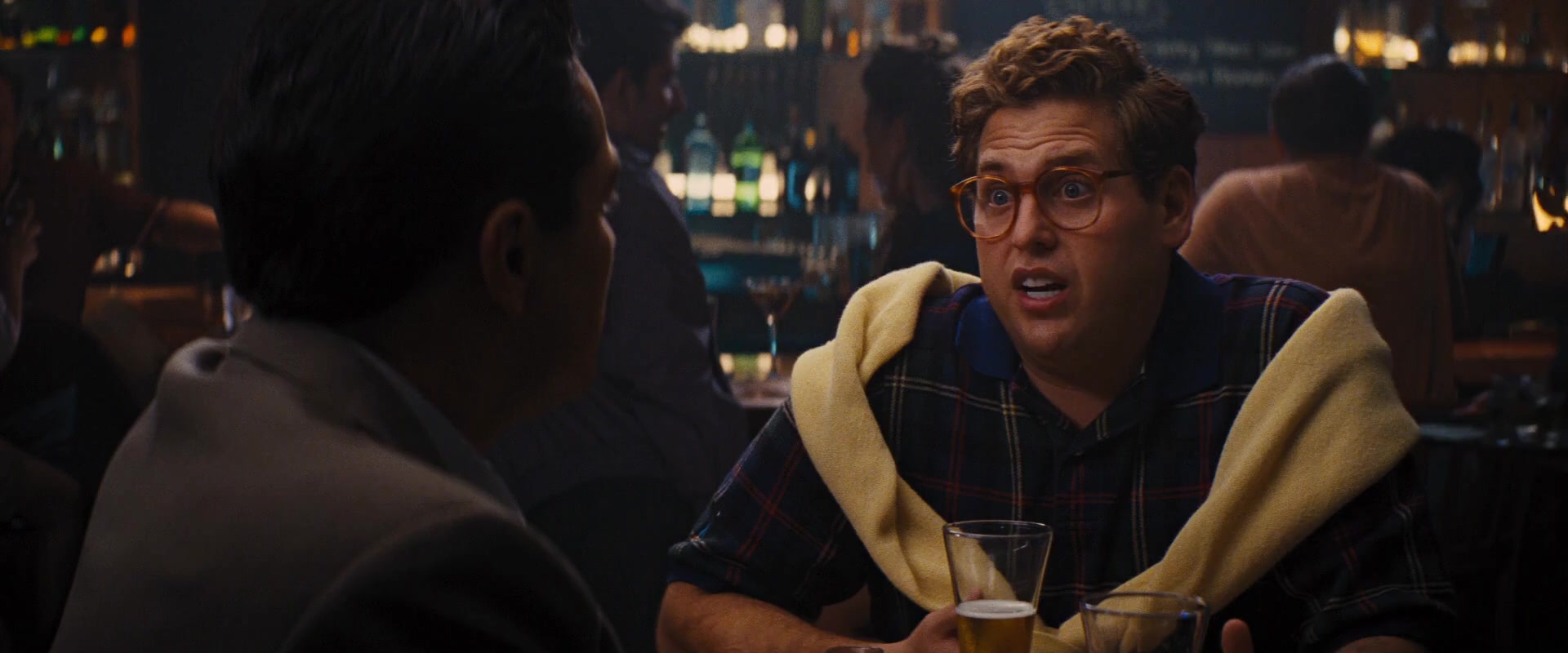} &
    \includegraphics[width=0.19\textwidth]{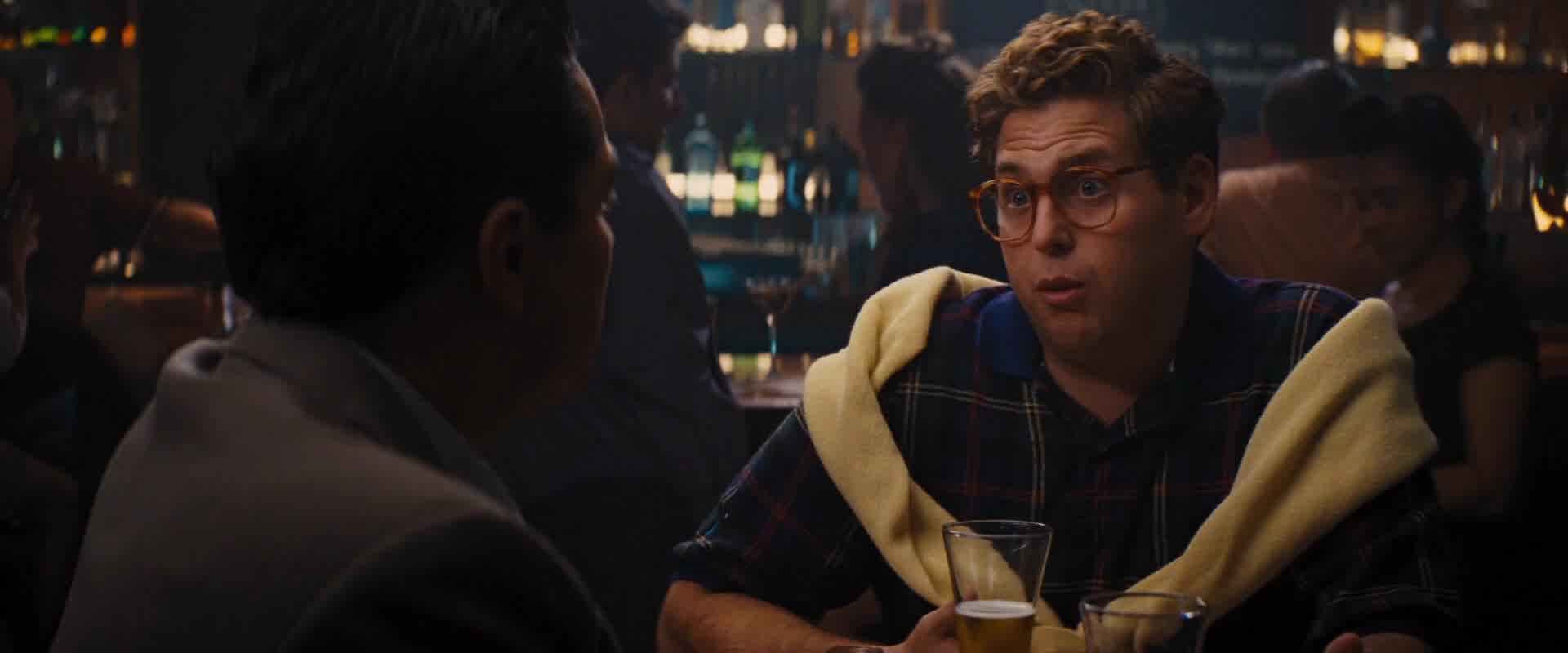} &
    \includegraphics[width=0.19\textwidth]{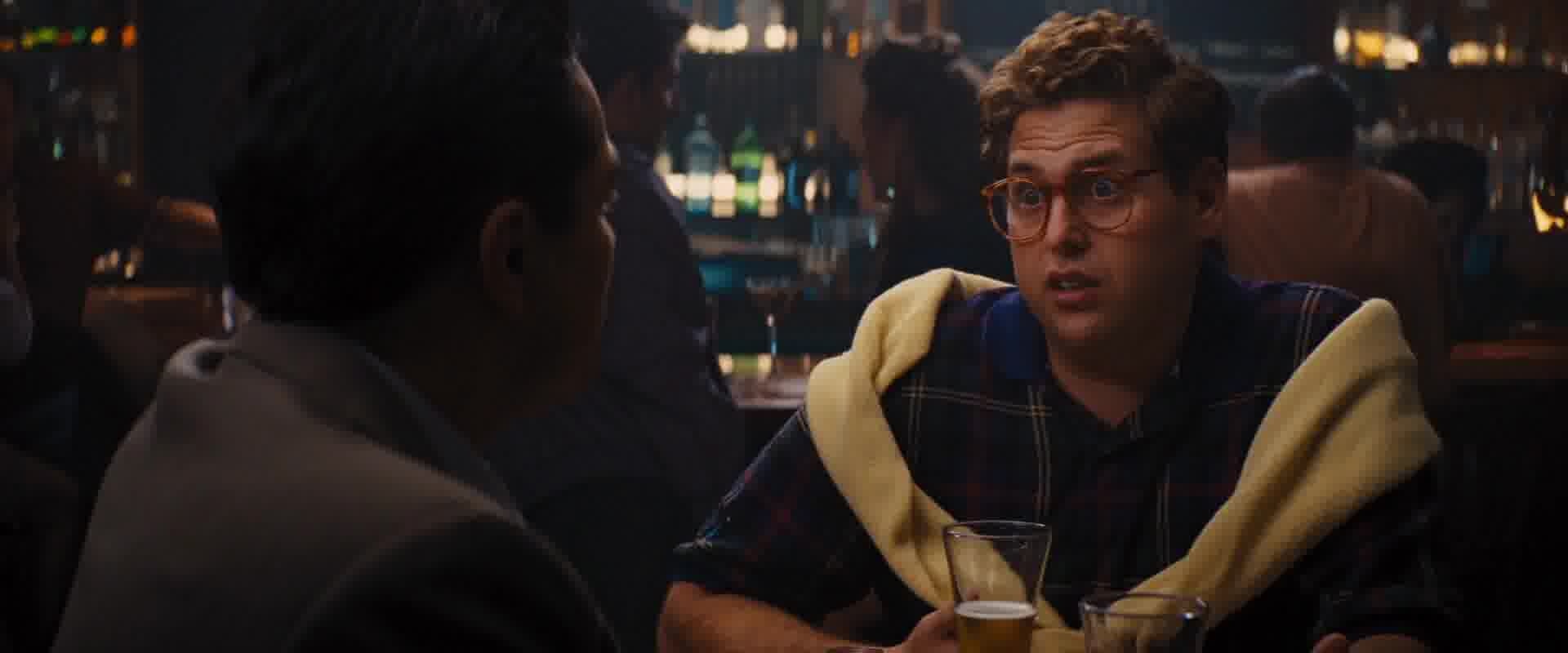} &
    \includegraphics[width=0.19\textwidth]{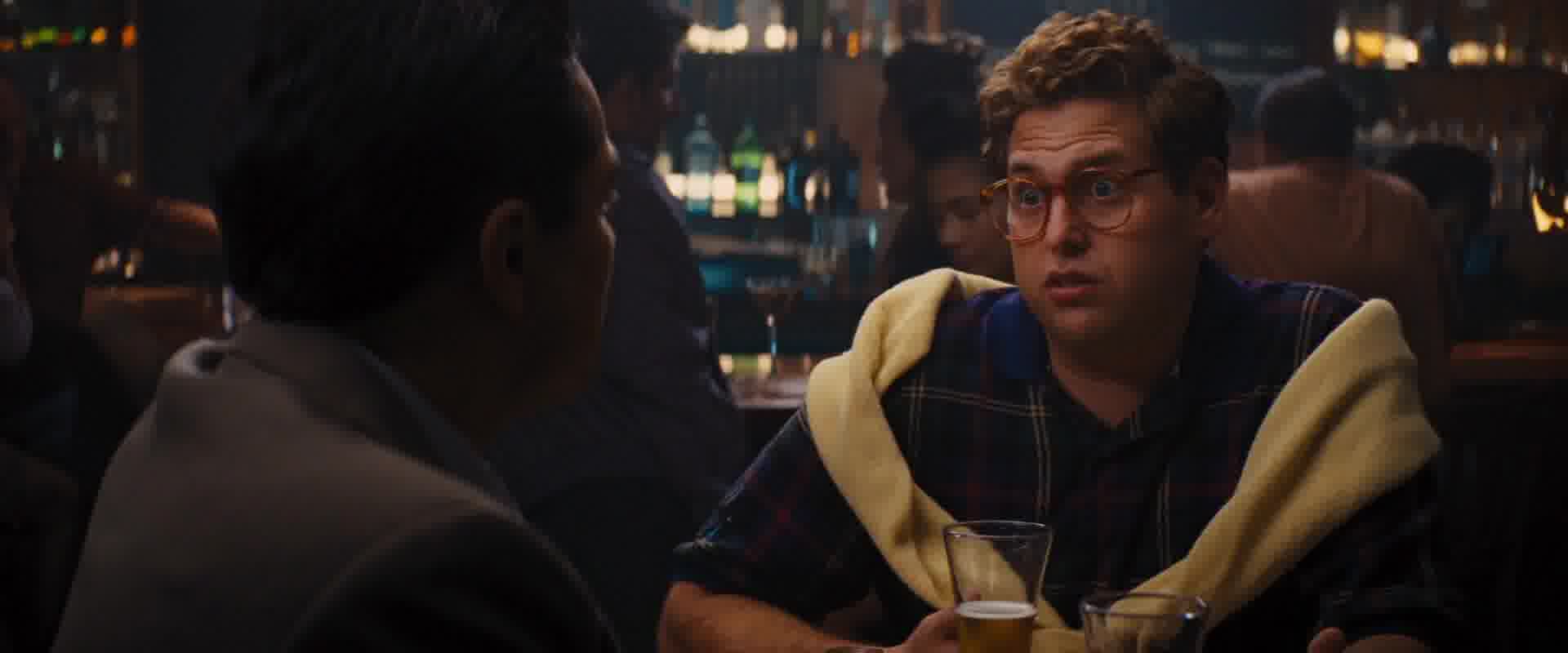}\\
    \small Frame 1 & \small Frame 3 & \small Frame 8 & \small Frame 13 & \small Frame 16 \\
  \end{tabular}
} \\ \hline
\bf Instruction&\bf Model Output\\ 
\hline
\multirow{1}{*}{Action Recognition}  & The main event captured in the video is a man sitting at a bar with a glass of beer, engaging in conversation with another man. The man is wearing glasses and is sitting in a dining area, surrounded by other people, possibly in a bar or restaurant. The scene also features a man wearing a towel, which is wrapped around his neck. \\ \hline
\multirow{1}{*}{Video Understanding}  &    In the scene, a man is sitting in a bar with a beer in his hand, engrossed in a conversation with another man. The two men are sitting at a table, with one of them holding a glass of wine. There are multiple bottles of beer visible in the background, placed on a table and around the bar area. A few other people can also be seen in the scene, possibly patrons or other bar visitors, scattered around the bar and interacting with the two main characters. The scene is set in a dimly lit environment with some lighting, providing a relaxed atmosphere for the men's conversation. \\
 \hline
Visual Question Answering &There are two people in the scene, one man and one woman. They are having a conversation at a bar, sitting at a table, and drinking. \\ \hline
Video Captioning & A man with glasses sitting at a table talking with another man.\\ \hline
\multirow{1}{*}{Object and Scene Recognition}   & In the scene, there is a man wearing glasses sitting in a bar, talking to another person, who is also sitting in a chair next to him. The man is holding a glass of beer, while the other person is holding a cup. \\ \hline
Commonsense Reasoning & The man is sitting in a bar with a glass of wine, engaging in conversation with another man, possibly discussing something. This action could have motivated the man to take this step for various reasons. It could be that he is seeking to socialize, share his thoughts or ideas, or simply engage in casual conversation with a friend or acquaintance.  \\ \hline
Spatial Understanding & Restaurant \\ \hline
Temporal Ordering & The scene depicted in the video shows a man sitting at a bar, engaging in a conversation with another man. The man is wearing glasses and is holding a glass in his hand, which suggests that he is drinking or holding a drink. They appear to be having a casual conversation, possibly discussing something of interest or sharing a story. \\ \hline
Video reasoning & In this video, there is an unusual scene where a man with glasses is having a conversation with another man in a bar, and they are both holding glasses. This is an unusual scene because it is not a typical social setting for two men to be having a conversation in a bar. They are sitting in a dark, dimly lit environment, which adds to the unusual nature of their interaction. \\ \hline
Narrative Understanding & The central conflict in the movie is the man sitting in a bar talking to another man, who is wearing glasses. The man in the bar is having a conversation with another man, possibly discussing something of interest. It is not specified what the topic of their conversation is, but it is likely related to the movie's storyline. \\ \hline
Emotion and Sentiment Analysis & The video shows a man sitting at a bar with a friend, talking and having a conversation, which suggests a social and relaxed atmosphere. The man is wearing glasses, and the scene is set in a dimly lit bar, which creates a more intimate and cozy atmosphere. \\ \hline
Global Appearance & Man with glasses and a towel \\ \hline
Linking Events &The scene depicted in the image shows a man wearing glasses sitting at a bar, talking to another person, possibly a friend, while holding a glass of beer. This early event sets the stage for the later development of a story or a situation that may unfold in the video.
\\ \hline
\end{tabular}
\end{center}
\end{table*}

%\vspace{25pt}
\begin{table*}[!ht]
\caption{Outputs from Qwen-2.5-VL for a sample video from the Movie10 dataset.}
\label{qwen_model_outputs_prompts}
\begin{center}
\scriptsize
\begin{tabular}{|l|p{10cm}|}
\hline
\multicolumn{2}{|c|}{
  \begin{tabular}{ccccc}
    \includegraphics[width=0.19\textwidth]{image_frames/ezgif-frame-001.jpg} &
    \includegraphics[width=0.19\textwidth]{image_frames/ezgif-frame-003.jpg} &
    \includegraphics[width=0.19\textwidth]{image_frames/ezgif-frame-008.jpg} &
    \includegraphics[width=0.19\textwidth]{image_frames/ezgif-frame-013.jpg} &
    \includegraphics[width=0.19\textwidth]{image_frames/ezgif-frame-016.jpg}\\
    \small Frame 1 & \small Frame 3 & \small Frame 8 & \small Frame 13 & \small Frame 16 \\
  \end{tabular}
} \\ \hline
\bf Instruction&\bf Model Output\\ 
\hline
\multirow{1}{*}{Action Recognition}  & The video appears to be a still from a movie or a scene from a film. It shows two individuals engaged in a conversation in what looks like a dimly lit bar or restaurant setting. The person on the left is wearing a suit and tie, holding a drink, and seems to be listening intently. The person on the right is gesturing with their hand, possibly emphasizing a point during the conversation. The atmosphere suggests a serious or intense discussion. \\ \hline
\multirow{1}{*}{Video Understanding}  &    The video depicts a scene from a movie or TV show, featuring two characters engaged in a conversation. The character on the left is dressed in a formal gray suit with a patterned tie and is holding a cigarette in his right hand. He appears to be listening intently to the other character, who is partially visible from behind. The setting seems to be an indoor environment, possibly a bar or a similar establishment, as suggested by the background elements such as a mirror and what looks like a bar counter.\\
 \\ \hline
Visual Question Answering & There are two people in the scene. One person is sitting at a table, wearing a suit with a patterned tie, and appears to be engaged in a conversation or interaction with another person who is partially visible from behind. The second person is facing away from the camera, and their attire includes a plaid shirt. The setting seems to be indoors, possibly in a bar or restaurant, given the background elements like the glassware and the dim lighting. \\ \hline
Video Captioning & A man in a suit is engaged in a conversation with another person in a dimly lit bar setting. \\\hline
\multirow{1}{*}{Object and Scene Recognition}   & In the video, there are two main individuals and several objects: \textbf{1. The Man in the Suit:} He is wearing a brown suit with a white shirt and a red paisley tie. He has dark hair styled neatly and is holding a glass of what appears to be a drink in his left hand. His expression seems serious or contemplative as he looks at the other person. \textbf{2. The Other Person:} Only the back of this individual's head is visible, suggesting they are facing away. \\ \hline
Commonsense Reasoning & In the video, the character on the left appears to be in a conversation with another person whose back is turned to the camera. The character's hand is extended towards the other person, possibly gesturing or pointing at something. This action could be motivated by several reasons: \textbf{1. Clarification or Explanation:} The character might be trying to explain something important or clarify a point in their conversation. \textbf{2. Emphasis: } The gesture could be used to emphasize a particular point they are making.\\ \hline
Spatial Understanding & The video appears to be from the movie ``The Wolf of Wall Street'', directed by Martin Scorsese and released in 2013. The setting looks like a bar or a lounge, which is consistent with the film's depiction of the characters' activities in the financial world. However, there are no specific landmarks visible in the image that would definitively identify the exact location within the movie. The scene is set in New York City, as the film is based on the real-life events. \\ \hline
Temporal Ordering & The image depicts a scene from a movie where two characters are engaged in a conversation. Here is a step-by-step description of the activity shown: \textbf{1. Setting:} The setting appears to be an indoor environment, possibly a bar or a restaurant, given the presence of a glass and what looks like a menu board in the background. \textbf{2. Characters:}   - The character on the left is wearing a suit with a patterned tie and is holding a glass in his hand. \\ \hline
Video reasoning & The video appears to be from a scene in a movie or TV show, featuring two characters engaged in a conversation. The setting looks like a bar or a similar social environment, with dim lighting and a relaxed atmosphere. What might be considered unusual is the character on the right, who seems to be gesturing with their hand while holding a drink, which could indicate they are explaining something or making a point during the conversation. However, without additional context, it's difficult to determine if there is anything particularly. \\ \hline
Narrative Understanding & The movie depicted in the video is ``The Wolf of Wall Street,'' directed by Martin Scorsese and starring Leonardo DiCaprio as Jordan Belfort. The film is based on the memoir of the same name by Belfort himself. \textbf{Main Storyline:} The story follows Jordan Belfort, a young stockbroker who rises to become one of the most successful and corrupt figures in the financial world during the late 1980s and early 1990s. \\ \hline
Emotion and Sentiment Analysis & The video depicts two individuals in what appears to be a dimly lit, possibly upscale setting, such as a bar or restaurant. The person on the left is dressed in a suit and tie, suggesting a formal or professional demeanor, while the person on the right is wearing a plaid shirt, which might indicate a more casual or relaxed attitude.  \\ \hline
Global Appearance & In the video provided, there is no indication of significant changes in the characters' appearances or outfit changes. The man on the left appears to be wearing a brown suit with a patterned tie and a white shirt. He has dark hair styled neatly. The person on the right is seen from behind, wearing a plaid shirt with a blue and red pattern. There are no visible changes in their attire or appearance within this single frame. \\ \hline
Linking Events & In the video provided, we see a scene from the movie ``The Wolf of Wall Street,'' where Leonardo DiCaprio's character, Jordan Belfort, is engaged in a conversation with another character. The setting appears to be a dimly lit bar or lounge, which is typical for scenes involving discussions about business deals or personal matters. An early event that could influence later developments in this scene might involve a previous interaction between Jordan and his current interlocutor.
\\ \hline
\end{tabular}
\end{center}
\end{table*}

\begin{table*}[!ht]
\caption{Outputs from Video-LLaVA for a sample clip from the Movie10 dataset.}
\label{video_llava_model_outputs_prompts}
\begin{center}
\scriptsize
\begin{tabular}{|l|p{10cm}|}
\hline
\multicolumn{2}{|c|}{
  \begin{tabular}{ccccc}
    \includegraphics[width=0.19\textwidth]{image_frames/ezgif-frame-001.jpg} &
    \includegraphics[width=0.19\textwidth]{image_frames/ezgif-frame-003.jpg} &
    \includegraphics[width=0.19\textwidth]{image_frames/ezgif-frame-008.jpg} &
    \includegraphics[width=0.19\textwidth]{image_frames/ezgif-frame-013.jpg} &
    \includegraphics[width=0.19\textwidth]{image_frames/ezgif-frame-016.jpg}\\
    \small Frame 1 & \small Frame 3 & \small Frame 8 & \small Frame 13 & \small Frame 16 \\
  \end{tabular}
} \\ \hline
\bf Instruction&\bf Model Output\\ 
\multirow{1}{*}{Action Recognition}  & The main events in the video include a man sitting at a bar, holding a beer, and talking to another man. The man holding the beer is wearing a plaid shirt and glasses. \\ \hline
\multirow{1}{*}{Video Understanding}  &    The video features a man sitting at a bar with a drink in his hand. He appears to be enjoying the drink and is engaged in conversation with someone else at the bar. The man is wearing glasses and a plaid shirt, giving him a casual and relaxed demeanor. The bar is well-lit, with a variety of bottles and glasses on display, indicating a lively and bustling atmosphere. In the background, there are other patrons.\\
 \hline
Visual Question Answering & There are three people in the scene. Two of them are sitting down, while the third man is standing up and holding a beer glass. \\ \hline
Video Captioning & A man in a blue and plaid shirt drinking beer at a bar. \\ \hline
\multirow{1}{*}{Object and Scene Recognition}   & In the scene, the main objects visible are a man sitting at a bar with a beverage in a glass and another man sitting next to him. There are also two other men sitting at the bar, and a woman standing nearby. The bar is well-stocked with various bottles and cups. The man sitting at the bar is wearing glasses, a blue plaid shirt, and a beige blanket. \\ \hline
Commonsense Reasoning & The character in the image is holding a beer glass and appears to be in a bar or a similar setting. The motivation for the character to take this action could be due to various reasons, such as relaxation, socialization, or simply enjoying a beverage during leisure time. The character may have chosen to have a beer to unwind after a long day, to connect with others in a social setting, or simply to indulge in a favorite beverage. \\ \hline
Spatial Understanding & The video is taken from a bar, as indicated by the presence of multiple bottles on the bar and a man holding a beer glass. The video also shows the man standing in front of a bar, suggesting the location of the video is a bar or similar establishment. \\ \hline
Temporal Ordering & The video depicts a man sitting at a bar, holding a beer in his hand. He is wearing a plaid shirt and glasses. The scene is set in a dark room, and there are other people in the background. The man is seen drinking from the beer glass, which is placed on the bar counter. The camera captures the man's facial expressions and body language, highlighting his relaxed and social atmosphere. \\ \hline
Video reasoning & The unusual aspect of this video is that it features a man in a plaid shirt and glasses, who is sitting in a bar and drinking a beer. The man is also wearing a beige scarf, which adds a distinctive touch to his appearance. However, the most striking feature of this video is that the man is not only drinking a beer, but he is also holding a cell phone. \\ \hline
Narrative Understanding & The central conflict in the movie is a man's desire for a drink at a bar, but he is unable to get one due to a bartender's refusal. The man's frustration and disappointment lead to him making a scene and causing a commotion at the bar. The conflict is ultimately resolved when the man is given a drink, and he leaves the bar with a smile. \\ \hline
Emotion and Sentiment Analysis & In the video, the man in the blue shirt and the woman are seen laughing and smiling while enjoying drinks and conversation. The man in the plaid shirt is seen drinking beer and looking down, while the man in the blue shirt is holding a beer glass. The video conveys a sense of relaxation, camaraderie, and enjoyment among the characters.  \\ \hline
Global Appearance & In the video, the main character, a man with a beard, is seen wearing a plaid shirt and a scarf. He is also seen wearing a different outfit in one of the scenes, which is not specified in the given information. The other man in the scene is wearing a dark shirt and glasses. There are also two other men in the background, one wearing a dark jacket and the other wearing a dark shirt. \\ \hline
Linking Events & The early event in the video is the man in the blue and white checkered shirt sitting at the bar with a beer in his hand. This scene introduces the setting and the man's presence at the bar, setting the stage for later developments. The man's interaction with other patrons, the woman, and the man in glasses adds depth and complexity to the narrative.
\\ \hline
\end{tabular}
\end{center}
\end{table*}

\begin{table*}[!ht]
\caption{Outputs from LLaVA-NeXT-Video for a sample clip from the Movie10 dataset.}
\label{llava_next_model_outputs_prompts}
\begin{center}
\scriptsize
\begin{tabular}{|l|p{10cm}|}
\hline
\multicolumn{2}{|c|}{
  \begin{tabular}{ccccc}
    \includegraphics[width=0.19\textwidth]{image_frames/ezgif-frame-001.jpg} &
    \includegraphics[width=0.19\textwidth]{image_frames/ezgif-frame-003.jpg} &
    \includegraphics[width=0.19\textwidth]{image_frames/ezgif-frame-008.jpg} &
    \includegraphics[width=0.19\textwidth]{image_frames/ezgif-frame-013.jpg} &
    \includegraphics[width=0.19\textwidth]{image_frames/ezgif-frame-016.jpg}\\
    \small Frame 1 & \small Frame 3 & \small Frame 8 & \small Frame 13 & \small Frame 16 \\
  \end{tabular}
} \\ \hline
\bf Instruction&\bf Model Output\\ 
\multirow{1}{*}{Action Recognition}  & The main event captured in the video is a conversation between two men at a bar, where one of them is wearing a red and white checkered shirt and glasses, while the other man is wearing a dark suit and tie. They are discussing a topic, possibly related to a movie, as suggested by the context. There are also other patrons in the background, and a bartender is visible in the foreground. \\ \hline
\multirow{1}{*}{Video Understanding}  &    The video features a scene from the movie \"The Hangover\" featuring actors Zach Galifianakis and Bradley Cooper. In the scene, Zach Galifianakis, portraying a character named Chip Baskin, is sitting at a bar with a group of people. He is seen holding a beer and appears to be in a state of disbelief as he looks on at the camera. \\
 \hline
Visual Question Answering & There are several people in the scene. The man with glasses is sitting at a bar, holding a drink, and appears to be in a conversation with someone. Another man is standing at the bar, and there is a woman sitting nearby. There is a bartender standing at the bar, and a person in the background who is partially visible, but it's unclear what their role is in the scene. \\ \hline
Video Captioning & Two men are seen having a conversation in a bar, with the man in the plaid shirt appearing confused or disbelieving the man in the suit.\\ \hline
\multirow{1}{*}{Object and Scene Recognition}   & In the scene, we can see a man in a suit sitting at a table with a woman and a bartender. There's also a man in a suit standing at a bar, a woman sitting at a table with a man, a woman standing at a bar, a man standing at a bar, and a man sitting at a table with a woman. \\ \hline
Commonsense Reasoning & In the video you provided, the character is seen holding up a beer bottle and making a face, which could be interpreted as a playful or humorous gesture, possibly to amuse others or to express a lighthearted or jovial moment. This action might be motivated by a desire to entertain, bond with others, or to simply enjoy a casual and relaxed moment. \\ \hline
Spatial Understanding & The video appears to have been taken from a scene in a movie, specifically ``Goodfellas'' featuring Robert De Niro. The scene is set in a bar, and the background includes a sign that says ``The Goodfella's.'' This is a reference to the character Robert De Niro's character, who is a bartender in the movie. The specific setting and the sign are consistent with the movie's depiction of the bar. \\ \hline
Temporal Ordering & The video depicts two men sitting at a bar, engaged in a conversation. One man is wearing a suit, while the other is dressed in a plaid shirt. They are both holding drinks and appear to be having a casual conversation. The setting suggests they might be at a social event, such as a bar or a party, where people gather to relax and enjoy drinks. \\ \hline
Video reasoning & In the video, there is a scene where a man is seen sitting at a bar with another man. The man in glasses appears to be engaged in a conversation with someone off-camera, but the camera is focused on the other man, who is not the focus of the scene. The man with glasses is holding a beer and appears to be in the middle of a discussion, but his face is not shown in the shot. \\ \hline
Narrative Understanding & The movie ``The Big Short'' is a biographical comedy-drama film about the life of stockbroker Michael Burry, who predicts the financial crisis of 20008. The central conflict of the story is the housing bubble and the subprime mortgage market, which Burry sees as unsustainable and warns his colleagues about, but they dismiss his ideas as unrealistic. \\ \hline
Emotion and Sentiment Analysis & The characters in the video express a range of emotions, from surprise and shock to amusement and even a bit of confusion. The video itself is quite humorous, with the man in a suit and tie appearing to be in a state of disbelief or shock as the other man, who is wearing a plaid shirt and a hoodie, is seen drinking from his beer bottle.  \\ \hline
Global Appearance & In the video, we can see a man with a beard who is wearing a suit at the beginning, but he is later seen in a more casual outfit with a plaid shirt. There are also other people in the scene, including a man with glasses and a red shirt, a man with a beard and a red shirt, and a woman in a red dress. \\ \hline
Linking Events & In the scene you've described, a man is shown sitting in a bar with a drink in front of him, looking at a TV. The TV shows a news segment about a man who has died, which is later revealed to be the man himself. This early event in the video influences the later developments because it sets up the narrative of the story, introducing the main character and establishing his presence in the scene.
\\ \hline
\end{tabular}
\end{center}
\end{table*}

\begin{table*}[!ht]
\caption{Outputs from LLaVA-OneVision Video for a sample clip from the Movie10 dataset.}
\label{llava_onevision_model_outputs_prompts}
\begin{center}
\scriptsize
\begin{tabular}{|l|p{10cm}|}
\hline
\multicolumn{2}{|c|}{
  \begin{tabular}{ccccc}
    \includegraphics[width=0.19\textwidth]{image_frames/ezgif-frame-001.jpg} &
    \includegraphics[width=0.19\textwidth]{image_frames/ezgif-frame-003.jpg} &
    \includegraphics[width=0.19\textwidth]{image_frames/ezgif-frame-008.jpg} &
    \includegraphics[width=0.19\textwidth]{image_frames/ezgif-frame-013.jpg} &
    \includegraphics[width=0.19\textwidth]{image_frames/ezgif-frame-016.jpg}\\
    \small Frame 1 & \small Frame 3 & \small Frame 8 & \small Frame 13 & \small Frame 16 \\
  \end{tabular}
} \\ \hline
\bf Instruction&\bf Model Output\\ \hline
\multirow{1}{*}{Action Recognition}  & The video captures a conversation between two individuals at a bar, with one person gesturing and speaking while the other listens. \\ \hline
\multirow{1}{*}{Video Understanding}  &    The video takes place in a lively bar setting, creating a warm and inviting atmosphere. The background features a well-lit bar counter adorned with various bottles and glasses, and several people can be seen in the background, adding to the bustling ambiance. In the foreground, two individuals are seated at a table. One person is dressed in a gray suit with a red and white patterned tie, while the other is wearing a blue and red plaid shirt with a yellow jacket draped over their shoulders. \\
 \hline
Visual Question Answering & There are two people in the scene, engaged in a conversation. \\ \hline
Video Captioning & A man with curly hair and glasses sits at a table in a bar, holding a glass of beer, and engages in a conversation.\\ \hline
\multirow{1}{*}{Object and Scene Recognition}   & The main objects and people visible in the scene are two men sitting at a table in a dimly lit, bustling bar or restaurant. One man is dressed in a gray suit and tie, and the other is wearing a plaid shirt with a beige sweater over it, holding a drink. The background features a bar with various bottles and glasses, and other patrons are visible in the distance. \\ \hline
Commonsense Reasoning & The character took this action to emphasize their point in the conversation, which could have been motivated by the seriousness of the discussion or the need to convey a specific message. \\ \hline
Spatial Understanding & The video is taken from a movie scene, and it shows a bar. \\ \hline
Temporal Ordering & The video takes place in a dimly lit bar with a lively background filled with people and bottles on shelves. Two individuals are seated at a table, engaged in a conversation. The person on the left is dressed in a gray suit with a red and white patterned tie and is gesturing with their hand, while the person on the right, wearing a blue and red plaid shirt with a yellow sweater draped over their shoulders and holding a glass of beer, listens attentively. \\ \hline
Video reasoning & The unusual aspect is that the scene remains static with minimal changes in the individuals' expressions and the background, suggesting a continuous conversation or interaction. \\ \hline
Narrative Understanding & The central conflict involves two characters in a bar, with the character in the plaid shirt and yellow sweater being advised or persuaded by the other character. The conflict is resolved through a conversation, leading to a change in the character's demeanor and posture. \\ \hline
Emotion and Sentiment Analysis & The characters express a range of emotions, including surprise, concern, and contemplation. The video creates a sense of intrigue and emotional depth, drawing the viewer into the characters' conversation. \\ \hline
Global Appearance & The character on the right initially has curly hair and is wearing a blue plaid shirt with a beige sweater, but later, the character's hair appears straight and they are dressed in a dark suit with a red tie. \\ \hline
Linking Events & The man in the plaid shirt and glasses is seen holding a glass of beer and engaged in a conversation with the man in the gray suit, which sets the stage for the unfolding dialogue and interaction.
\\ \hline
\end{tabular}
\end{center}
\end{table*}

\begin{table*}[!ht]
\caption{Outputs from VideoChat-R1 for a sample clip from the Movie10 dataset.}
\label{videochat_r1_model_outputs_prompts}
\begin{center}
\scriptsize
\begin{tabular}{|l|p{10cm}|}
\hline
\multicolumn{2}{|c|}{
  \begin{tabular}{ccccc}
    \includegraphics[width=0.19\textwidth]{image_frames/ezgif-frame-001.jpg} &
    \includegraphics[width=0.19\textwidth]{image_frames/ezgif-frame-003.jpg} &
    \includegraphics[width=0.19\textwidth]{image_frames/ezgif-frame-008.jpg} &
    \includegraphics[width=0.19\textwidth]{image_frames/ezgif-frame-013.jpg} &
    \includegraphics[width=0.19\textwidth]{image_frames/ezgif-frame-016.jpg}\\
    \small Frame 1 & \small Frame 3 & \small Frame 8 & \small Frame 13 & \small Frame 16 \\
  \end{tabular}
} \\ \hline
\bf Instruction&\bf Model Output\\ 
\multirow{1}{*}{Action Recognition}  & The video captures a scene from what appears to be a social setting, possibly a bar or restaurant. Two individuals are engaged in conversation. One person is wearing a plaid shirt and has a blanket draped over their shoulders, suggesting they might be cold or seeking comfort. The other individual is dressed in a suit and tie, indicating a more formal attire. The background shows other patrons and bar stools, contributing to the ambiance of a casual gathering.  \\ \hline
\multirow{1}{*}{Video Understanding}  &    The video depicts a scene from what appears to be a movie or TV show, set in a dimly lit bar or lounge. Two individuals are seated at a table, engaged in conversation. The person on the left is wearing a dark suit and tie, while the individual on the right is dressed in a plaid shirt with a yellow scarf draped over their shoulders. Both appear to be holding drinks, suggesting they might be enjoying a casual evening out. \\
 \hline
Visual Question Answering & There are two people in the scene. They appear to be engaged in a conversation at a bar or restaurant. One person is wearing a plaid shirt and has a blanket draped over their shoulders, while the other person is dressed in a suit. The setting suggests a casual social interaction. \\ \hline
Video Captioning & Two individuals engage in conversation at a bar, surrounded by dim lighting and patrons.\\ \hline
\multirow{1}{*}{Object and Scene Recognition}   & The video shows two individuals seated at a bar, engaged in conversation. The person on the left is wearing a dark suit and appears to be listening attentively. The person on the right is wearing a plaid shirt with a yellowish blanket draped over their shoulders and is holding a drink. The background features a bar setting with various bottles and glasses, indicating a social or casual environment. The lighting suggests an indoor setting, likely during the evening. \\ \hline
Commonsense Reasoning & In the video, one character appears to be engaged in a conversation with another character at what looks like a bar or a similar social setting. The character on the right seems to be leaning forward slightly, which might indicate that they are listening intently or trying to emphasize a point during their conversation. The motivation for this action could be: 1. \textbf{Engagement}: The character might be trying to show interest or engagement in the conversation by leaning closer. \\ \hline
Spatial Understanding & The video appears to be from a movie or TV show set in a bar or restaurant. The setting includes a bar counter with bottles and glasses, suggesting it could be a scene from a film or series that takes place in a social or dining establishment. However, without additional context or recognizable landmarks, it's difficult to pinpoint the exact location or landmark shown in the video. It could be from a variety of films or shows that feature such settings. \\ \hline
Temporal Ordering & The video depicts two individuals seated at a bar, engaged in conversation. The person on the left is wearing a suit and appears to be listening attentively. The individual on the right is dressed casually in a plaid shirt with a blanket draped over their shoulders, suggesting they might be cold or seeking comfort. They both have drinks in front of them, indicating they are likely enjoying a social setting. The background shows other patrons and bar stools, contributing to the ambiance of a lively bar environment. \\ \hline
Video reasoning & The video appears to be from a scene in a movie or TV show, likely set in a bar or similar social setting. The unusual aspect here is the presence of a large, animated hand with a glove that resembles a cartoonish character, which seems out of place in a real-world setting like a bar. This suggests it might be a special effect or a prop used for comedic or dramatic purposes within the context of the film. \\ \hline
Narrative Understanding & The video appears to be from a scene in the movie ``The Hangover," where the characters are at a bar, seemingly engaged in a conversation. The central conflict in this movie revolves around a group of friends who go on a bachelor party to Las Vegas but end up getting lost and having various mishaps. \\ \hline
Emotion and Sentiment Analysis & The video depicts two individuals engaged in what appears to be an intense conversation at a bar. The person on the right seems to be speaking with a sense of urgency or concern, as indicated by their open mouth and expressive eyes. The other individual is listening attentively, suggesting a moment of serious discussion. Overall, the scene conveys a sense of tension and engagement. The setting, with its dim lighting and crowded bar, adds to the atmosphere of a private yet public interaction.  \\ \hline
Global Appearance & The video shows two characters sitting at a bar. The character on the left is wearing a dark suit and tie, while the character on the right is wearing a plaid shirt with a yellow blanket draped over their shoulders. There are no significant changes in their appearances or outfits throughout the video; they remain in these respective outfits for the duration of the frames provided. \\ \hline
Linking Events & In the video, two individuals are seated at a bar, engaged in conversation. The setting suggests a casual and possibly intimate interaction, which could imply that they are discussing something personal or significant. This early event of them sitting together and talking could influence later developments by establishing a foundation for their relationship or the context of their conversation. For instance, if this is part of a larger narrative, the initial interaction might set up themes such as friendship, trust, or conflict that will unfold in subsequent scenes.
\\ \hline
\end{tabular}
\end{center}
\end{table*}

\begin{table*}[!ht]
\caption{Outputs from Qwen-2.5 Audio for a sample audio from the Movie10  dataset (Wolf of wallstreet).}
\label{audio_instruct_model_outputs_prompts_qwen2.5}
\begin{center}
\scriptsize
\begin{tabular}{|l|p{10cm}|}
\hline
\multicolumn{2}{|c|}{
  \begin{tabular}{c}
    \includegraphics[width=0.25\textwidth]{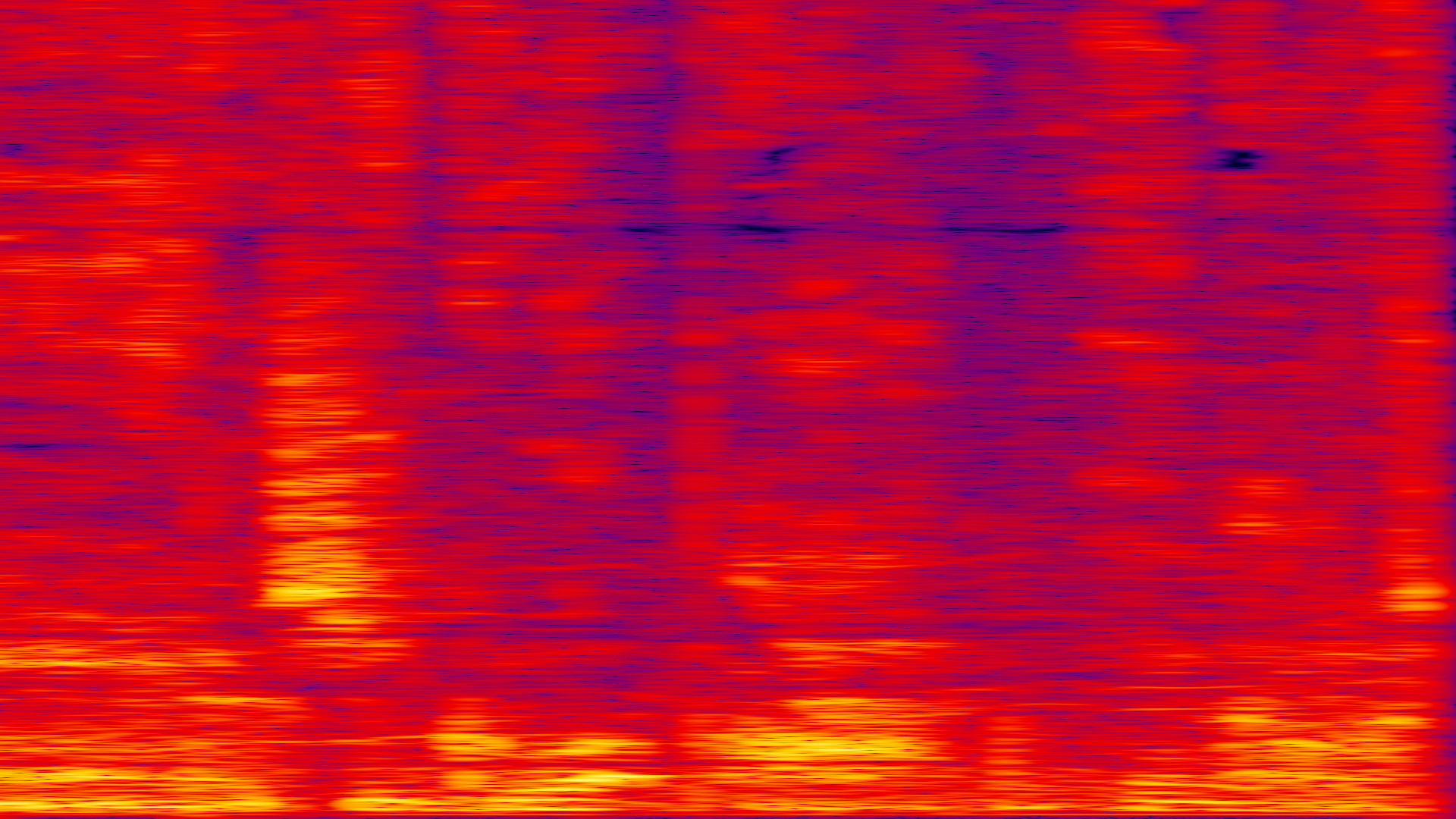} \\
    \small Spectrogram  \\
  \end{tabular}
} \\ \hline
\bf Instruction&\bf Model Output\\ 
\hline
\multirow{1}{*}{Audio Understanding}  & There's a male singing in English with an angry mood, saying 'And she grew up hot. \\ \hline
\multirow{1}{*}{Audio Comprehension}  &   They are talking and laughing. \\
 \hline
Audio Captioning &The audio is about someone's upbringing. \\ \hline
% Video Captioning & A man with glasses sitting at a table talking with another man.\\ \hline
\multirow{1}{*}{Sound Event Detection}   & There are no background sounds or music, only speech. A male voice speaks in English with the phrase 'And she grew up hot. \\ \hline
Speaker Identification & A man speaking in English with an angry mood said 'And she grew up hot, you know?  \\ \hline
\end{tabular}
\end{center}
\end{table*}

\begin{table*}[!ht]
\caption{Outputs from Kimi-Audio for a sample audio from the Movie10  dataset (Wolf of wallstreet)}
\label{audio_instruct_model_outputs_prompts_kimi}
\begin{center}
\scriptsize
\begin{tabular}{|l|p{10cm}|}
\hline
\multicolumn{2}{|c|}{
  \begin{tabular}{c}
    \includegraphics[width=0.25\textwidth]{image_frames/spectrogram.png} \\
    \small Spectrogram  \\
  \end{tabular}
} \\ \hline
\bf Instruction&\bf Model Output\\ 
\hline
\multirow{1}{*}{Audio Understanding}  &  The audio contains a speech where kids are talking by the door in English with a female voice and a neutral mood.\\ \hline
\multirow{1}{*}{Audio Comprehension}  &   The people are talking and there is music playing in the background. \\
 \hline
Audio Captioning & A man is speaking loudly and aggressive.\\ \hline
% Video Captioning & A man with glasses sitting at a table talking with another man.\\ \hline
\multirow{1}{*}{Sound Event Detection}   &  There is no speech in this audio.\\ \hline
Speaker Identification &  The speaker in the audio is a Mandarin-speaking male, aged between 16-25 years old, with an angry mood.\\ \hline
\end{tabular}
\end{center}
\end{table*}

\FloatBarrier

\section{Correlation Between Instruction Semantics and Model Representations}
\label{app:instruction_semantics_representaions}

We conducted a comprehensive study of the robustness of semantic similarity.

\begin{figure*}[!ht]
    \centering
    \includegraphics[width=0.49\linewidth]{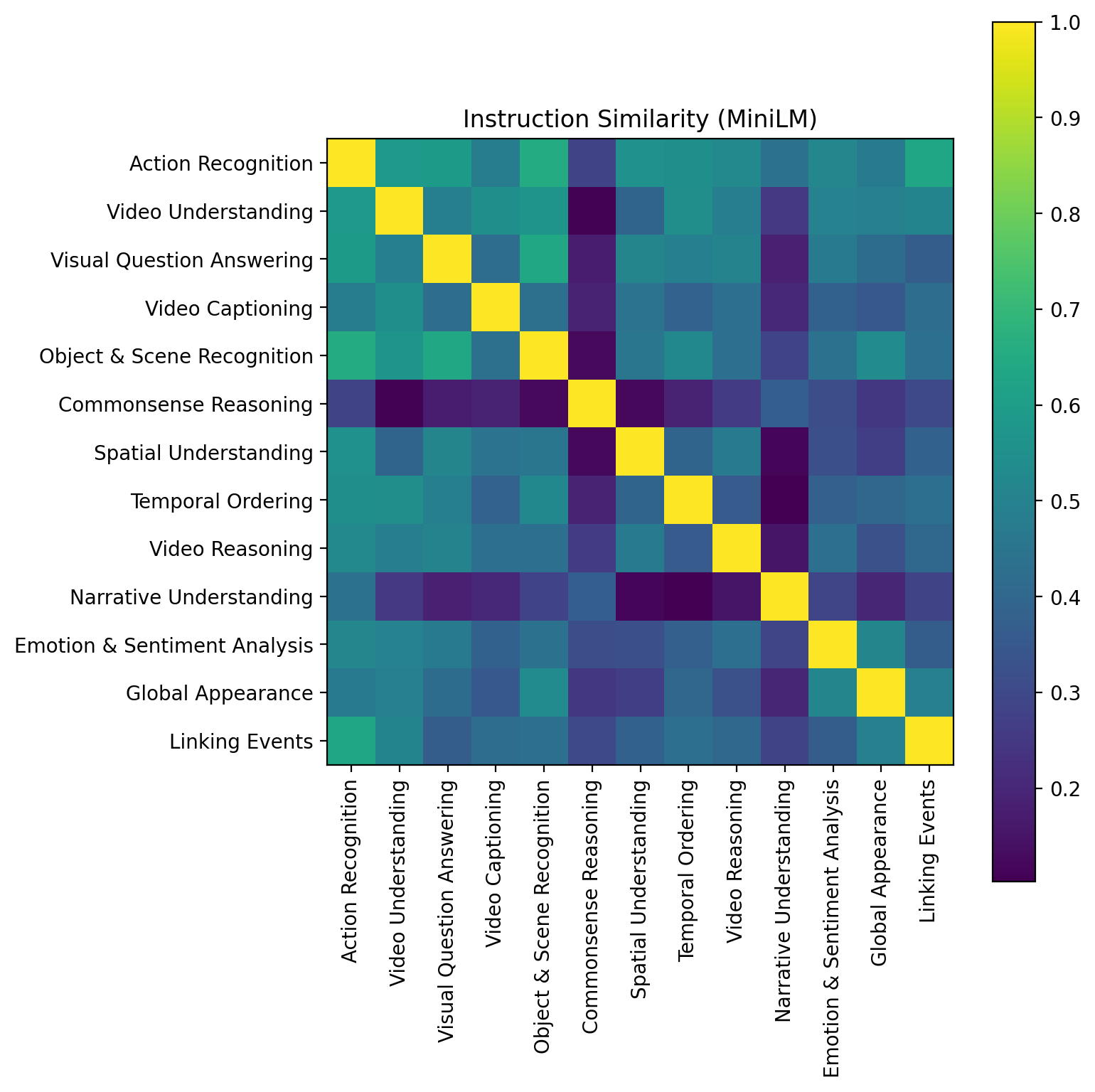}
    \includegraphics[width=0.49\linewidth]{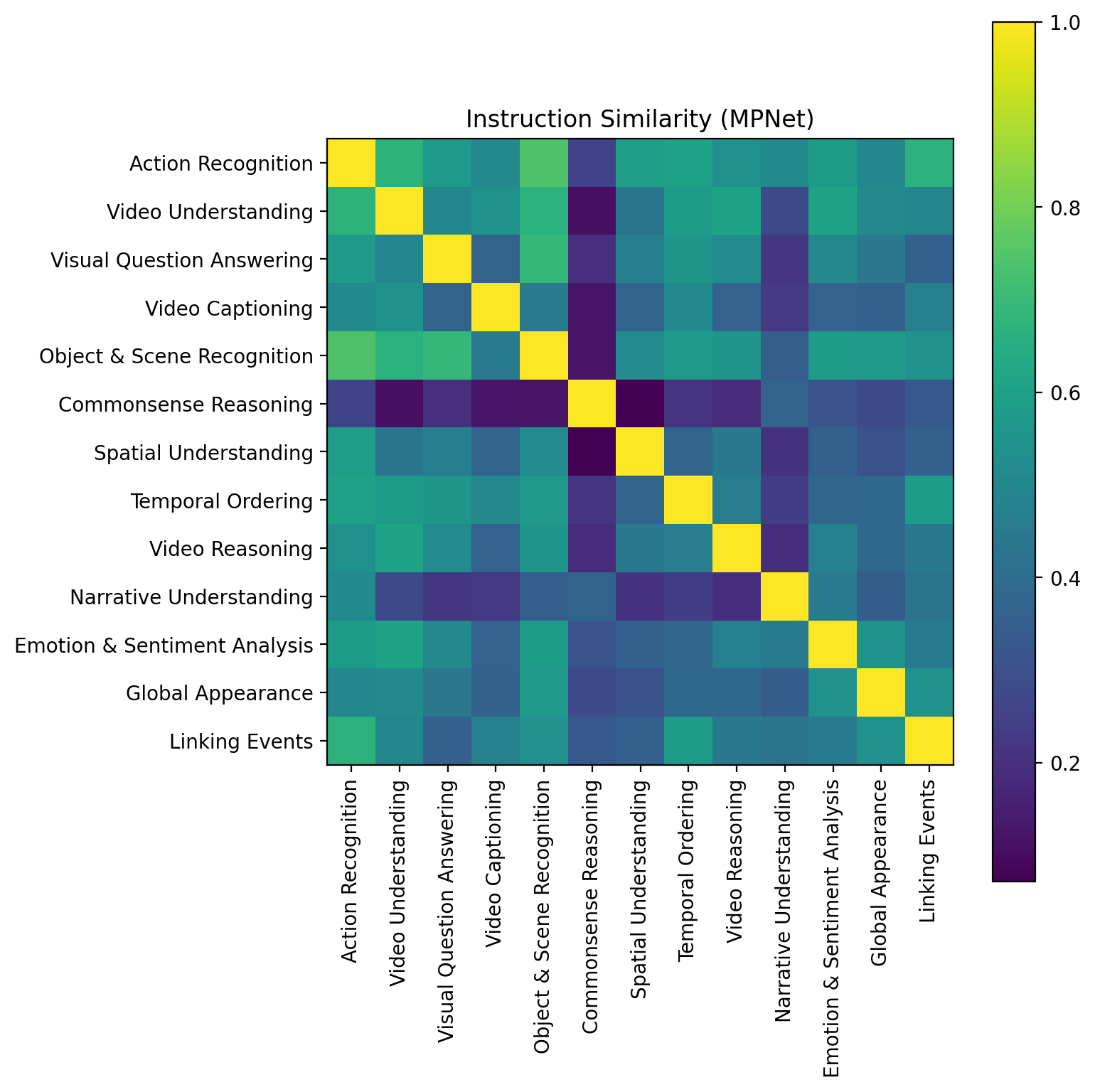}
    \caption{Instruction semantic similarity using text-embeddings.}
    \label{fig:instruction_similarity_sbert}
\end{figure*}

\begin{figure*}[!ht]
    \centering
    \includegraphics[width=0.95\linewidth]{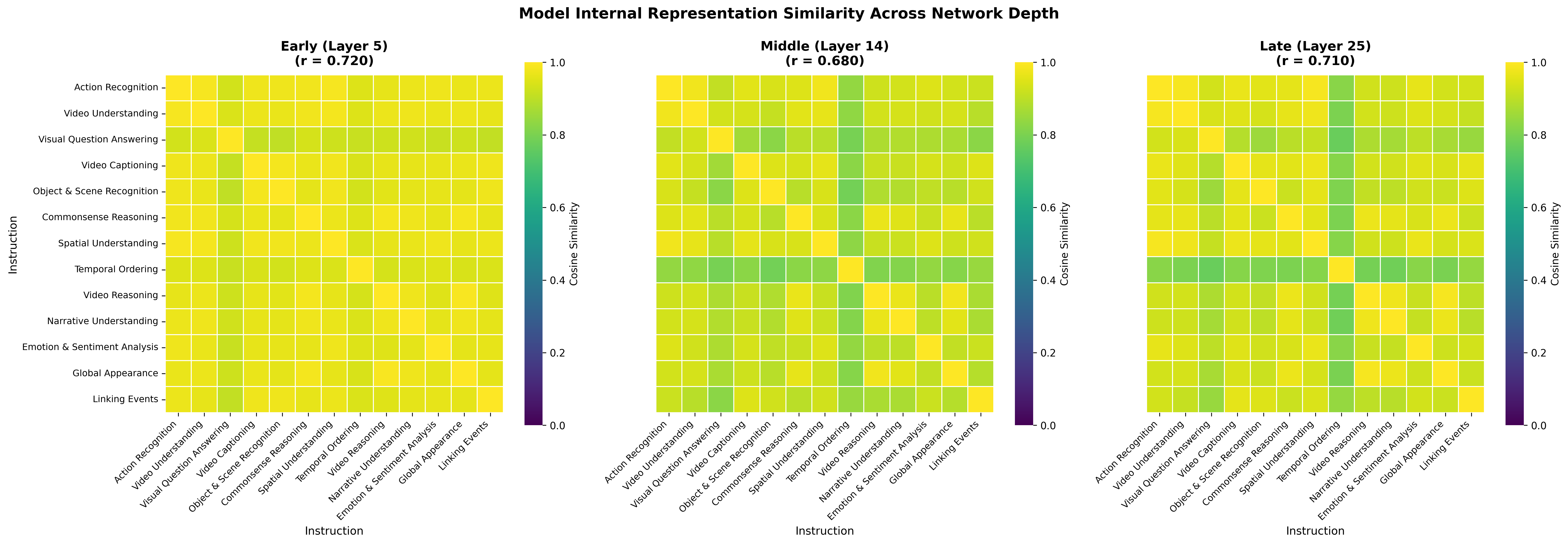}
    \caption{Qwen-2.5-VL-7B-Instruct: Model internal representational similarity across layer depth.}
    \label{fig:instruction_similarity_representations}
\end{figure*}

\noindent\textbf{Semantic similarity measurement.} To validate that our task-specific instruction set contains semantically similar pairs, we perform following: (i) Using 13 video task-specific instructions, we first computed pairwise semantic similarity using two independent text embedding models (all-MiniLM-L6-v2~\citep{reimers-2019-sentence-bert} and MPNet~\citep{reimers-2019-sentence-bert}). Both models produced highly consistent semantic similarity matrices (Pearson r = 0.94 between MiniLM and MPNet embeddings). (ii) Captures fine-grained semantic relationships between instruction texts. (iii) Example pairs: ``Describe the video'' vs. ``Caption the video'' (high semantic similarity).
Fig.~\ref{fig:instruction_similarity_sbert} shows pairwise semantic similarity between 13 instruction prompts computed using MiniLM embeddings (left) and MPNet (right). From Fig.~\ref{fig:instruction_similarity_sbert} (left), we observe that the semantic similarity ranges from 0.15 to 0.85 (mean: 0.42 $\pm$ 0.18), with multiple pairs of high-similarity identified ($>$0.60):(Action Recognition vs. Video Understanding (0.68), Video Understanding vs. Visual Question Answering (0.65), Object \& Scene Recognition vs. Action Recognition (0.68)). We also observe low semantic similarity pairs: (Commonsense Reasoning vs. Most others, Commonsense Reasoning vs. other tasks, Spatial Understanding vs. Emotional/Narrative). This confirms our instruction set contains the ``semantically similar or equivalent instructions'' for testing fine-grained task distinctions.

\noindent\textbf{Internal representations of the IT MLLM.} To measure how the model internally processes different task-specific instructions, we extracted hidden language states from the Qwen-2.5-VL-7B-Instruct model across all processing layers. We perform the following: (i) Extracted language hidden states across all 29 layers for each task-specific instruction. (ii) Used the same video input with varying instructions to isolate instruction effects. (iii) Analyzed three key layers: Early (Layer 5), Middle (Layer 14), and Late (Layer 25). (iv) While we present detailed results for three representative layers, we computed correlations and clustering metrics across all 29 layers to ensure findings are not artifacts of specific layer selection.

Fig.~\ref{fig:instruction_similarity_representations} shows pairwise similarity between 13 instruction prompts computed using three key layers. From Fig.~\ref{fig:instruction_similarity_representations}, we observe that layer-specific differentiation patterns emerge: (i) In early layers, relatively uniform representation similarity suggests that initial encoding stage has limited task specialization, (ii) In middle layers, increased variation in similarity patterns indicates evidence of task-specific transformations, and (iii) In later layers, pattern similar to early layer but with subtle differences implies refinement of task-specific representations. Overall, the quantitative analysis from Table 1 reveal that middle layer shows greatest differentiation, while early and late layers show convergent patterns. While, we observe clear structure in model representations, a critical question remains: Do these patterns reflect semantic similarity or functional task requirements? To answer this, we next compare model representations with semantic similarity computed from text embeddings. 

\noindent\textbf{Correlation analysis: semantic similarity vs. model representations.}
For each layer, we computed correlations between: Semantic similarity (from text embeddings) and Model representation similarity (from hidden states) as follows: 
For each layer L in [1, 2, $\cdots$, 29]:
(i) Extract upper triangle of semantic similarity matrix (78 pairs),
(ii) Extract upper triangle of model similarity matrix (78 pairs),
(iii) Compute Pearson correlation between the two sets,
(iv) Compute Spearman correlation for robustness, and 
(v) Test significance against random baseline

We compute both (a) the semantic similarity between instructions using text embedding models, and (b) the similarity of internal model representations (e.g., cosine similarity between layer vectors). A strong positive correlation between these measures would demonstrate that instruction tuning enhances the model's sensitivity to fine-grained semantic distinctions. Contrary to the hypothesis that surface semantic similarity should drive internal similarity, as shown in Table~\ref{tab:correlation_depth}, we observe near-zero to modestly positive coupling (Pearson r $\approx$ 0.04–0.18; Spearman $\rho$ $\approx$ 0.15–0.27), with several mid/late layers show uncorrected Spearman p$<$0.05, but no layer survives FDR correction across 29 layers (Benjamini–Hochberg). This pattern is highly consistent throughout the network depth, indicating that instruction tuning does not enhance semantic sensitivity; instead, IT representations show task-conditioned clustering

\begin{table*}[t]
\centering
\scriptsize
\caption{Correlation across network depth.
Coupling between instruction semantic similarity and instruction-conditioned representation similarity.}
\label{tab:correlation_depth}
\begin{tabular}{|l|c|c|c|c|}
\hline
\textbf{Layer range} &
\textbf{Pearson $r$ (mean $\pm$ sd)} &
\textbf{p-value range} &
\textbf{Spearman $\rho$ (mean $\pm$ sd)} &
\textbf{Significance} \\
\hline
Early (L1--L10) &
$0.066 \pm 0.015$ &
$0.497$--$0.717$ &
$0.184 \pm 0.022$ &
No individual layer sig.\ after FDR \\
Middle (L11--L20) &
$0.126 \pm 0.009$ &
$0.219$--$0.330$ &
$0.199 \pm 0.011$ &
Some uncorrected $p < 0.05$; none FDR-sig \\
Late (L21--L28) &
$0.149 \pm 0.018$ &
$0.109$--$0.240$ &
$0.247 \pm 0.013$ &
Several uncorrected $p < 0.05$; none FDR-sig \\
Overall (L1--L29) &
$0.113 \pm 0.041$ &
$0.109$--$0.717$ &
$0.207 \pm 0.034$ &
No FDR-significant effects \\
\hline
\end{tabular}
\end{table*}

\noindent\textbf{Statistical validation (Mantel permutation).}
We assess whether instruction semantics predict representation similarity using a Mantel permutation test (10,000 label shuffles) per layer, with FDR correction across layers. Pearson r and Spearman $\rho$ are the raw matrix correlations; Mantel z and q report significance. 
From Table~\ref{tab:mantel_test}, we find that Mantel tests show no significant association between semantic and representation spaces across layers, indicating representations are not organized by surface semantic similarity.

\begin{table}[!ht]
\centering
\scriptsize
\caption{\textbf{Mantel permutation test (semantic vs.\ representation similarity).} 
10{,}000 label permutations; FDR applied across layers.}
\label{tab:mantel_test}
\begin{tabular}{|l|c|c|c|c|}
\hline
\textbf{Layer} &
\textbf{Pearson $r$} &
\textbf{Spearman $\rho$} &
\textbf{Mantel $z$} &
\textbf{$q$-value (FDR)} \\
\hline
L5 (early)   & 0.077 & 0.210 & 0.65 & 0.482 \\
L14 (middle) & 0.130 & 0.201 & 1.12 & 0.376 \\
L25 (late)   & 0.143 & 0.237 & 1.23 & 0.338 \\
\hline
\end{tabular}
\end{table}

\noindent\textbf{Clustering Quality Analysis.}
We perform clustering analysis to determine whether the model organizes instructions by semantic similarity or functional task categories.

\begin{table}[!ht]
\centering
\scriptsize
\caption{\textbf{Task-based vs. Semantic Clustering Alignment.} 
$\Delta = \text{Task} - \text{Semantic}$.}
\label{tab:functional_semantic}
\begin{tabular}{|l|c|c|c|}
\hline
\textbf{Layer} &
\textbf{ARI (Func)} &
\textbf{ARI (Sem)} &
\textbf{$\Delta$ ARI} \\
\hline
L5  & $-0.003$ & $-0.132$ & $+0.129$ \\
L14  & $+0.010$ & $-0.132$ & $+0.142$ \\
L25 & $+0.042$ & $-0.132$ & $+0.174$ \\
\hline
\end{tabular}
\end{table}

Note:  ARI $>$ 0 indicates clustering aligns with task-based labels, and $<$ 0 for semantic further indicates that functional labels align better with representation geometry.

From Table~\ref{tab:functional_semantic}, we make the following key findings: (i) Task-based categories consistently outperform semantic clusters across all metrics and layers. (ii) ARI advantage: $\Delta$ = +0.129 to +0.174 (strong functional alignment). (iii) Middle layer specialization: mid/later layers shows strongest functional differentiation (ARI $\Delta$ = +0.423)

\noindent\textbf{Progressive functional specialization:} ARI advantage increases with depth, demonstrating task-specific organization strengthens in later layers. We focus on ARI rather than Silhouette because ARI measures alignment with ground-truth task categories (what instruction tuning teaches), while Silhouette measures cluster compactness (which may reflect pre-trained semantic structure). Overall, instruction tuning is associated with task-conditioned clustering (ARI evidence) while preserving semantic structure (Silhouette evidence).

\begin{figure}[t]
    \centering
    \includegraphics[width=\linewidth]{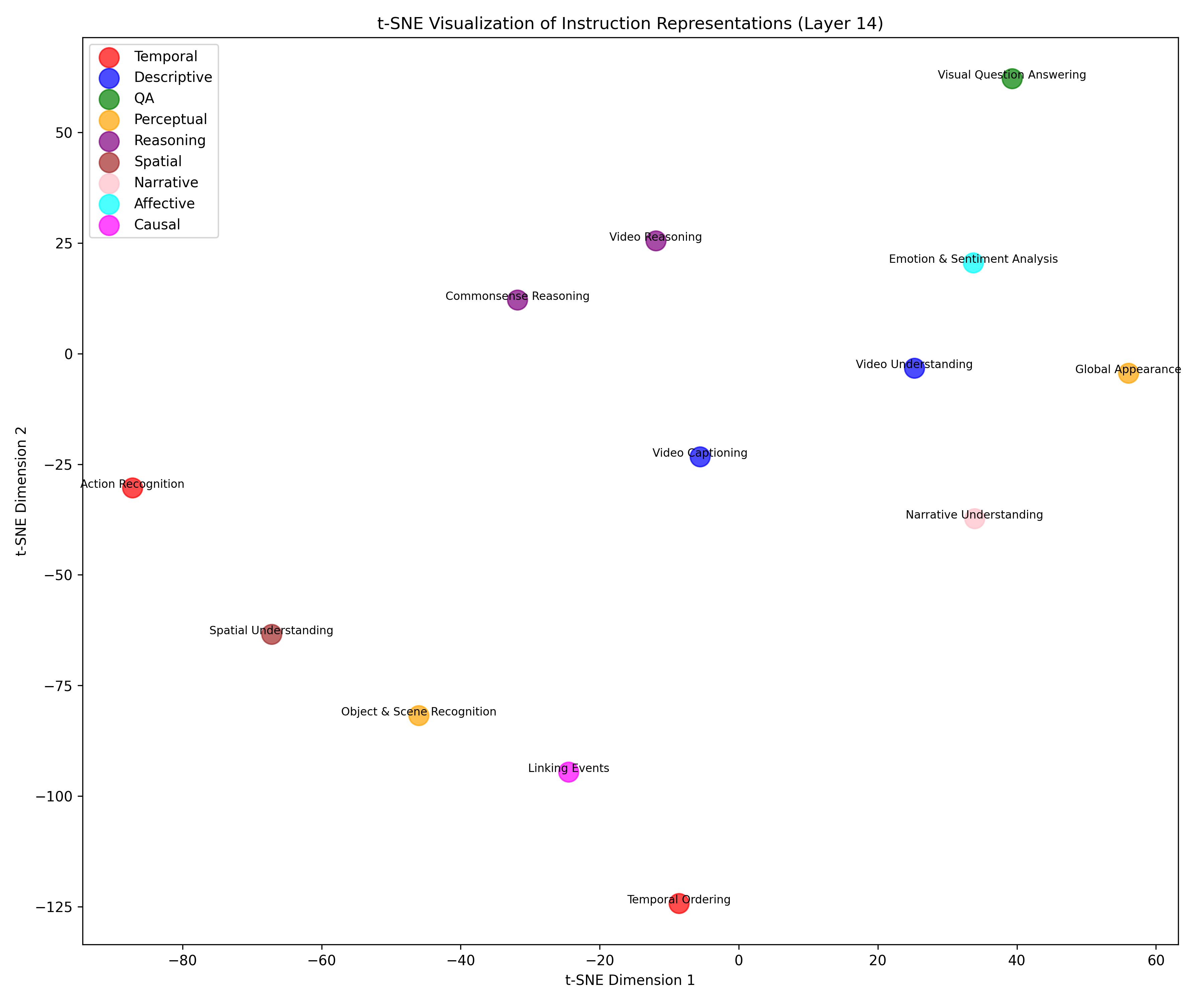}
    \caption{Clustering instruction task-specific representations.}
    \label{fig:clustering_instrcution_categories}
\end{figure}

\noindent\textbf{Visualization: t-SNE Projections (illustrative only)}
We performed t-SNE dimensionality reduction to visualize how the model organizes instruction task-specific representations in each layer (see Fig.~\ref{fig:clustering_instrcution_categories}, attached). We emphasize that t-SNE is used for visualization only; conclusions rely on ARI/silhouette and Mantel statistics. From Fig.~\ref{fig:clustering_instrcution_categories}, we make the following observations:
(1) Instructions cluster by task-based labels rather than semantic similarity. (2) Clear spatial separation between: (i) Perceptual prompts(object recognition, appearance) $\rightarrow$ Right region, (ii) Reasoning prompts(commonsense, video reasoning) $\rightarrow$ Center region, (iii) Temporal-causal prompts(action recognition, event linking) $\rightarrow$ Bottom region, (iv) Descriptive prompts(video understanding, captioning, narrative) $\rightarrow$ Distributed pattern.
(3) Semantically similar instructions (e.g., "Describe video" vs. "Caption video") are spatially separated. (4) Functional categories show consistent grouping patterns across layers.

\begin{table*}[!ht]
\centering
\scriptsize
\caption{IT vs. ICL models: Representational similarity and clustering metrics. *** p$<$0.001, FDR-significant across 29 layers; ns = not significant after FDR.}
\label{}
\resizebox{\textwidth}{!}{\begin{tabular}{|l |l |c |c |c |l |l|}
\hline
\textbf{Metric} & \textbf{Model} & \textbf{Layer 5} & \textbf{Layer 14} & \textbf{Layer 25} & \textbf{Winner} & \textbf{Magnitude} \\
\hline

\multirow{2}{*}{\textbf{Semantic correlation}} 
 & IT (Qwen-2.5-VL-7B-Instruct)    & Pearson $r = 0.077$ (ns) & 0.130 (ns) & 0.143 (ns) & --- & Weak \\
 & ICL (Qwen-2.5-VL-7B) & \textbf{0.690***} & \textbf{0.680***} & \textbf{0.777***} & \textbf{ICL} & $\sim$4.2$\times$ stronger \\

% \hline

% \multirow{2}{*}{\textbf{Silhouette score (cluster cohesion)}} 
%  & IT $\Delta$ (Func $-$ Sem)  & $-0.102$ & $-0.171$ & $-0.172$ & \textbf{Semantic} & Weak \\
%  & ICL $\Delta$ (Func $-$ Sem) & $-0.157$ & $-0.252$ & $-0.234$ & \textbf{Semantic} & Moderate--Strong \\

\hline

\multirow{2}{*}{\textbf{Adjusted Rand Index (label alignment)}}
 & IT $\Delta$ (Func $-$ Sem)  & \textbf{+0.129} & \textbf{+0.142} & \textbf{+0.174} & \textbf{Functional} & Weak \\
 & ICL $\Delta$ (Func $-$ Sem) & $-0.442$ & $-0.291$ & $-0.349$ & \textbf{Semantic} & Very Strong \\

\hline
\end{tabular}}
\end{table*}

\section{Representational Differences Between IT and ICL Models}
\label{app:instrcution_tuning_icl}

ICL models can follow zero-shot prompting i.e. in-context prompt, whereas Instruction-tuning adds a supervised signal. Below we describe the representational differences observed between IT and ICL models.

To understand the difference in the working of IT and ICL models, we perform additional analysis. We compared instruction-tuned (IT) and in-context learning (ICL) models to identify differences in representational structure. For the 13 tasks, we first compute a 13×13 semantic-similarity matrix using MiniLM embeddings. We then perform Representational Similarity Analysis (RSA) by computing correlations between the upper triangles of the 13×13 semantic-similarity matrix and the corresponding representation-similarity matrix (same videos, same pipeline), per layer.

We find that for the instruction-tuned model (Qwen-2.5-VL-7B-Instruct), the correlation between instruction semantic similarity and internal representation similarity is weak across layers. (e.g., best layer L28: Pearson r=0.183, p=0.109; Spearman $\rho$=0.266, p=0.018); no layer remains significant after FDR correction across 29 layers. This indicates that IT representations are less coupled to surface prompt semantics than ICL representations, consistent with task-conditioned representational structure that diverges from instruction-text wording.

% \textcolor{blue}{\noindent
% % \textbf{Stable task routing vs. wording sensitivity.}
% For the instruction-tuned model (Qwen-2.5-VL-7B-Instruct), the correlation between instruction semantic similarity and internal representation similarity is weak across layers. (e.g., best layer L28: Pearson r=0.183, p=0.109; Spearman $\rho$=0.266, p=0.018); no layer remains significant after FDR across 29 layers. This supports a function-driven geometry rather than surface wording.}

ICL (Qwen-2.5-VL-7B): RSA correlation r$\approx$0.78 $\rightarrow$ strong coupling to prompt semantics $\rightarrow$ shallow text-proximal matching
IT (Qwen-2.5-VL-7B-Instruct): RSA correlation r=0.183 (best layer, ns after FDR correction) $\rightarrow$ weak coupling to prompt semantics $\rightarrow$ task-conditioned representational structure that diverges from instruction-text wording

% For Qwen-2.5-Omni-7B (ICL), the correlation between MiniLM-based instruction semantic similarity and model representation similarity is high (e.g., r$\approx$0.78), indicating that prompt wording strongly drives internal states, consistent with more shallow, text-proximal matching.

% \textcolor{blue}{Note: correlations are computed between the upper triangles of the 13$\times$13 semantic-similarity matrix and the corresponding representation-similarity matrix (same videos, same pipeline), per layer.}

% \textcolor{blue}{Legend: }

\begin{itemize}
    \item Adjusted Rand Index (ARI) measures agreement between predicted clusters and ground-truth labels by counting pairwise agreements, adjusted for chance, with values ranging from -1 (worse than random) to +1 (perfect agreement). Positive ARI $\rightarrow$ indicates clustering aligns with task-based grouping, Negative ARI $\rightarrow$ clustering aligns with semantic-based grouping~\citep{hubert1985comparing}.
\end{itemize}

\noindent\textbf{Key findings: ICL vs IT.}
We compared instruction-tuned (IT) and in-context learning (ICL) models to identify differences in representational structure. We find that ICL models show stronger coupling to prompt semantics (RSA correlation r=0.78 vs. 0.14 for IT, a 4.2$\times$ difference, p$<$0.001). IT models show stronger task-conditioned clustering across layers as measured by ARI ($\Delta$=+0.13 to +0.17 across layers), with the effect strengthening progressively with depth (L5: +0.129 $\rightarrow$ L25: +0.174).

% We find that ICL models show higher semantic organization with semantic correlation: r=0.78 vs 0.14 (IT), a 4.2x advantage (p$<$0.001). IT models show emerging functional organization but with weak effects: Functional ARI advantage: $\Delta$=+0.13 to +0.17 across layers, progressive strengthening: L5: +0.129 → L25: +0.174 (increasing with depth).

ICL MLLMs behaves wording-sensitive (strong semantic-representation coupling), whereas IT representations are less coupled to surface wording and show task-conditioned representational structure that strengthens with depth. This is consistent with prior findings that instruction tuning leads to task-specific representations supporting downstream task performance~\citep{zhu2025internvl3}, and we extend this by showing that these representations also yield improved brain alignment.

\FloatBarrier

\section{Effectiveness of instruction-tuned video MLLMs vs audio MLLMs vs multimodal vs unimodal representations for various brain regions}
\label{app:MultimodalityResults}

\begin{figure*}[!ht]
\centering
    \includegraphics[width=0.95\linewidth]{images/cbar_mllms_new1.pdf}\\
\includegraphics[width=0.3\linewidth]{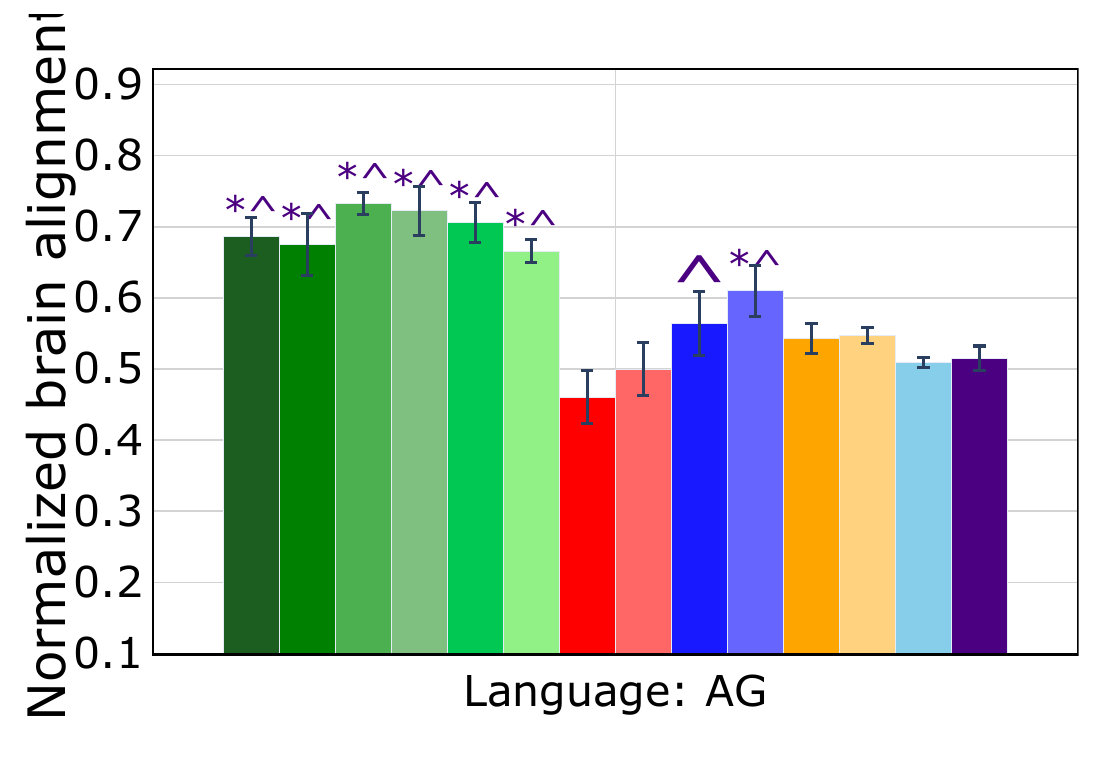}
\includegraphics[width=0.3\linewidth]{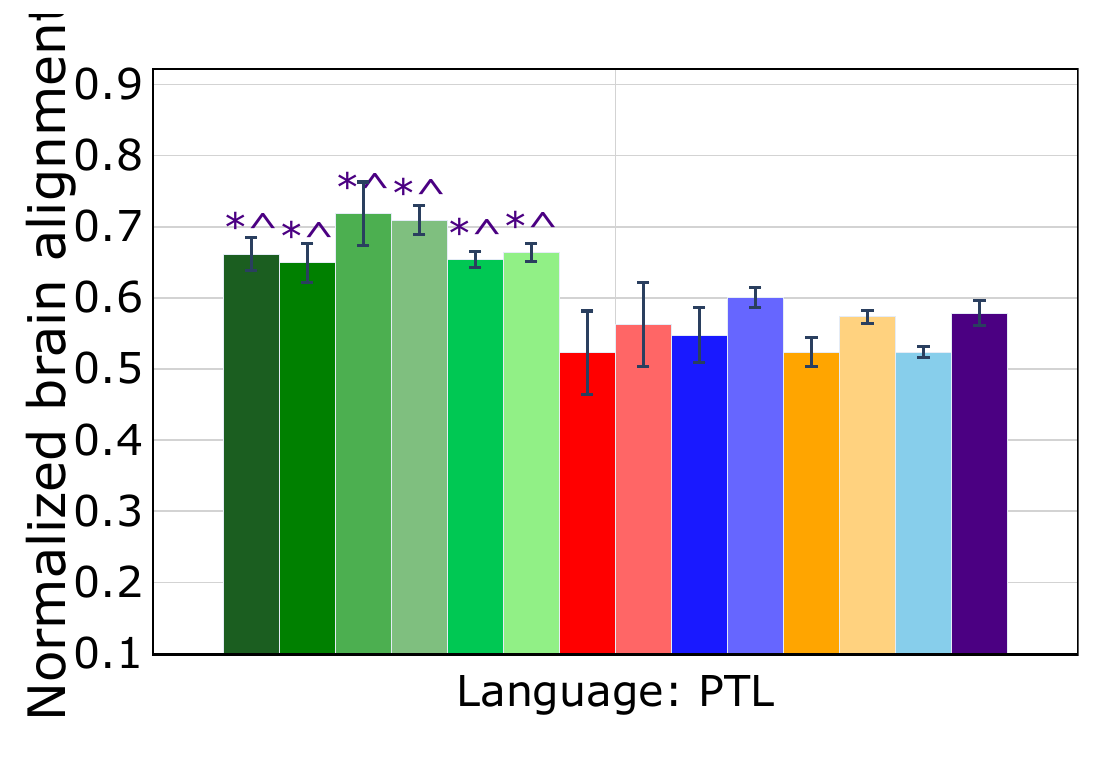}
\includegraphics[width=0.3\linewidth]{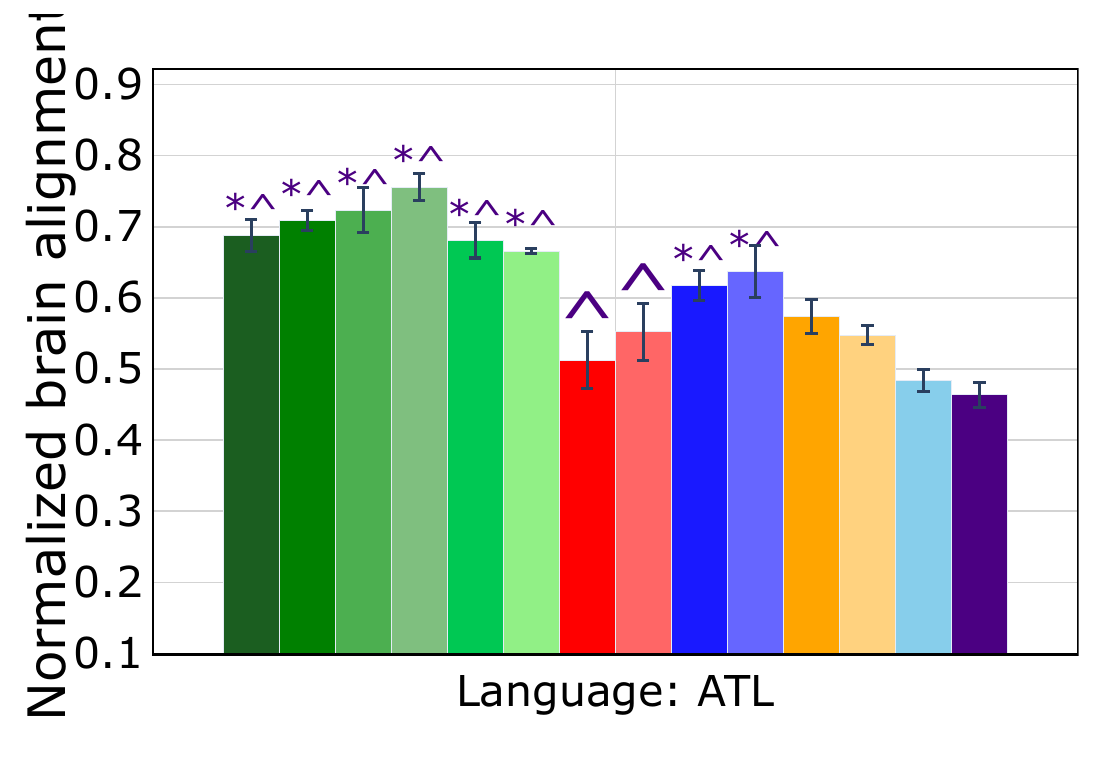}
\includegraphics[width=0.3\linewidth]{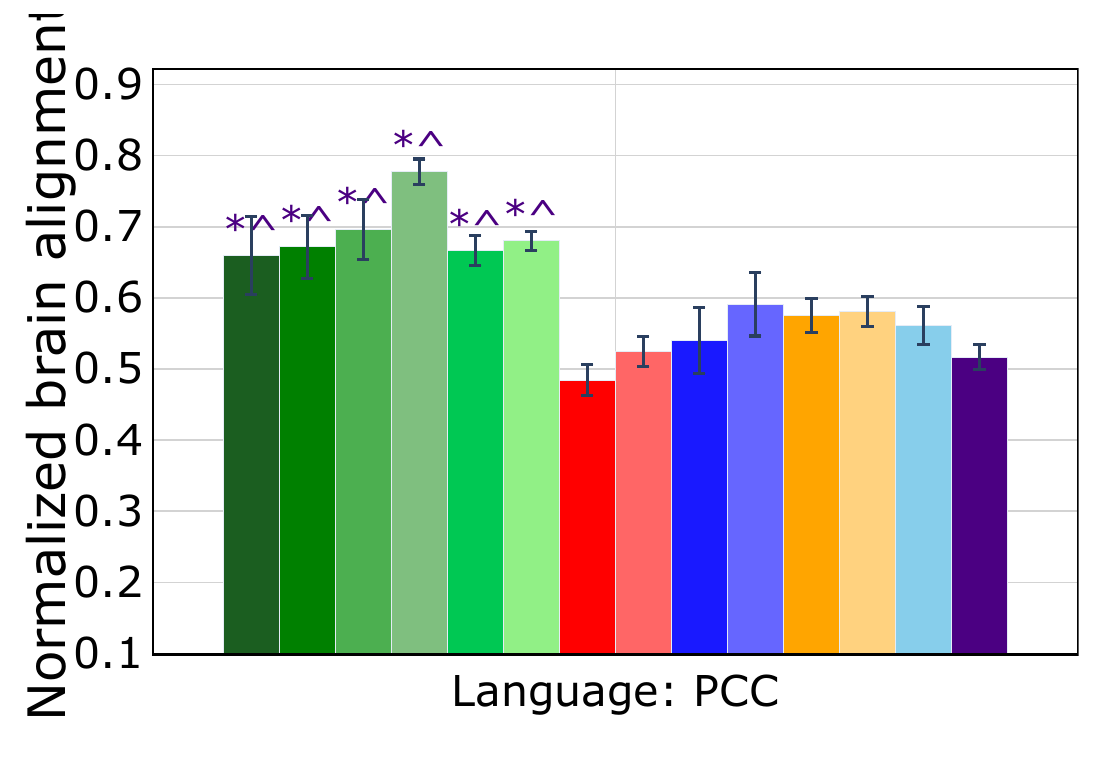}
\includegraphics[width=0.3\linewidth]{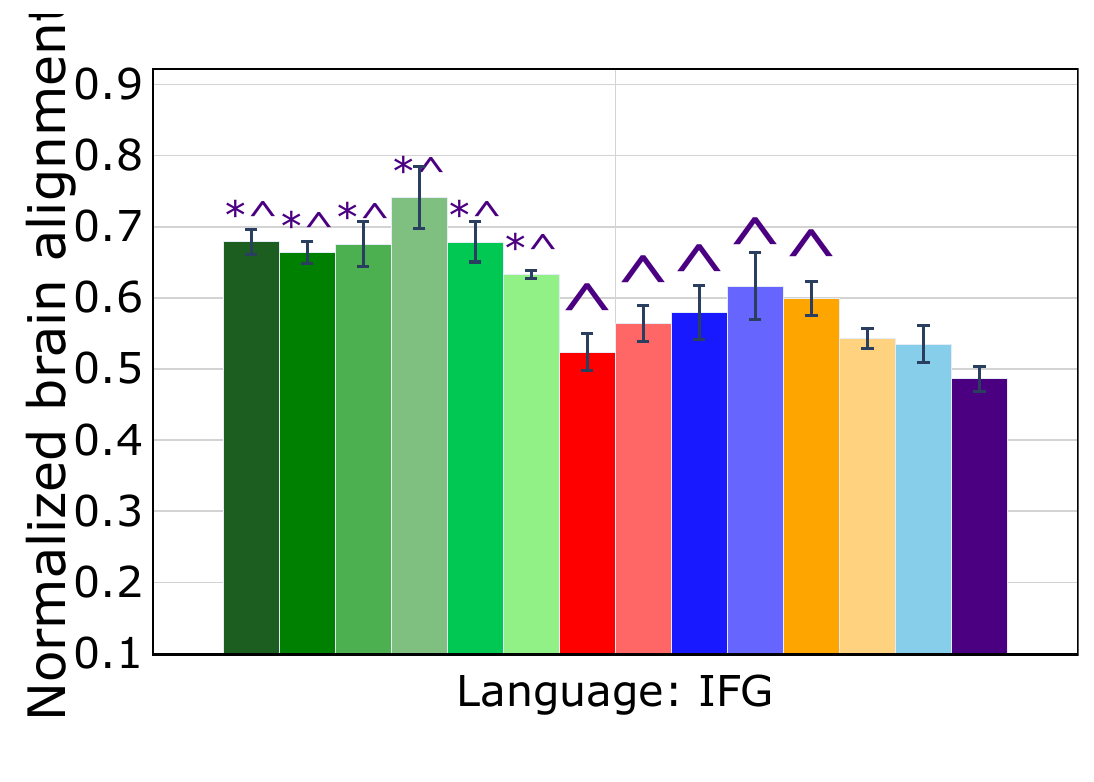}
\includegraphics[width=0.3\linewidth]{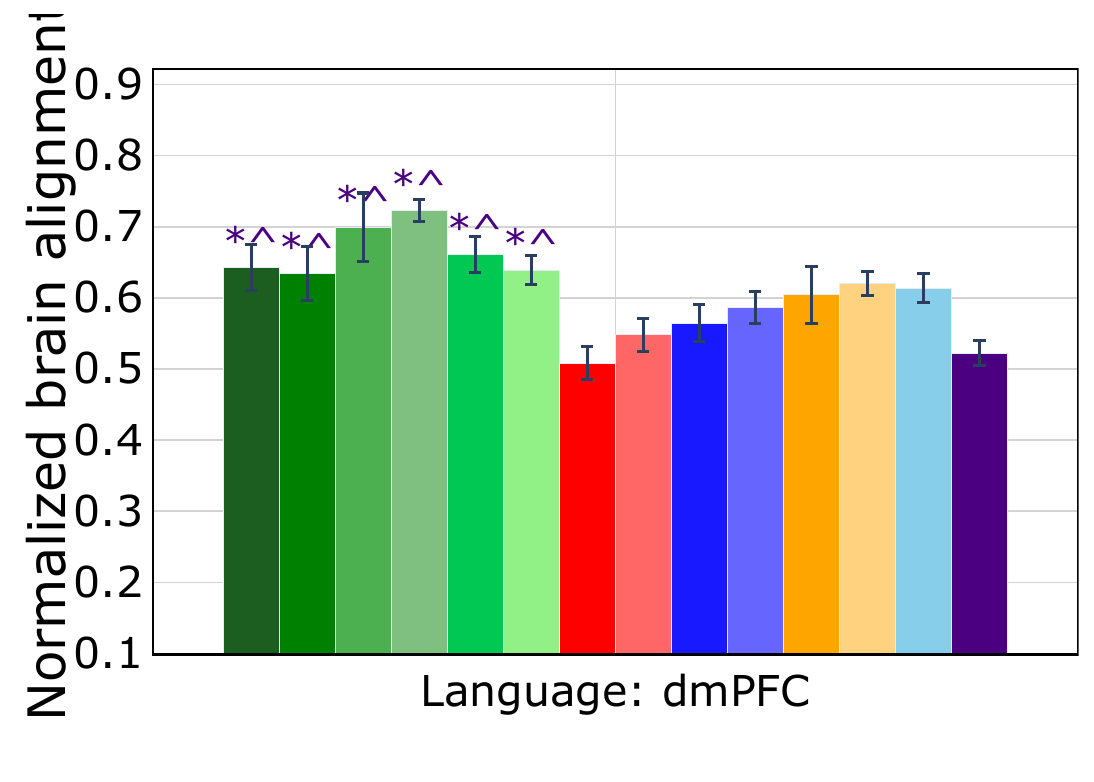}
\includegraphics[width=0.3\linewidth]{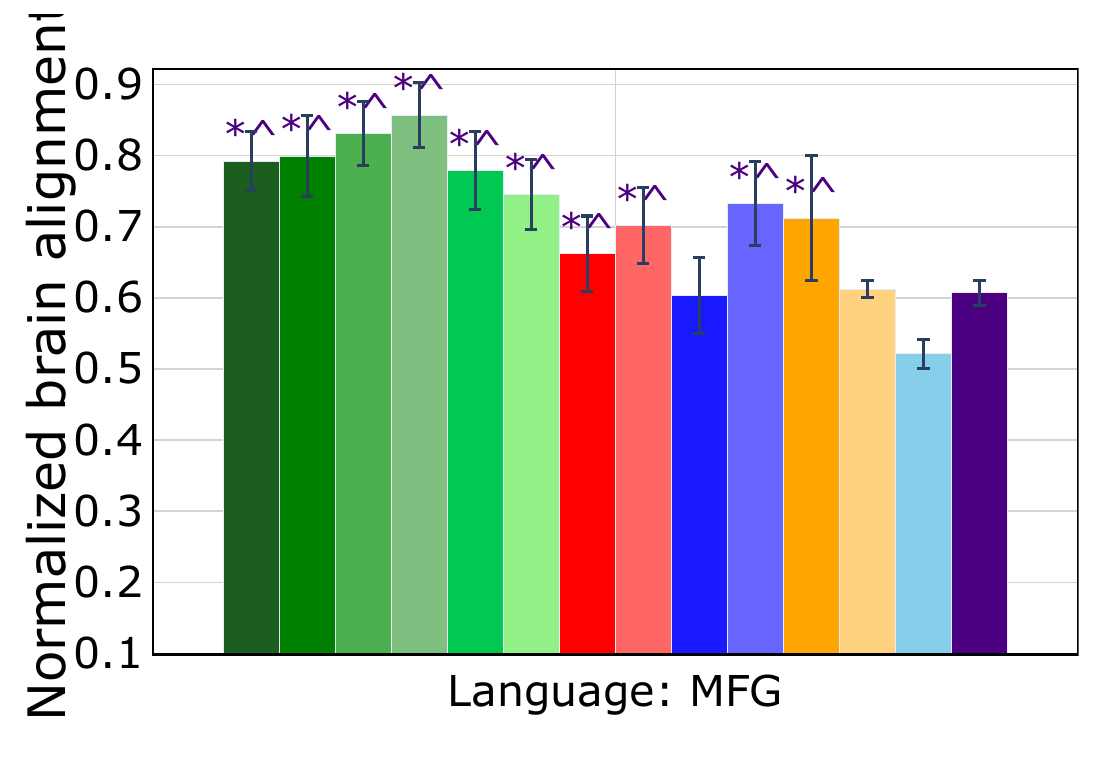}
\includegraphics[width=0.3\linewidth]{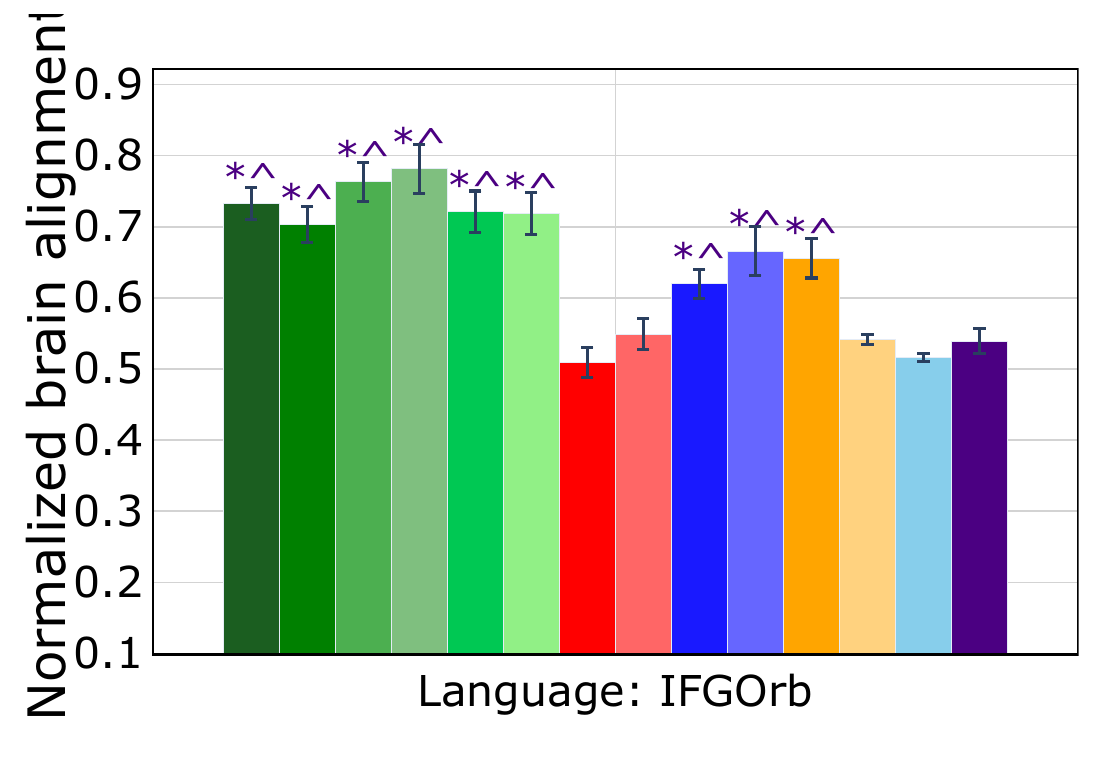}
\vspace{-0.5cm}
\caption{Average normalized brain alignment of instruction-tuned video MLLMs vs instruction-tuned audio MLLMs vs multimodal and unimodal models across several ROIs (AG, ATL, PTL, IFG, MFG, IFGOrb, PCC and dmPFC) of language region. Error bars indicate the standard error of the mean across participants. $*$ implies that instruction-tuned MLLM embeddings are significantly better than multimodal models and $\wedge$ means that instruction-tuned MLLM embeddings are significantly better unimodal models with p$\leq 0.05$.
}
\label{fig:language_results_subrois}
\end{figure*}

\begin{figure*}[!ht]
\centering
    \includegraphics[width=0.95\linewidth]{images/cbar_mllms_new1.pdf}\\
\includegraphics[width=0.3\linewidth]{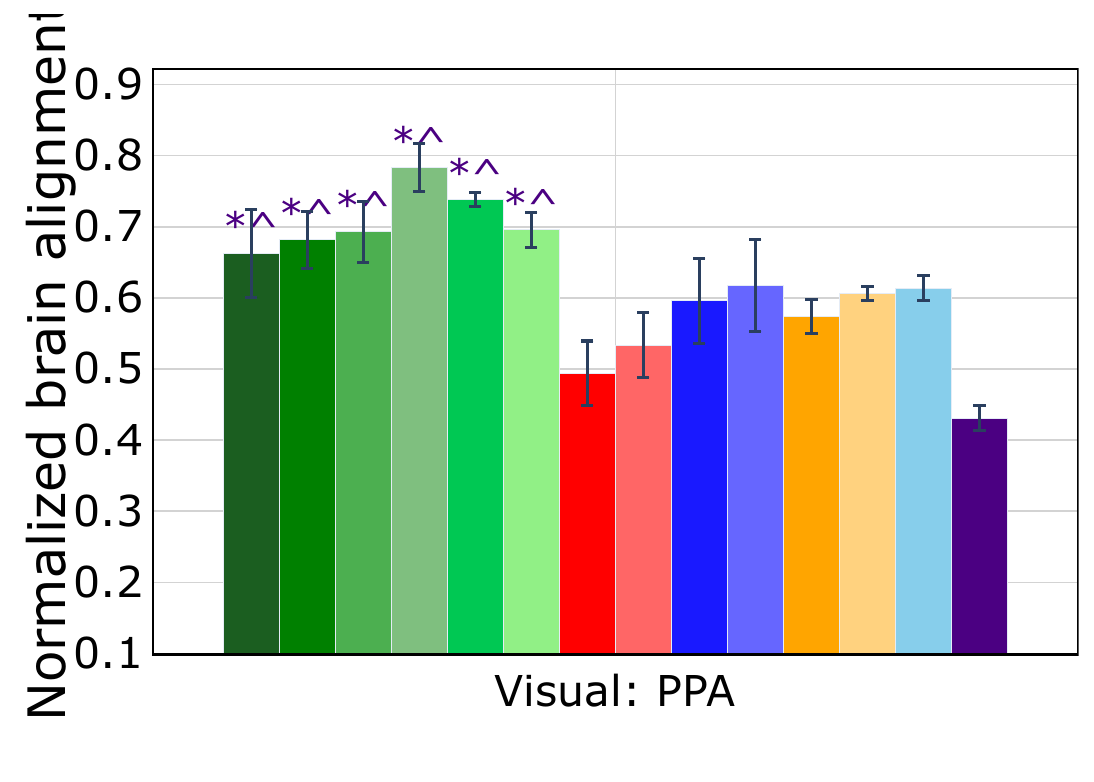}
\includegraphics[width=0.3\linewidth]{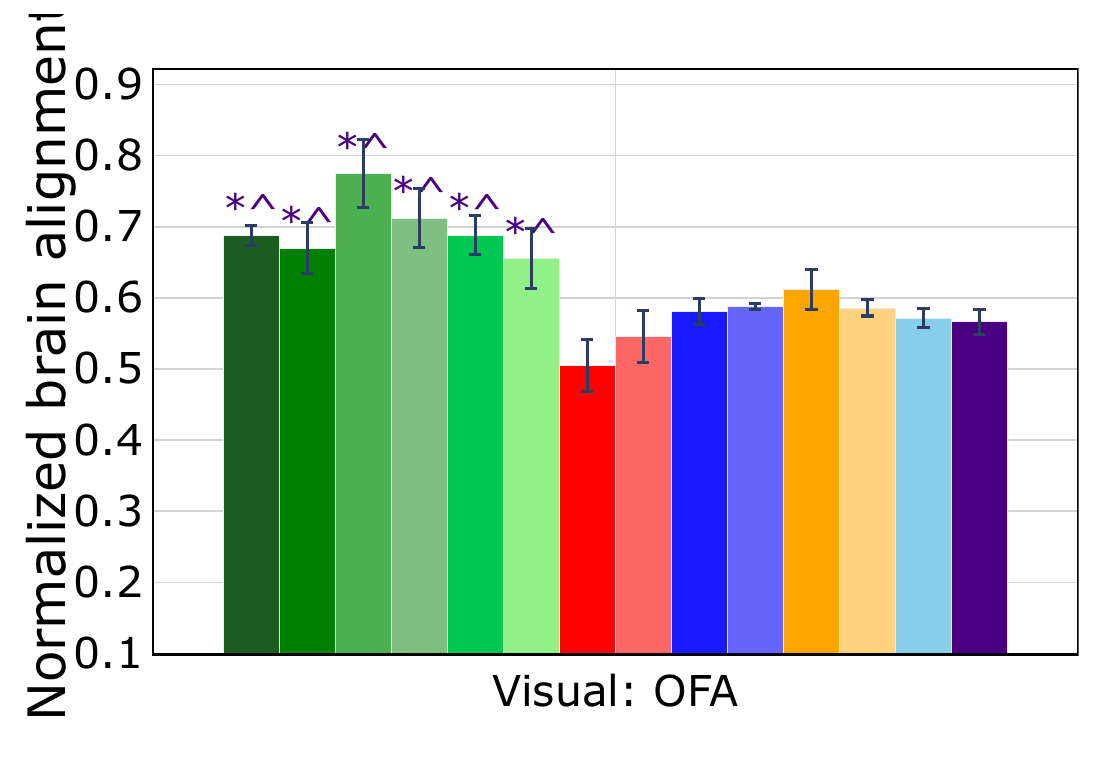}
\includegraphics[width=0.3\linewidth]{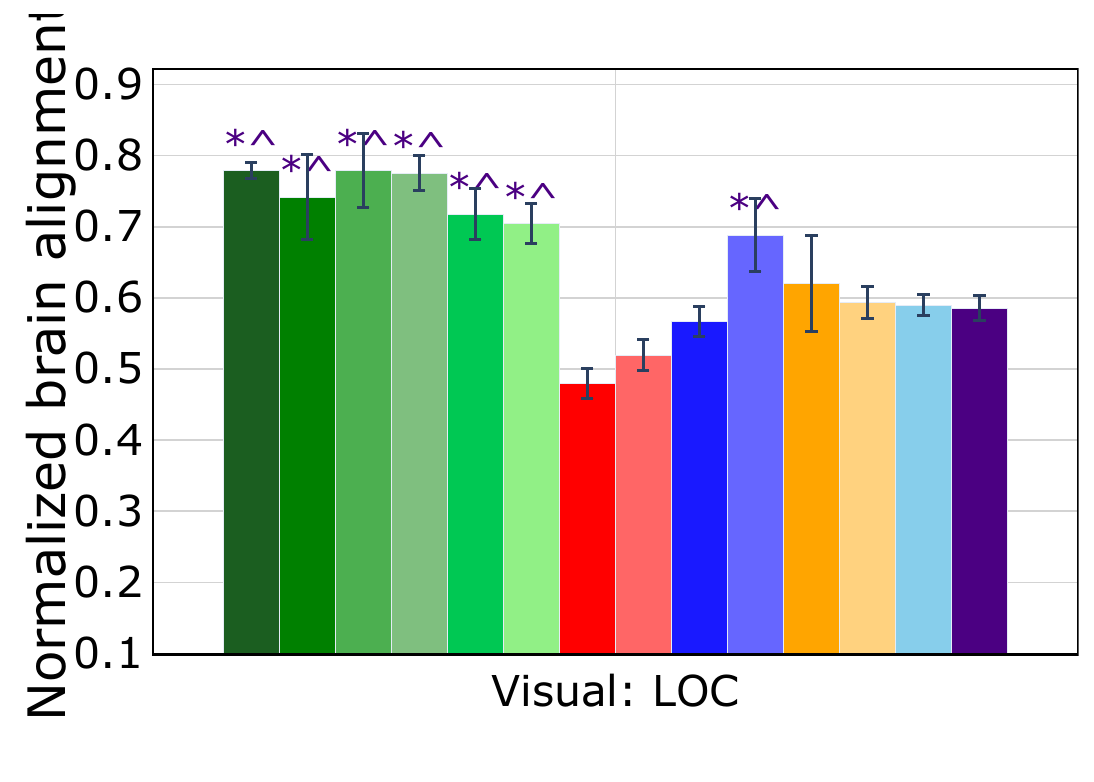}
\includegraphics[width=0.3\linewidth]{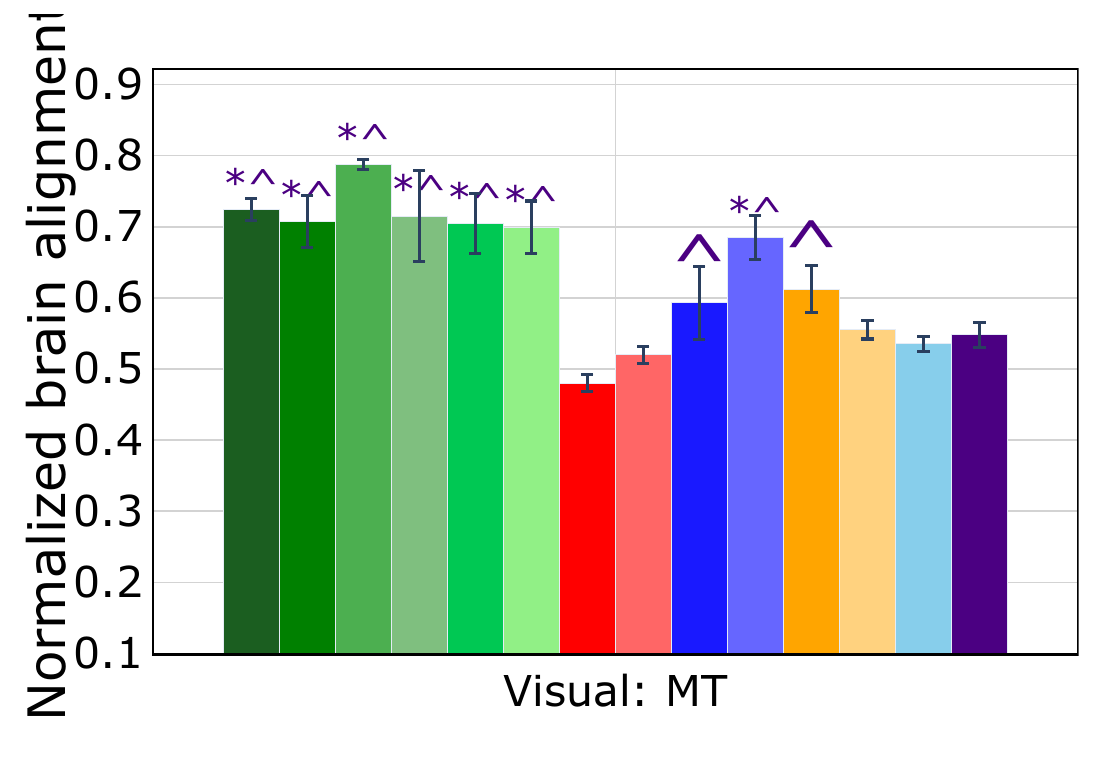}
\includegraphics[width=0.3\linewidth]{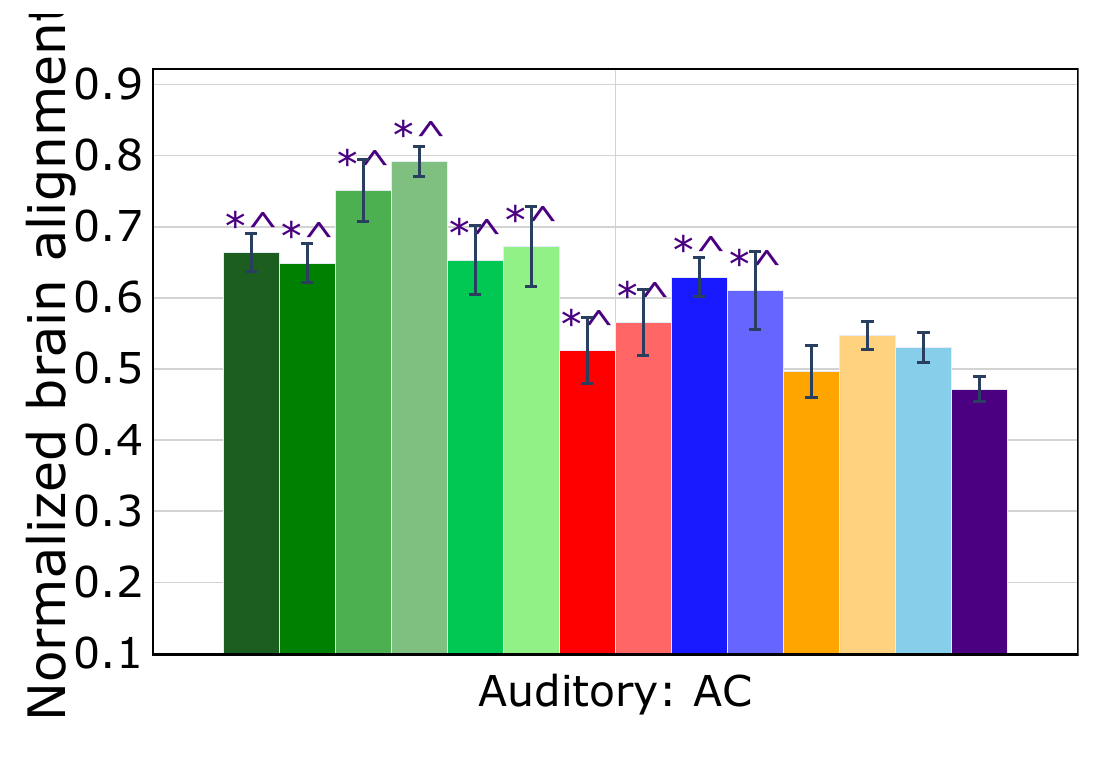}
 \includegraphics[width=0.3\linewidth]{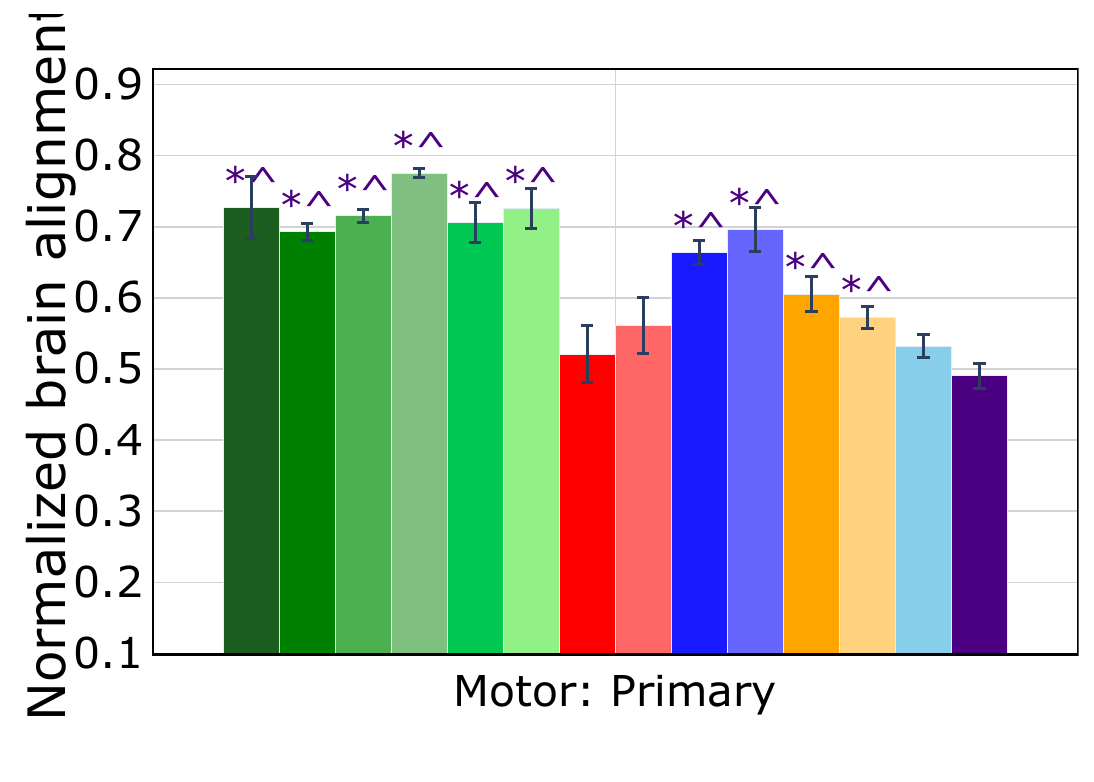}
 \includegraphics[width=0.3\linewidth]{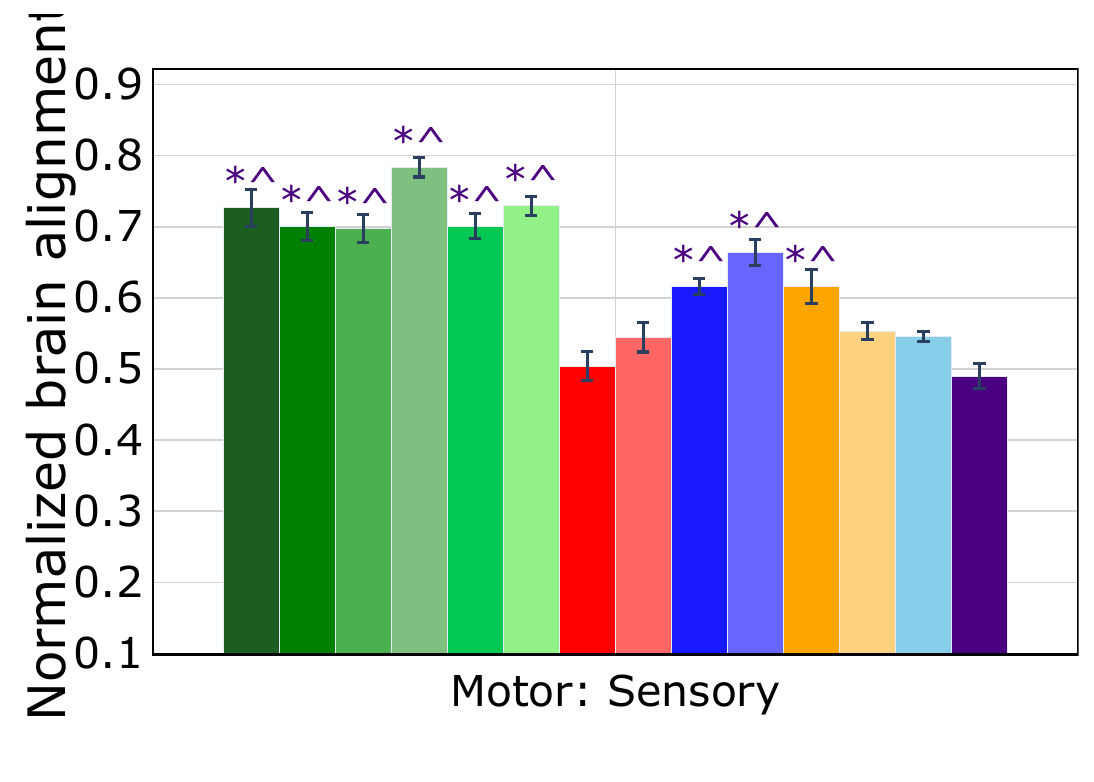}
\vspace{-0.5cm}
\caption{Average normalized brain alignment of instruction-tuned video MLLMs vs instruction-tuned audio MLLMs vs multimodal and unimodal models across several ROIs of visual cortex (PPA, OFA, LOC, MT), Auditory cortex (AC), and Motor Area (PMA and SMA). Error bars
indicate the standard error of the mean across participants. $*$ implies that instruction-tuned MLLM embeddings are significantly better than multimodal models and $\wedge$ means that instruction-tuned MLLM embeddings are significantly better unimodal models with p$\leq 0.05$.
 }
\label{fig:visual_speech_results_subrois}
\end{figure*}

Fig.~\ref{fig:language_results_subrois} show average normalized brain alignment of instruction-tuned video MLLMs vs instruction-tuned audio MLLMs vs multimodal and unimodal models across several ROIs (AG, ATL, PTL, IFG, MFG, IFGOrb, PCC and dmPFC) of language region. Fig.~\ref{fig:visual_speech_results_subrois} show the same for visual, auditory and motor regions.

\begin{figure*}[t]
    \centering
     \includegraphics[width=0.85\linewidth]{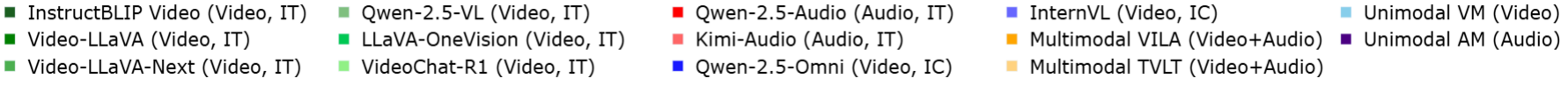}\\
     %\vspace{-0.1cm}
      \includegraphics[width=0.22\linewidth]{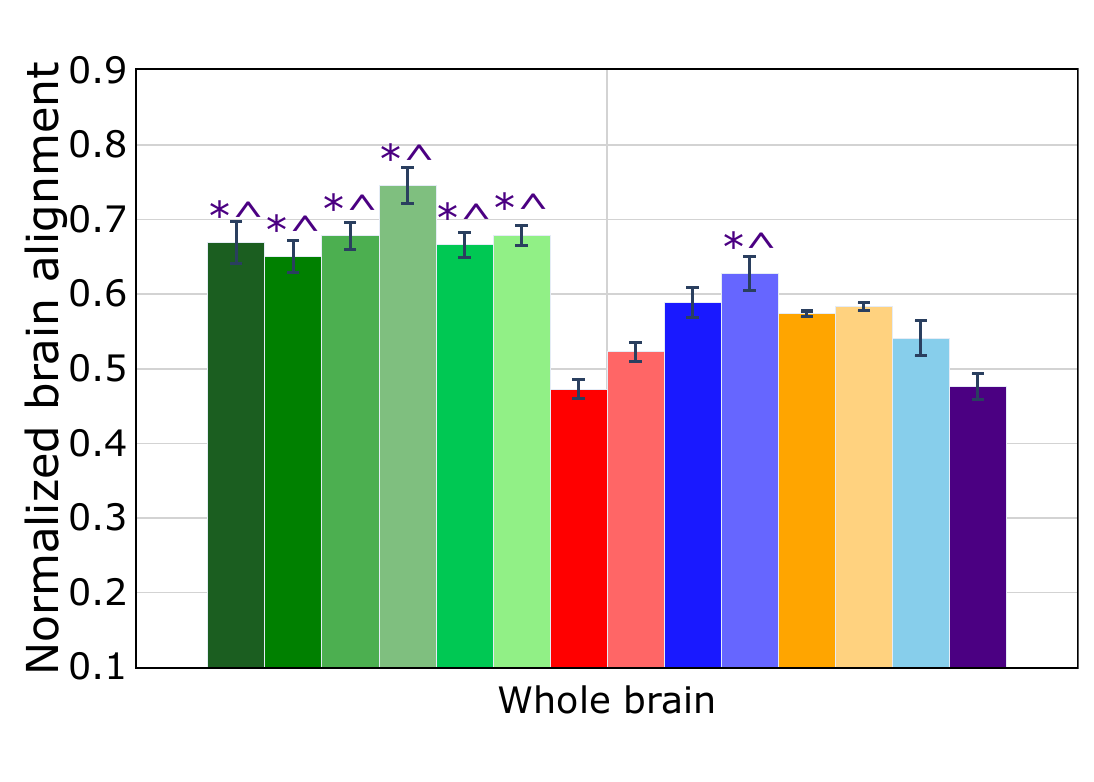}
    % % % \hspace{-10pt}
    \includegraphics[width=0.22\linewidth]{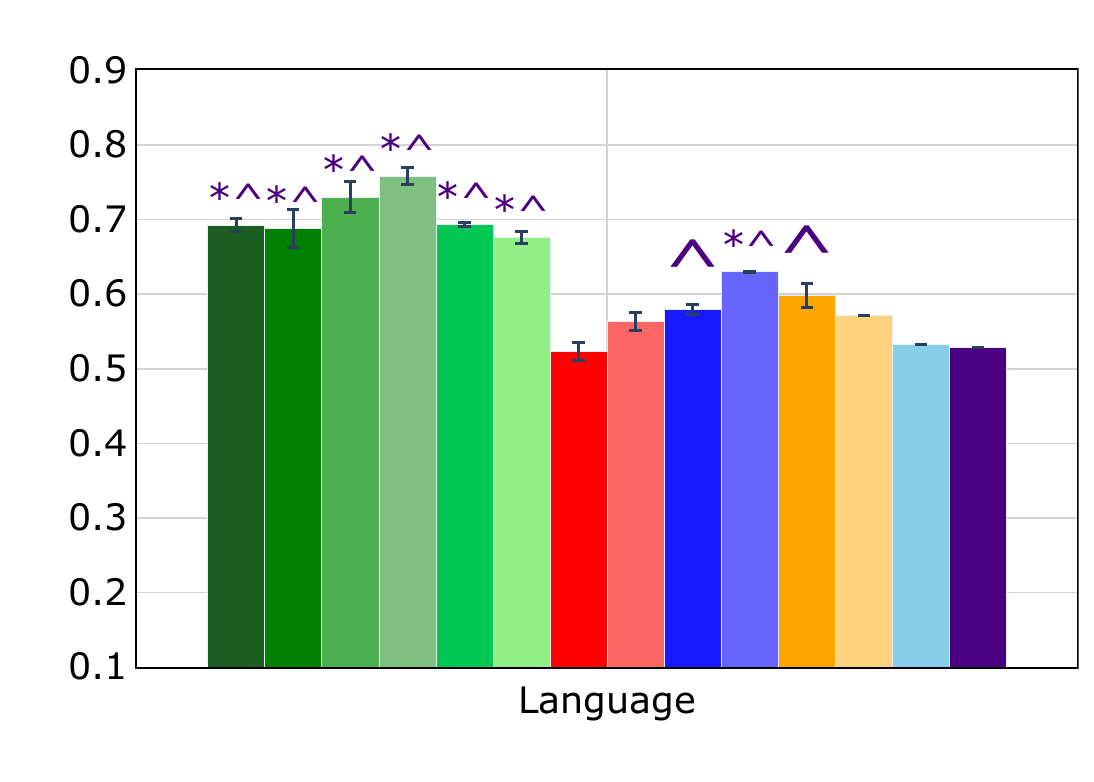}
     \includegraphics[width=0.22\linewidth]{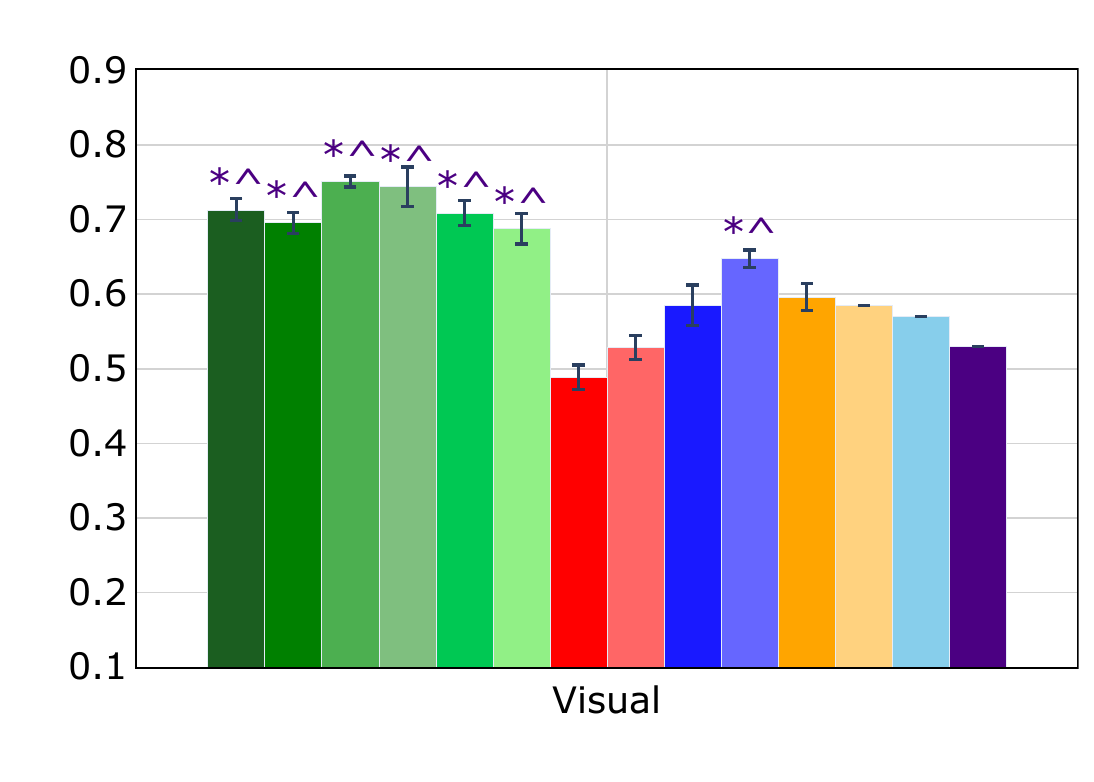}
     \includegraphics[width=0.22\linewidth]{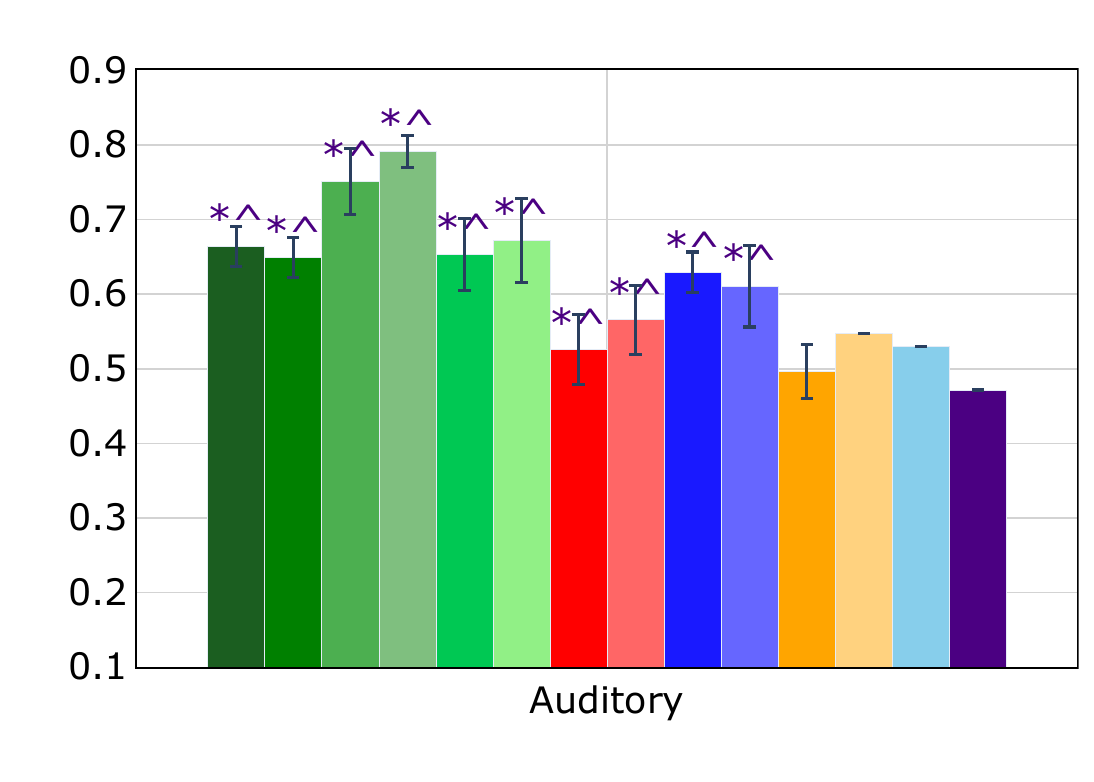}
    \vspace{-0.4cm}
    \caption{Average normalized brain alignment of instruction-tuned video MLLMs vs instruction-tuned audio MLLMs vs in-context learning video MLLMs vs multimodal and unimodal models across whole brain, language, visual and auditory regions. Error bars indicate the standard error of the mean across participants. $*$ implies that instruction-tuned MLLM embeddings are significantly better than multimodal models and $\wedge$ means that instruction-tuned MLLM embeddings are significantly better unimodal models with p$\leq 0.05$. IT: Instruction-tuned, IC: In-context learning
    % \sri{Can we mention about this p-value somewhere in Section 4 or Appendix or near Wilcox-signed test?}
    % \textcolor{green}{\textbf{$\times$}}
    % $*$ and $\wedge$ imply that instruction-tuned MLLM embeddings are significantly better than multimodal and unimodal models with p$\leq 0.05$.  
    }
    \label{fig:normalized_brain_alignment_allmodels}
\end{figure*}

\section{Inclusion of other Video MLLMs}
\label{app:other_videomllm_kimi}

\begin{figure*}[!ht]
    \centering
     \includegraphics[width=0.95\linewidth]{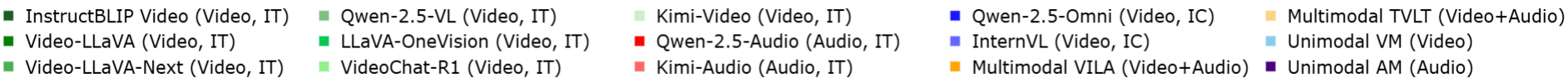}\\
      \includegraphics[width=0.24\linewidth]{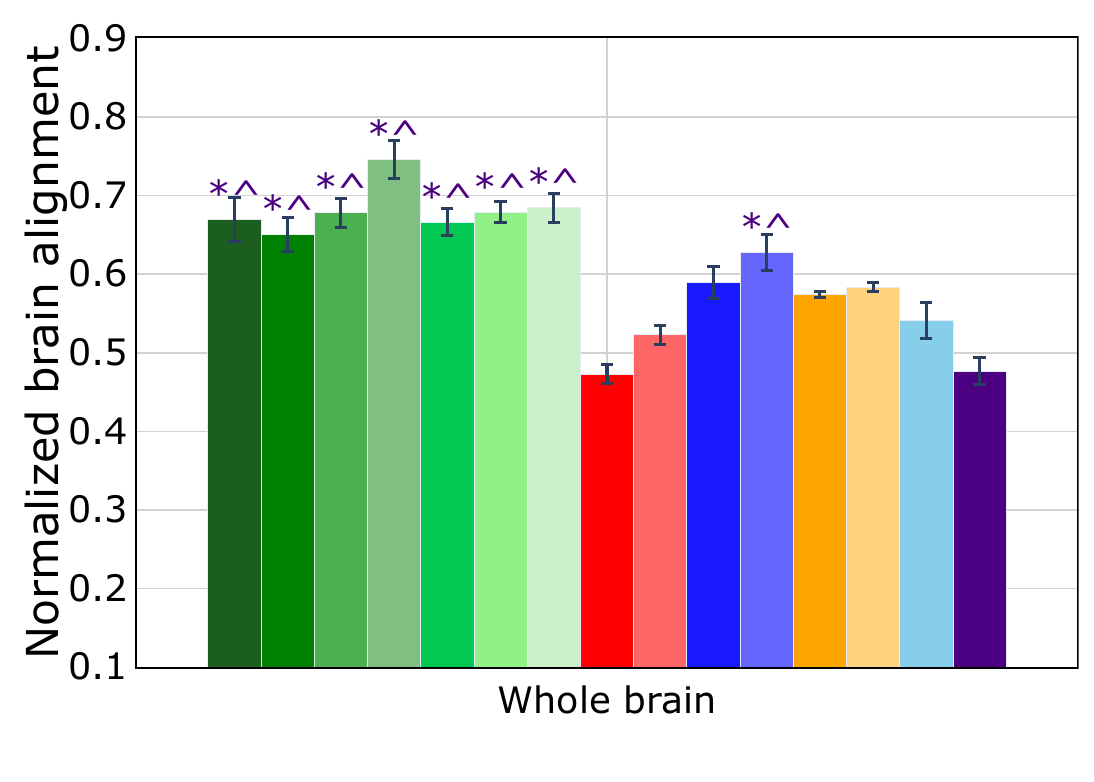}
    % % % \hspace{-10pt}
    \includegraphics[width=0.24\linewidth]{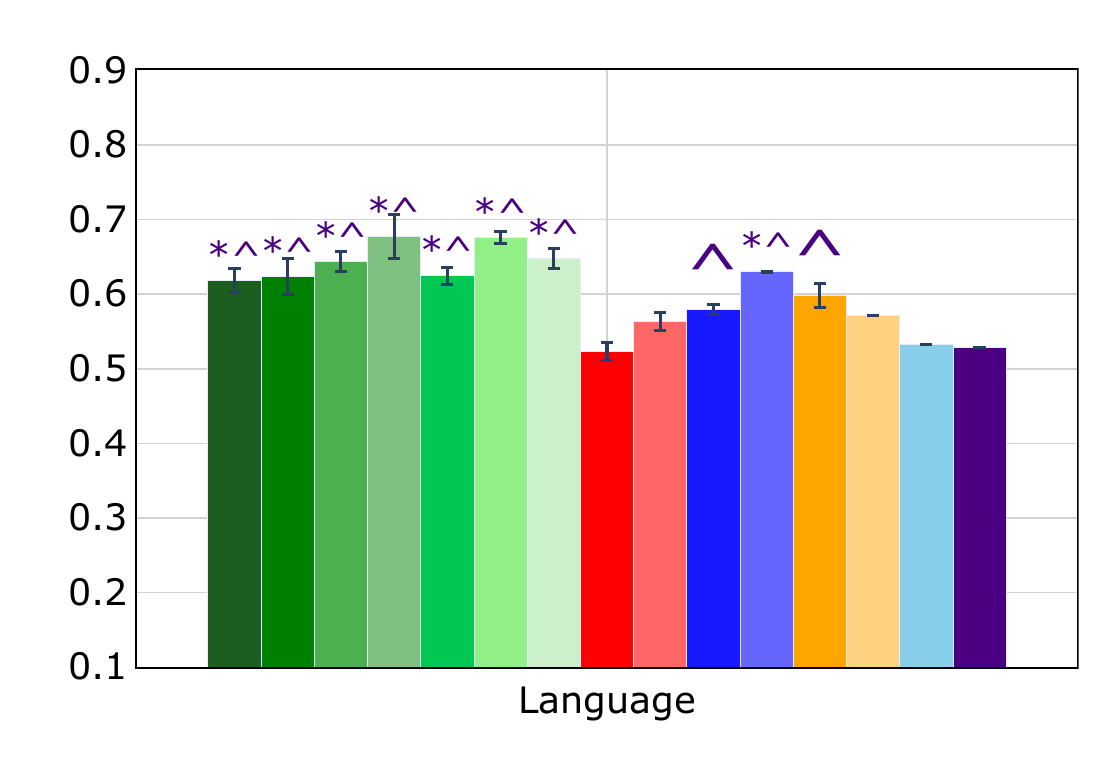}
     \includegraphics[width=0.24\linewidth]{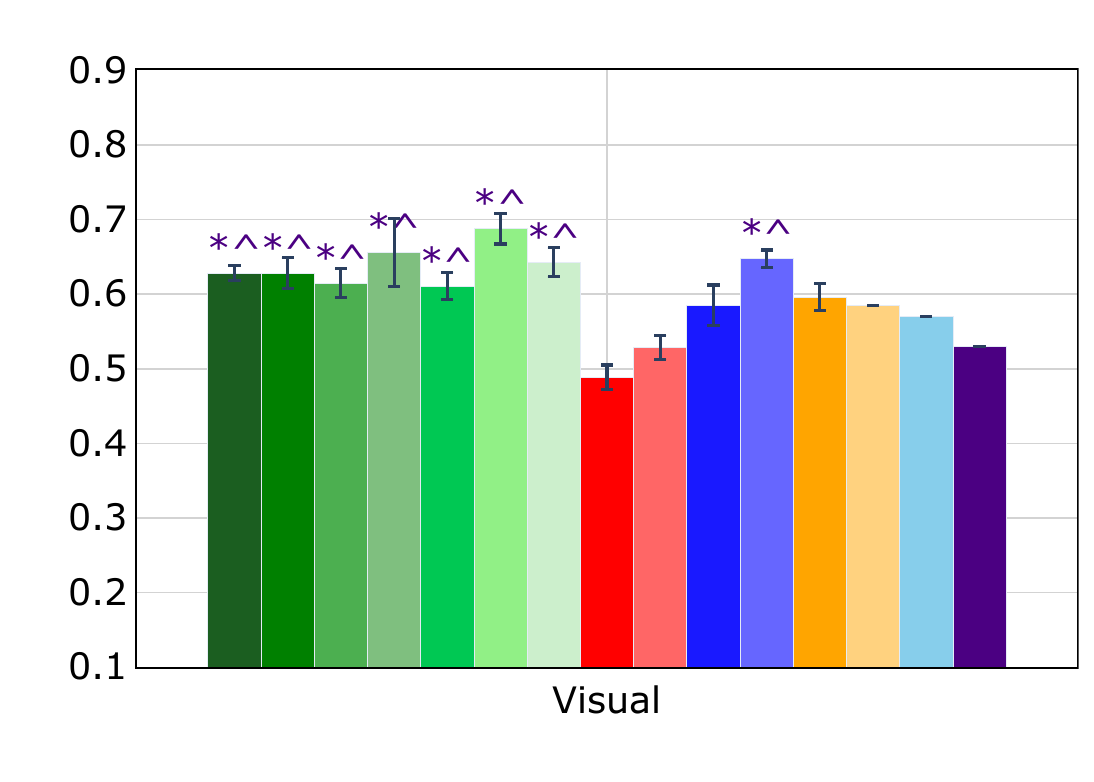}
     \includegraphics[width=0.24\linewidth]{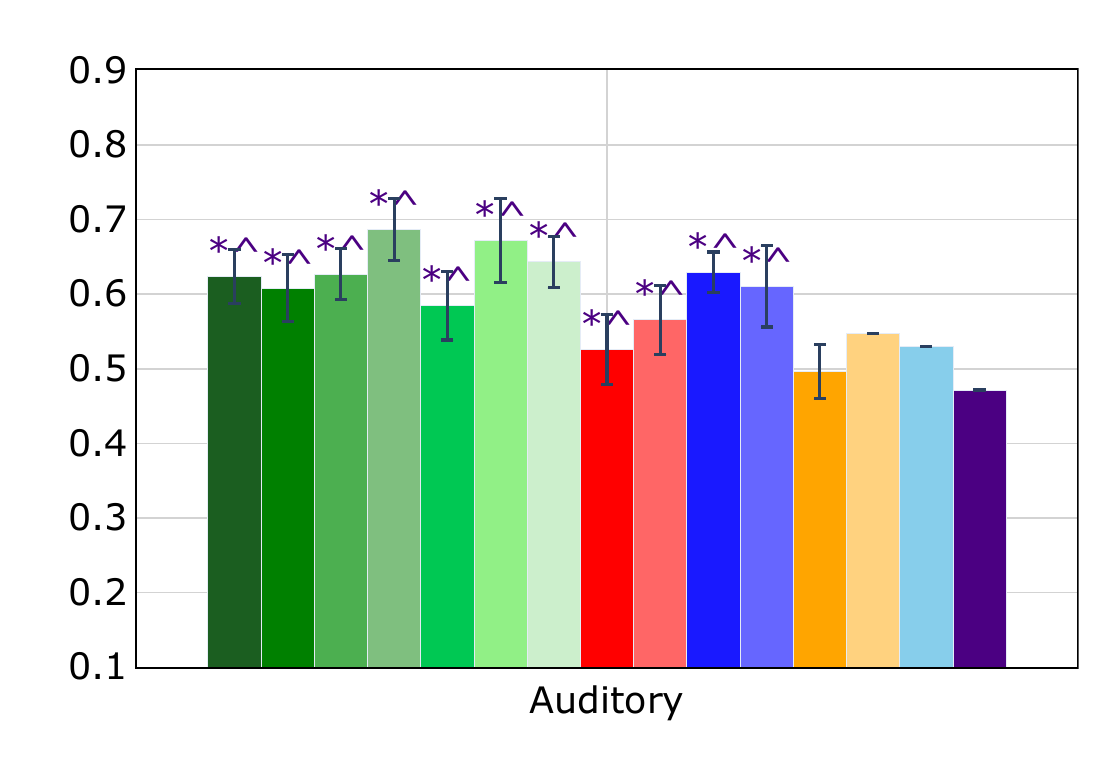}
    \vspace{-0.3cm}
    \caption{Average normalized brain alignment of instruction-tuned video MLLMs vs instruction-tuned audio MLLMs vs in-context learning video MLLMs vs multimodal and unimodal models across whole brain, language, visual and auditory regions. Error bars indicate the standard error of the mean across participants. $*$ implies that instruction-tuned MLLM embeddings are significantly better than multimodal models and $\wedge$ means that instruction-tuned MLLM embeddings are significantly better unimodal models with p$\leq 0.05$.} 
    % \sri{Can we mention about this p-value somewhere in Section 4 or Appendix or near Wilcox-signed test?}
    % \textcolor{green}{\textbf{$\times$}}
    % $*$ and $\wedge$ imply that instruction-tuned MLLM embeddings are significantly better than multimodal and unimodal models with p$\leq 0.05$.  
\label{fig:normalized_brain_alignment_allmodels_kimivideo}
\end{figure*}

We have added Kimi-VL~\citep{team2025kimi} to our evaluation and ran it through the same brain-encoding pipeline (identical preprocessing, instruction prompts, voxel-wise mapping, and normalization). The results are now reported in Fig.~\ref{fig:normalized_brain_alignment_allmodels_kimivideo}. We observe that the instruction-tuned Kimi-VL model demonstrates similar encoding performance to the other six instruction-tuned video MLLMs.

%\textcolor{blue}{For SEED-1.5-VL~\citep{guo2025seed1}, we could not find an official, publicly available checkpoint; only a demo/API was accessible.}
% Due to this reason, we could not include SEED-1.5-VL under the same controlled conditions.}

\begin{figure*}[!ht]
    \centering
     \includegraphics[width=0.49\linewidth]{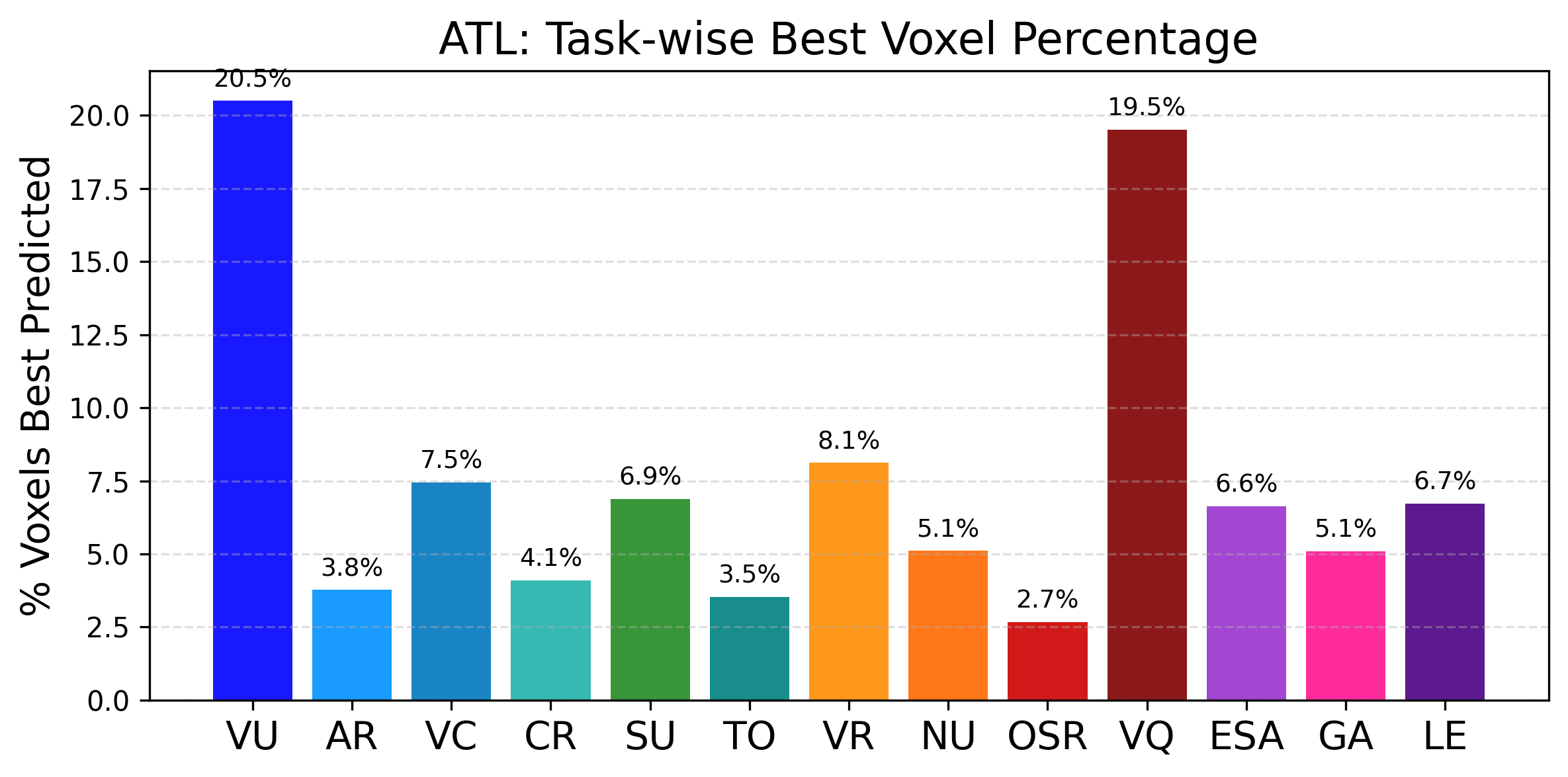}
     \includegraphics[width=0.49\linewidth]{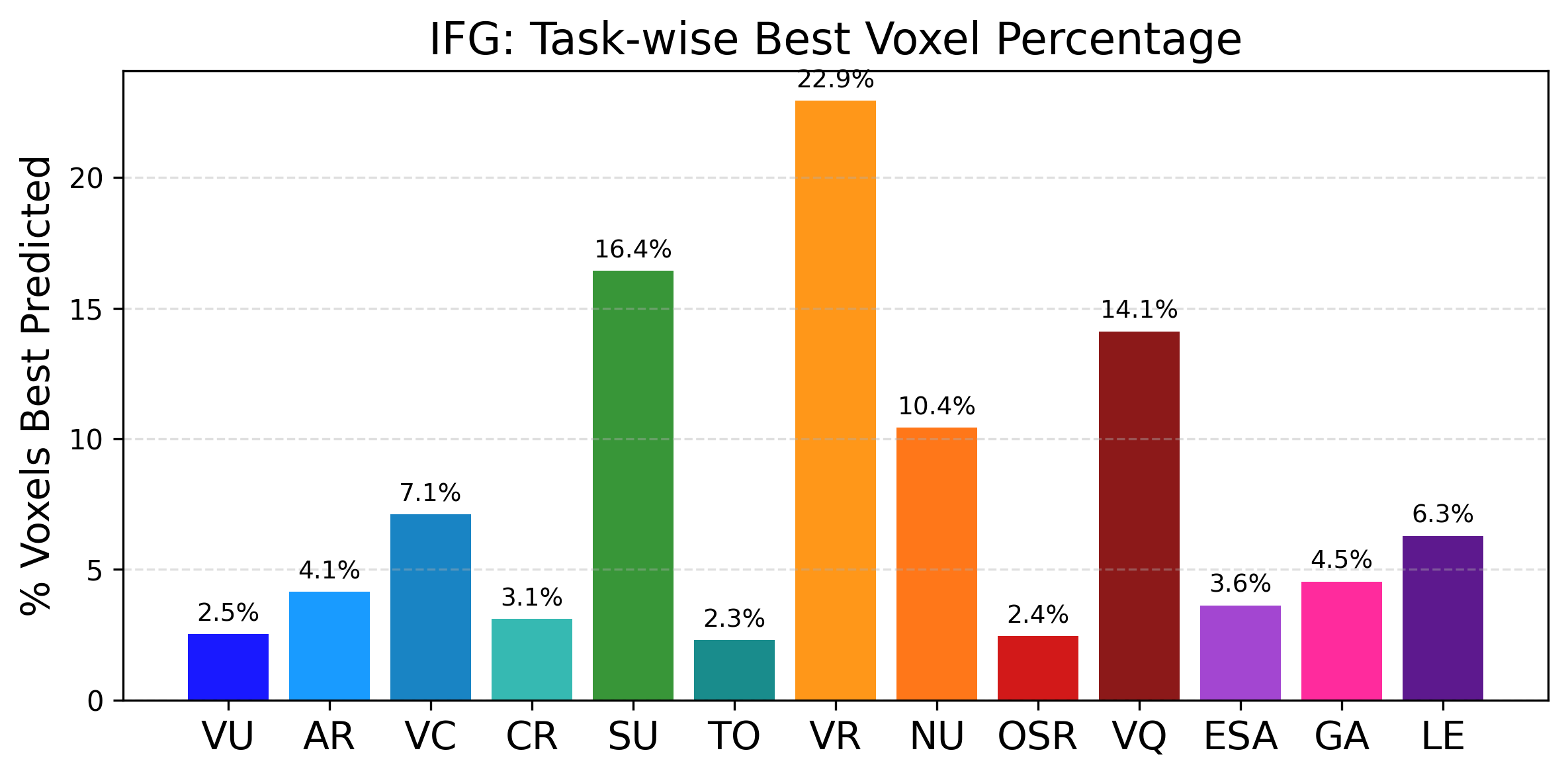}
      \includegraphics[width=0.49\linewidth]{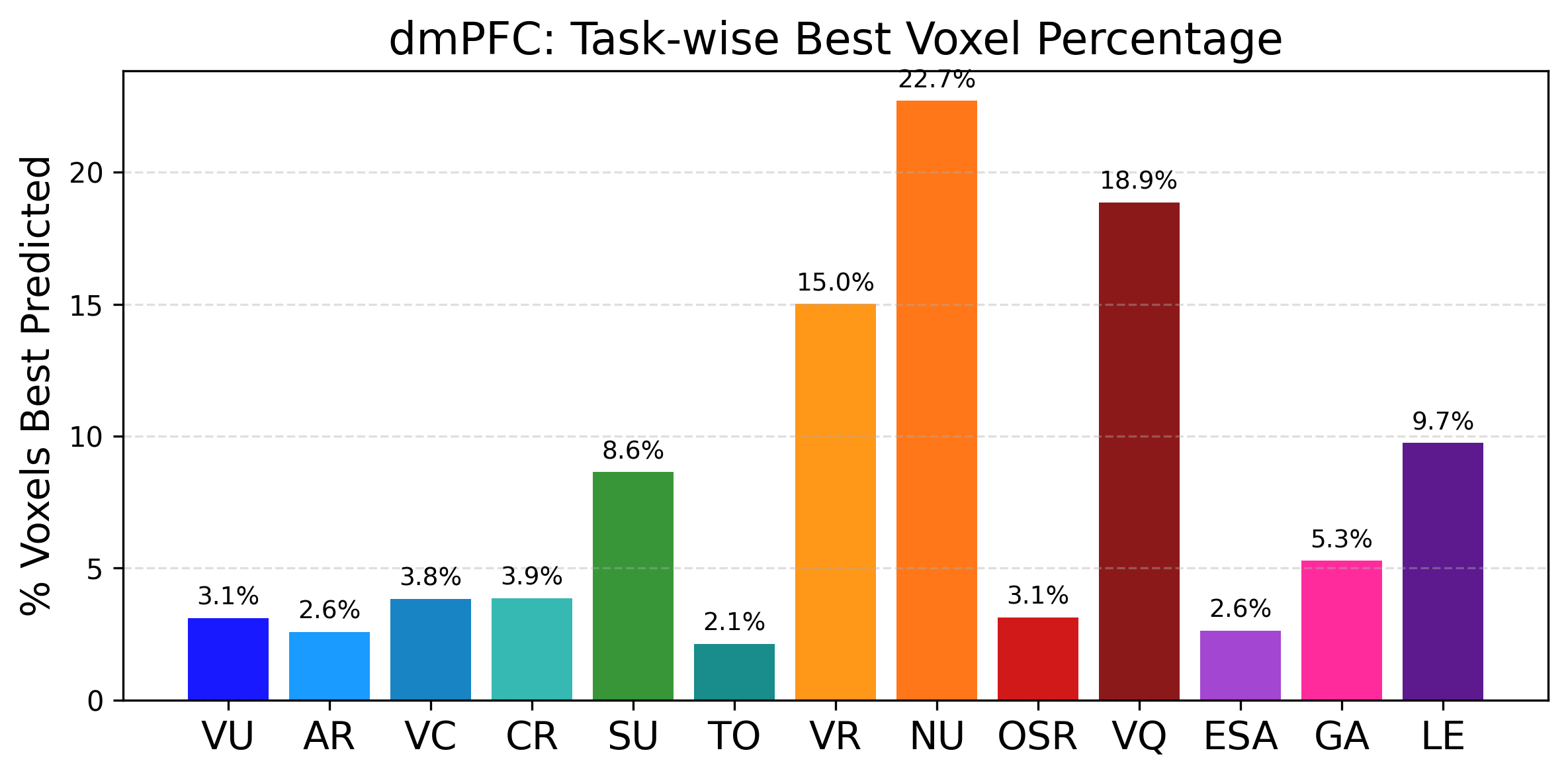}
      \includegraphics[width=0.49\linewidth]{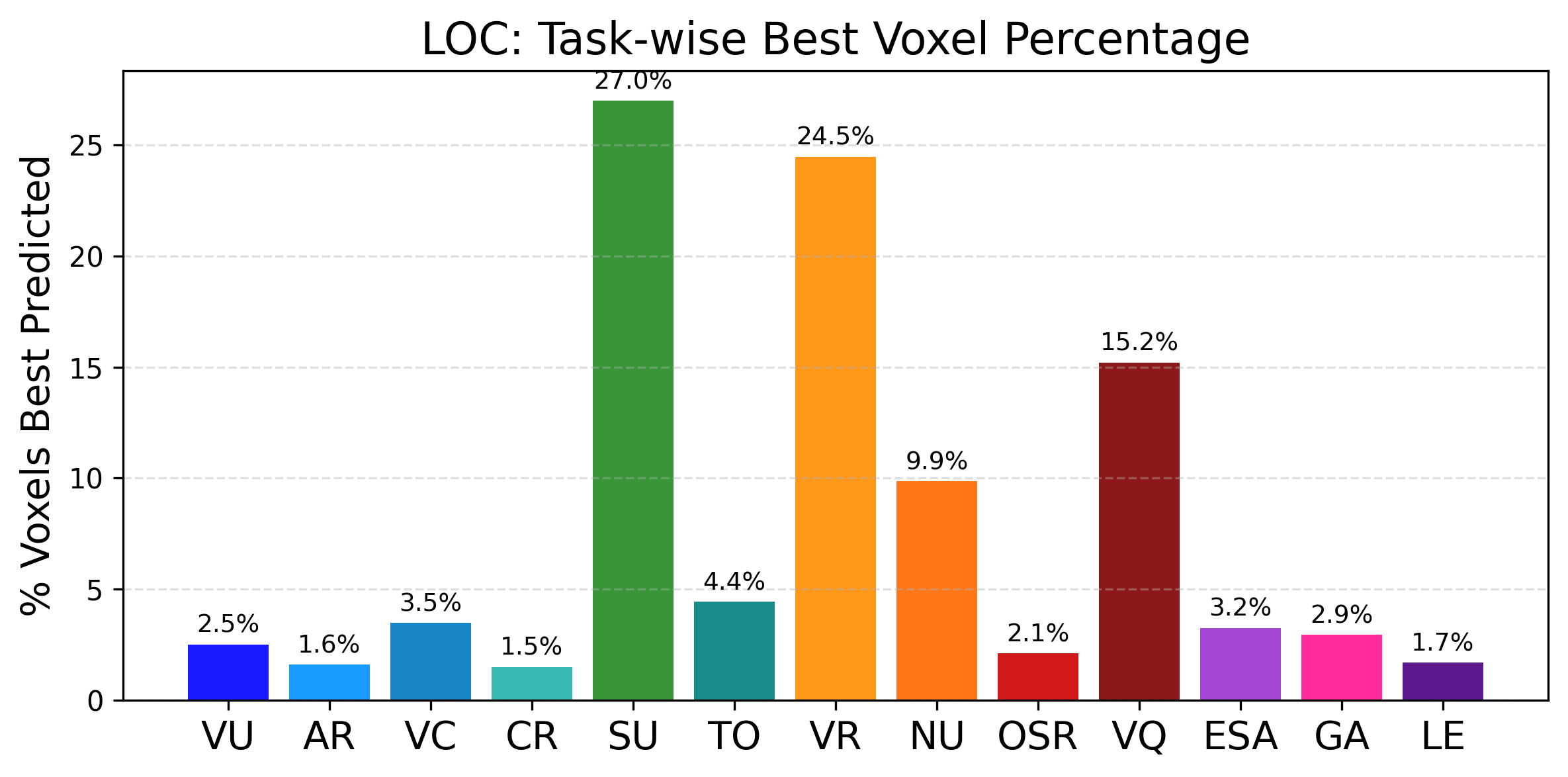}
      \includegraphics[width=0.325\linewidth]{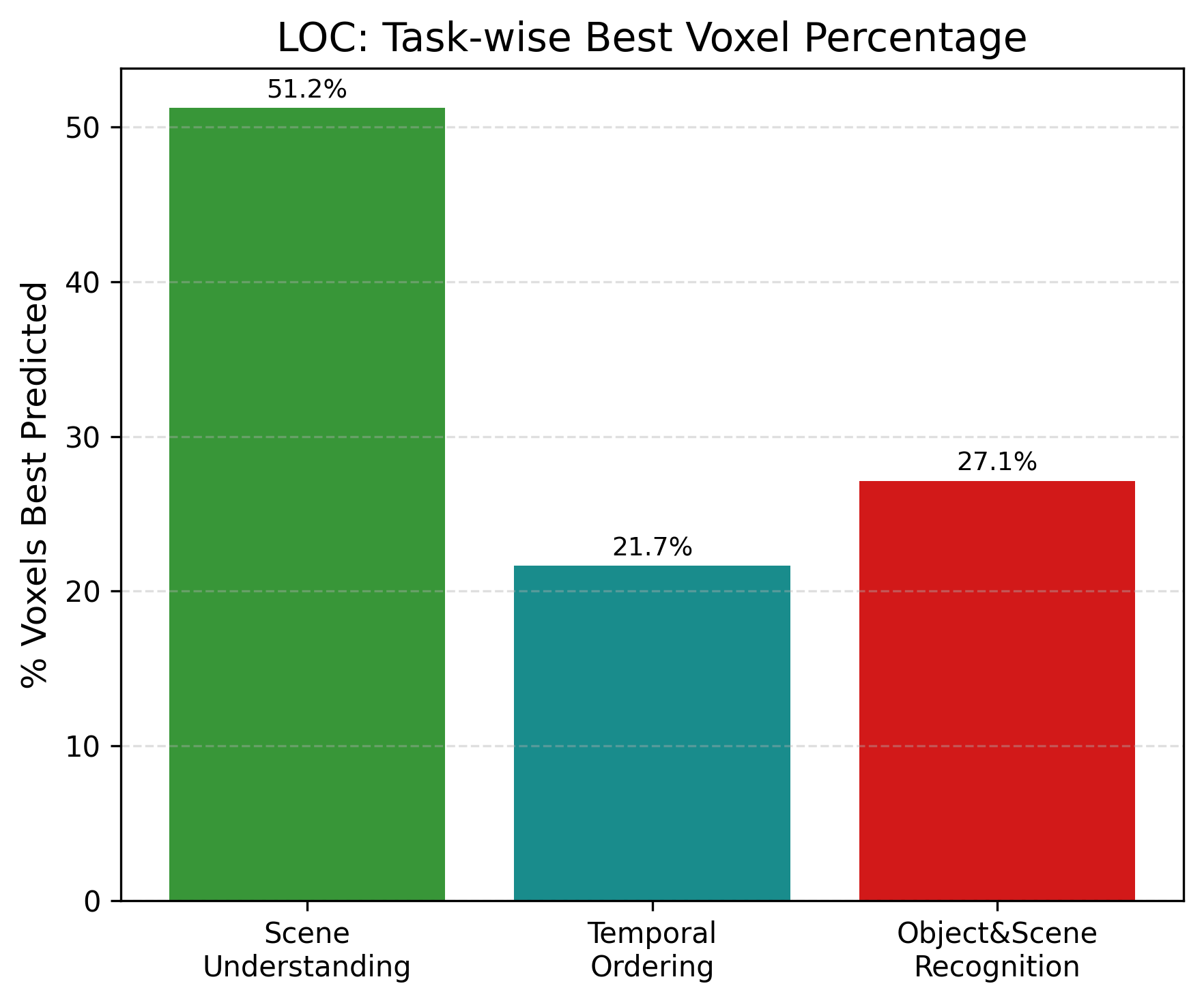}
      \includegraphics[width=0.325\linewidth]{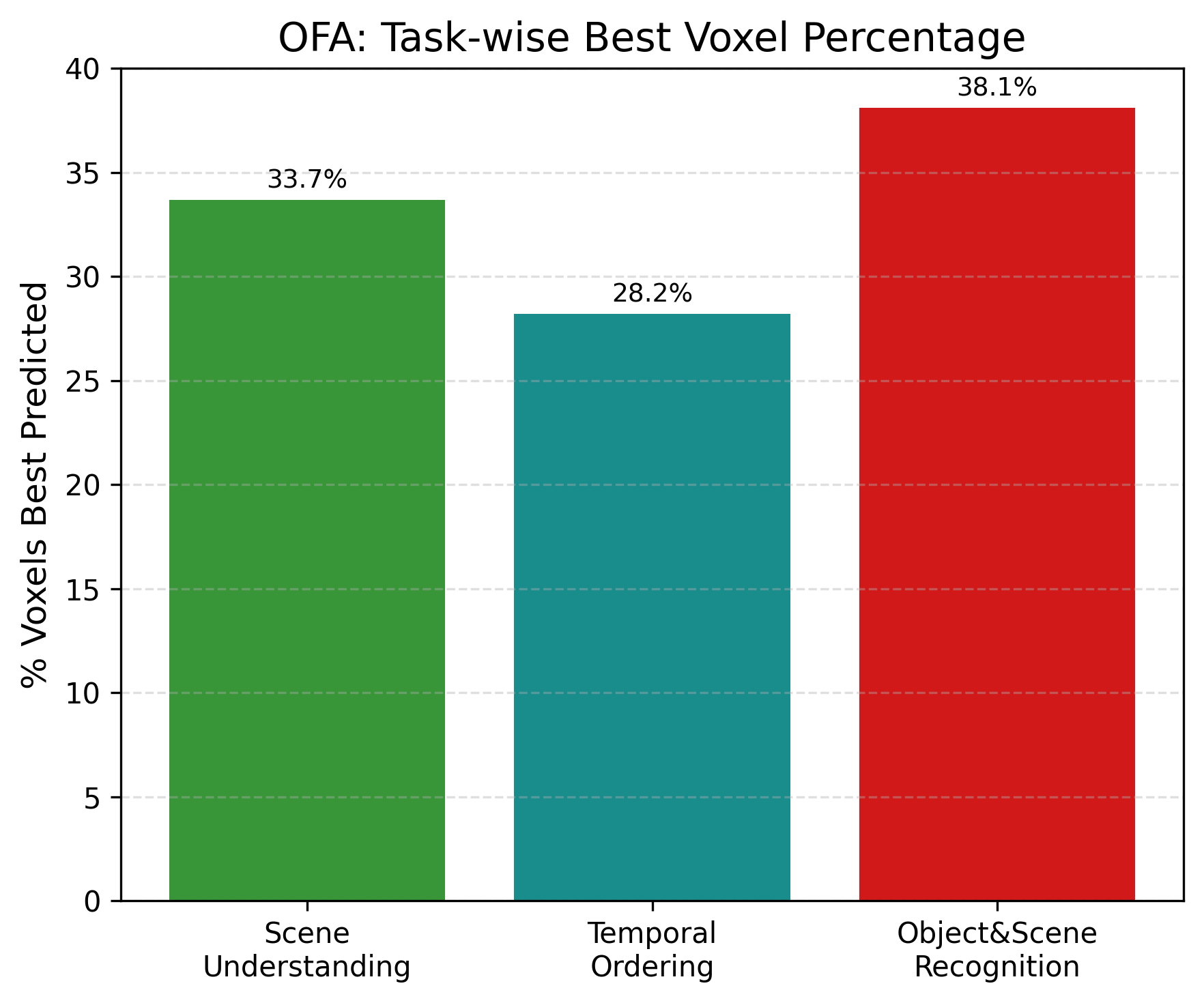}
      \includegraphics[width=0.325\linewidth]{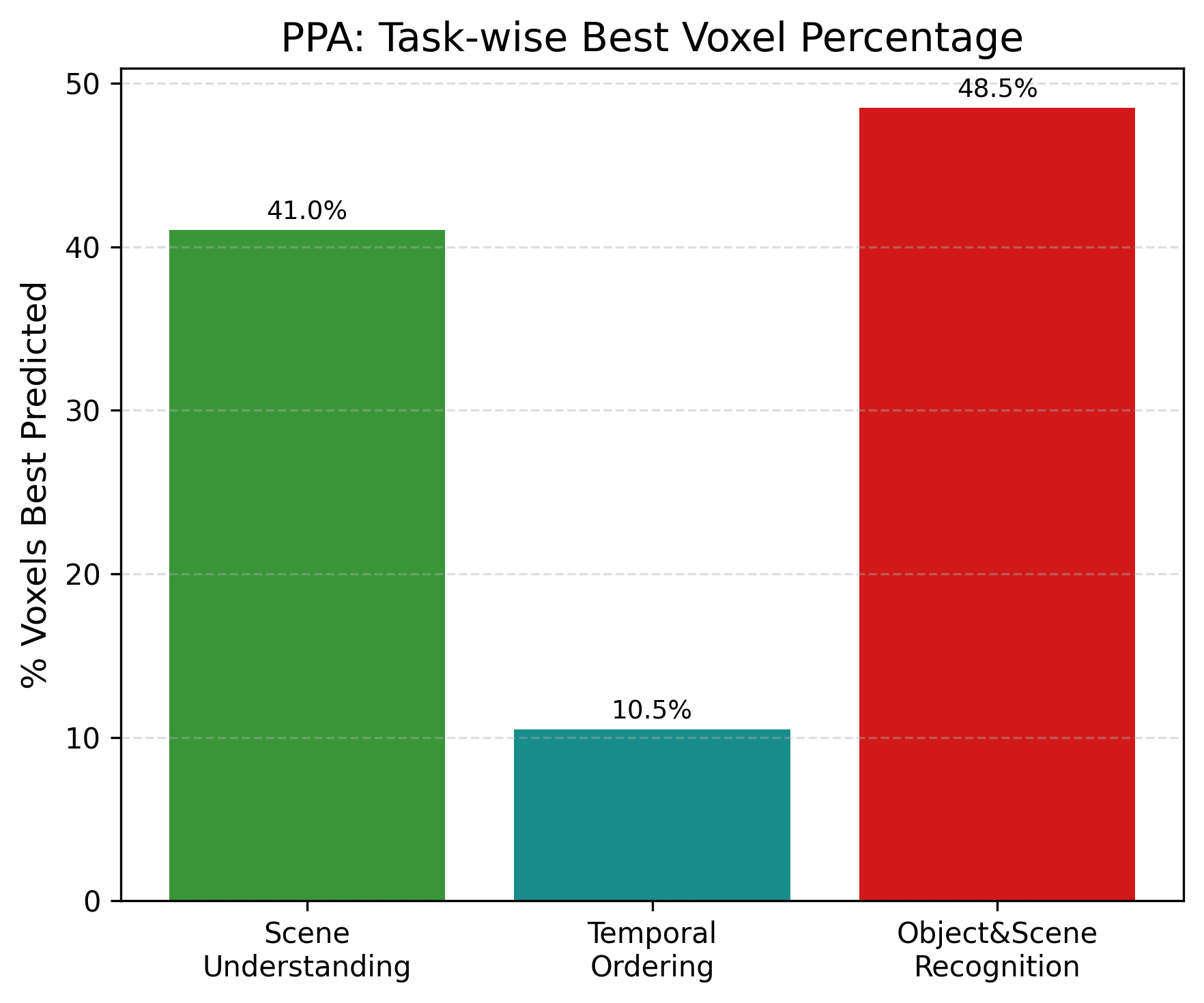}
     %\vspace{-0.1cm}
    %   \includegraphics[width=0.22\linewidth]{images/whole_brain_normalized_alignment_video_audio_barplot.pdf}
    % % % % \hspace{-10pt}
    % \includegraphics[width=0.22\linewidth]{images/Language_brain_normalized_alignment_barplot.pdf}
    %  \includegraphics[width=0.22\linewidth]{images/Visual_brain_normalized_alignment_barplot.pdf}
    %  \includegraphics[width=0.22\linewidth]{images/Auditory_brain_normalized_alignment_barplot.pdf}
    %\includegraphics[width=\linewidth]{images/barplots2.pdf}\
    \vspace{-0.4cm}
    \caption{Percentage of best predicted voxels per task for an instruction-tuned video MLLM across three language ROIs: ATL, IFG and dmPFC.
    }
    \label{fig:qwen_percentage_voxels_remaining_rois}
\end{figure*}

Fig.~\ref{fig:qwen_percentage_voxels_remaining_rois} shows task-wise best-voxel fractions in key
higher-order language and visual ROIs to assess how instructions differentially engage them.
As shown in Fig.~\ref{fig:qwen_percentage_voxels_remaining_rois}, restricting the analysis to spatial-visual tasks (Scene Understanding, Temporal Ordering, Object \& Scene Recognition) reveals clean dissociations across early visual ROIs. PPA is dominated by Object \& Scene Recognition (48.5\%) and Scene Understanding (41.0\%), consistent with its established role in scene and place processing. OFA shows a balanced profile led by Object \& Scene Recognition (38.1\%), consistent with its role in low-level object recognition including faces. LOC is dominated by Scene Understanding (51.2\%), a richer task than bare recognition, consistent with LOC's role in object and scene processing under spatial layout reasoning. These findings align with classical neuroanatomical findings on functional specialization across early visual cortex.

\begin{figure*}[!ht]
    \centering
     \includegraphics[width=0.49\linewidth]{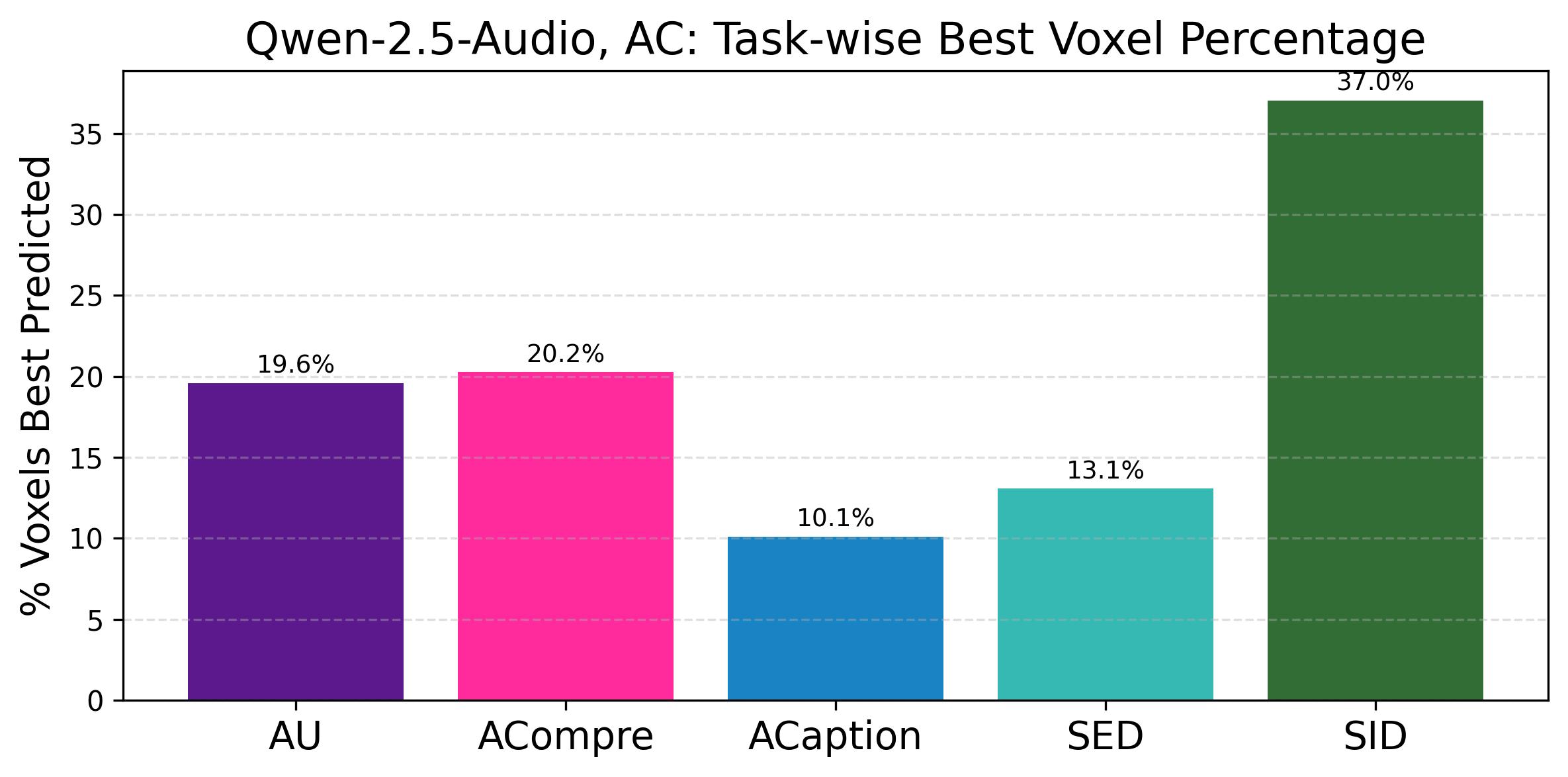}
     \includegraphics[width=0.49\linewidth]{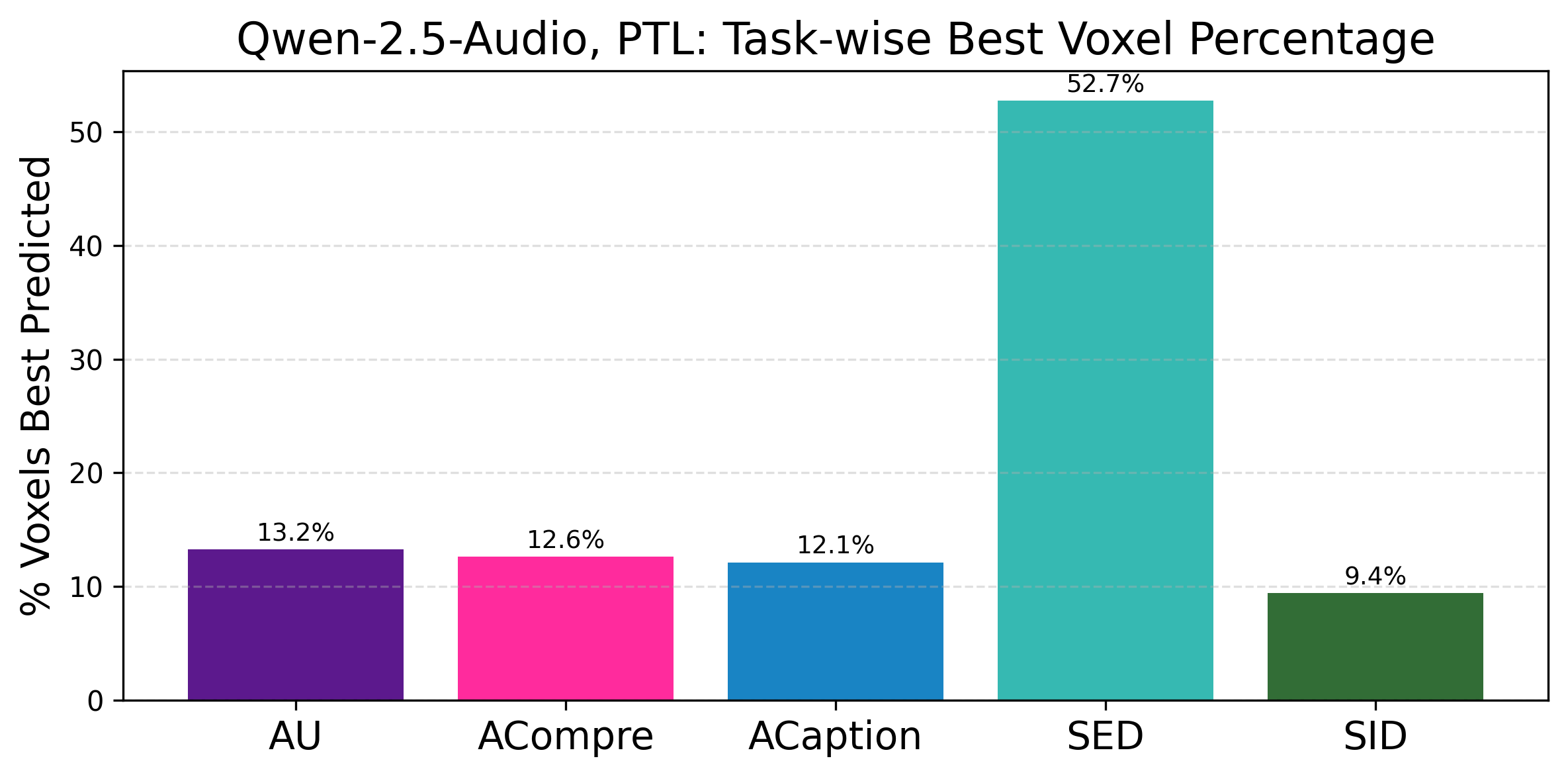}
      \includegraphics[width=0.49\linewidth]{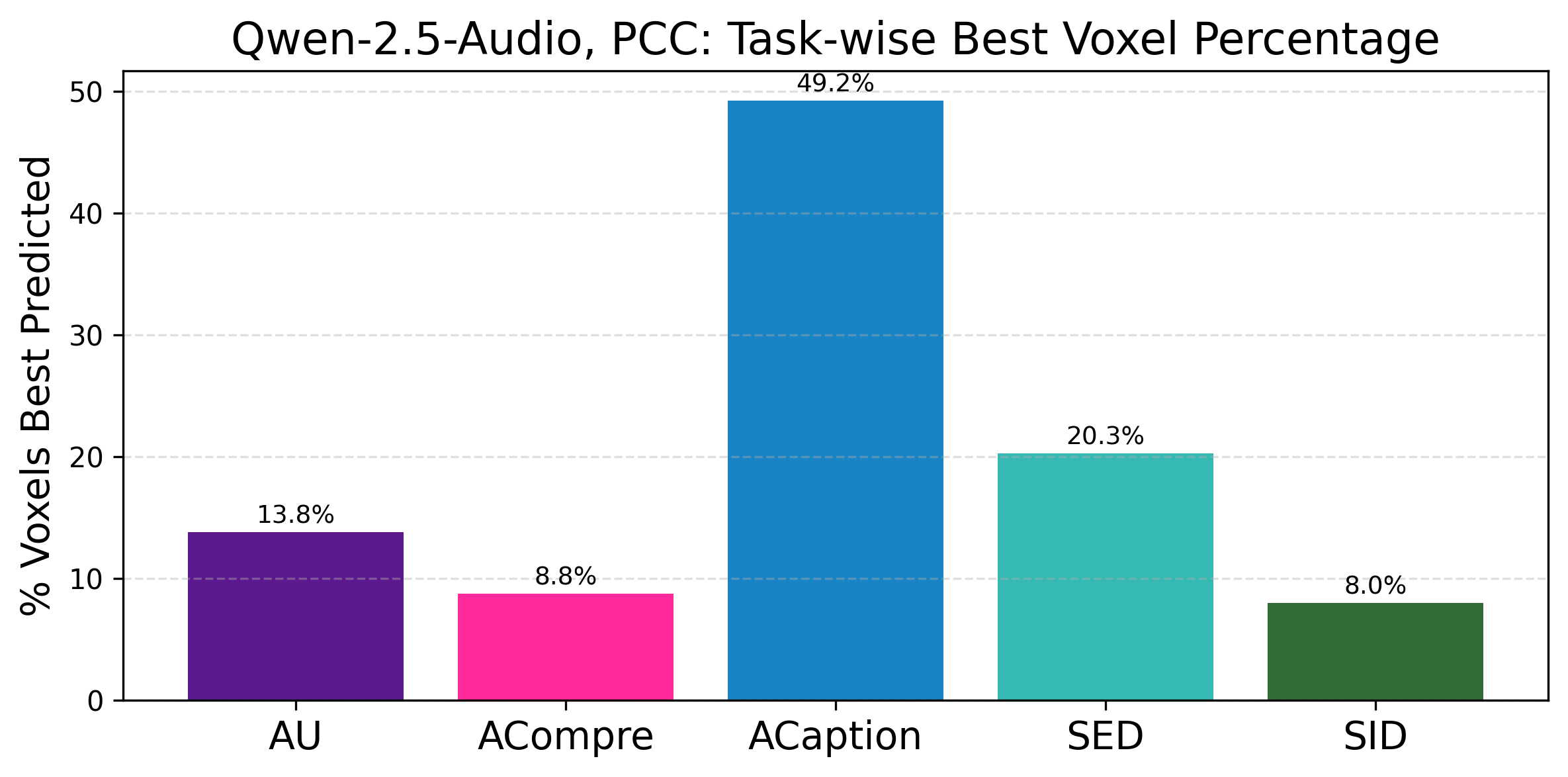}
      % \includegraphics[width=0.49\linewidth]{images/LOC_task_performance.png}
      % \includegraphics[width=0.325\linewidth]{images/LOC_task_performance_spatial_to_osr.png}
      % \includegraphics[width=0.325\linewidth]{images/OFA_task_performance_spatial_to_osr.png}
      % \includegraphics[width=0.325\linewidth]{images/PPA_task_performance_spatial_to_osr.png}
     %\vspace{-0.1cm}
    %   \includegraphics[width=0.22\linewidth]{images/whole_brain_normalized_alignment_video_audio_barplot.pdf}
    % % % % \hspace{-10pt}
    % \includegraphics[width=0.22\linewidth]{images/Language_brain_normalized_alignment_barplot.pdf}
    %  \includegraphics[width=0.22\linewidth]{images/Visual_brain_normalized_alignment_barplot.pdf}
    %  \includegraphics[width=0.22\linewidth]{images/Auditory_brain_normalized_alignment_barplot.pdf}
    %\includegraphics[width=\linewidth]{images/barplots2.pdf}\
    \vspace{-0.4cm}
    \caption{Percentage of best predicted voxels per task for an instruction-tuned audio MLLM across three language ROIs: AC, PTL and PCC.
    }
    \label{fig:qwen_percentage_voxels_remaining_rois_audio}
\end{figure*}

\FloatBarrier

\section{Contrasting Instruction-tuned video MLLMs with In-context learning video MLLMs}
\label{app:instruction_video_incontext_flatmaps}
\begin{figure*}[!ht]
\begin{minipage}{\textwidth}
\centering \includegraphics[width=0.55\linewidth]{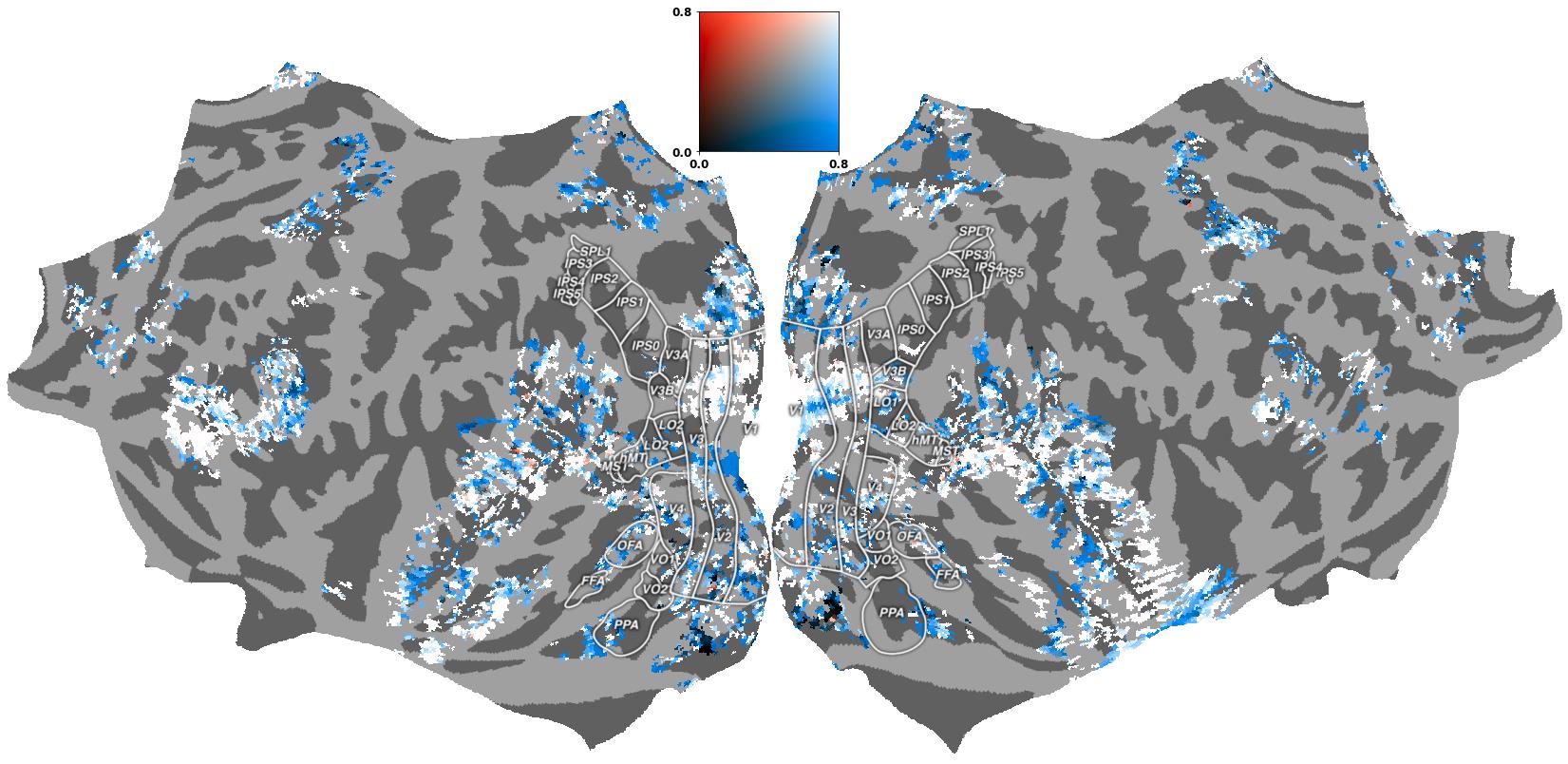}
    \\(a) Subject-01 \\
\end{minipage}
\begin{minipage}{\textwidth}
\centering
    \includegraphics[width=0.55\linewidth]{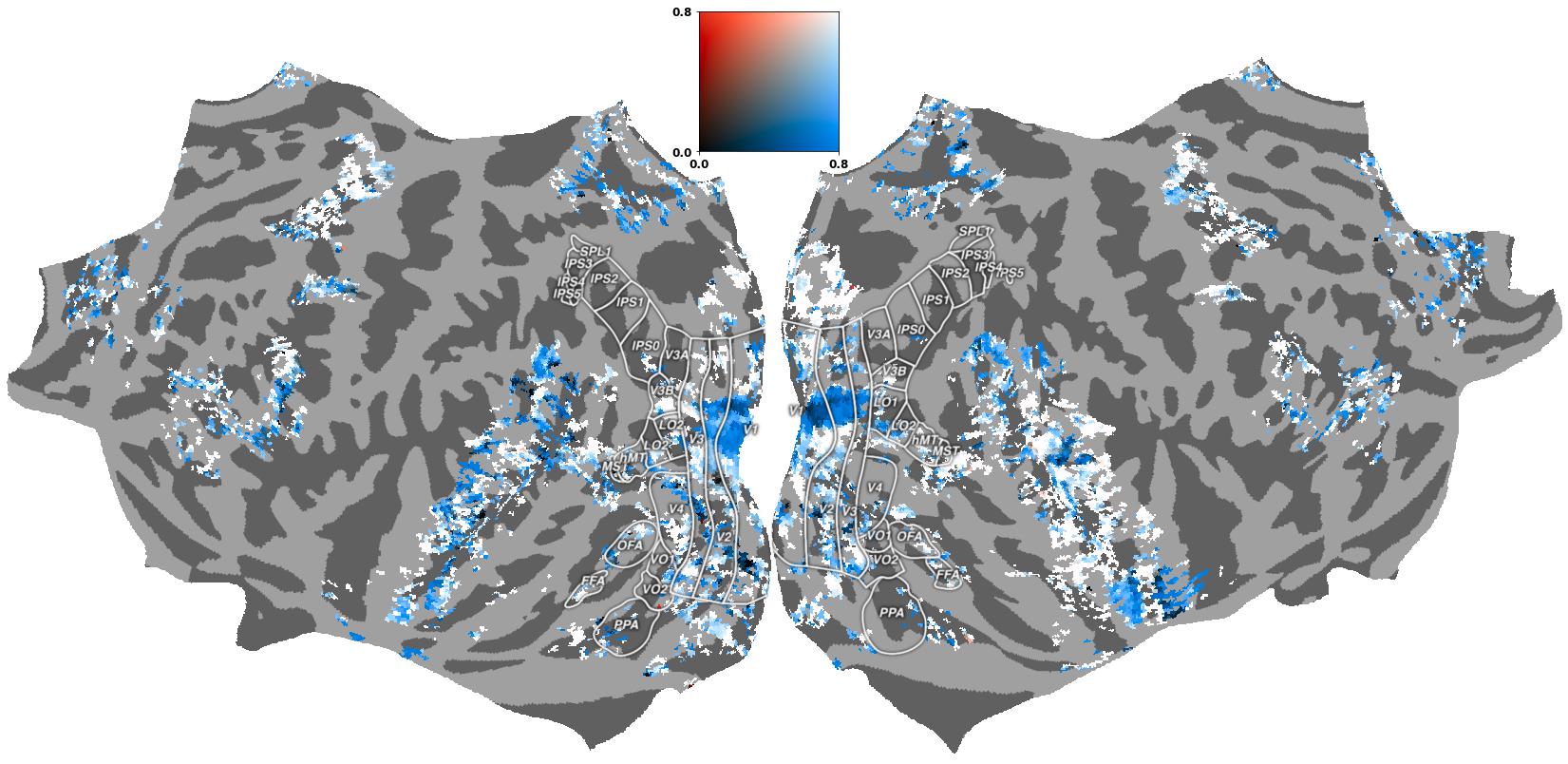}
    \\(b) Subject-02 \\
\end{minipage}
\begin{minipage}{\textwidth}
\centering
    \includegraphics[width=0.55\linewidth]{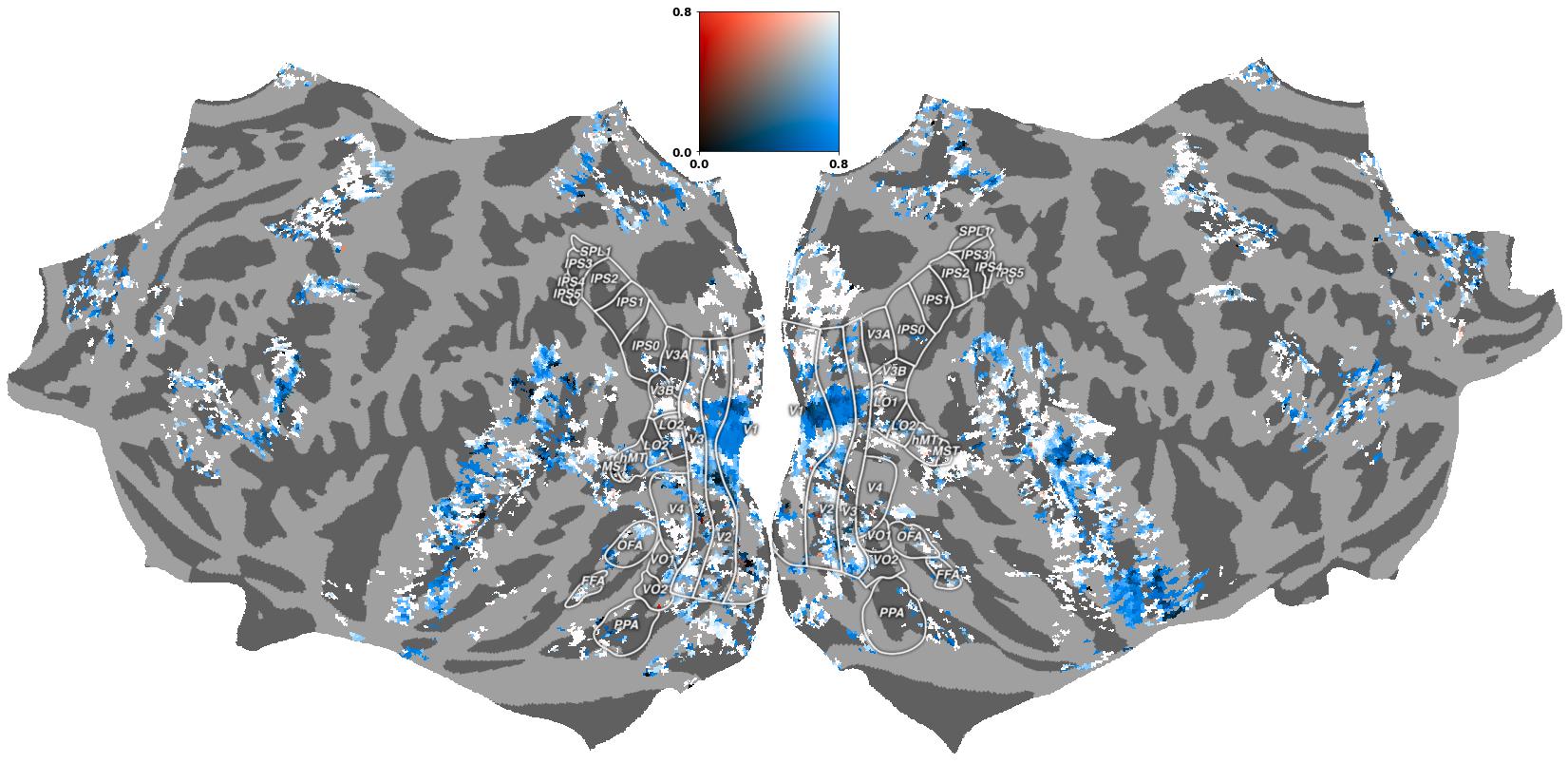}
    \\(c) Subject-03 \\
\end{minipage}
\begin{minipage}{\textwidth}
\centering
    \includegraphics[width=0.55\linewidth]{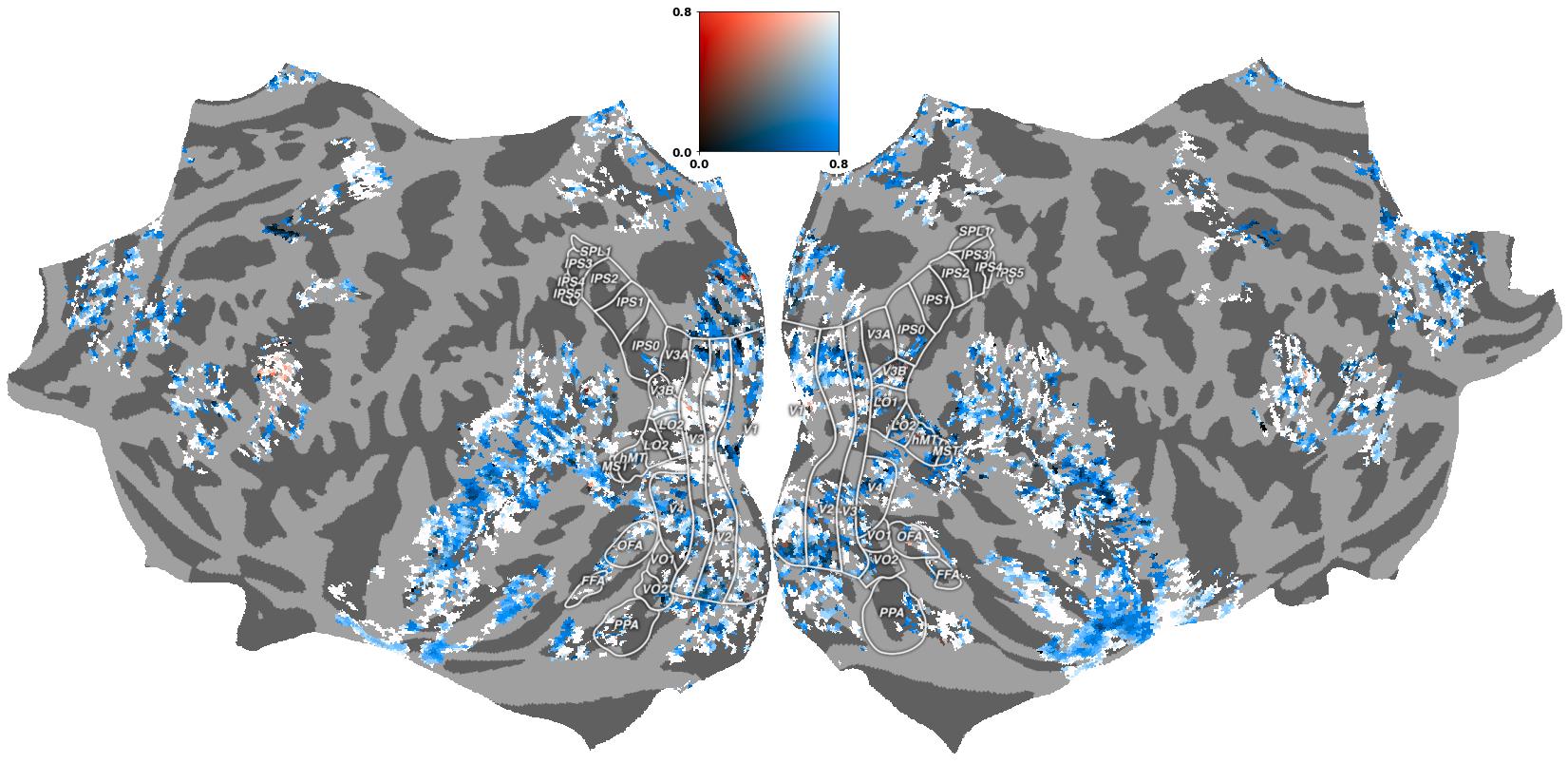}
    \\(d) Subject-05 \\
\end{minipage}
    \caption{Qwen-2.5-VL vs. InternVL: Contrast of estimated cross-subject prediction accuracy for all participants for the naturalistic movie watching. Pearson correlation scores for each voxel in each subject are projected onto the subject’s flattened cortical surface.
    \textcolor{cyan}{Blue} and \textcolor{red}{Red} voxels depict higher prediction accuracy estimates during instruction-tuned video MLLM and in-context learning video MLLM (InternVL), respectively. Voxels that have similar cross-subject prediction accuracy appear white. Here, middle frontal gyrus (MFG), inferior frontal gyrus (IFG),  inferior frontal gyrus orbital (IFGOrb), angular gyrus (AG), and lateral temporal cortex (LTC) are late language regions, EVC denotes early visual cortex and AC denotes auditory cortex.}
    \label{fig:qwenvl_internvl_2dmap}
\end{figure*}

\begin{figure*}[!ht]
\begin{minipage}{\textwidth}
\centering \includegraphics[width=0.55\linewidth]{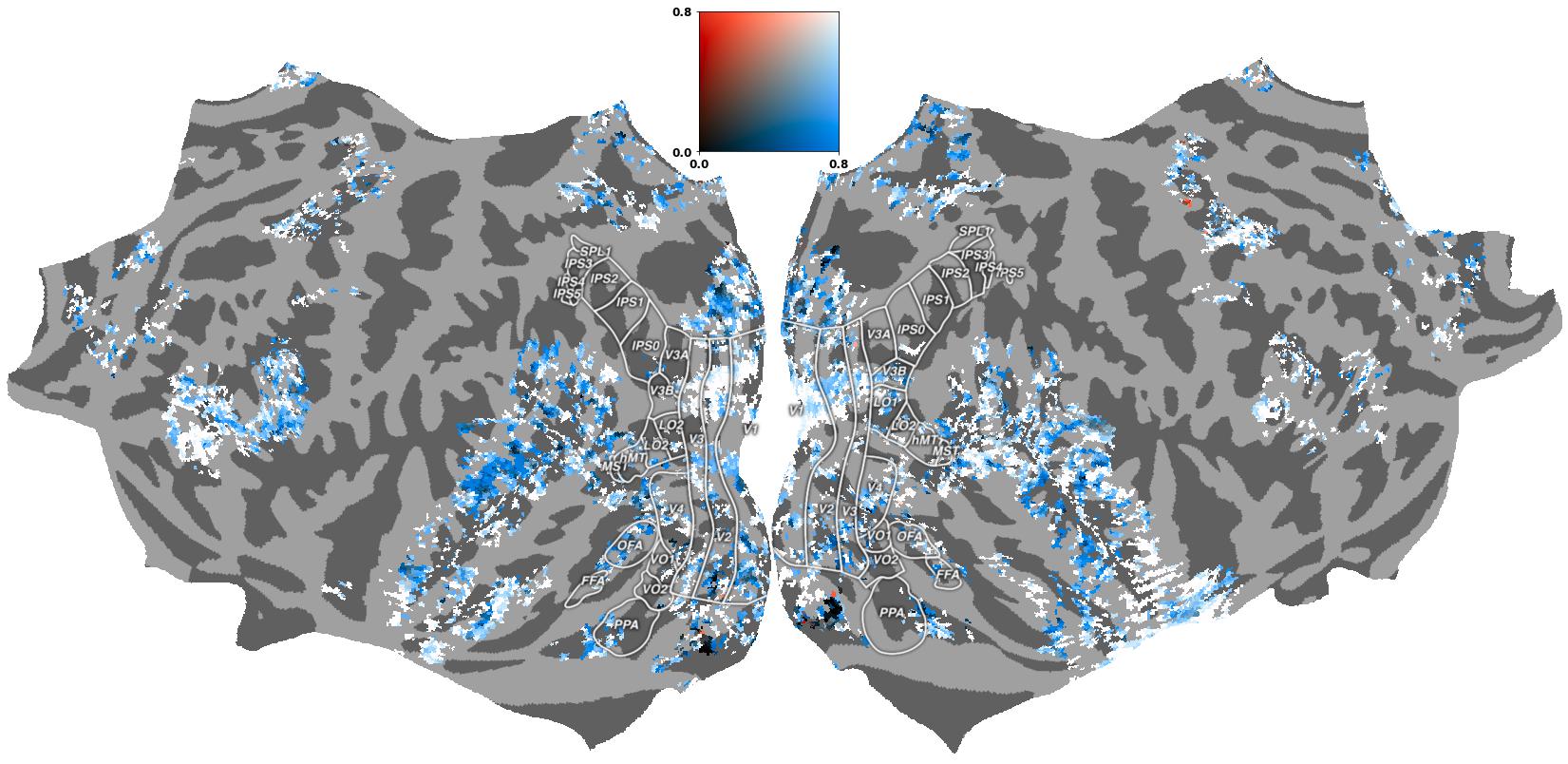}
    \\(a) Subject-01 \\
\end{minipage}
\begin{minipage}{\textwidth}
\centering
    \includegraphics[width=0.55\linewidth]{images/qwen_internvl_2dcolormap_sub2.jpg}
    \\(b) Subject-02 \\
\end{minipage}
\begin{minipage}{\textwidth}
\centering
    \includegraphics[width=0.55\linewidth]{images/qwen_omni_2dcolormap_sub2.jpg}
    \\(c) Subject-03 \\
\end{minipage}
\begin{minipage}{\textwidth}
\centering
    \includegraphics[width=0.55\linewidth]{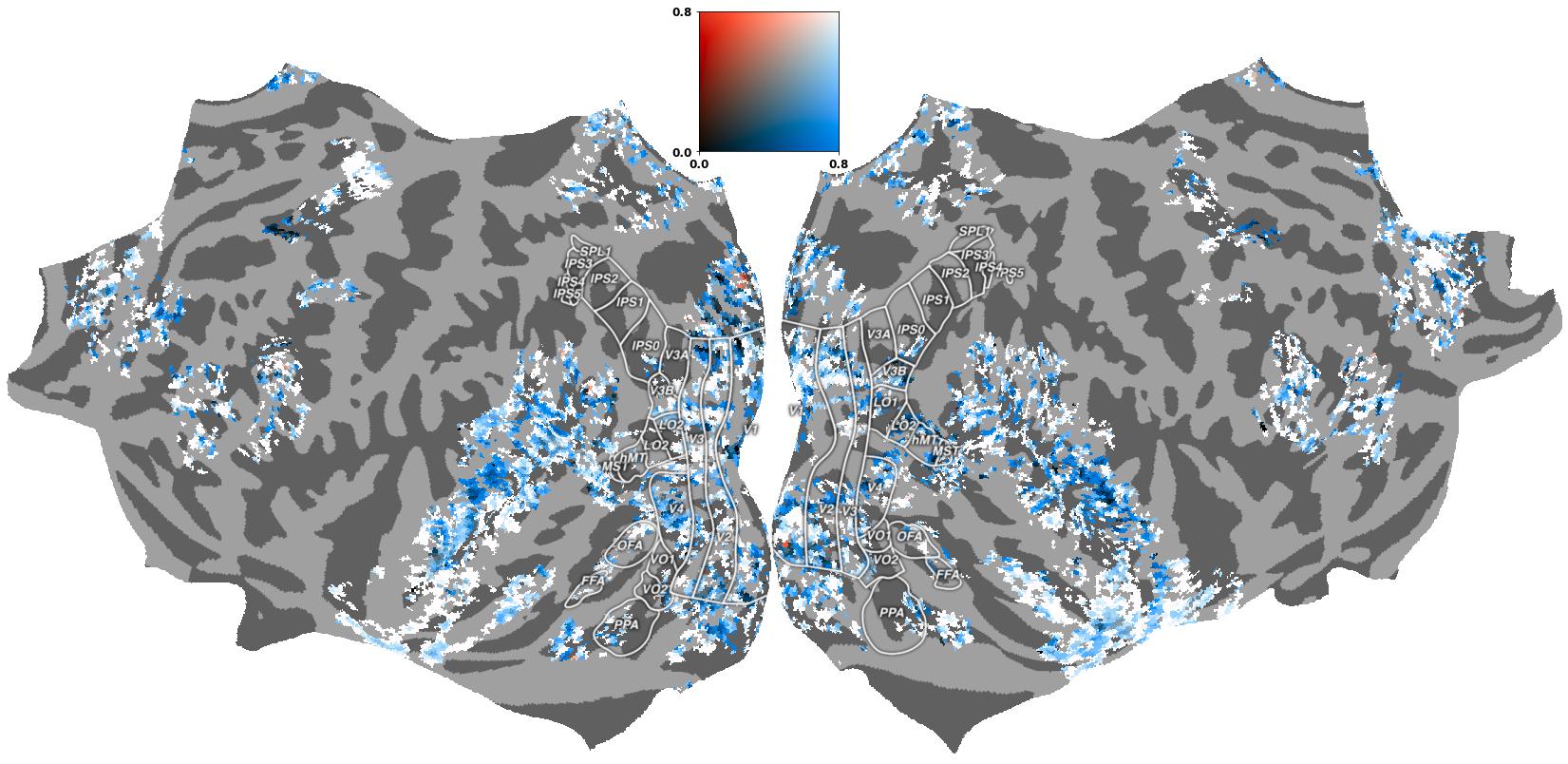}
    \\(d) Subject-05 \\
\end{minipage}
    \caption{Qwen-2.5-VL vs. Qwen-2.5-Omni: Contrast of estimated cross-subject prediction accuracy for all participants for the naturalistic movie watching. Pearson correlation scores for each voxel in each subject are projected onto the subject’s flattened cortical surface.
    \textcolor{cyan}{Blue} and \textcolor{red}{Red} voxels depict higher prediction accuracy estimates during instruction-tuned video MLLM and in-context learning video MLLM (Qwen-2.5-Omni), respectively. Voxels that have similar cross-subject prediction accuracy appear white. Here, middle frontal gyrus (MFG), inferior frontal gyrus (IFG),  inferior frontal gyrus orbital (IFGOrb), angular gyrus (AG), and lateral temporal cortex (LTC) are late language regions, EVC denotes early visual cortex and AC denotes auditory cortex.}
    \label{fig:qwenvl_omni_2dmap}
\end{figure*}

We present contrast of brainmaps to display the average normalized brain alignment across voxels for the instruction-tuned video MLLMs versus the in-context learning video MLLMs in Figures~\ref{fig:qwenvl_internvl_2dmap}, and ~\ref{fig:qwenvl_omni_2dmap}. The results show that instruction-tuned video MLLMs consistently achieve significantly higher alignment across all brain voxels.

\section{Contrasting Instruction-tuned video MLLMs with non-instruction-tuned multimodal}
\label{app:instruction_video_noninstruction_flatmaps}
\begin{figure*}[!ht]
    \centering
\begin{minipage}{\textwidth}
\centering
    \includegraphics[width=0.55\linewidth]{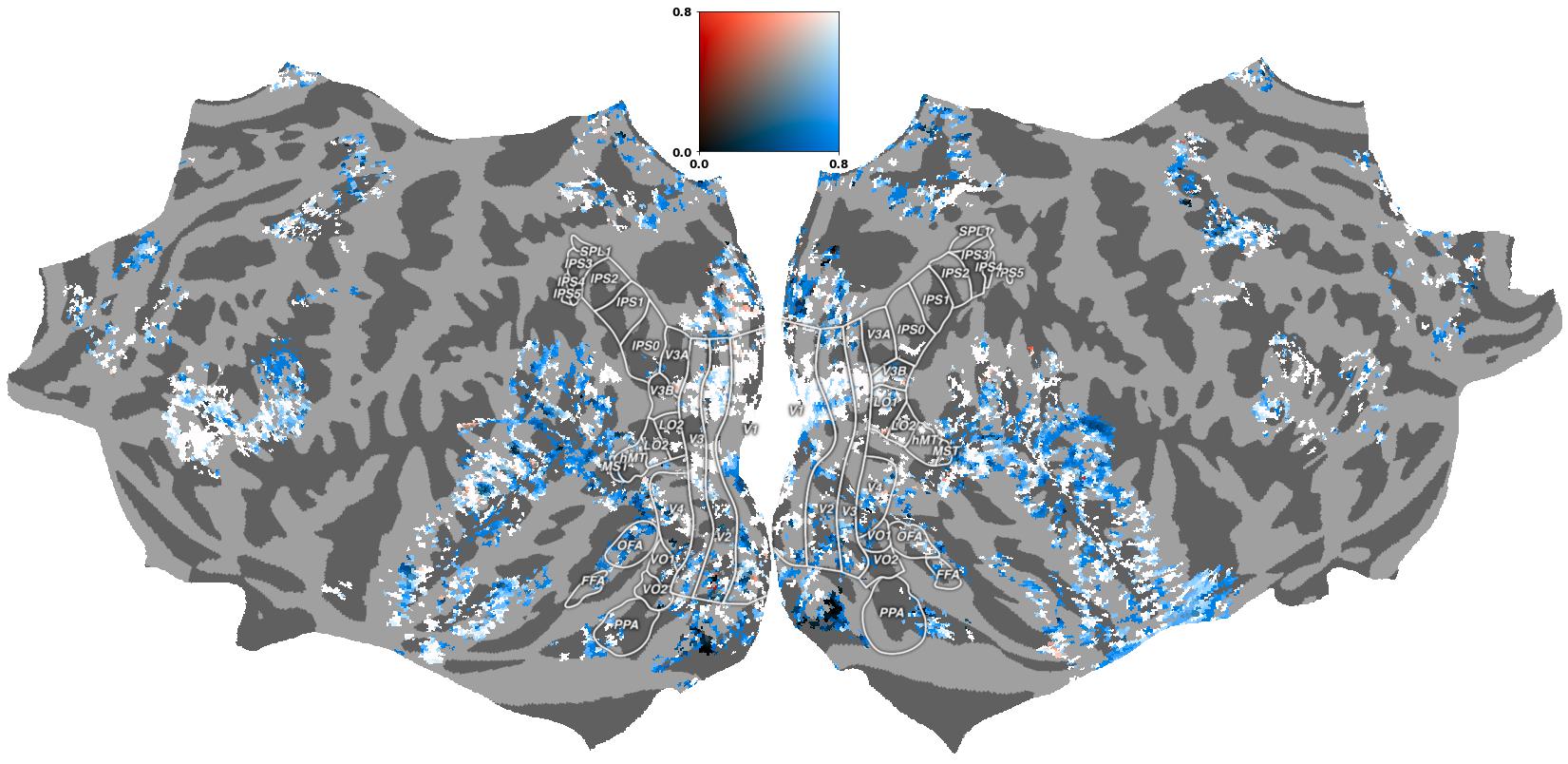}
    \\(a) Subject-01 \\
\end{minipage}
\begin{minipage}{\textwidth}
\centering
    \includegraphics[width=0.55\linewidth]{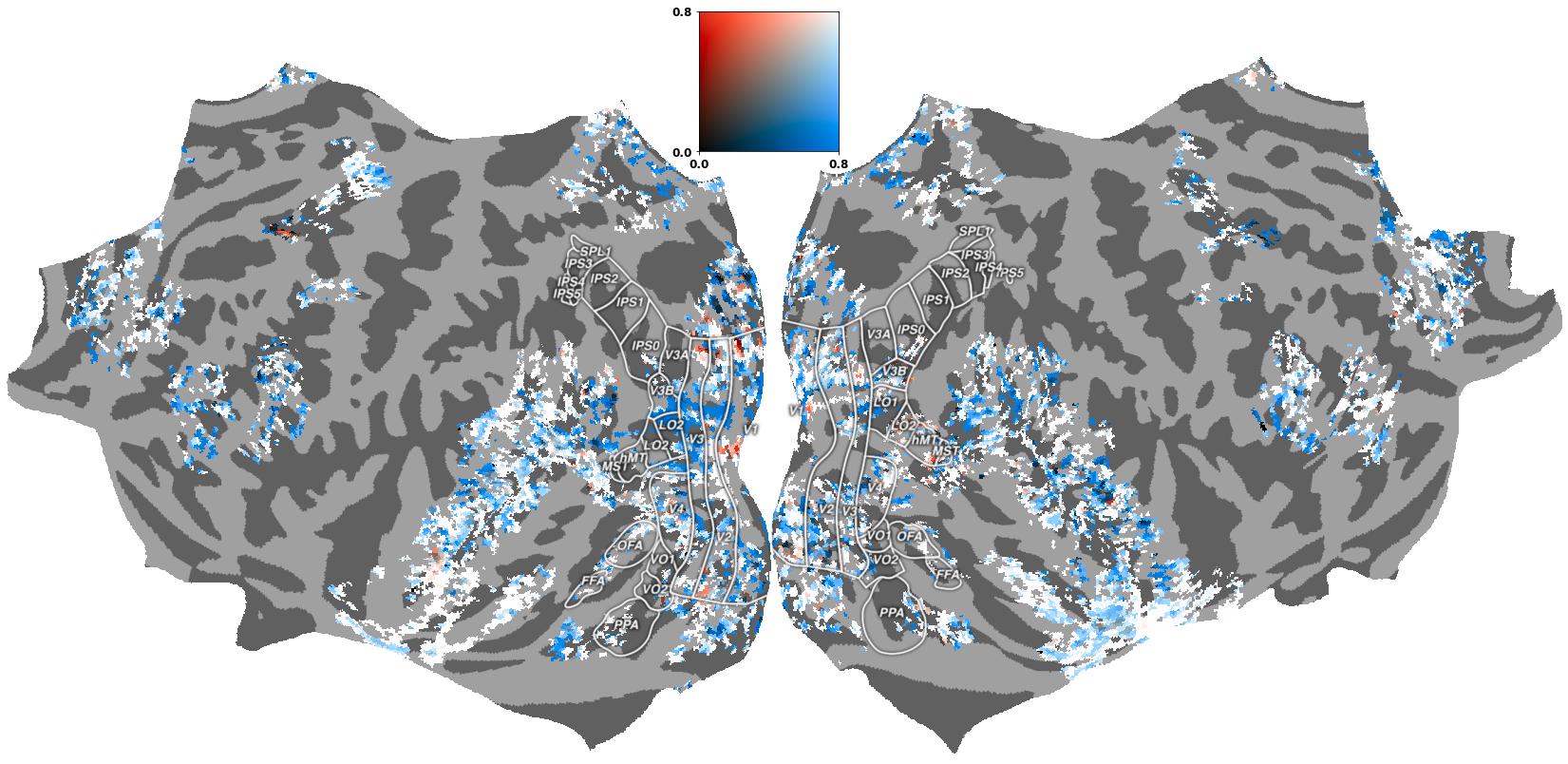}
    \\(c) Subject-02 \\
\end{minipage}
\begin{minipage}{\textwidth}
\centering
    \includegraphics[width=0.55\linewidth]{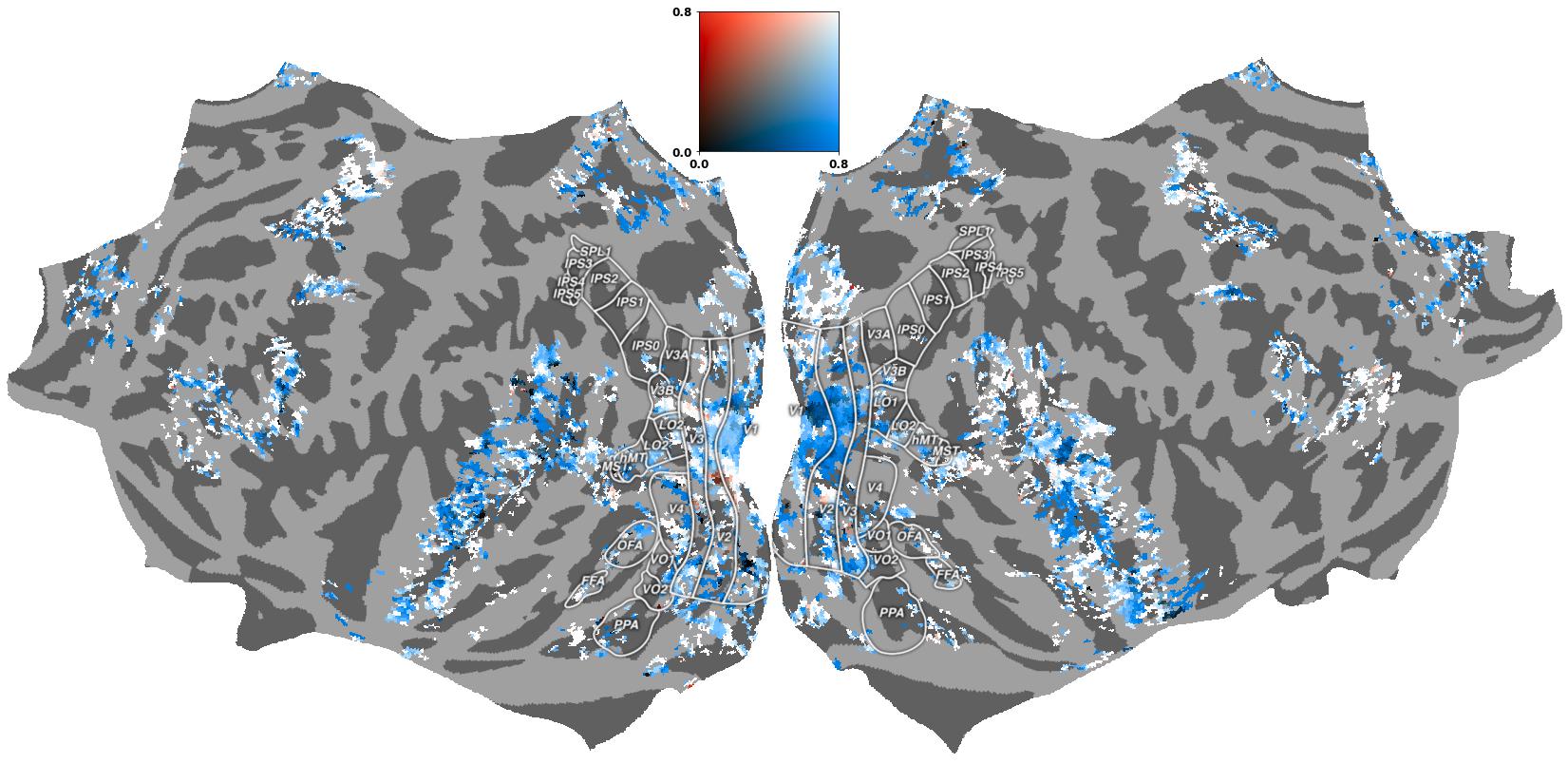}
    \\(c) Subject-03 \\
\end{minipage}
\begin{minipage}{\textwidth}
\centering
    \includegraphics[width=0.55\linewidth]{images/qwen_vila_2dcolormap_sub3.jpg}
    \\(d) Subject-05 \\
\end{minipage}
    \caption{Qwen-2.5-VL vs. VILA: Contrast of estimated cross-subject prediction accuracy for all participants for the naturalistic movie watching. Pearson correlation scores for each voxel in each subject are projected onto the subject’s flattened cortical surface. \textcolor{cyan}{Blue} and \textcolor{red}{Red} voxels depict higher prediction accuracy estimates during instruction-tuned video MLLM and multimodal VILA, respectively. Voxels that have similar cross-subject prediction accuracy appear white.  Here, middle frontal gyrus (MFG), inferior frontal gyrus (IFG),  inferior frontal gyrus orbital (IFGOrb), angular gyrus (AG), and lateral temporal cortex (LTC) are late language regions, EVC denotes early visual cortex and AC denotes auditory cortex.}
\label{fig:instruction_tuned_video_multimodal_brainmaps_vila}
\end{figure*}

\begin{figure*}[!ht]
\begin{minipage}{\textwidth}
\centering \includegraphics[width=0.55\linewidth]{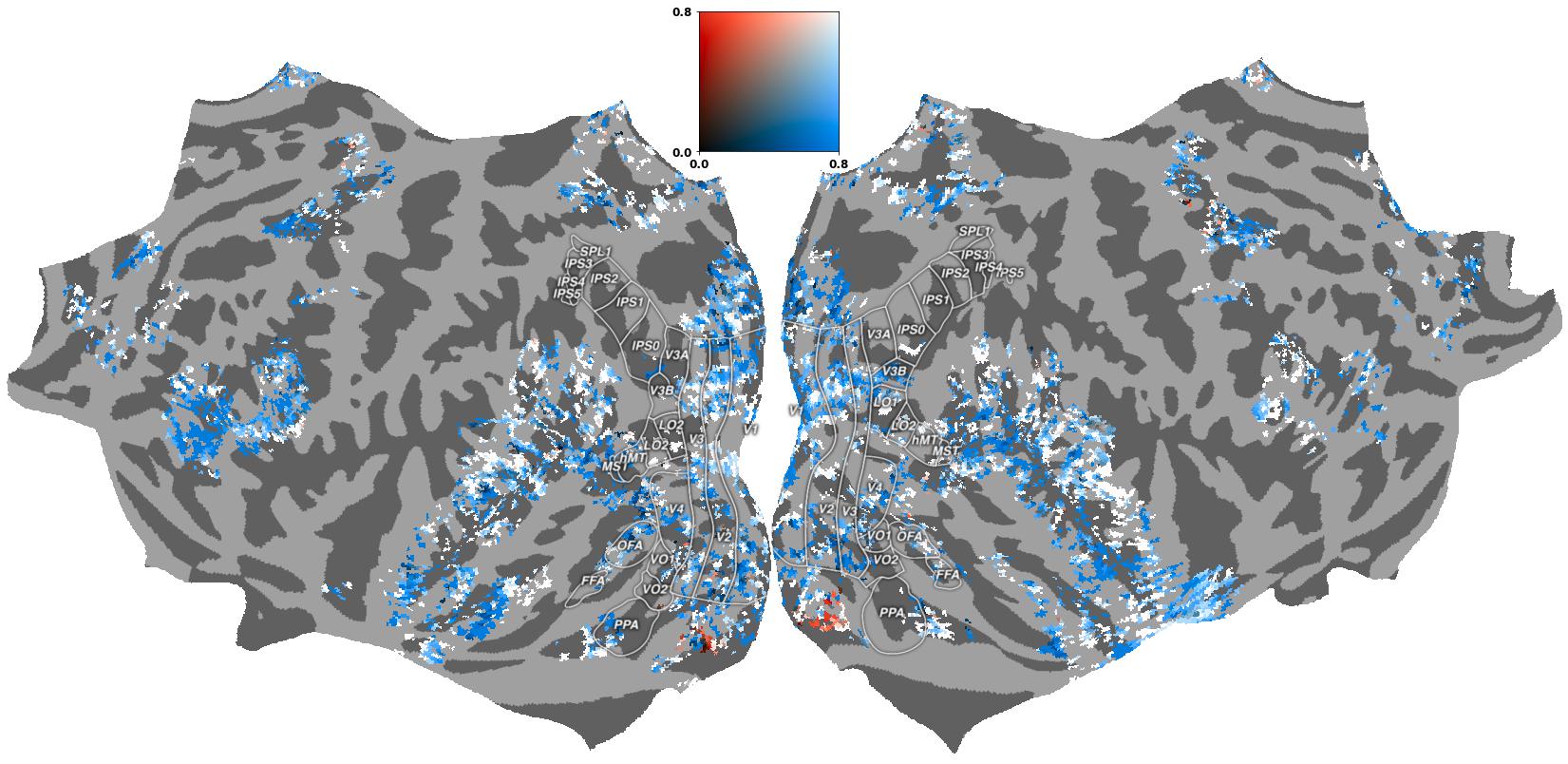}
    \\(a) Subject-01 \\
\end{minipage}
\begin{minipage}{\textwidth}
\centering
    \includegraphics[width=0.55\linewidth]{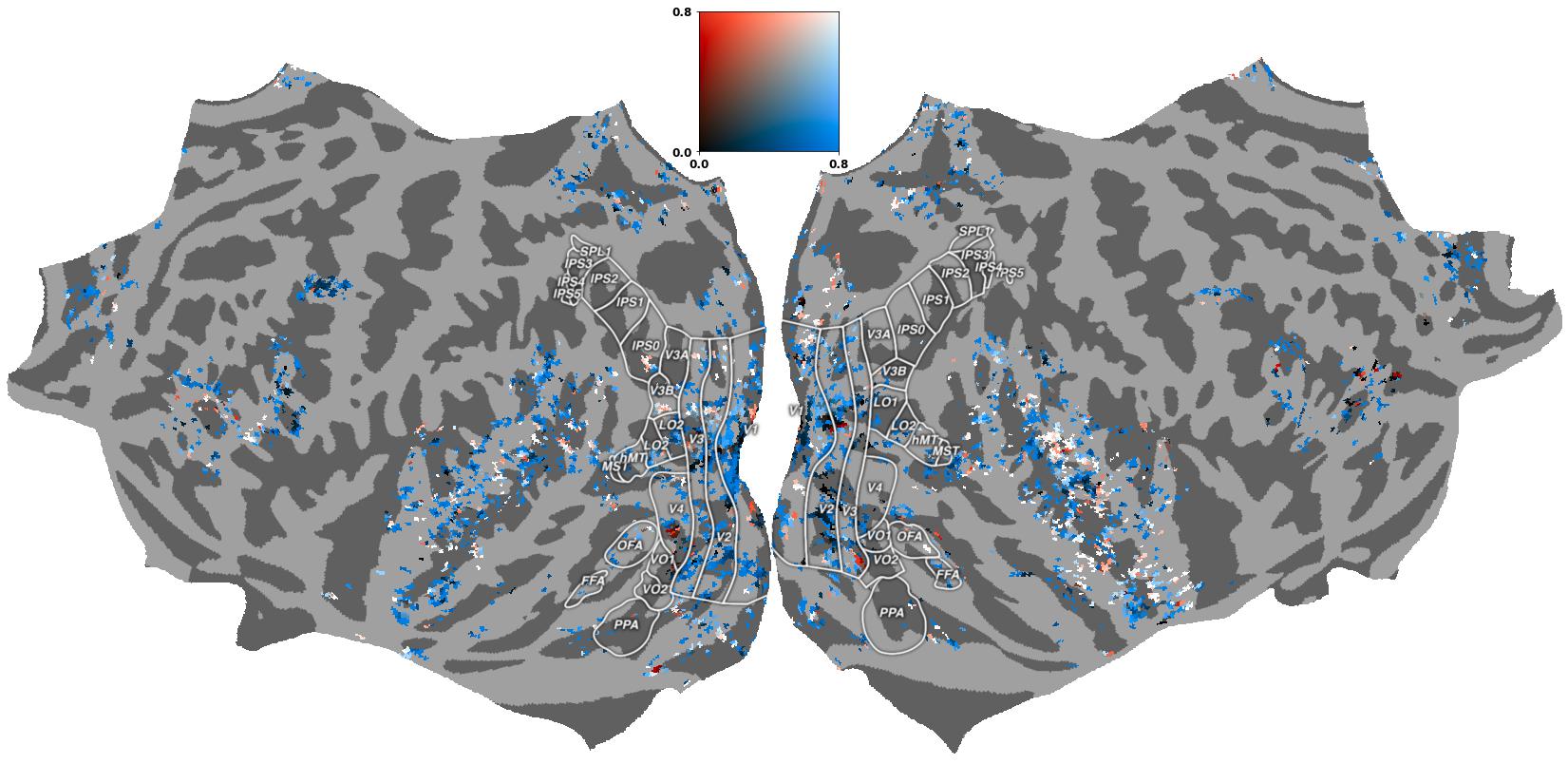}
    \\(b) Subject-02 \\
\end{minipage}
\begin{minipage}{\textwidth}
\centering
    \includegraphics[width=0.55\linewidth]{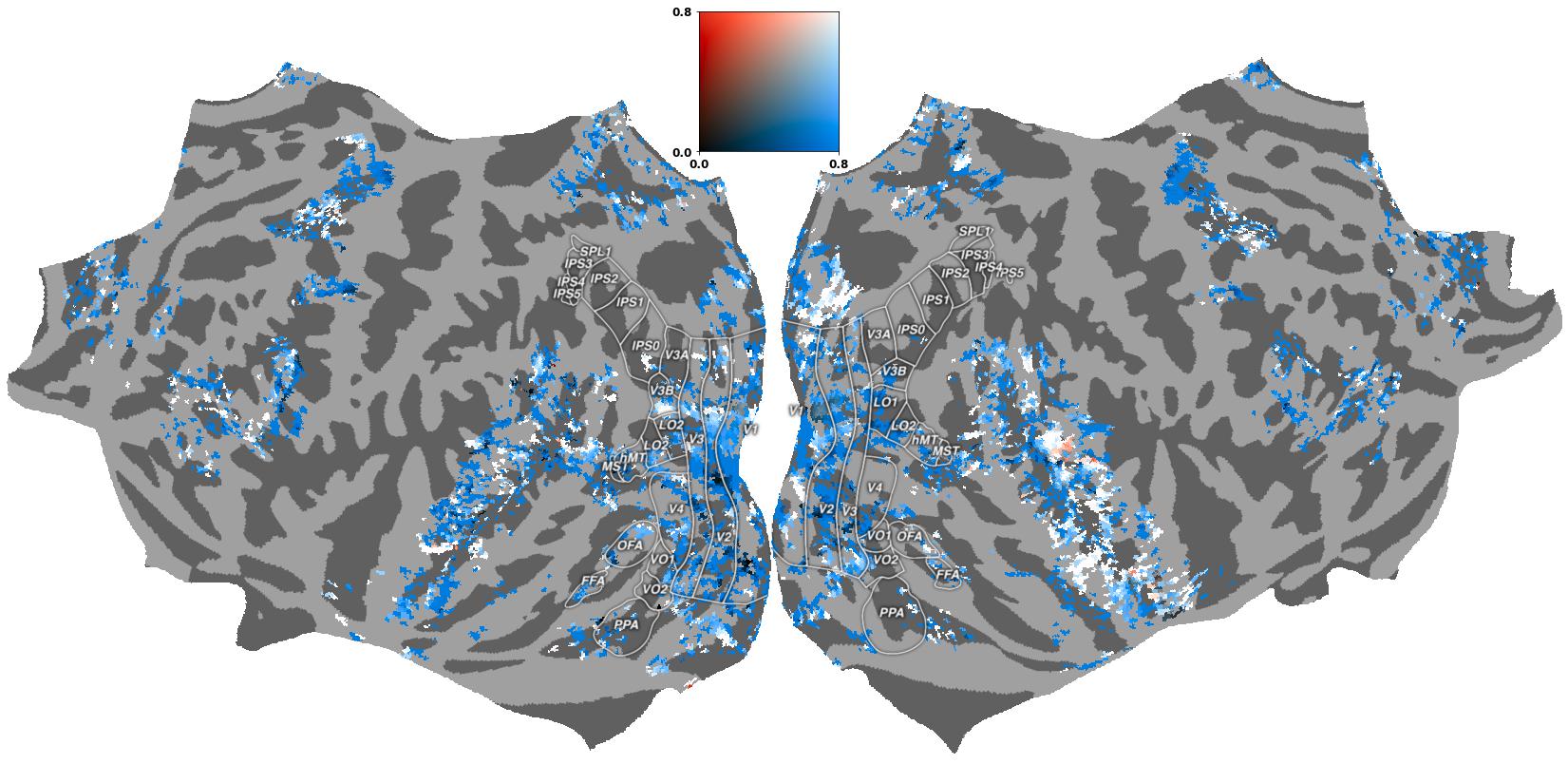}
    \\(c) Subject-03 \\
\end{minipage}
\begin{minipage}{\textwidth}
\centering
    \includegraphics[width=0.55\linewidth]{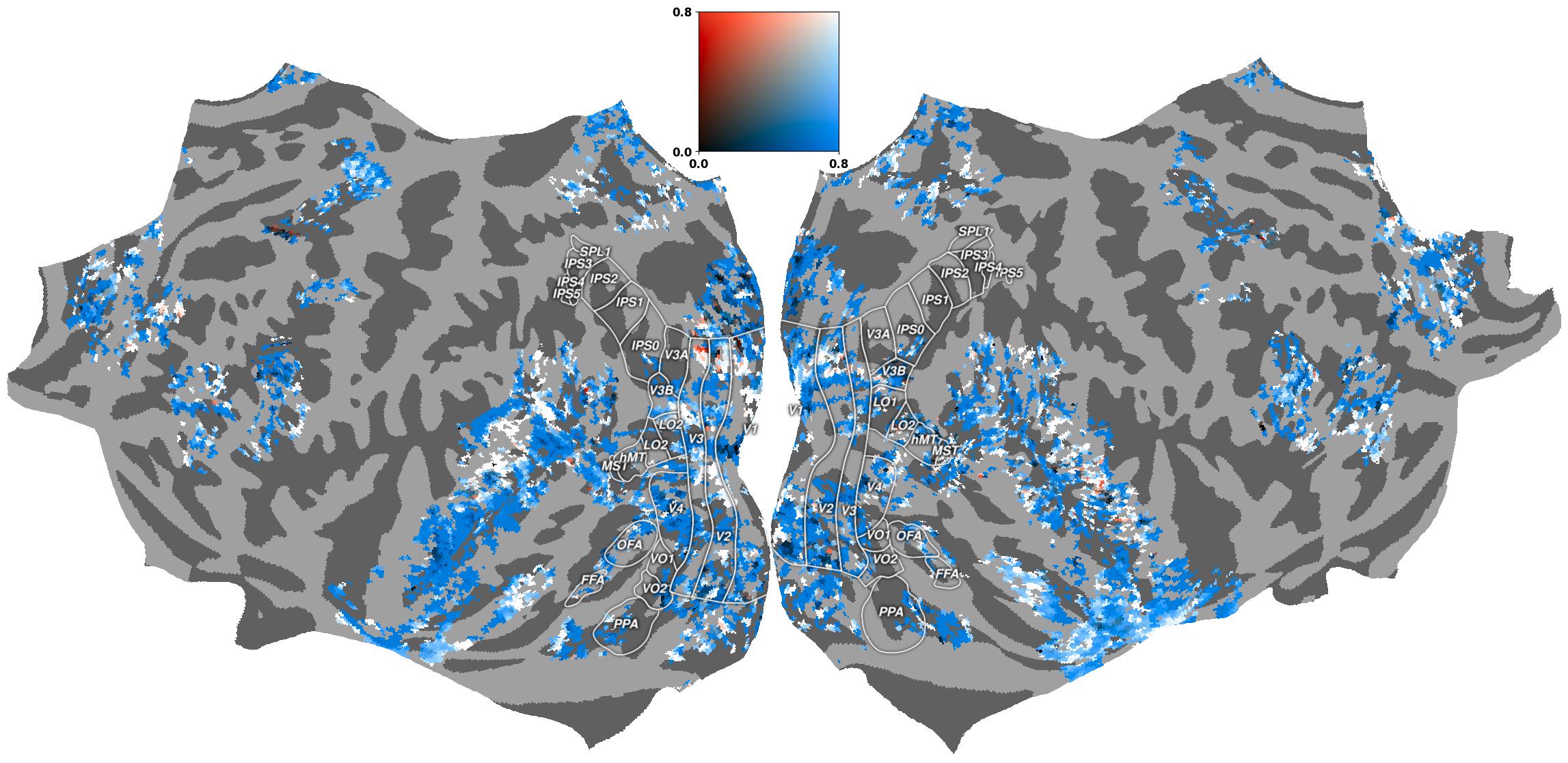}
    \\(d) Subject-05 \\
\end{minipage}
    \caption{Qwen-2.5-VL vs. TVLT: Contrast of estimated cross-subject prediction accuracy for all participants for the naturalistic movie watching. Pearson correlation scores for each voxel in each subject are projected onto the subject’s flattened cortical surface.
    \textcolor{cyan}{Blue} and \textcolor{red}{Red} voxels depict higher prediction accuracy estimates during instruction-tuned video MLLM and multimodal TVLT, respectively. Voxels that have similar cross-subject prediction accuracy appear white. Here, middle frontal gyrus (MFG), inferior frontal gyrus (IFG),  inferior frontal gyrus orbital (IFGOrb), angular gyrus (AG), and lateral temporal cortex (LTC) are late language regions, EVC denotes early visual cortex and AC denotes auditory cortex.}
    \label{fig:qwenvl_tvlt_2dmap}
\end{figure*}

\begin{figure*}[!ht] 
\centering
\begin{minipage}{\textwidth}
\centering \includegraphics[width=0.55\linewidth]{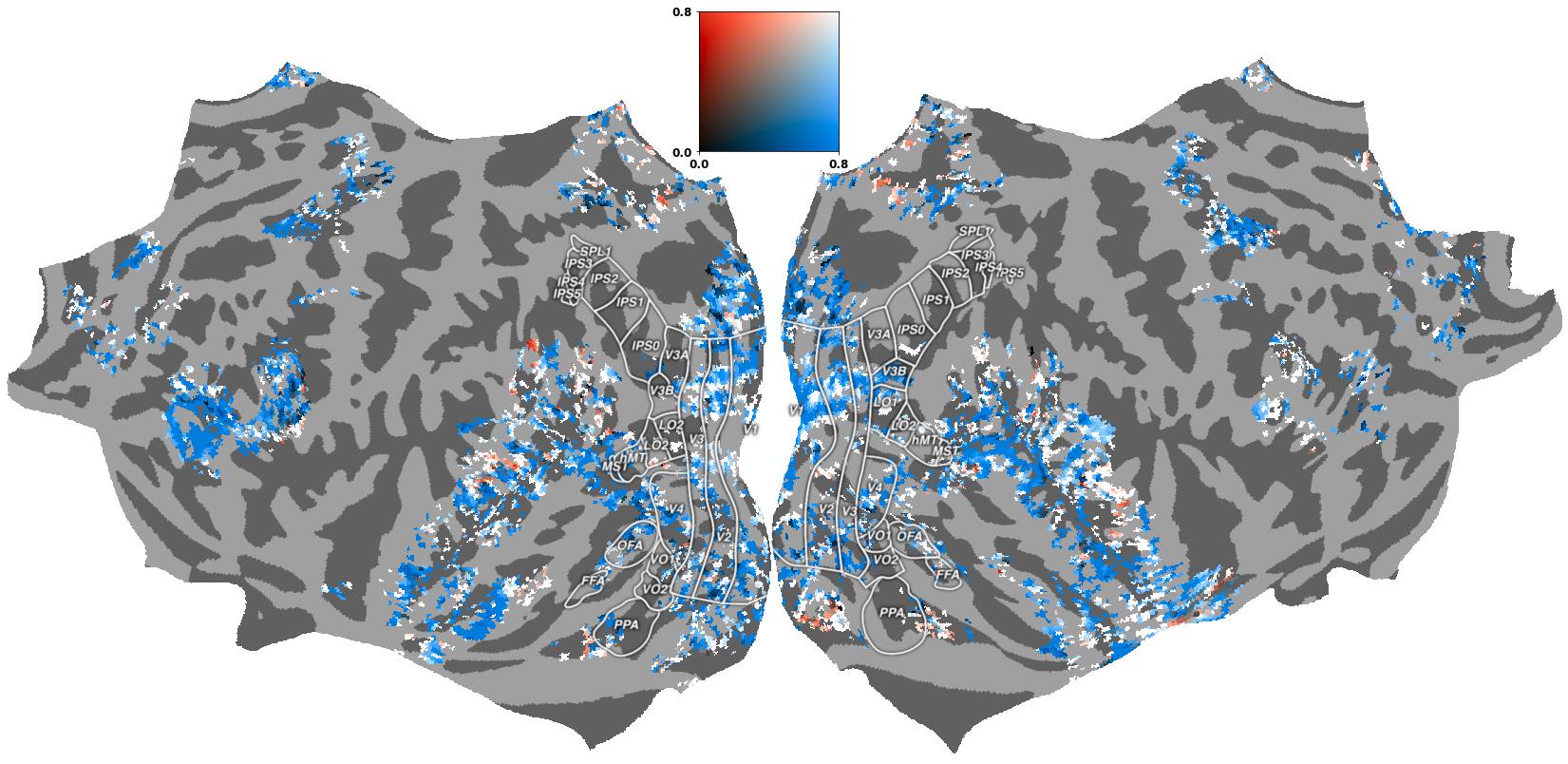}
    \\(a) Subject-01 \\
\end{minipage}
\begin{minipage}{\textwidth}
\centering
    \includegraphics[width=0.55\linewidth]{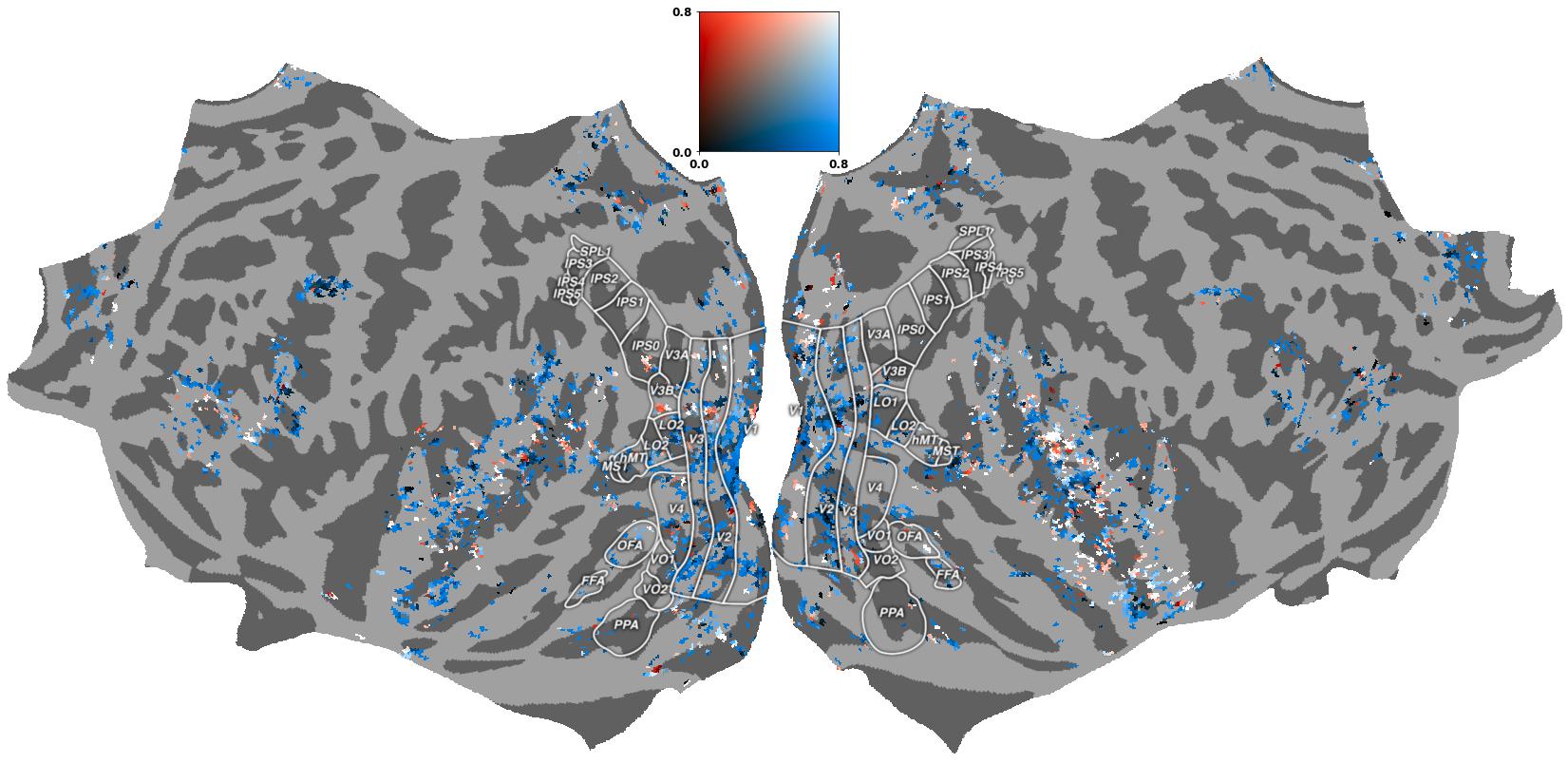}
    \\(b) Subject-02 \\
\end{minipage}
\begin{minipage}{\textwidth}
\centering
    \includegraphics[width=0.55\linewidth]{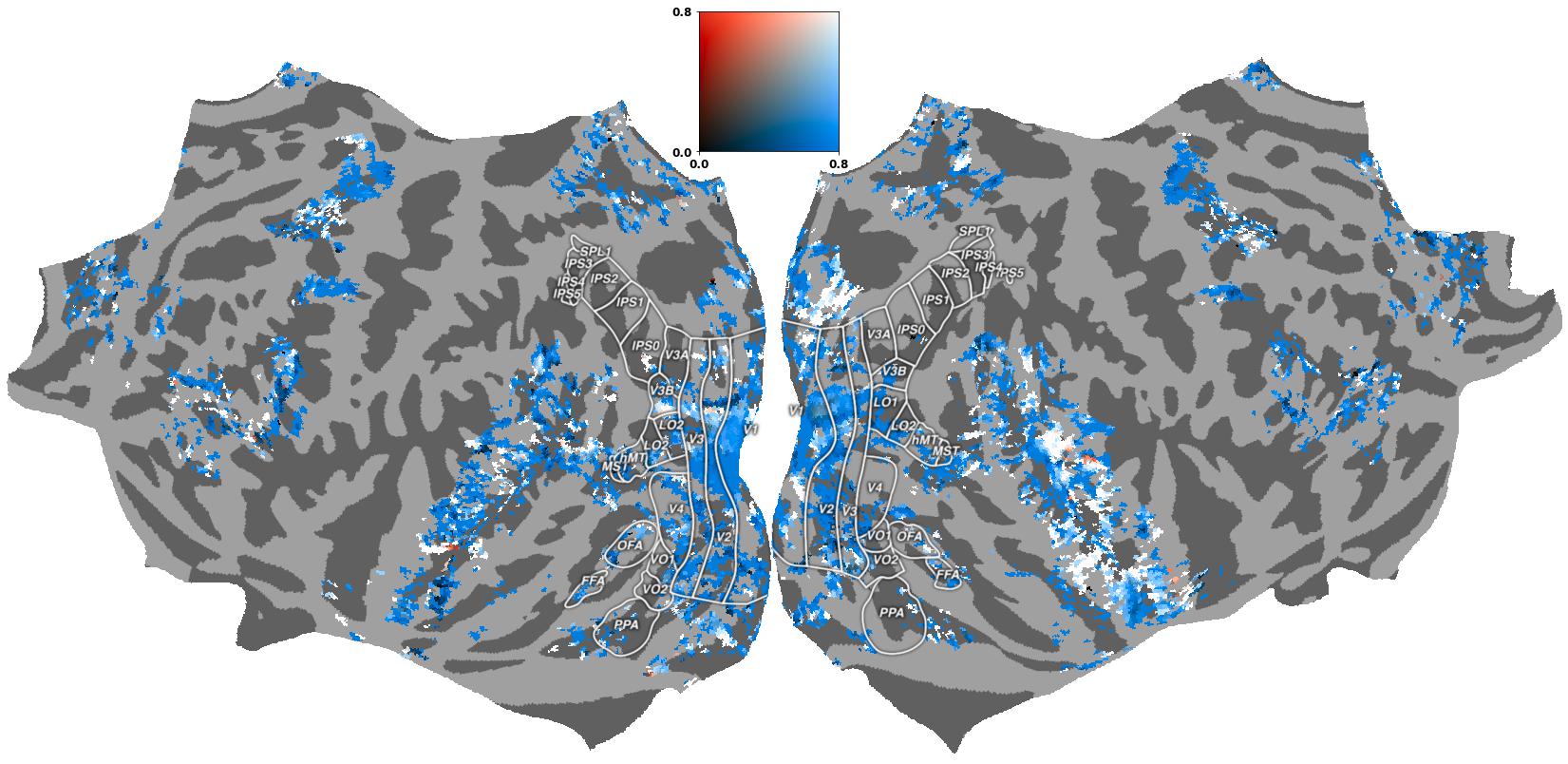}
    \\(c) Subject-03 \\
\end{minipage}
\begin{minipage}{\textwidth}
\centering
    \includegraphics[width=0.55\linewidth]{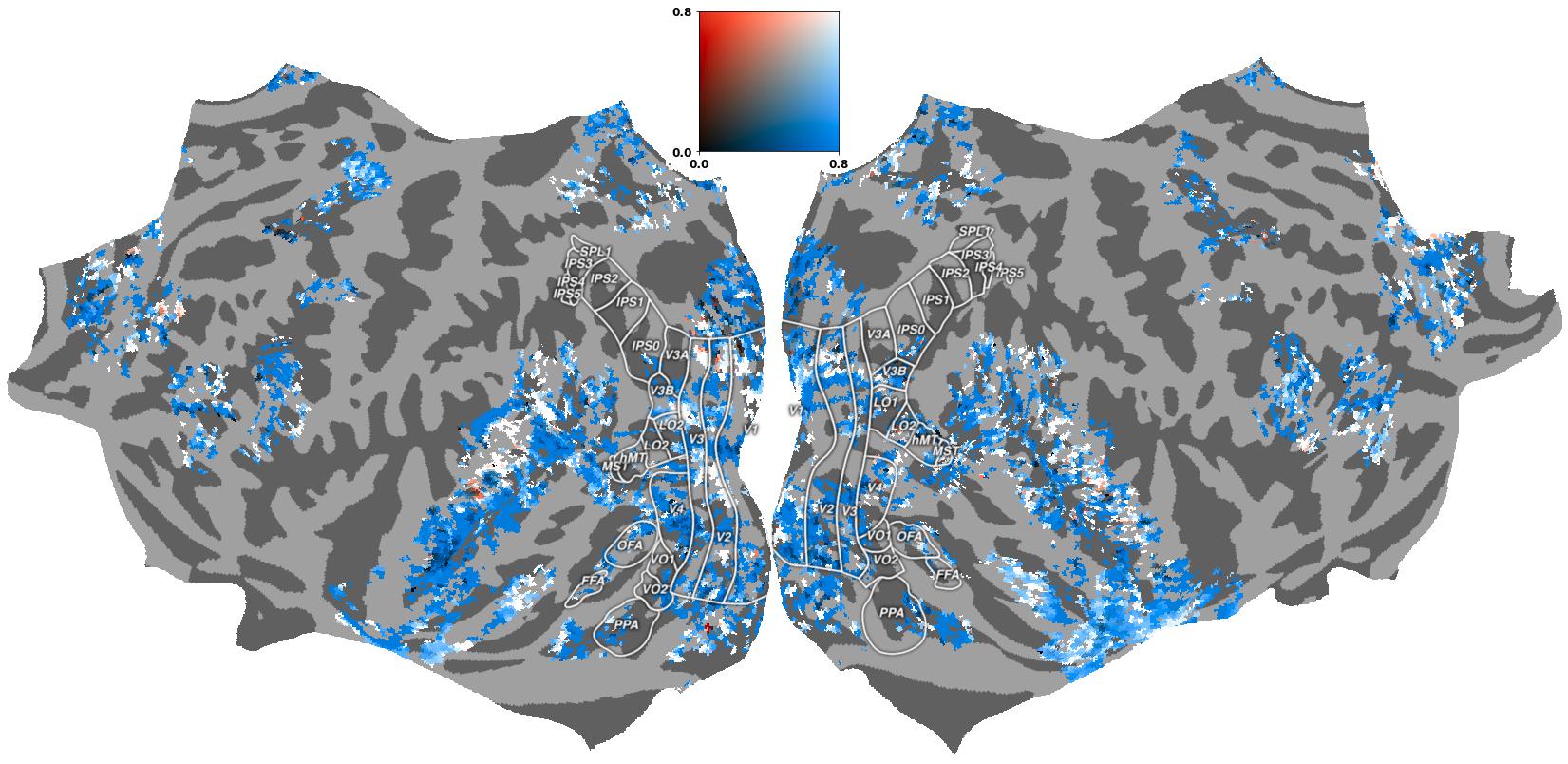}
    \\(d) Subject-05 \\
\end{minipage}
\caption{InstructBLIPVideo vs. TVLT: Contrast of estimated cross-subject prediction accuracy for all participants for the naturalistic movie watching. Pearson correlation scores for each voxel in each subject are projected onto the subject’s flattened cortical surface. \textcolor{cyan}{Blue} and \textcolor{red}{Red} voxels depict higher prediction accuracy estimates during instruction-tuned video MLLM and multimodal TVLT, respectively. Voxels that have similar cross-subject prediction accuracy appear white.
}
\label{fig:instructblip_tvlt_2dmap}
\end{figure*}

\begin{figure*}[!ht] 
\centering
\begin{minipage}{\textwidth}
\centering
    \includegraphics[width=0.55\linewidth]{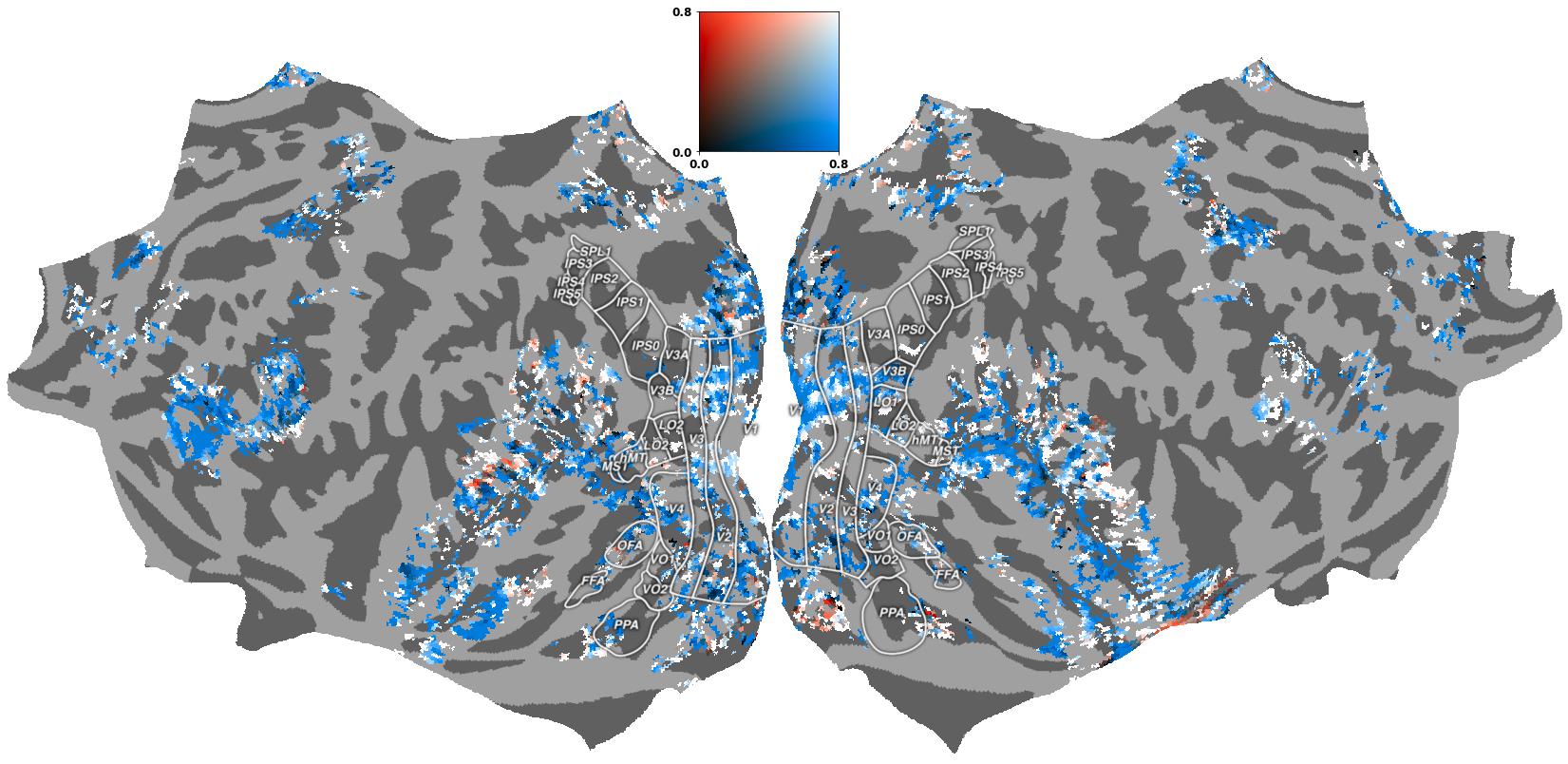}
    \\(a) Subject-01 \\
\end{minipage}
\begin{minipage}{\textwidth}
\centering
    \includegraphics[width=0.55\linewidth]{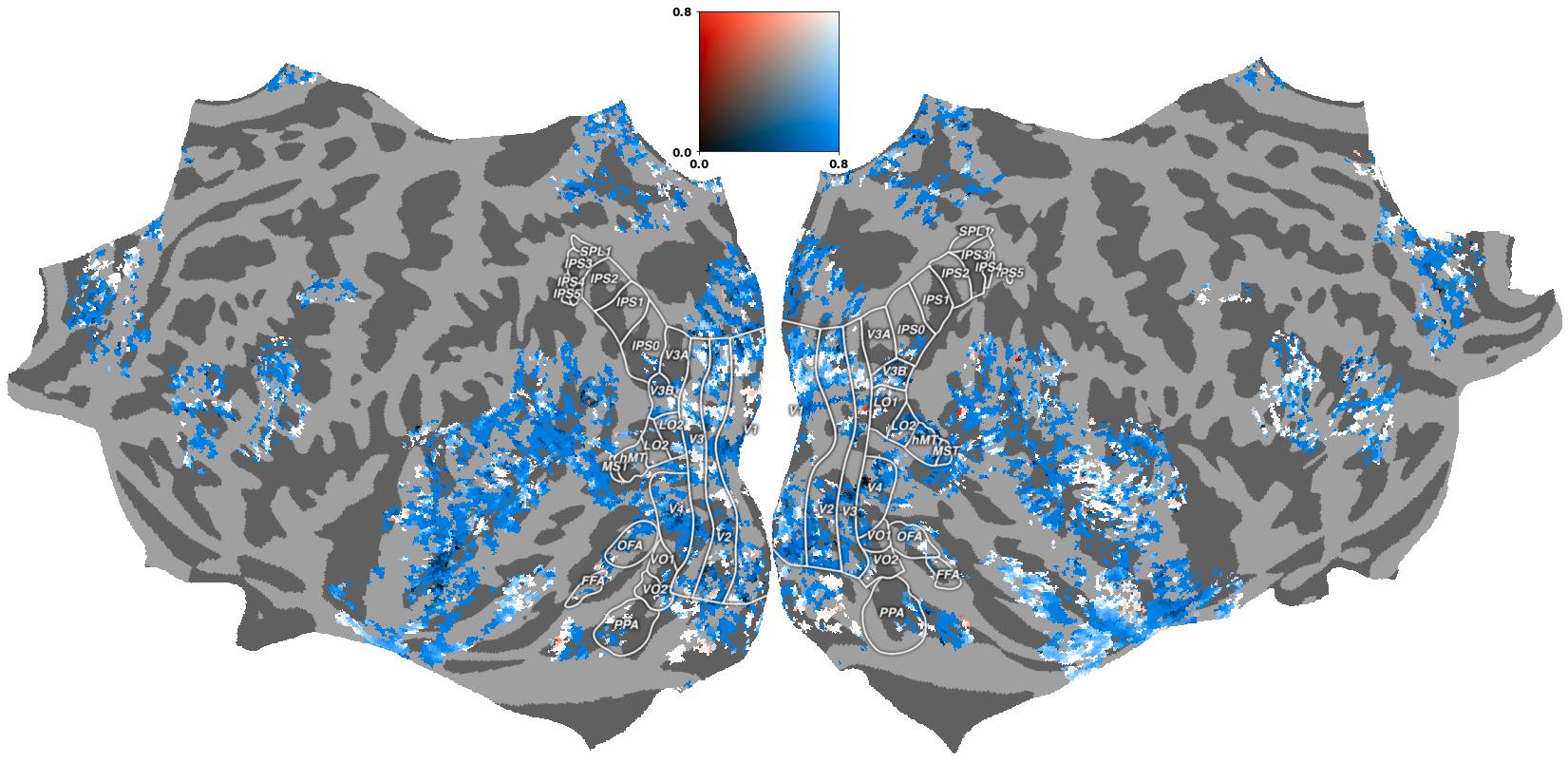}
    \\(b) Subject-02 \\
\end{minipage}
\begin{minipage}{\textwidth}
\centering
    \includegraphics[width=0.55\linewidth]{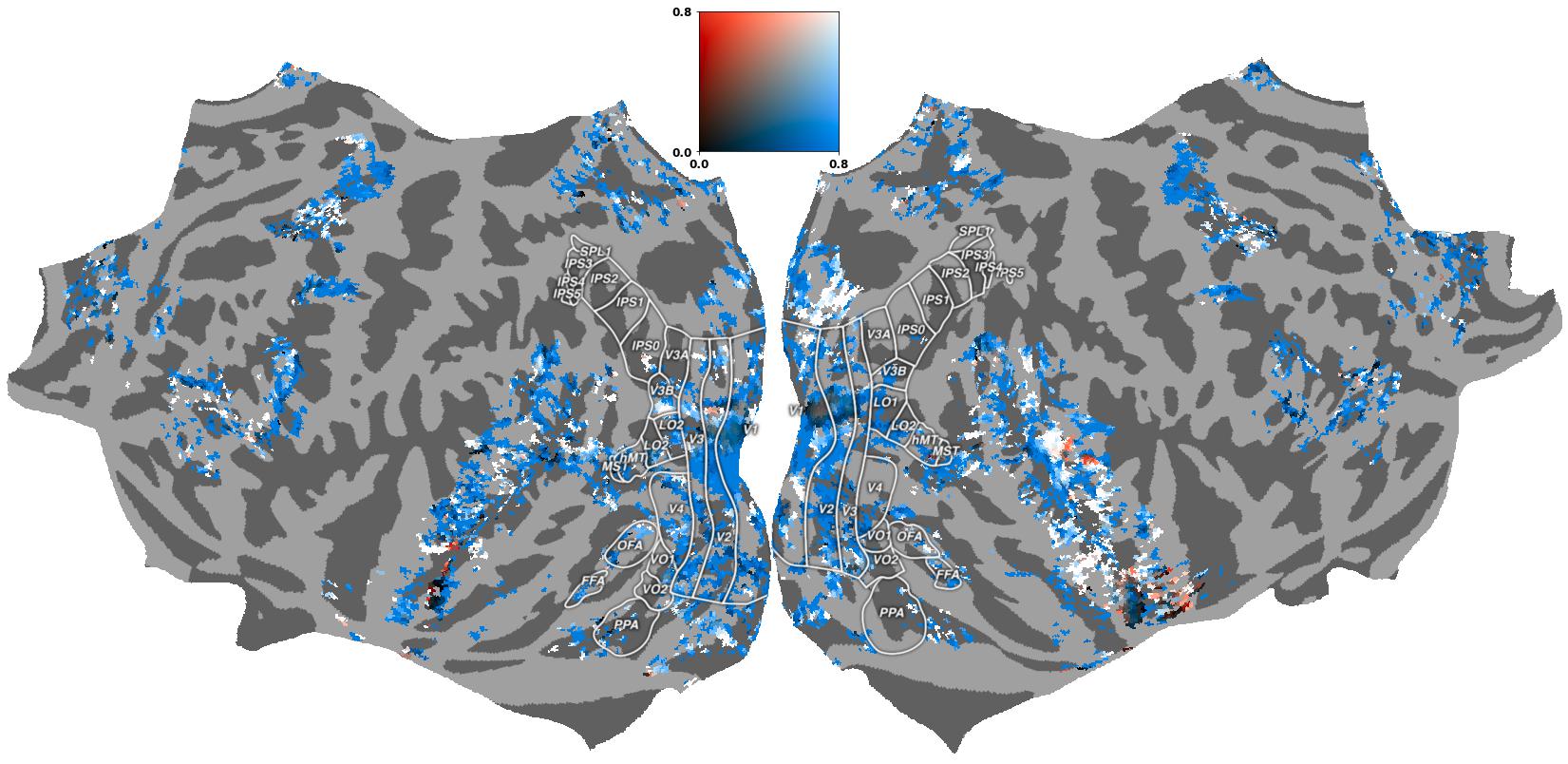}
    \\(c) Subject-03 \\
\end{minipage}
\begin{minipage}{\textwidth}
\centering
    \includegraphics[width=0.55\linewidth]{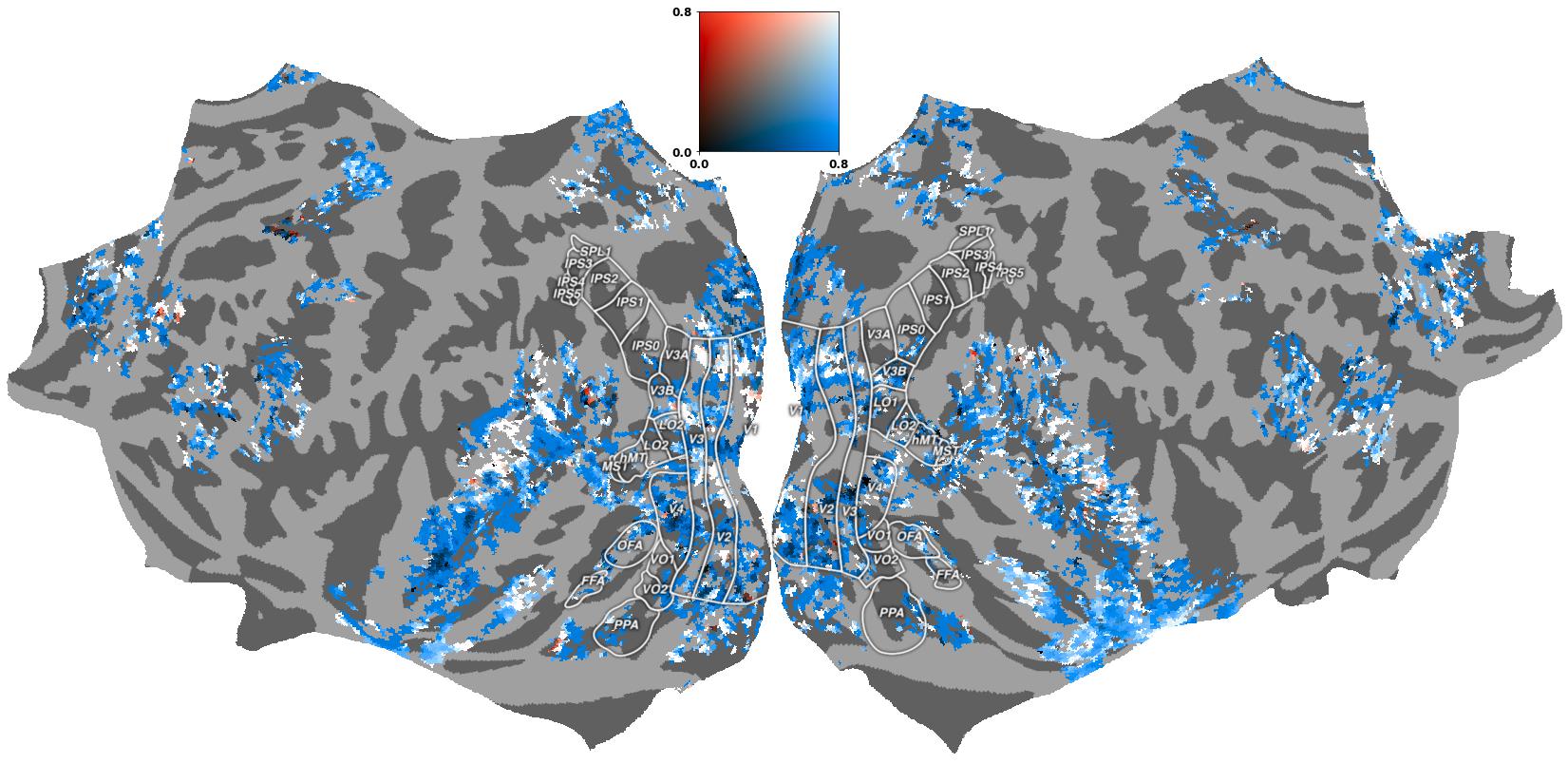}
    \\(d) Subject-05 \\
\end{minipage}
\caption{Video-LLaVA vs. TVLT: Contrast of estimated cross-subject prediction accuracy for all participants for the naturalistic movie watching. Pearson correlation scores for each voxel in each subject are projected onto the subject’s flattened cortical surface. \textcolor{cyan}{Blue} and \textcolor{red}{Red} voxels depict higher prediction accuracy estimates during instruction-tuned video MLLM and multimodal TVLT, respectively. Voxels that have similar cross-subject prediction accuracy appear white.
}
\label{fig:languagebind_tvlt_2dmap}
\end{figure*}

\begin{figure*}[!ht] 
\centering
\begin{minipage}{\textwidth}
\centering
    \includegraphics[width=0.55\linewidth]{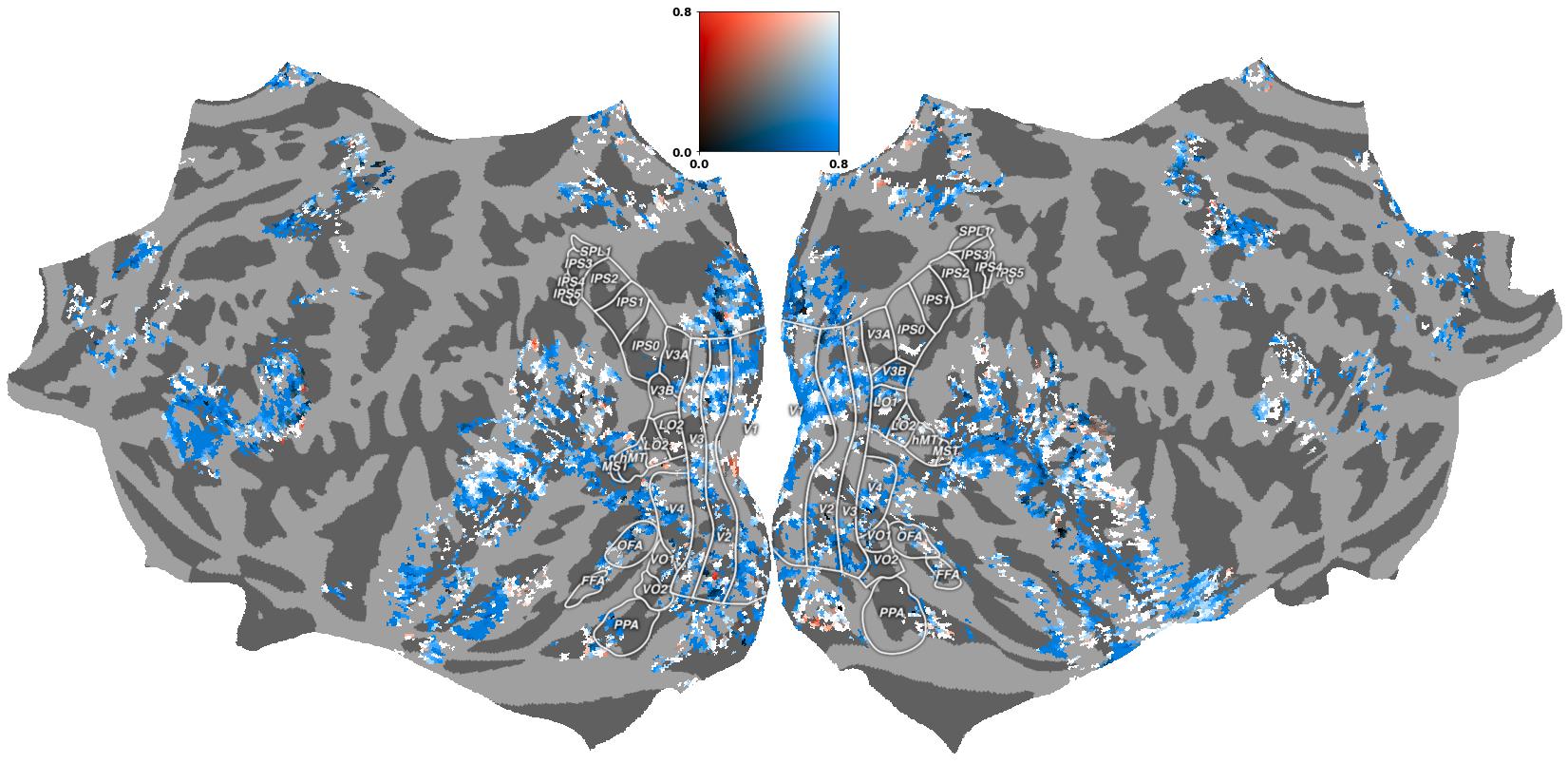}
    \\(a) Subject-01 \\
\end{minipage}
\begin{minipage}{\textwidth}
\centering
    \includegraphics[width=0.55\linewidth]{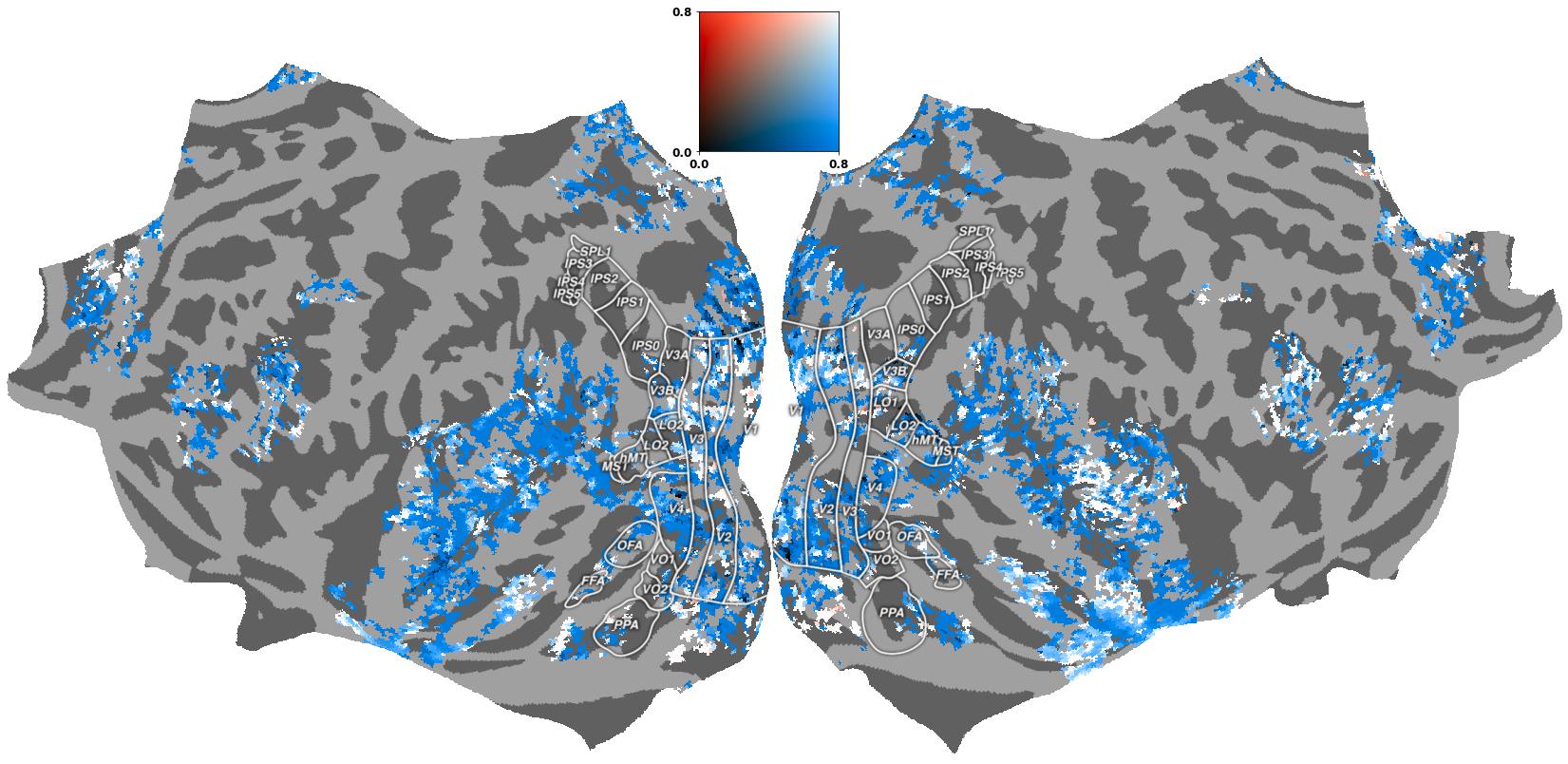}
    \\(c) Subject-02 \\
\end{minipage}
\begin{minipage}{\textwidth}
\centering
    \includegraphics[width=0.55\linewidth]{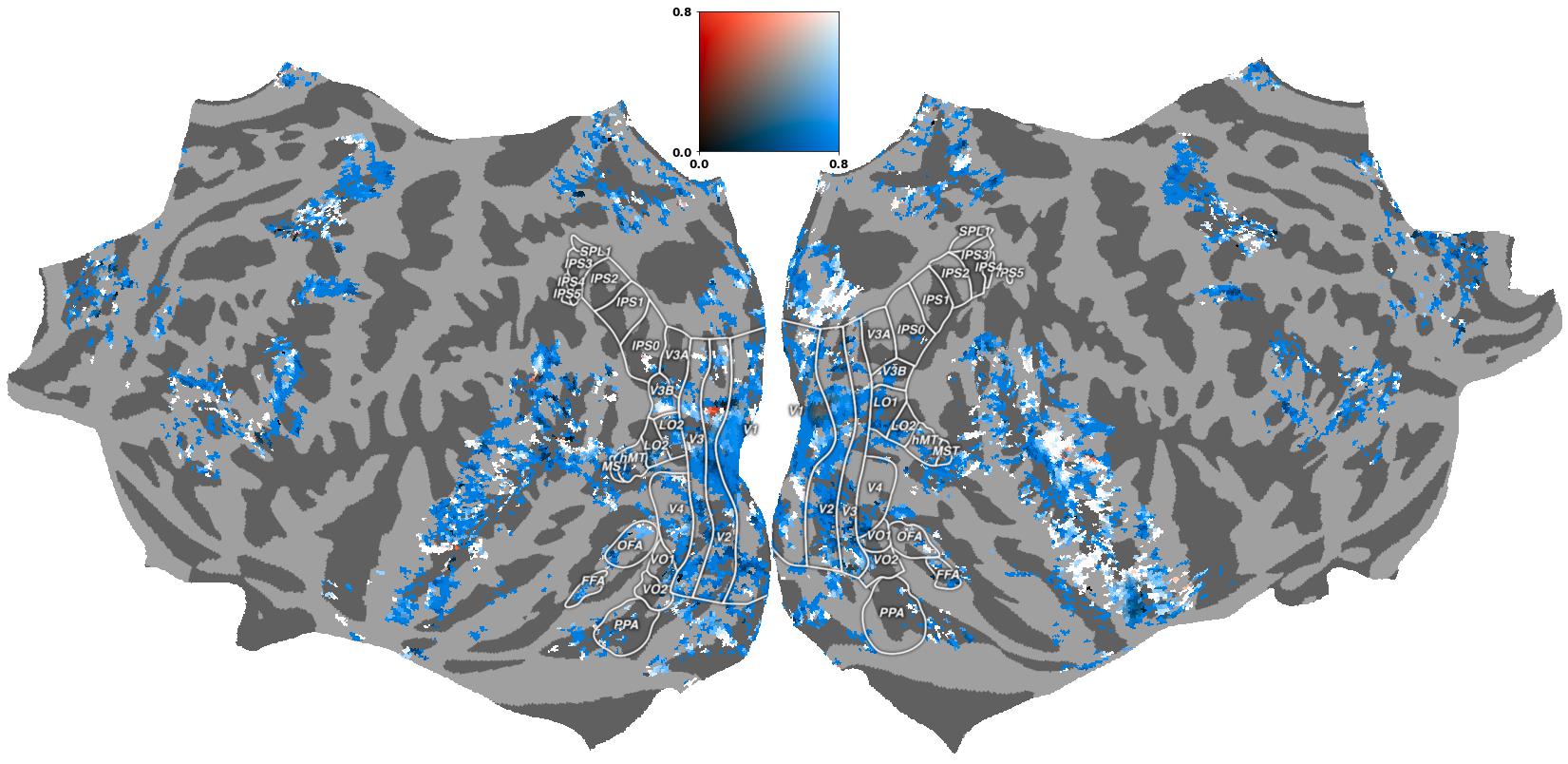}
    \\(c) Subject-03 \\
\end{minipage}
\begin{minipage}{\textwidth}
\centering
    \includegraphics[width=0.55\linewidth]{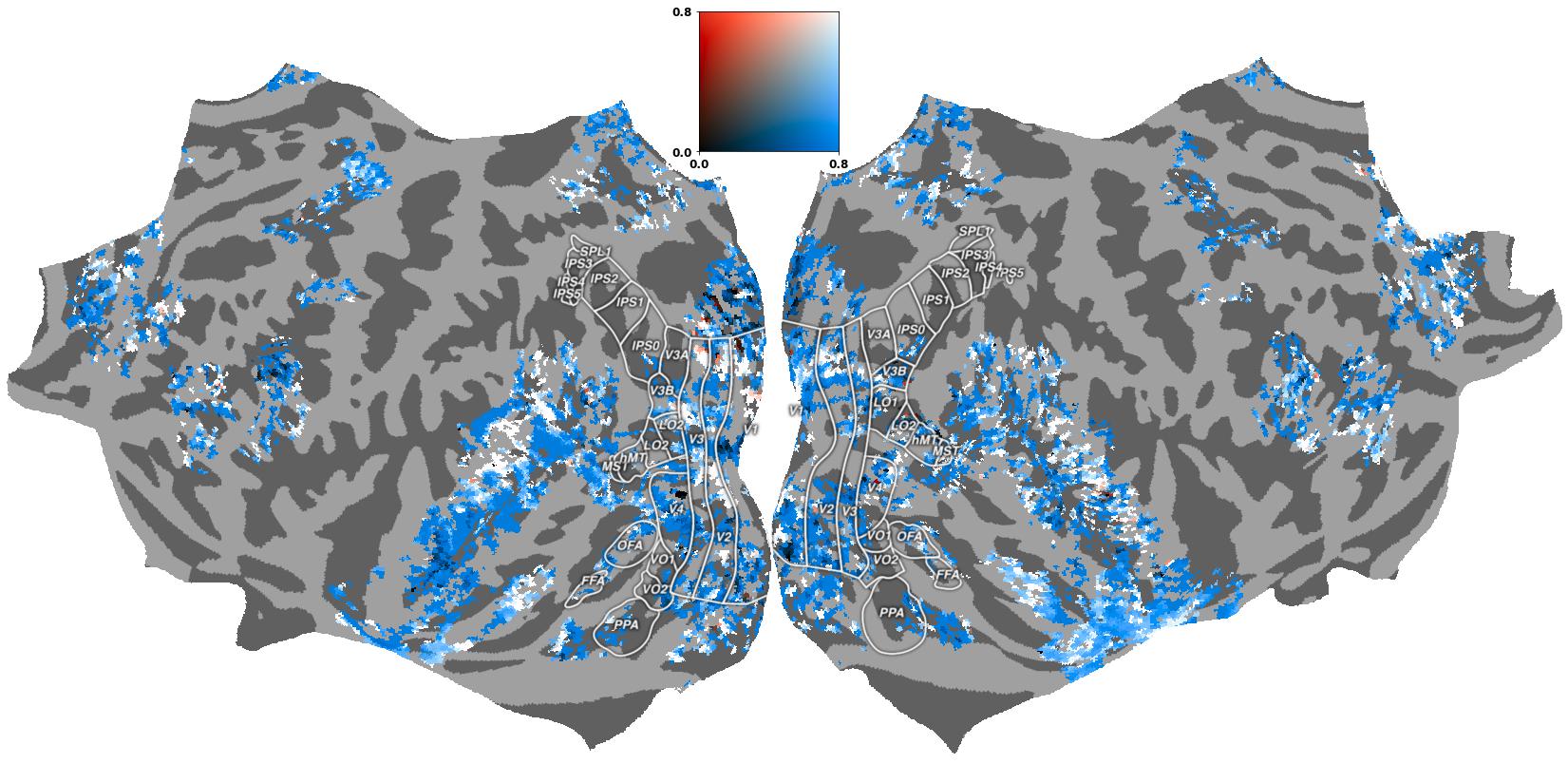}
    \\(d) Subject-05 \\
\end{minipage}
\caption{LLaVA-NeXT-Video vs. TVLT: Contrast of estimated cross-subject prediction accuracy for all participants for the naturalistic movie watching. Pearson correlation scores for each voxel in each subject are projected onto the subject’s flattened cortical surface. \textcolor{cyan}{Blue} and \textcolor{red}{Red} voxels depict higher prediction accuracy estimates during instruction-tuned video MLLM and multimodal TVLT, respectively. Voxels that have similar cross-subject prediction accuracy appear white.
}
\label{fig:llavanext_tvlt_2dmap}
\end{figure*}

\begin{figure*}[!ht]
    \centering
\begin{minipage}{\textwidth}
\centering
    \includegraphics[width=0.55\linewidth]{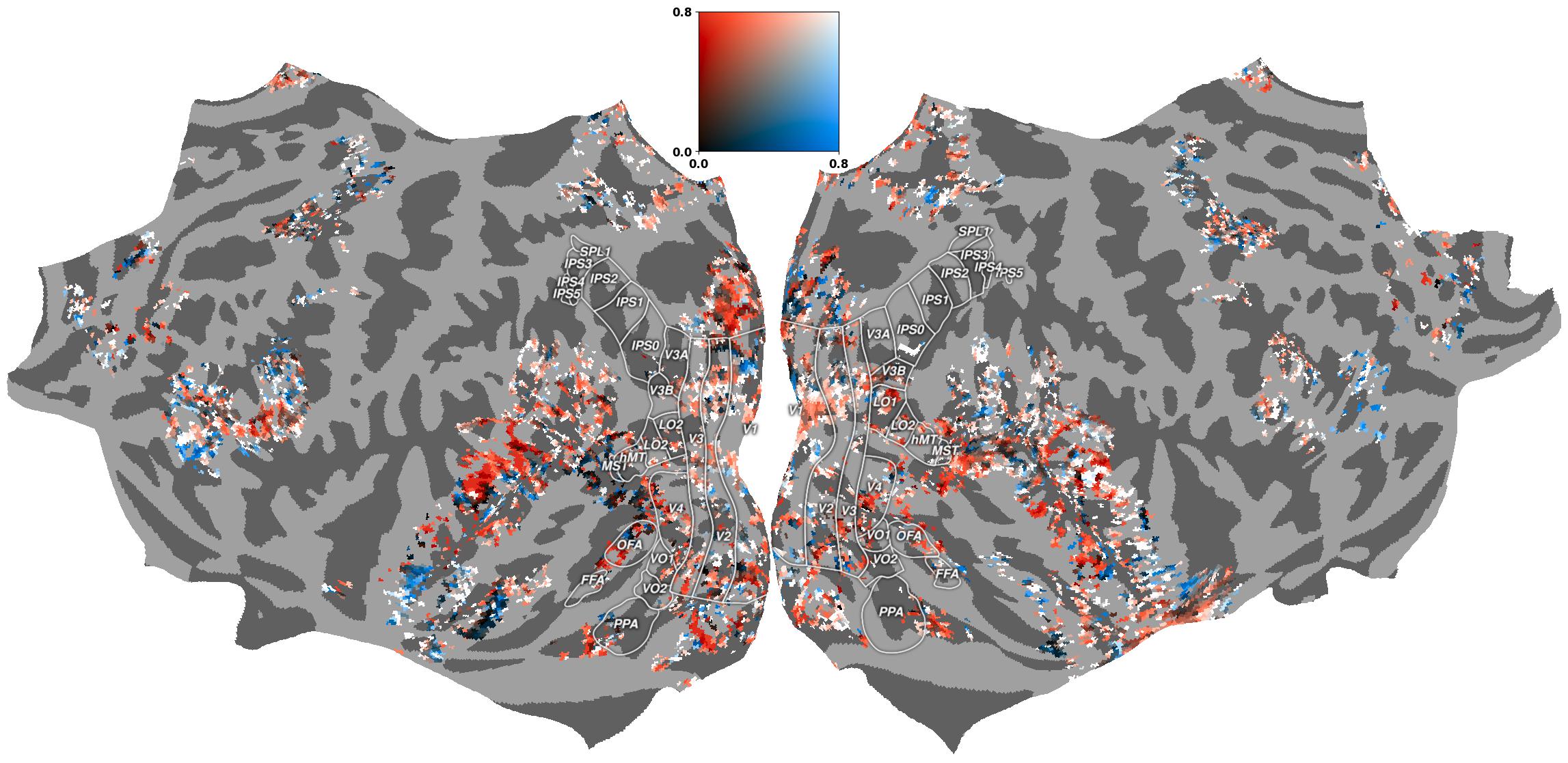}
    \\(a) Subject-01 \\
\end{minipage}
\begin{minipage}{\textwidth}
\centering
    \includegraphics[width=0.55\linewidth]{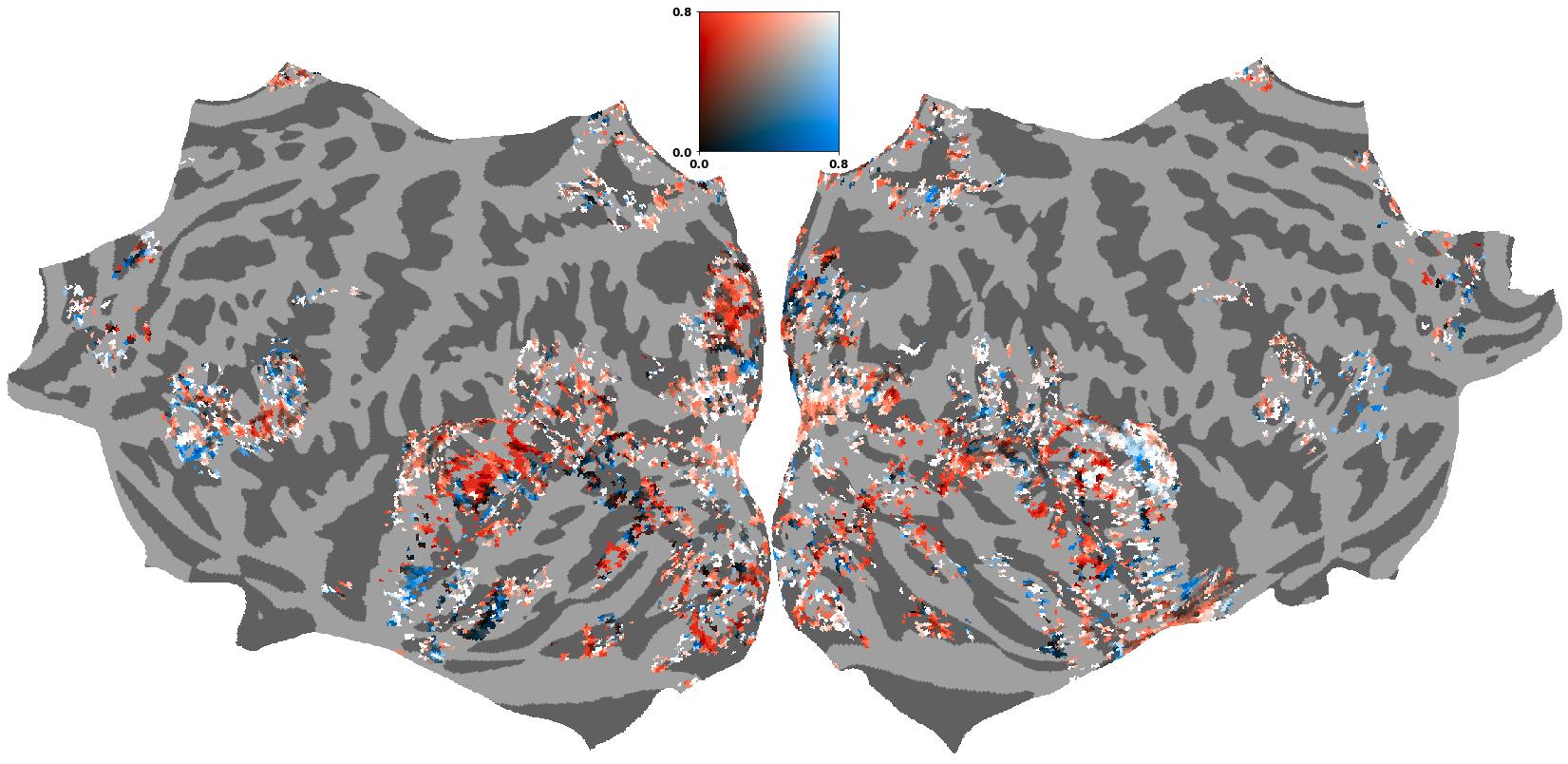}
    \\(c) Subject-02 \\
\end{minipage}
\begin{minipage}{\textwidth}
\centering
    \includegraphics[width=0.55\linewidth]{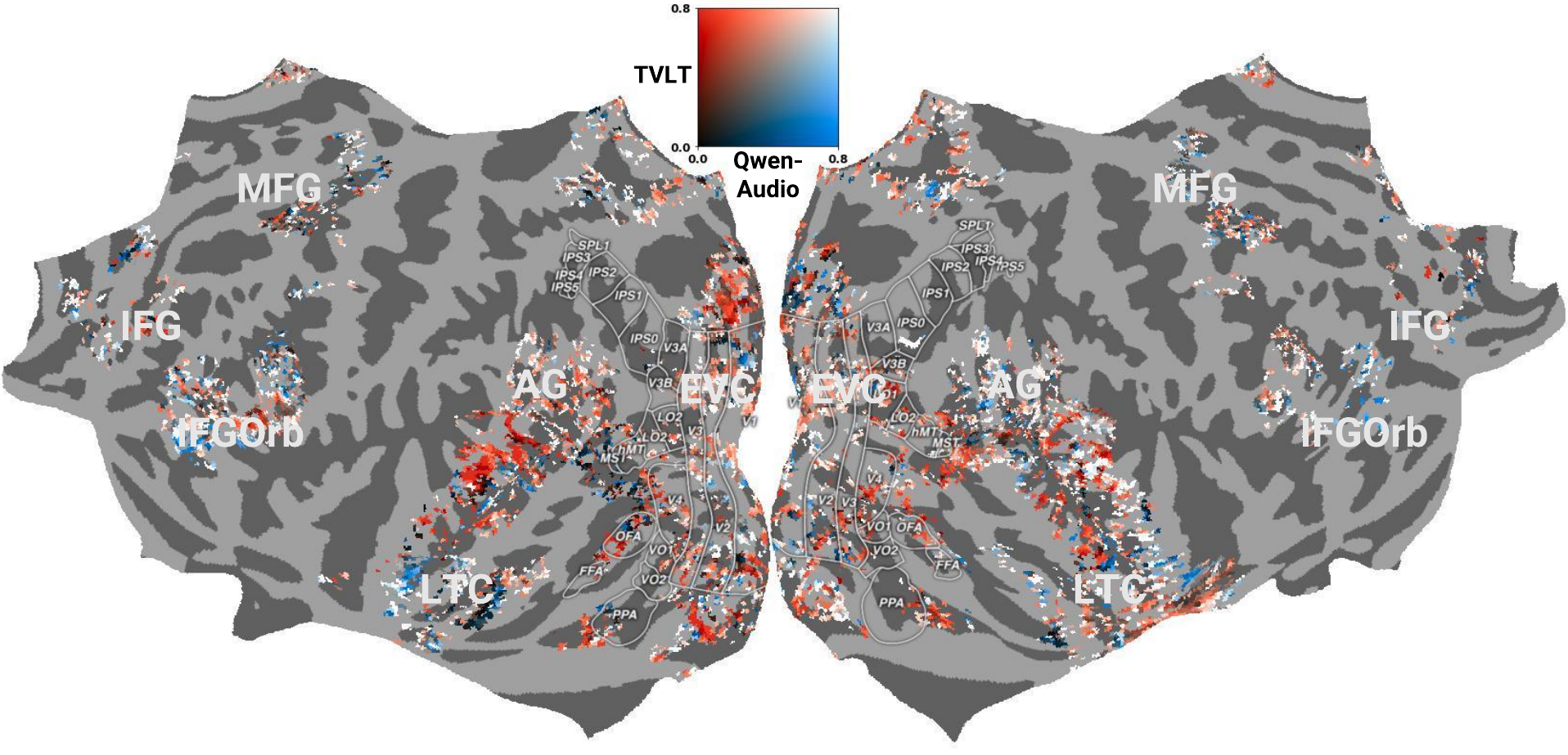}
    \\(c) Subject-03 \\
\end{minipage}
\begin{minipage}{\textwidth}
\centering
    \includegraphics[width=0.55\linewidth]{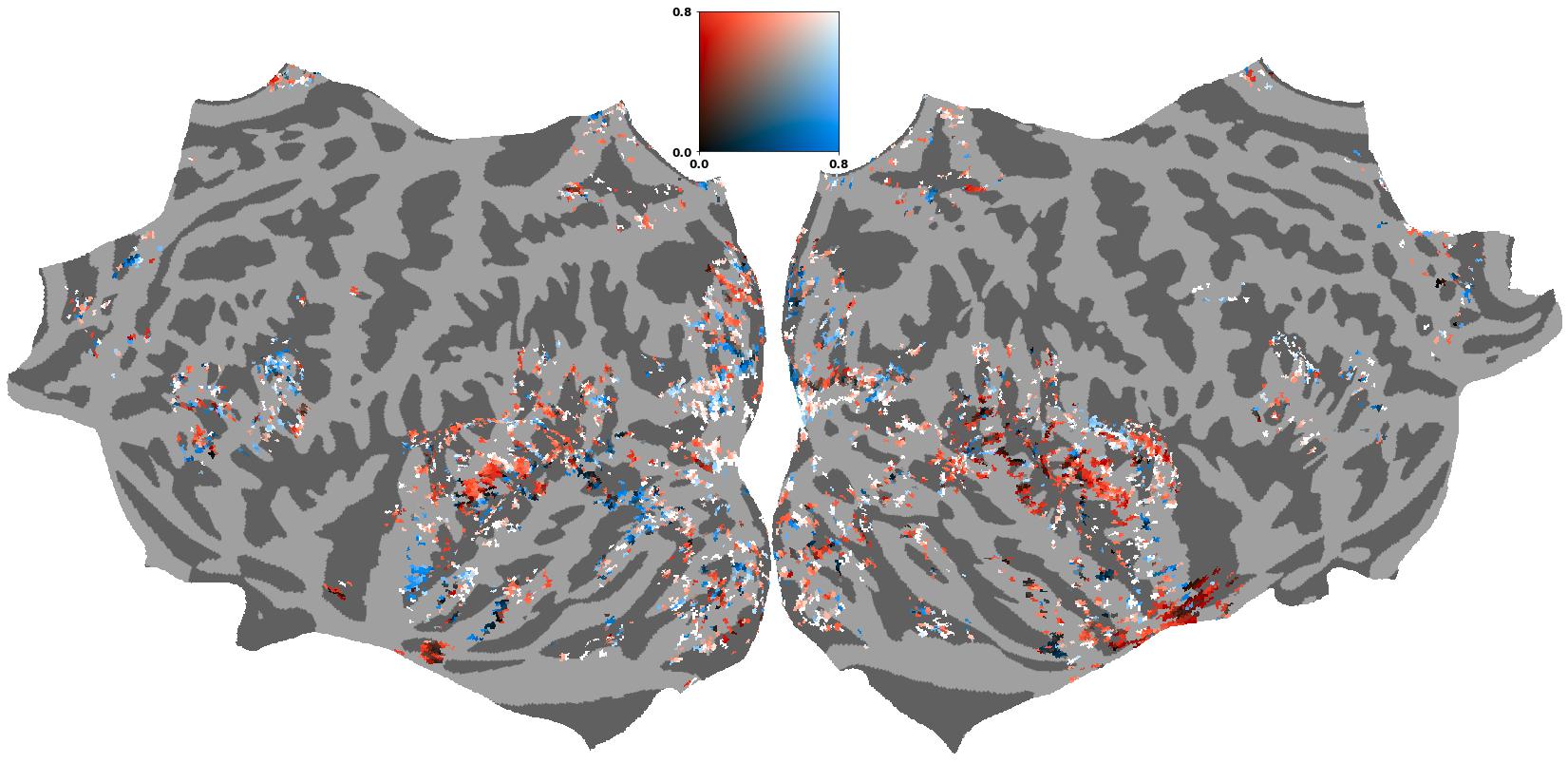}
    \\(d) Subject-05 \\
\end{minipage}
    \caption{Qwen-Audio vs. TVLT: Contrast of estimated cross-subject prediction accuracy for all participants for the naturalistic movie watching. Pearson correlation scores for each voxel in each subject are projected onto the subject’s flattened cortical surface. \textcolor{cyan}{Blue} and \textcolor{red}{Red} voxels depict higher prediction accuracy estimates during instruction-tuned audio MLLM and multimodal TVLT, respectively. Voxels that have similar cross-subject prediction accuracy appear white.  Here, middle frontal gyrus (MFG), inferior frontal gyrus (IFG),  inferior frontal gyrus orbital (IFGOrb), angular gyrus (AG), and lateral temporal cortex (LTC) are late language regions, EVC denotes early visual cortex and AC denotes auditory cortex.}
\label{fig:instruction_tuned_audio_multimodal_brainmaps_qwenaudio}
\end{figure*}

\begin{figure*}[!ht]
    \centering
\begin{minipage}{\textwidth}
\centering
    \includegraphics[width=0.55\linewidth]{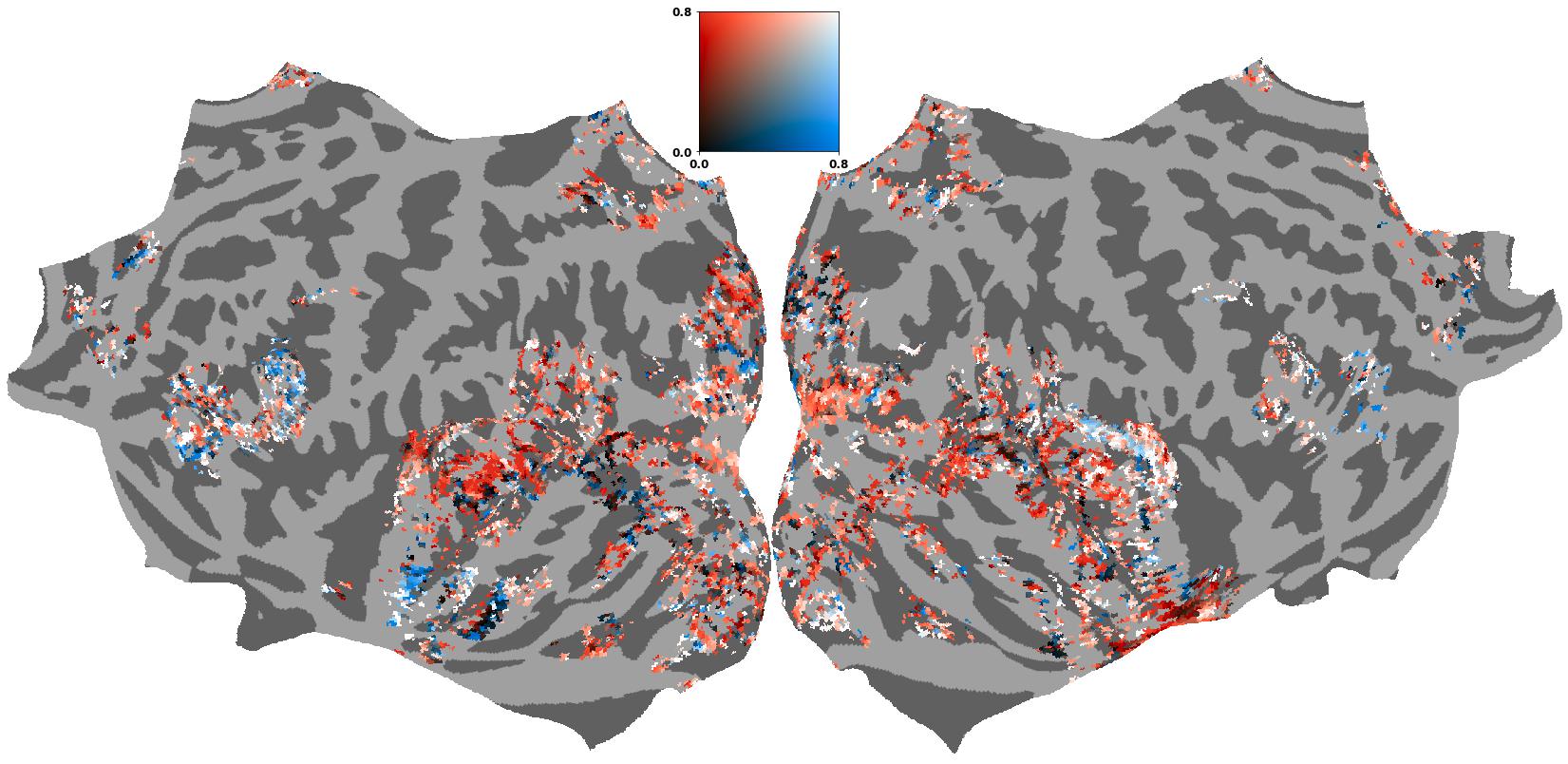}
    \\(a) Subject-01 \\
\end{minipage}
\begin{minipage}{\textwidth}
\centering
    \includegraphics[width=0.55\linewidth]{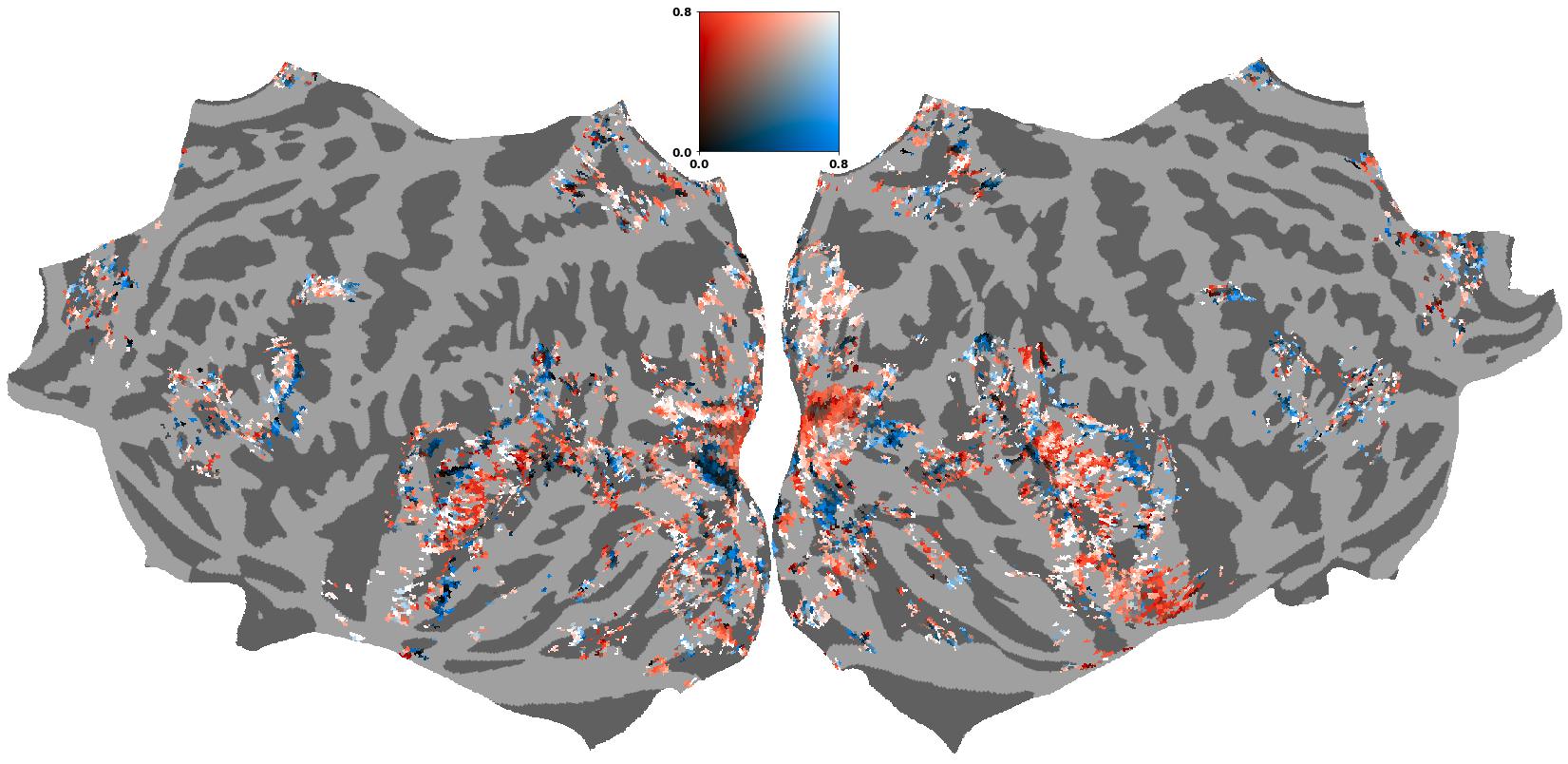}
    \\(c) Subject-02 \\
\end{minipage}
\begin{minipage}{\textwidth}
\centering
    \includegraphics[width=0.55\linewidth]{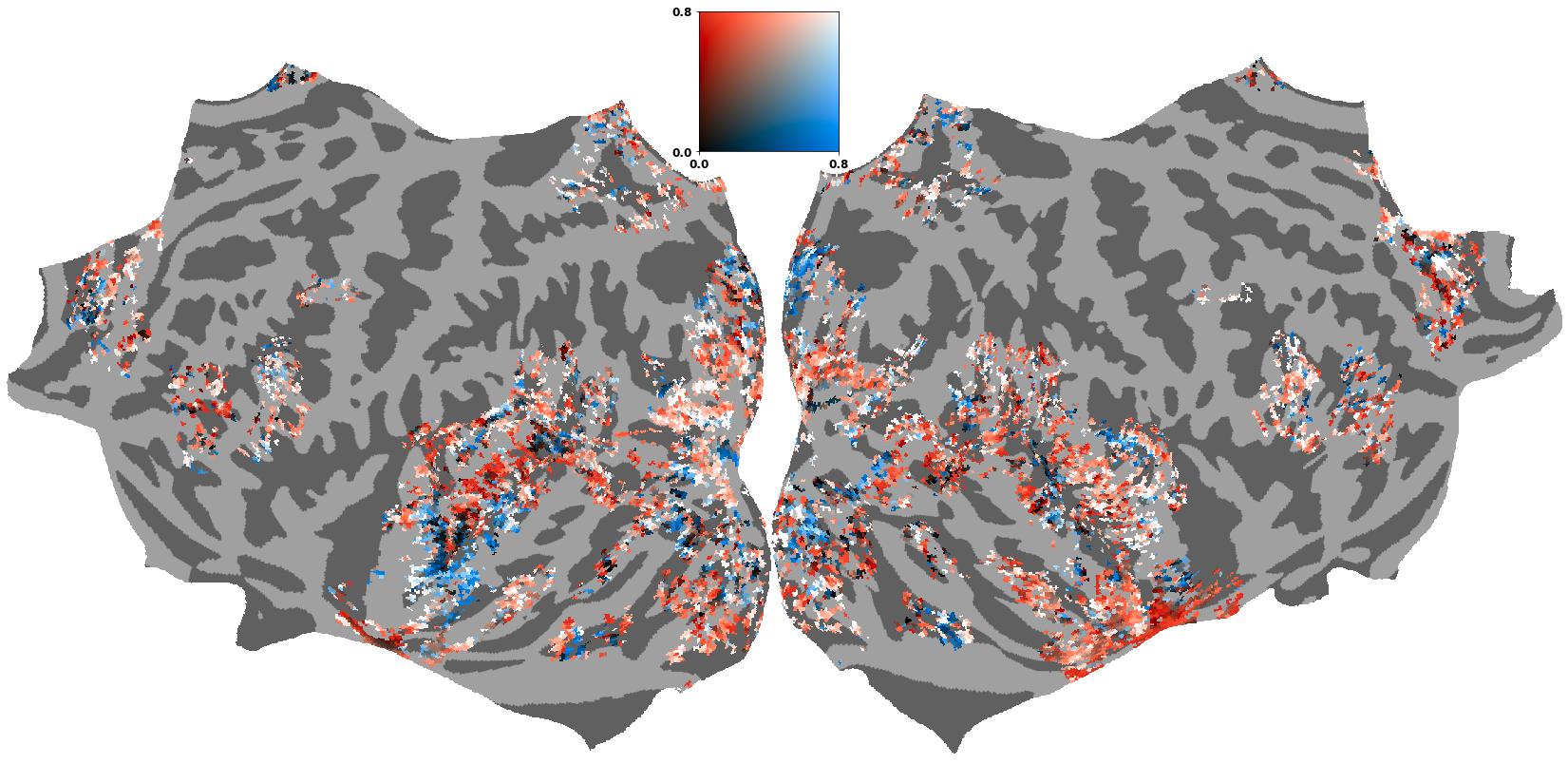}
    \\(c) Subject-03 \\
\end{minipage}
\begin{minipage}{\textwidth}
\centering
    \includegraphics[width=0.55\linewidth]{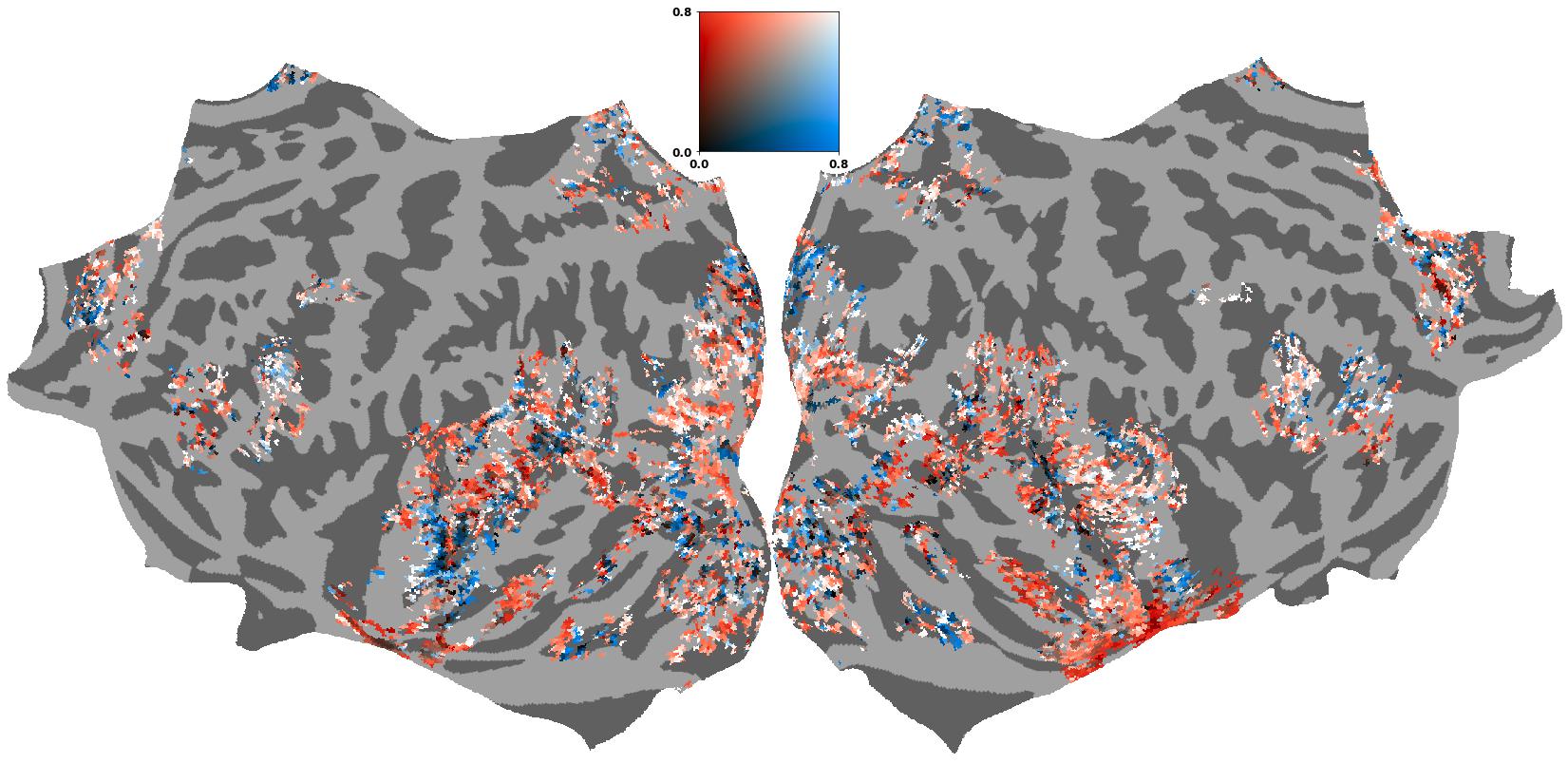}
    \\(d) Subject-05 \\
\end{minipage}
    \caption{Kimi-Audio vs. TVLT: Contrast of estimated cross-subject prediction accuracy for all participants for the naturalistic movie watching. Pearson correlation scores for each voxel in each subject are projected onto the subject’s flattened cortical surface. \textcolor{cyan}{Blue} and \textcolor{red}{Red} voxels depict higher prediction accuracy estimates during instruction-tuned audio MLLM and multimodal TVLT, respectively. Voxels that have similar cross-subject prediction accuracy appear white.  Here, middle frontal gyrus (MFG), inferior frontal gyrus (IFG),  inferior frontal gyrus orbital (IFGOrb), angular gyrus (AG), and lateral temporal cortex (LTC) are late language regions, EVC denotes early visual cortex and AC denotes auditory cortex.}
\label{fig:instruction_tuned_audio_multimodal_brainmaps_Kimiaudio}
\end{figure*}

We present contrast of brainmaps to display the average normalized brain alignment across voxels for the instruction-tuned video MLLMs versus the non-instruction-tuned multimodal models VILA and TVLT in Figures~\ref{fig:instruction_tuned_video_multimodal_brainmaps_vila},~\ref{fig:qwenvl_tvlt_2dmap},~\ref{fig:instructblip_tvlt_2dmap},~\ref{fig:languagebind_tvlt_2dmap}, and ~\ref{fig:llavanext_tvlt_2dmap}. The results show that instruction-tuned video MLLMs consistently achieve significantly higher alignment across all brain voxels. However, Figures~\ref{fig:instruction_tuned_audio_multimodal_brainmaps_qwenaudio} and~\ref{fig:instruction_tuned_audio_multimodal_brainmaps_Kimiaudio} reveal clear differences between audio MLLMs and multimodal models: the prediction performance of audio MLLMs lacks brain-relevant semantic information compared to multimodal models.

\FloatBarrier

\section{Brain Maps for Task-specific instructions}
\label{app:taskSpecificDetailedBrainMaps}
\begin{figure*}[!ht]
    \centering
\begin{minipage}{\textwidth}
\centering
    \includegraphics[width=0.8\linewidth]{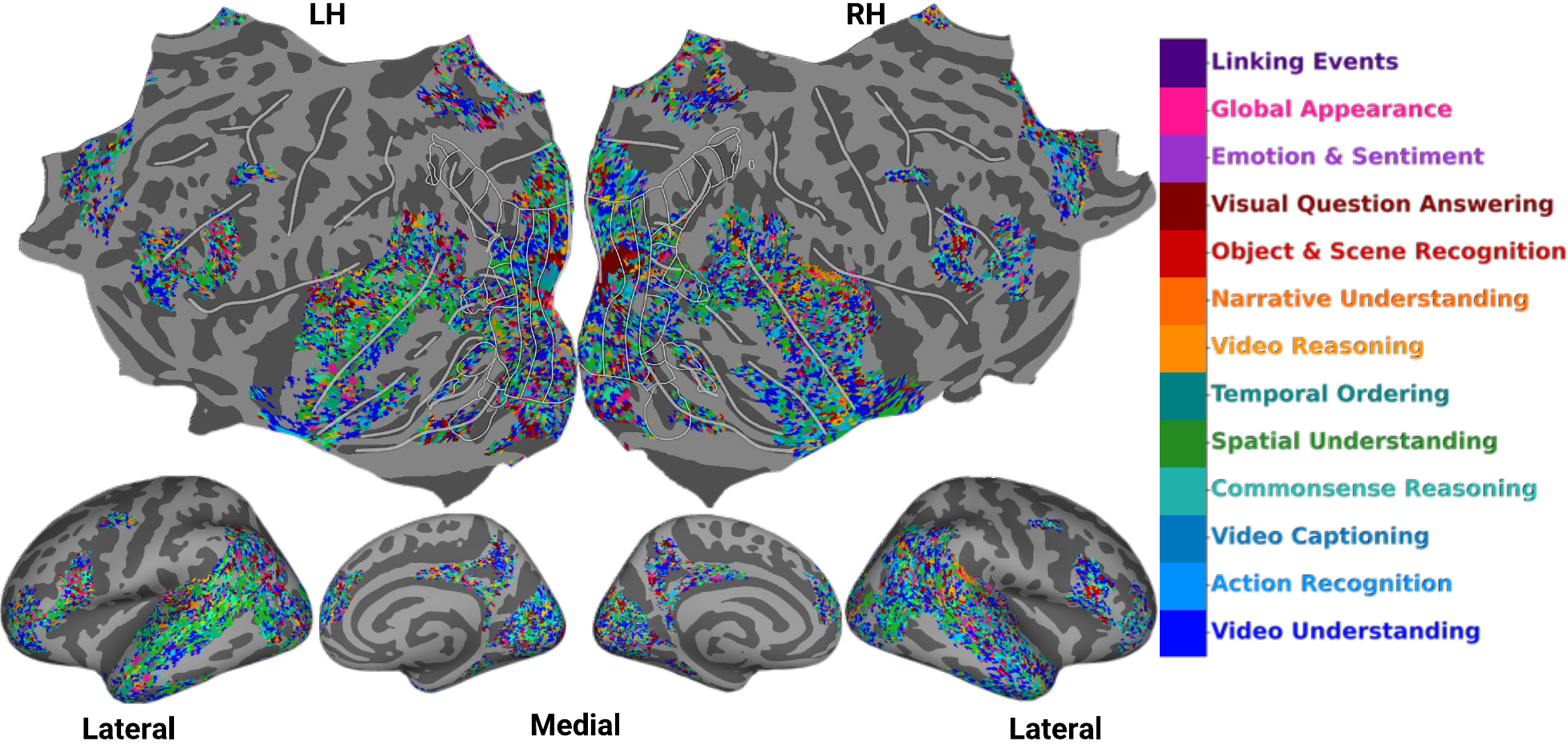}
    \\(a) InstructBLIPVideo \\
\end{minipage}
\hfill
\begin{minipage}{\textwidth}
\centering
    \includegraphics[width=0.8\linewidth]{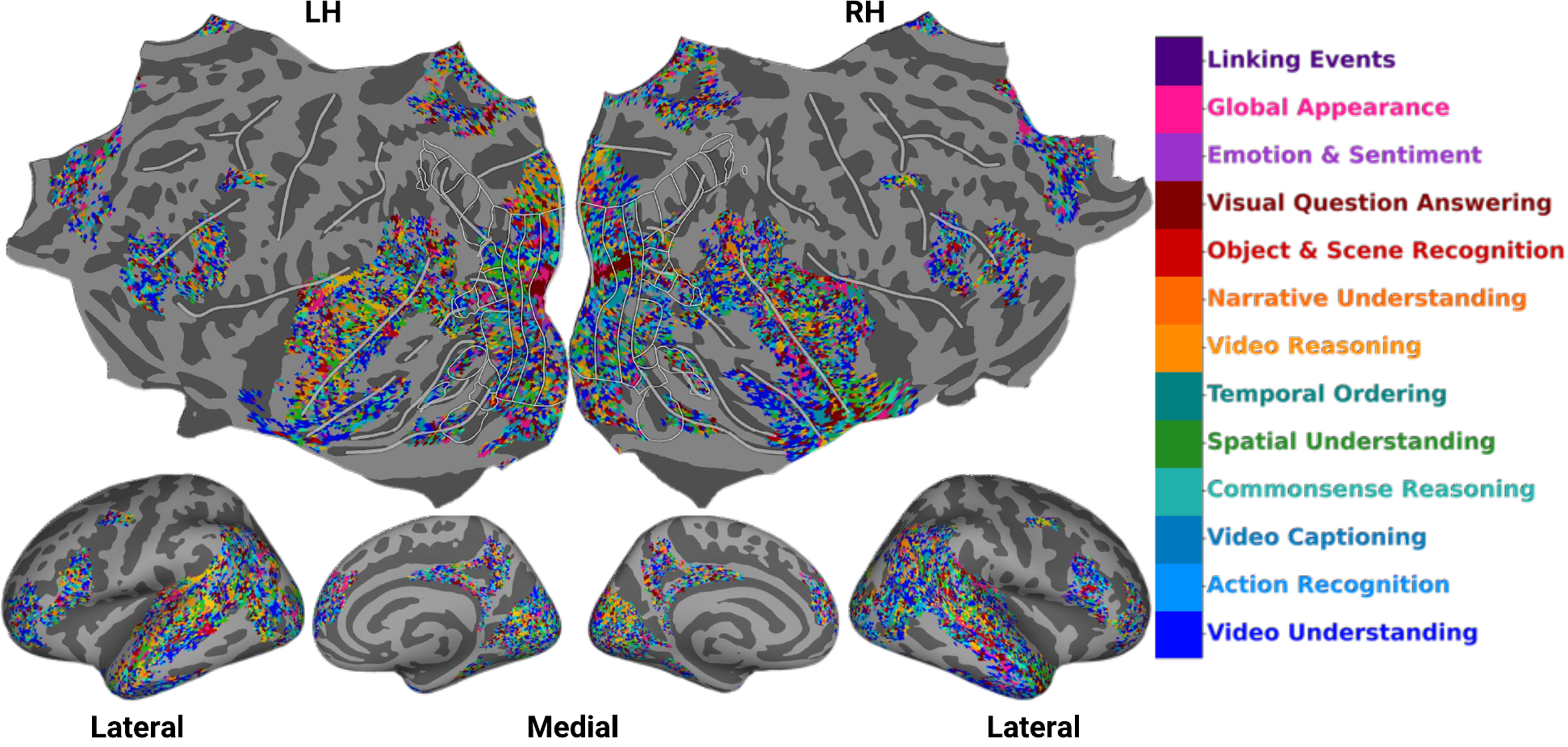}
    \\(b) Video-LLaVA \\
\end{minipage}
\hfill
\begin{minipage}{\textwidth}
\centering
    \includegraphics[width=0.8\linewidth]{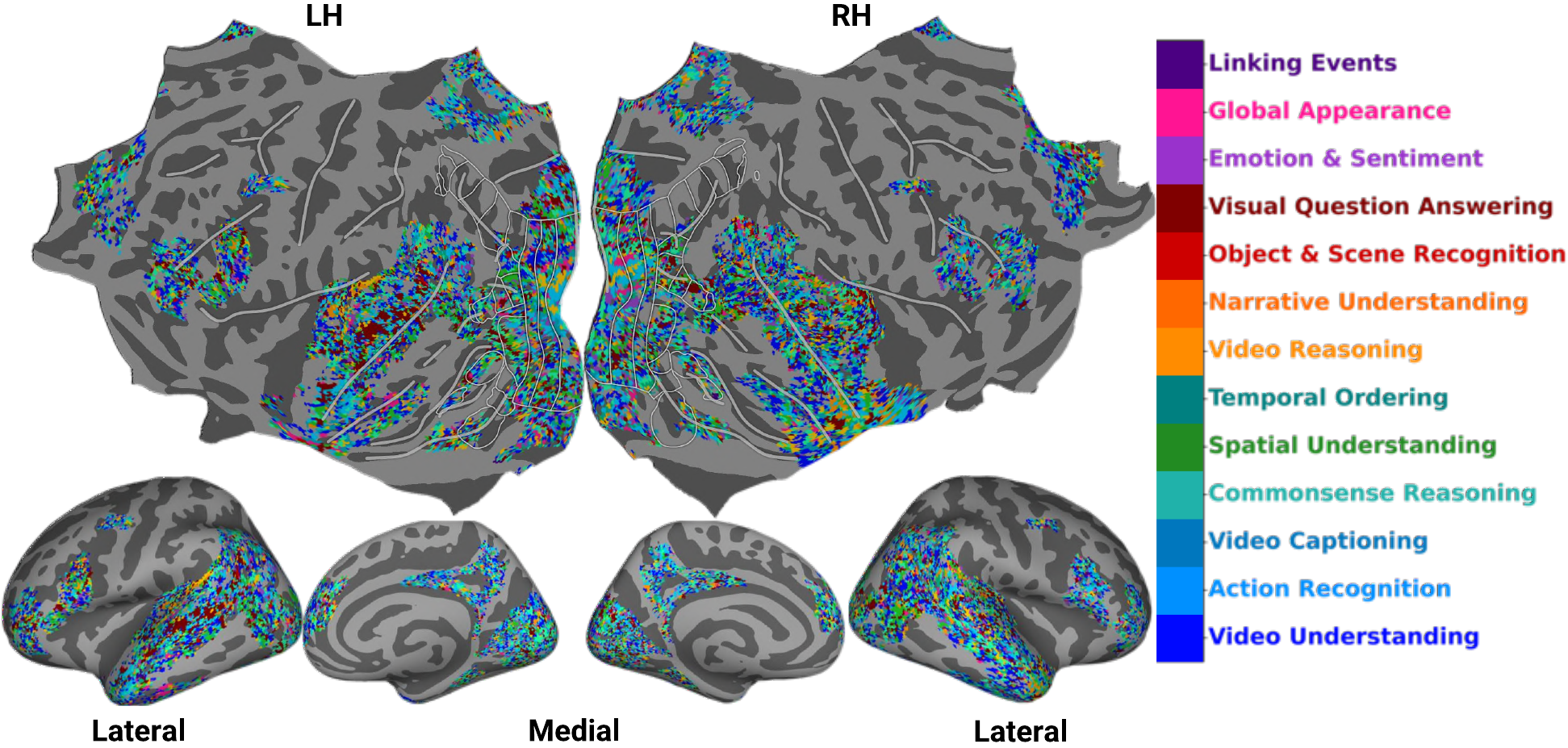}
    \\(c) LLaVA-NeXT-Video \\
\end{minipage}
    \caption{Each voxel is color coded with the instruction (out of 13) that led to the highest normalized brain alignment. The color bar highlights color codes for each instruction.
    % , with applied hex color codes for the 10 task instructions, 
    The voxels are projected onto the flattened cortical surface averaged across all 4 subjects for 3 video MLLM (InstructBLIPVideo, Video-LLaVA and LLaVA-NeXT-Video).}    \label{fig:video_instruction_brainmap_instructblip}
\end{figure*}

\begin{figure*}[!ht]
    \begin{minipage}{\textwidth}
\centering
    \includegraphics[width=0.72\linewidth]{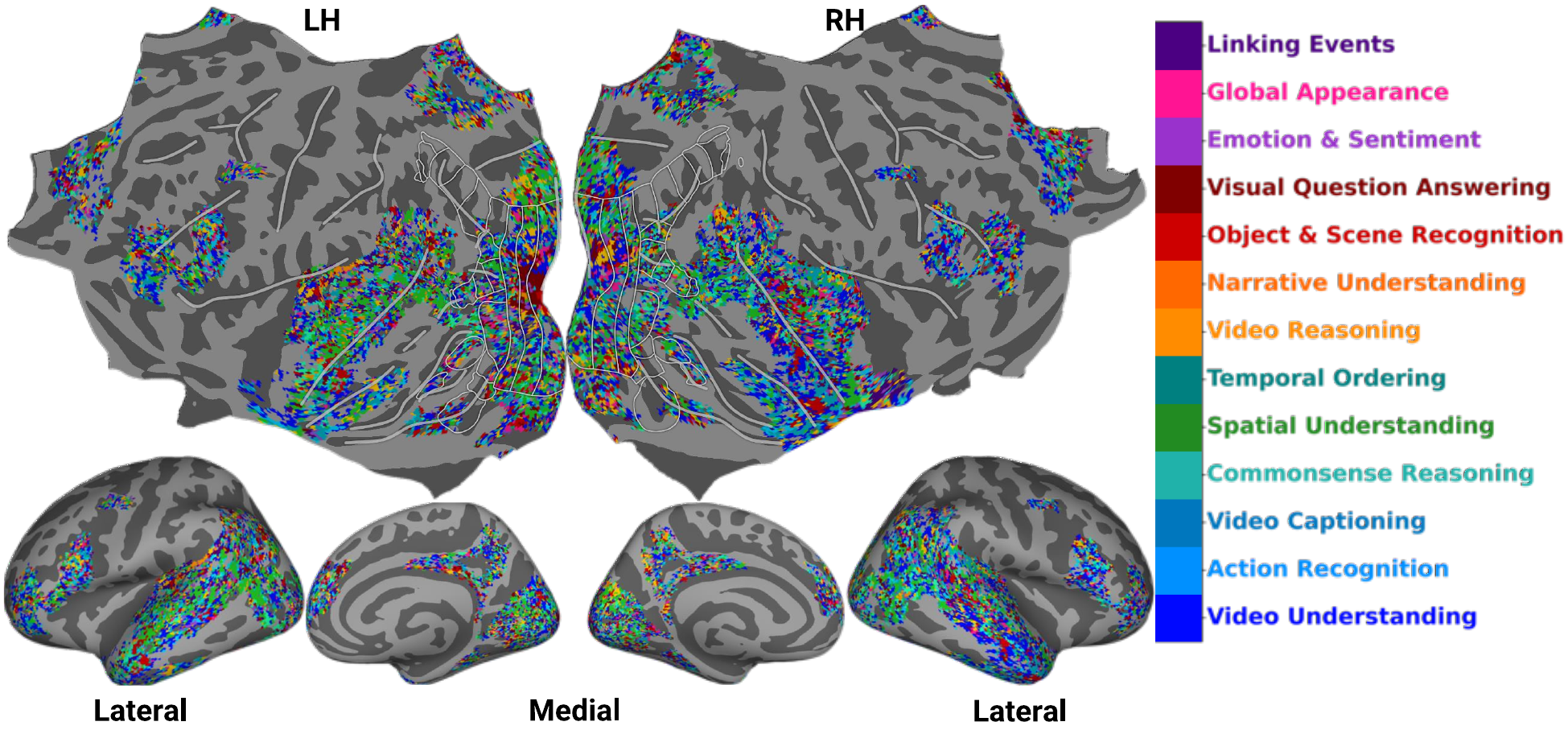}
    \\(a) LLaVA-OneVision \\
\end{minipage}
\hfill
\begin{minipage}{\textwidth}
\centering
    \includegraphics[width=0.72\linewidth]{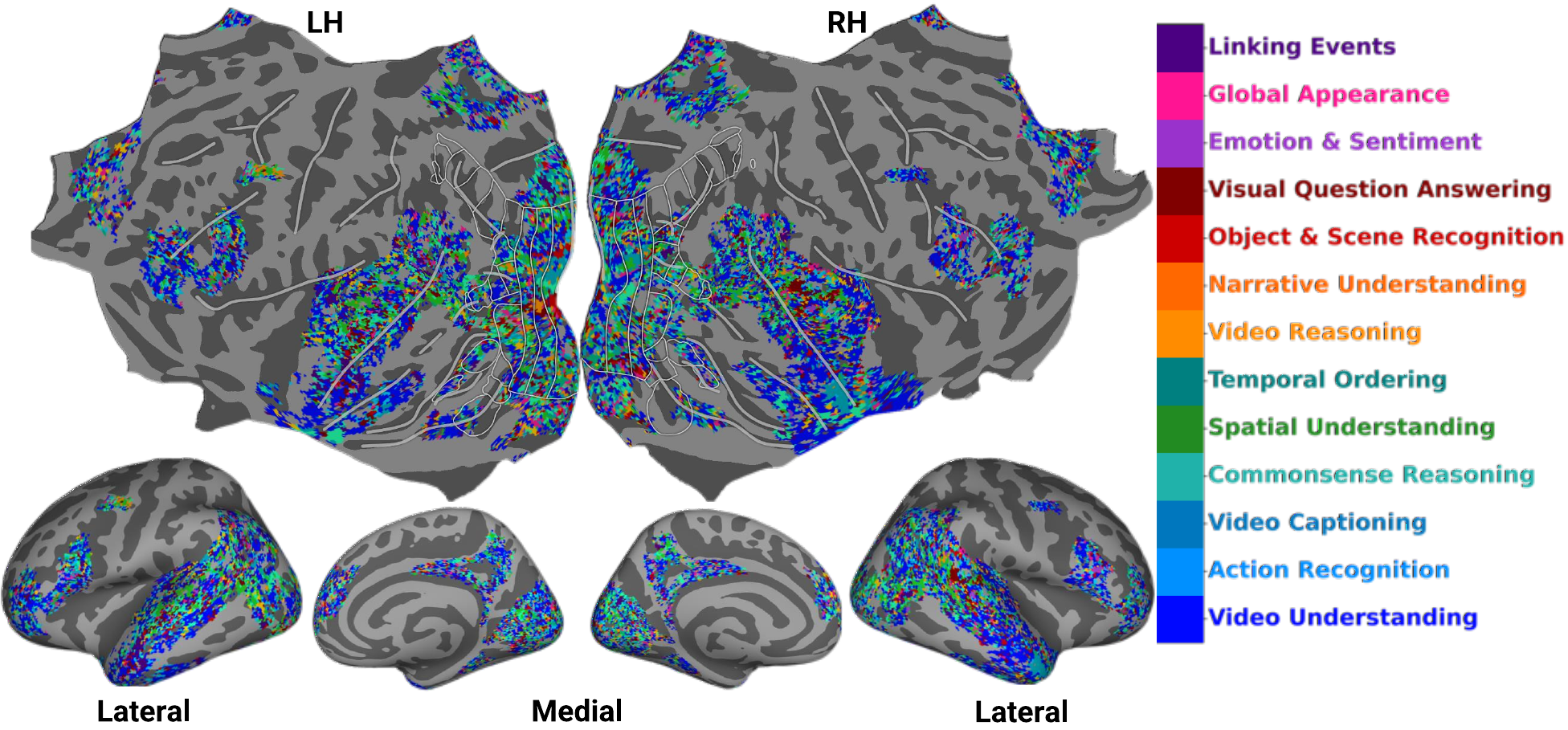}
    \\(b) VideoChat-R1 \\
\end{minipage}
    \caption{Each voxel is color coded with the instruction (out of 13) that led to the highest normalized brain alignment. The color bar highlights color codes for each instruction.
    % , with applied hex color codes for the 10 task instructions, 
    The voxels are projected onto the flattened cortical surface averaged across all 4 subjects for 2 video MLLM (LLaVA-OneVision, VideoChat-R1).}
    \label{fig:video_instruction_brainmap_llavaonevision}
\end{figure*}

Figures~\ref{fig:video_instruction_brainmap_instructblip} and~\ref{fig:video_instruction_brainmap_llavaonevision} show brain maps for InstructBLIPVideo, Video-LLaVA, LLaVA-NeXT-Video, LLaVA-OneVision and VideoChat-R1 for video tasks for average normalized brain predictivity across subjects where the voxel color codes are projected onto the flattened cortical surface of the `fsaverage' subject. The color-scheme corresponding to each instruction is also reported. We make the following observations: (i) Video understanding exhibits the strongest alignment across the whole brain.
(ii) Tasks such as spatial understanding, narrative understanding, and visual question answering show higher alignment in language-related regions, including the angular gyrus, posterior temporal lobe, and visual regions.
(iii) Higher-order language regions in the frontal cortex are predominantly identified by the video understanding task, with a smaller proportion of voxels also activated by video reasoning and temporal ordering tasks. 

\begin{figure*}[!ht]
\centering
    \includegraphics[width=0.72\linewidth]{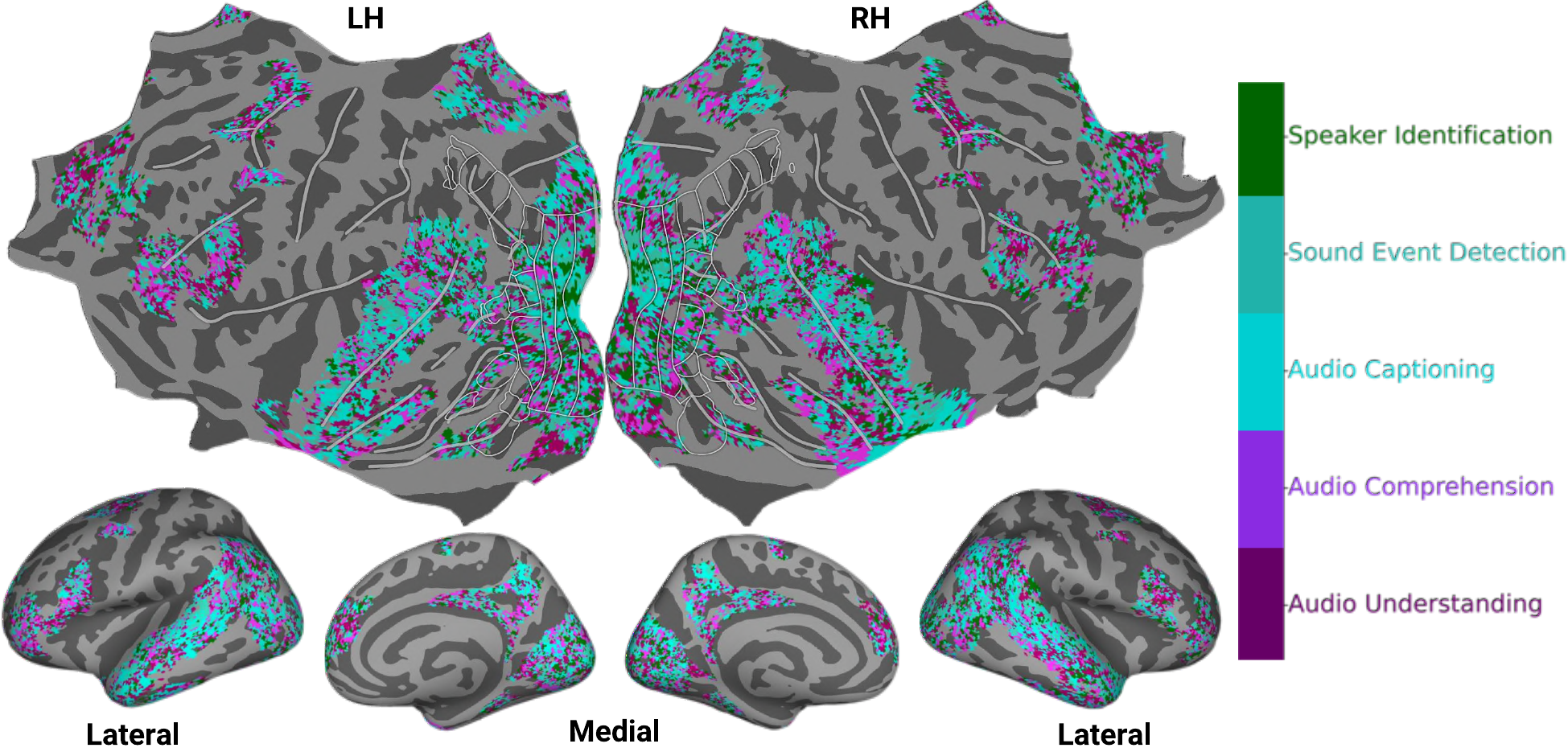}
    \caption{Kimi-Audio: Each voxel is color-coded with the instruction (out of 5) that led to the highest normalized brain alignment. The color bar highlights color codes for each instruction. The voxels are projected onto the flattened cortical surface of average across subjects on `fsaverage' surface.}
    \label{fig:audio_instruction_brainmap_kimi}
\end{figure*}

Fig.~\ref{fig:audio_instruction_brainmap_kimi} shows brainmap for audio instruction-tuned MLLM (Kimi-Audio) where the predictions are average across subjects. Here, the voxel color codes are projected onto the flattened cortical surface of the `fsaverage' subject. The figure shows a clear distinction between different audio tasks. 

% \begin{figure}[t]
%     \centering
%     \includegraphics[width=0.9\linewidth]{images/video_tasks_instruction_brainmap_videollava.pdf}
%     \caption{Each voxel is color coded with the instruction (out of 13) that led to the highest normalized brain alignment. The color bar highlights color codes for each instruction.
%     % , with applied hex color codes for the 10 task instructions, 
%     The voxels are projected onto the flattened cortical surface of a representative subject (S1) for video MLLM (Video-LLaVA).}
%     \label{fig:video_instruction_brainmap_videollava}
% \end{figure}

% \begin{figure}[t]
%     \centering
%     \includegraphics[width=0.9\linewidth]{images/video_tasks_instruction_brainmap_llavanext.pdf}
%     \caption{Each voxel is color coded with the instruction (out of 13) that led to the highest normalized brain alignment. The color bar highlights color codes for each instruction.
%     % , with applied hex color codes for the 10 task instructions, 
%     The voxels are projected onto the flattened cortical surface of a representative subject (S1) for video MLLM (LLaVA-Next).}
%     \label{fig:video_instruction_brainmap_llavanext}
% \end{figure}

\section{Brain Maps showing Layer-wise Details for Video Instruction-based MLLMs}
\label{app:layerwiseDetails}
\begin{figure*}[t]
\centering
\begin{minipage}{0.55\textwidth}
\centering
    \includegraphics[width=\linewidth]{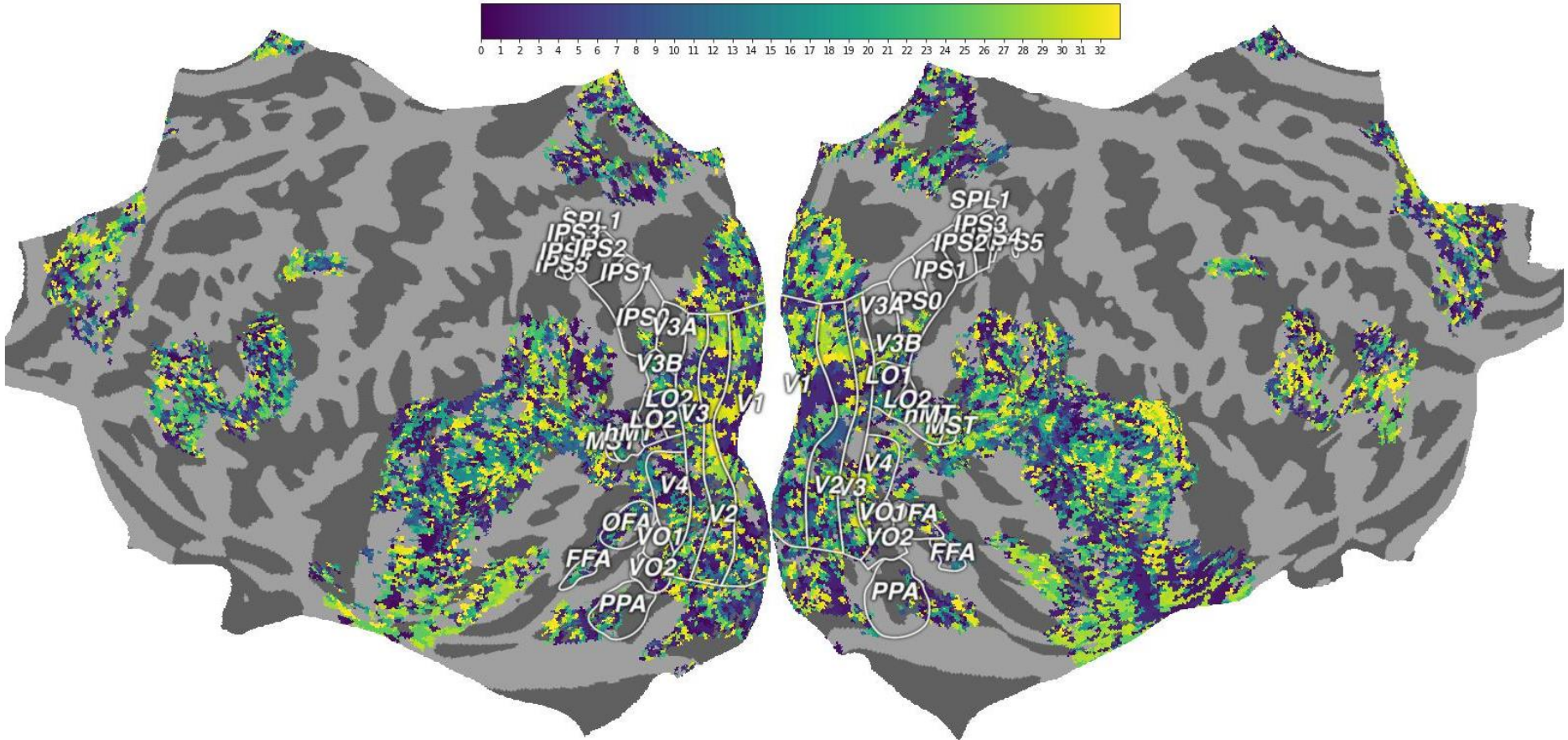}
    \\(a) InstructBLIPVideo\\
\end{minipage}
\hfill
\begin{minipage}{0.55\textwidth}
\centering
    \includegraphics[width=\linewidth]{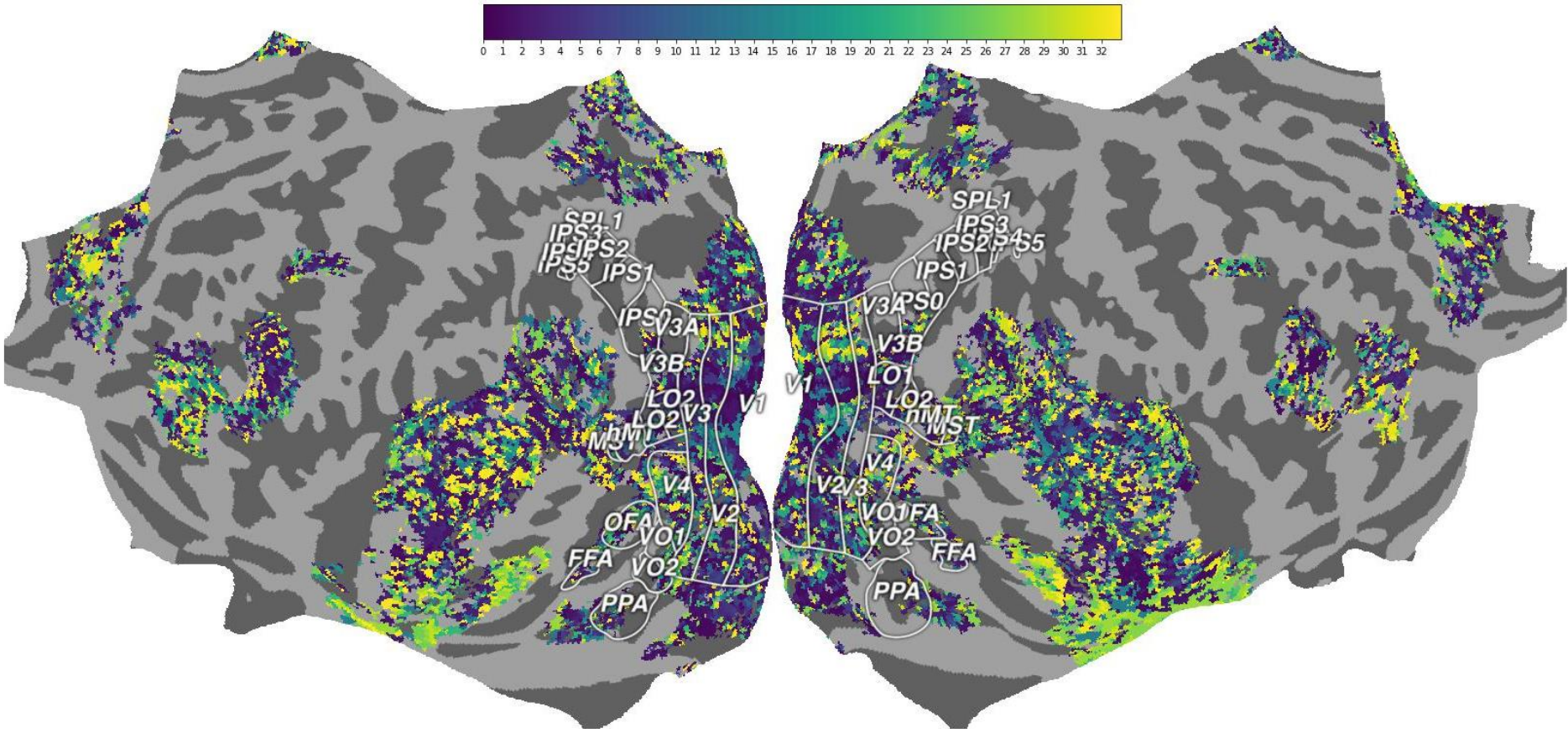}
    \\(b) Video-LLaVA\\
\end{minipage}
\hfill
\begin{minipage}{0.55\textwidth}
\centering
    \includegraphics[width=\linewidth]{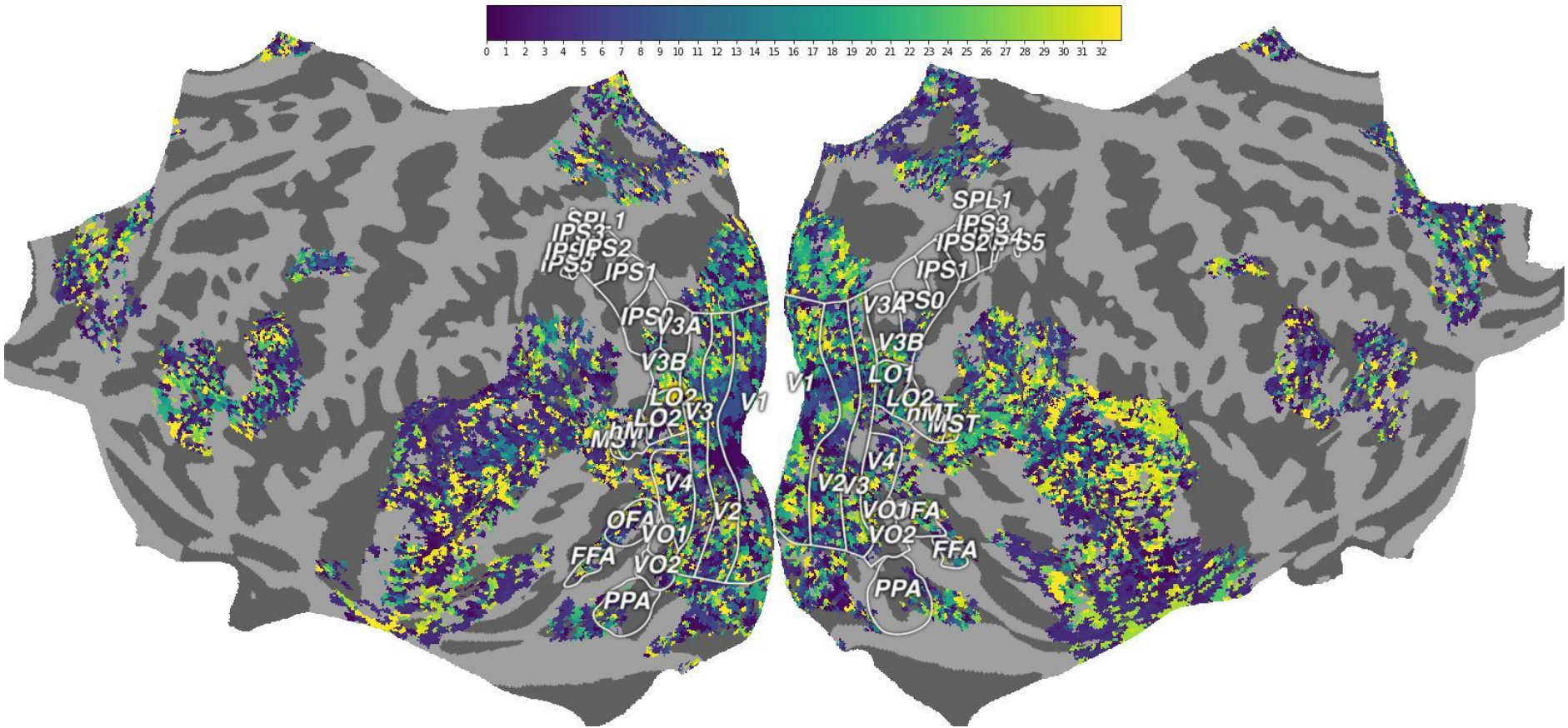}
    \\(c) LLaVa-NeXT-Video\\
\end{minipage}
\hfill
\begin{minipage}{0.55\textwidth}
\centering
    \includegraphics[width=\linewidth]{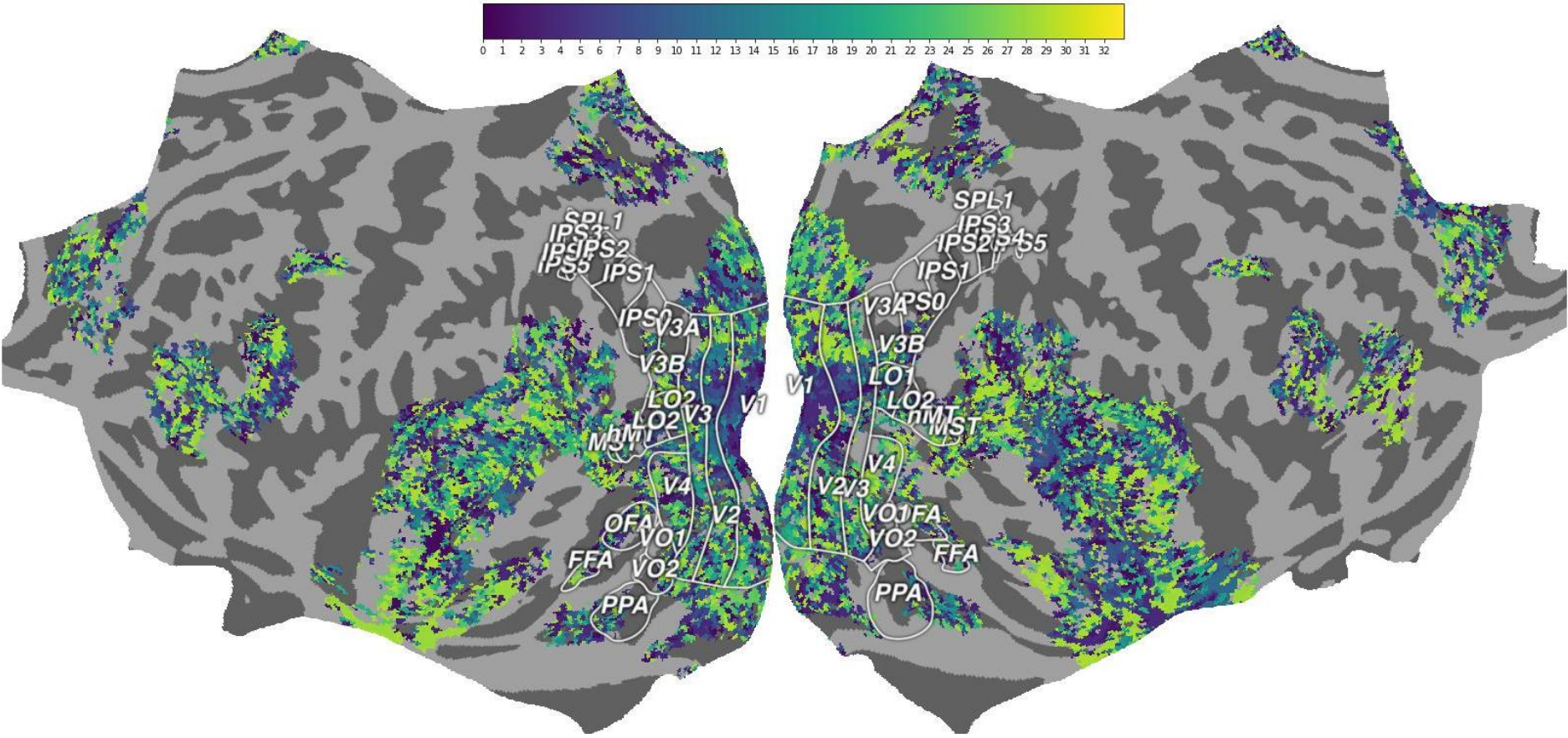}
    \\(d) LLaVA-OneVision\\
\end{minipage}
    \caption{Each voxel is color coded with the video MLLM layer number (out of 33) that led to the highest
normalized brain alignment. The color bar highlights color codes for each layer. The voxels are projected onto the flattened cortical surface of average across all 4 subjects on `fsaverage' surface for four MLLMs.}
    \label{fig:layers_instruction_brainmap}
\end{figure*}

To examine whether IT-MLLMs reflect the brain's hierarchy of information processing across layers, we analyze the voxels as follows. For each voxel, we select the layer that results in the
highest normalized brain alignment and apply a color code for the 29/33 layers across the various MLLMs. Fig.~\ref{fig:layers_instruction_brainmap} presents brain maps for four video MLLMs, where the voxels with their corresponding color codes are projected onto the flattened cortical surface of the `fsaverage' subject. 

\begin{figure*}[!ht]
\centering
\begin{minipage}{0.5\textwidth}
\centering
    \includegraphics[width=\linewidth]{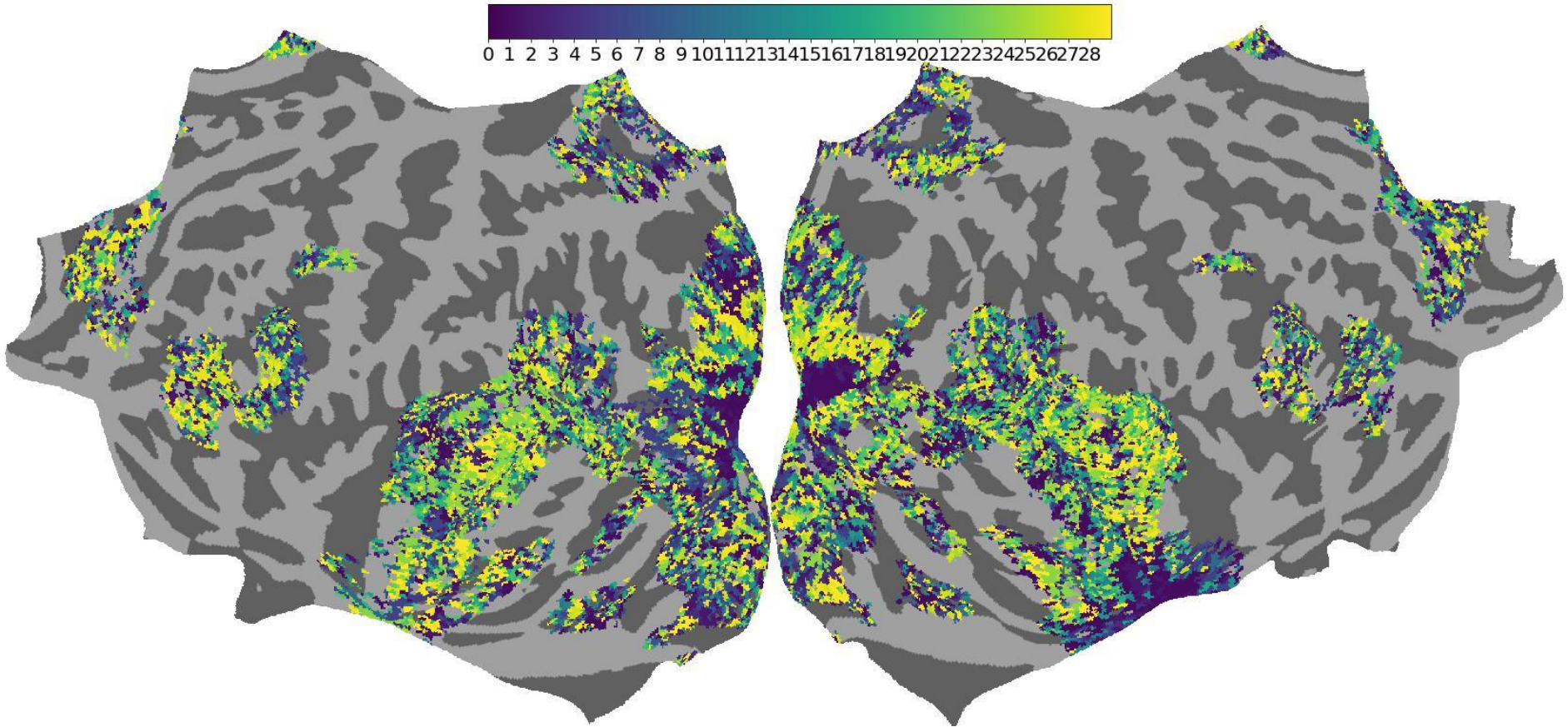}
    \\(a) Qwen-2.5-VL-Audio\\
\end{minipage}
    \caption{Each voxel is color coded with the video MLLM layer number (out of 29) that led to the highest
normalized brain alignment. The color bar highlights color codes for each layer. The voxels are projected onto the flattened cortical surface of average across all 4 subjects on `fsaverage' surface for four MLLMs.}
    \label{fig:layers_instruction_brainmap_audio}
\end{figure*}

\section{Validation of ``hierarchical correspondence'' between model layers and brain regions}
\label{app:layer_hierarchy_qwen}

\begin{figure*}[!ht] 
\centering
\begin{minipage}{0.49\textwidth}
\centering
    \includegraphics[width=0.9\linewidth]{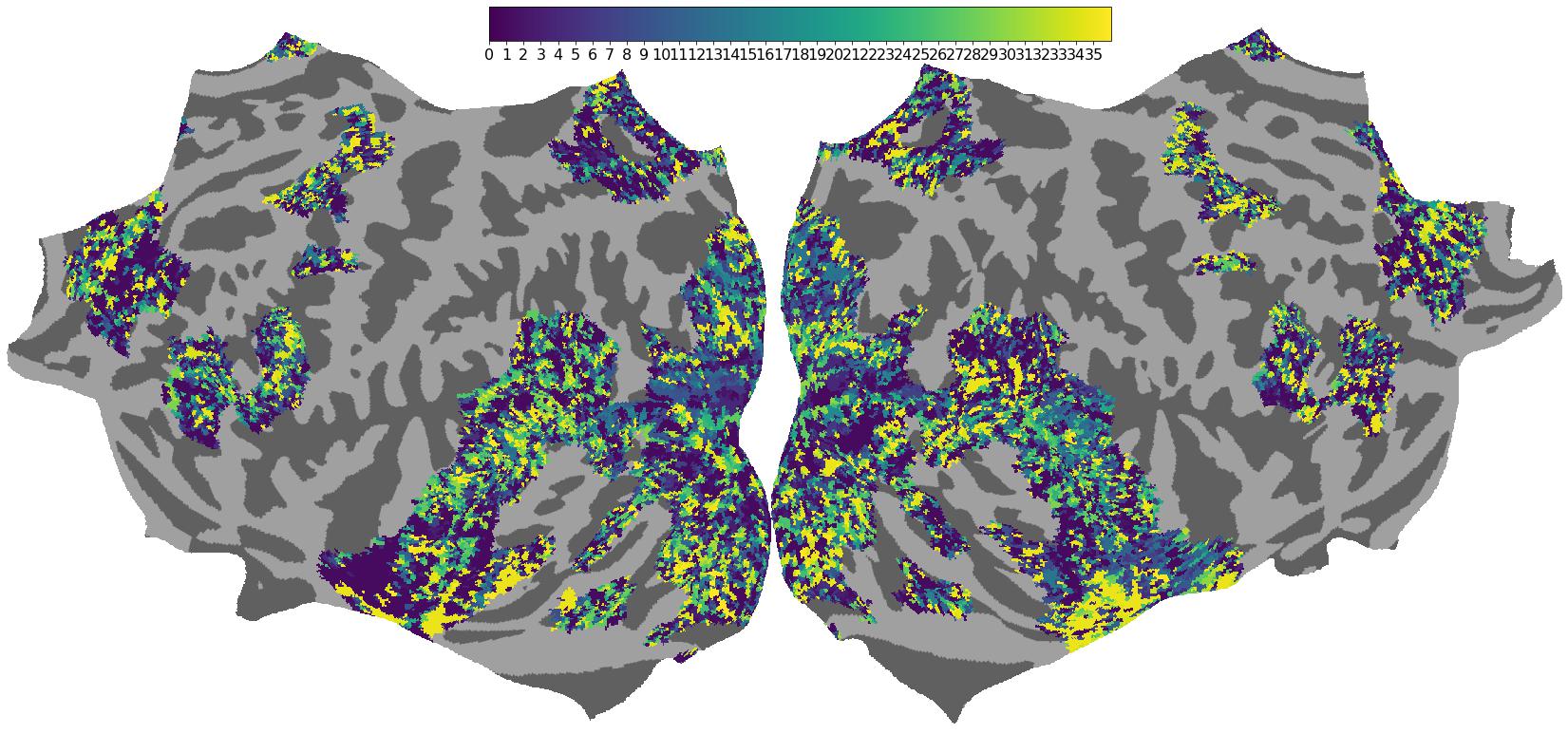}
    \\(a) Qwen-2.5-VL-3B Instruct \\
\end{minipage}
\begin{minipage}{0.49\textwidth}
\centering
    \includegraphics[width=0.9\linewidth]{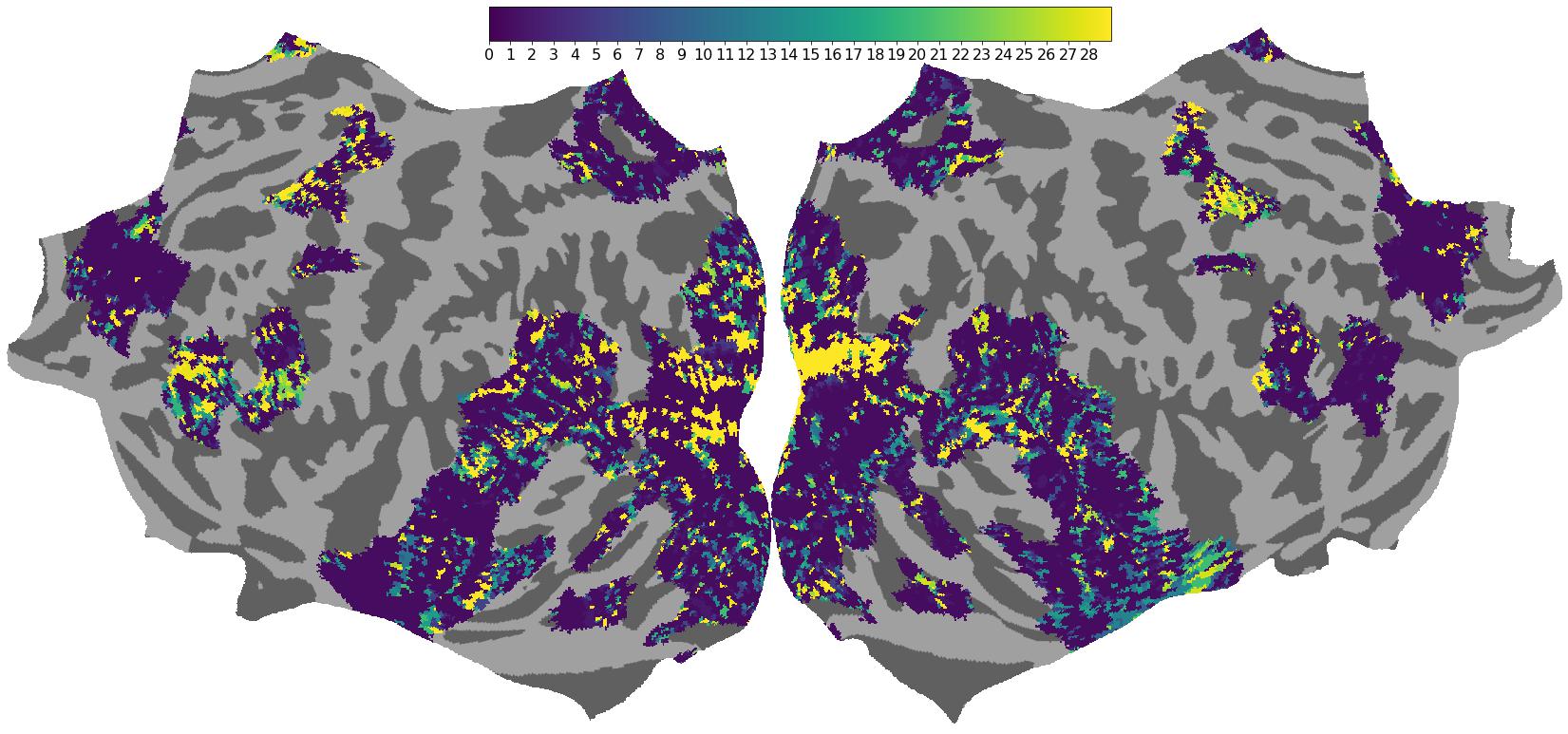}
    \\(b) Qwen-2.5-VL-7B Instruct non-natural language prompt \\
\end{minipage}
\caption{ (a) Qwen-2.5-VL-3B Instruct and (b) Qwen-2.5-VL-7B Instruct non-natural language prompt (layer-wise alignment): Each voxel is color
coded with the MLLM layer number (out of 29) that led to the highest normalized brain alignment.
The color bar highlights color codes for each layer. The voxels are projected onto the flattened
cortical surface of average across subjects on ‘fsaverage’ surface.}
\label{fig:qwen-2.5-vl_hierarchy_3b_random}
\end{figure*}

We compare hierarchical correspondence between model layers and brain regions in two settings: (i) Scaling the instruction-tuned model with correct video prompt: Qwen-2.5-vl-3B Instruct vs. Qwen-2.5-vl-7B Instruct, (ii) an instruction-tuned model with a non-natural language prompt, e.g.: \#\#\#\#\# \#\#\#\# \#\#\#\# \#\#\#\#\# this does not provide any meaning while passing video as input. Similar to Qwen-2.5-VL-7B Instruct model, we perform qualitative analysis by considering each voxel is color coded with the MLLM layer number (out of 29) that led to the highest normalized brain alignment. Fig~\ref{fig:qwen-2.5-vl_hierarchy_3b_random} shows the resulting brain maps for the 3B-Instruct model (a) and for the 7B-Instruct model with the non-language prompt (b). We make the following observations: (i) 
Non-language control (random prompt): There is no hierarchy of information processing observed across layers, where the brain prediction is dominated by early layers across cortex
(ii) Instruction condition (3B model): The hierarchical pattern still remains the same i.e., a systematic early to mid/late gradient across cortex, i.e., the same hierarchical pattern we observed with the larger model.

\begin{figure*}[!ht]
    \centering
    \includegraphics[width=0.6\linewidth]{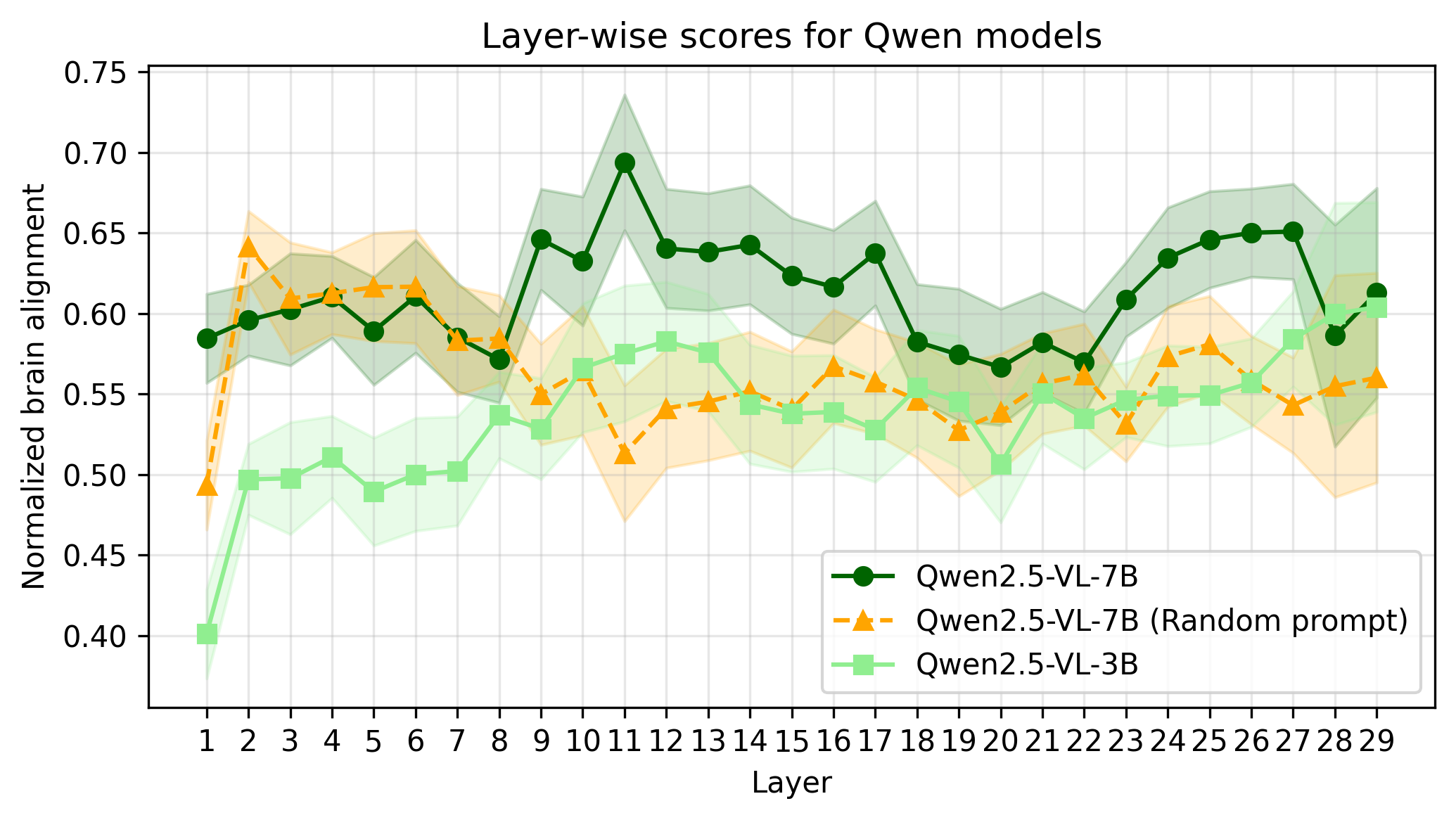}
    \caption{Layerwise normalized brain alignment for Qwen models.}
    \label{fig:qwen_layer_hieararchy}
\end{figure*}

We also perform quantitative analysis to show a layerwise normalized brain alignment across three models, as shown in Fig.~\ref{fig:qwen_layer_hieararchy}. Instruction prompts (Qwen-2.5-VL-7B; green solid) show a clear hierarchy: alignment rises from early layers and is strongest in mid to late layers. A non-language control (same model, orange dashed) flattens toward early layers, indicating that the hierarchy depends on instruction semantics rather than generic prompting. The smaller 3B model (light-green squares) exhibits the same shape with lower amplitude, demonstrating scale robustness of the hierarchy. Overall, we observe a significant shift in preferred layer under instruction vs. non-language, where a positive layer-trend is only observed for valid natural language instructions.

\FloatBarrier

\section{Self-Controlled Experiments: Comparing Models Before and After Instruction Tuning}
\label{app:before_after_tuning}

Comparing the same backbone before and after instruction tuning is the most direct way to reduce architectural confounds when assessing differences in brain alignment. Similar to InternVL-8B, we use same Qwen-2.5 family and the same scale (7B): Qwen-2.5-VL-7B-Instruct and Qwen-2.5-VL-7B models, so that both models receive the same visual input. This keeps architecture family, parameter size, and modality comparable, and highlights the effect of instruction tuning.  For both models, we issue the same instruction (``Describe the video'') and extract the resulting instruction-specific representations across layers. This analysis shows how instruction tuning on the same model category changes layer-wise representations and improves brain alignment.

%To clarify, we have already reported the corresponding comparison in the original draft. For clarity, we provided the same result in the paper that compares the same model category with instruction-tuned and pretrained versions. 

% We use the pretrained InternVL-8B model already reported in the paper and compare it to its instruction-tuned variant InternVL-8B-Instruct. For both models, we issue the same instruction (``Describe the video'') and extract the resulting instruction-specific representations across layers. This analysis shows how instruction tuning on the same model category changes layer-wise representations and improves brain alignment.

\noindent\textbf{Qwen-2.5-VL: before vs. after instruction tuning:} Using brain predictions across layers for Qwen-2.5-VL-7B-Instruct and Qwen-2.5-VL-7B models, we compute per-layer $\Delta$ alignment as

$\Delta$ alignment per layer = Normalized brain alignment (Qwen-2.5-VL-7B-Instruct) - Normalized Brain alignment (Qwen-2.5-VL-7B)

\begin{figure*}[!ht] 
\centering
\begin{minipage}{\textwidth}
\centering
\includegraphics[width=0.6\linewidth]{images/qwen_alignment_curves.pdf}
    \includegraphics[width=0.6\linewidth]{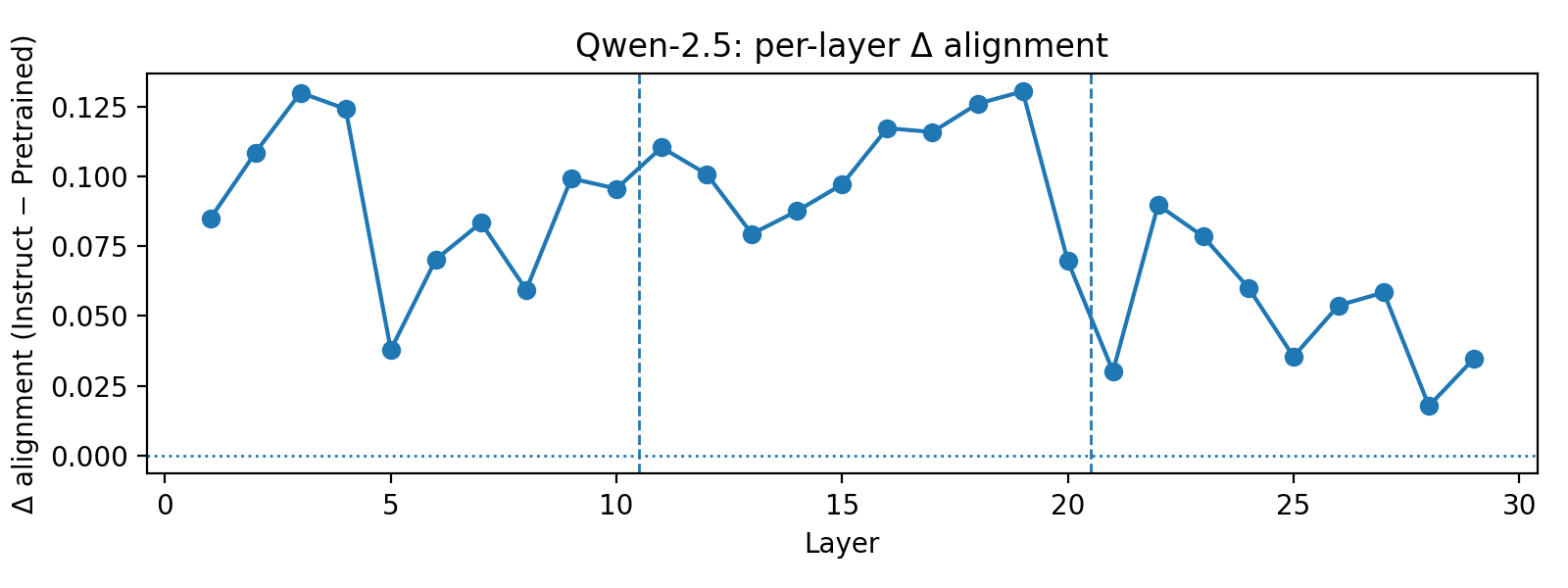}
\end{minipage}
\vspace{-0.5cm}
\caption{Qwen-2.5-VL: (top) Normalized brain alignment was computed before vs. after instruction tuning: Using brain predictions across layers for Qwen-2.5-VL-7B-Instruct and Qwen-2.5-VL-7B models. (bottom) $\Delta$ alignment per layer:= Normalized brain alignment (Qwen-2.5-VL-7B-Instruct) -  Normalized Brain alignment (Qwen-2.5-VL-7B)
}.
\label{fig:qwen2.5_vl_omni}
\end{figure*}

We make the following observations: (i) All layers improve (all $\Delta > $0), (ii) Depth trend: $\Delta$ decreases with depth (early $>$ mid $>$ late): early 0.177, mid 0.156, late 0.139; Spearman $\rho$ = -0.77, p $\ll$ 0.001. (ii) The center-of-mass of the improvement is near Layer 14/29 (normalized 0.47). Thus, while the absolute alignment peak under instruction occurs in mid to late layers (Fig.~\ref{fig:qwen2.5_vl_omni}), the incremental benefit over the pretrained baseline is strongest early and remains positive throughout depth.

% \noindent\textbf{InternVL: before vs. after instruction tuning:} Using brain predictions across layers for InternVL-8B-Instruct and InternVL-8B (pretrained) models, we compute per-layer $\Delta$ alignment, as shown in Fig.~\ref{fig:internvl_before_after}.  We make the following observations: (i) Similar to Qwen series, all layers positive: min $\Delta$ = 0.0557 (L5), max $\Delta$ = 0.1396 (L1), (ii)  Depth trend: gains increase into mid/late (early 0.0934 $\pm$ 0.0250, mid 0.0961 $\pm$ 0.0128, late 0.1100 $\pm$ 0.0104); Spearman $\rho \approx$ +0.34, p $\approx$ 0.07. (iii) The center-of-mass of the improvement is near Layer 15/29 (normalized 0.535). This implies that Instruction tuning yields mid/late-layer improvements for InternVL.

Table.~\ref{tab:self_controlled_delta} shows quantitative analysis of $\Delta$ alignment across families, we make following observations: (i) instruction tuning preserves the same alignment hierarchy (mid$\rightarrow$late layers peak) and shifts preferred processing toward later layers (positive shift in preferred layer), (ii) the locus of the gain is family-specific: Qwen shows larger early-layer increases, whereas InternVL shows mid/late-layer increases—consistent with non-uniform representational shifts rather than a uniform shift.

\begin{figure*}[!ht] 
\centering
\includegraphics[width=0.6\linewidth]{images/internvl_alignment_curves.pdf}
    \includegraphics[width=0.6\textwidth]{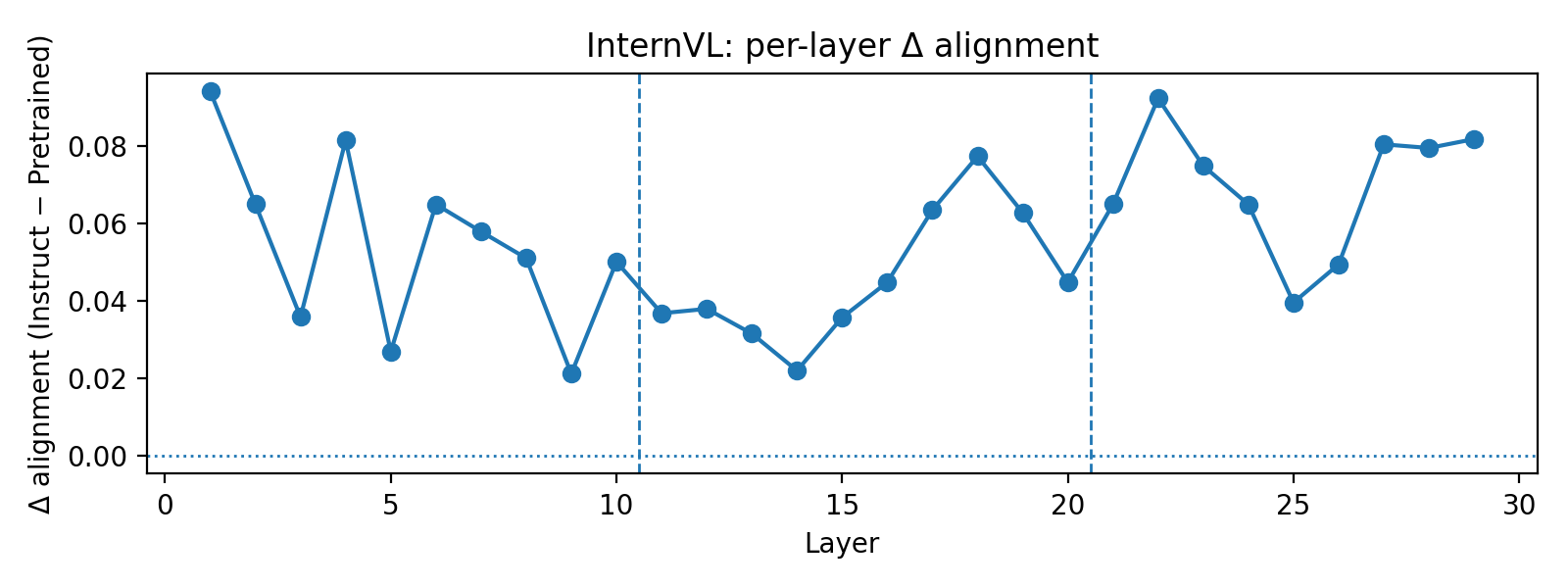}
\vspace{-0.2cm}
\caption{InternVL: Normalized brain alignment was computed before vs. after instruction tuning: Using brain predictions across layers for InternVL-8B-Instruct and InternVL-8B models.}
\label{fig:internvl_before_after}
\end{figure*}

\begin{table*}[!ht]
\centering
\scriptsize
\tabcolsep1pt
\caption{Self-controlled $\Delta$ alignment (Instruction $-$ Pretrained).
Early = Layers 1--10, Mid = 11--20, Late = 21--29 (mean$\pm$SD). 
COM is center-of-mass of $\Delta$ across layers (normalized by depth).
$\rho$ is Spearman correlation between layer index and $\Delta$.}
\label{tab:self_controlled_delta}
\resizebox{\textwidth}{!}{\begin{tabular}{|l|c|c|c|c|c|c|c|c|c|c|}
\hline
\textbf{Model} & 
\textbf{Early $\Delta$ } &
\textbf{Mid $\Delta$ } &
\textbf{Late $\Delta$ } &
\textbf{COM (norm)} &
\textbf{$\rho$ (layer,$\Delta$)} &
\textbf{p-value} &
\textbf{Min $\Delta$ (layer)} &
\textbf{Max $\Delta$ (layer)} &
\textbf{All $\Delta{>}0$} \\
\hline
Qwen-2.5-VL-7B &
$0.177\pm0.020$ &
$0.156\pm0.015$ &
$0.139\pm0.023$ &
0.474 &
$-0.767$ &
$<1\mathrm{e}{-6}$ &
$0.088\ (\mathrm{L}29)$ &
$0.207\ (\mathrm{L}1)$ &
\checkmark \\
InternVL-8B &
$0.093\pm0.025$ &
$0.096\pm0.013$ &
$0.110\pm0.010$ &
0.535 &
$+0.337$ &
0.074 &
$0.056\ (\mathrm{L}5)$ &
$0.140\ (\mathrm{L}1)$ &
\checkmark \\
\hline
\end{tabular}}
\end{table*}

\section{Details of Semantic Task Group Analysis}
\label{app:semTaskGroupAnalysis}
\begin{figure*}[!ht]
    \centering
    \includegraphics[width=0.49\linewidth]{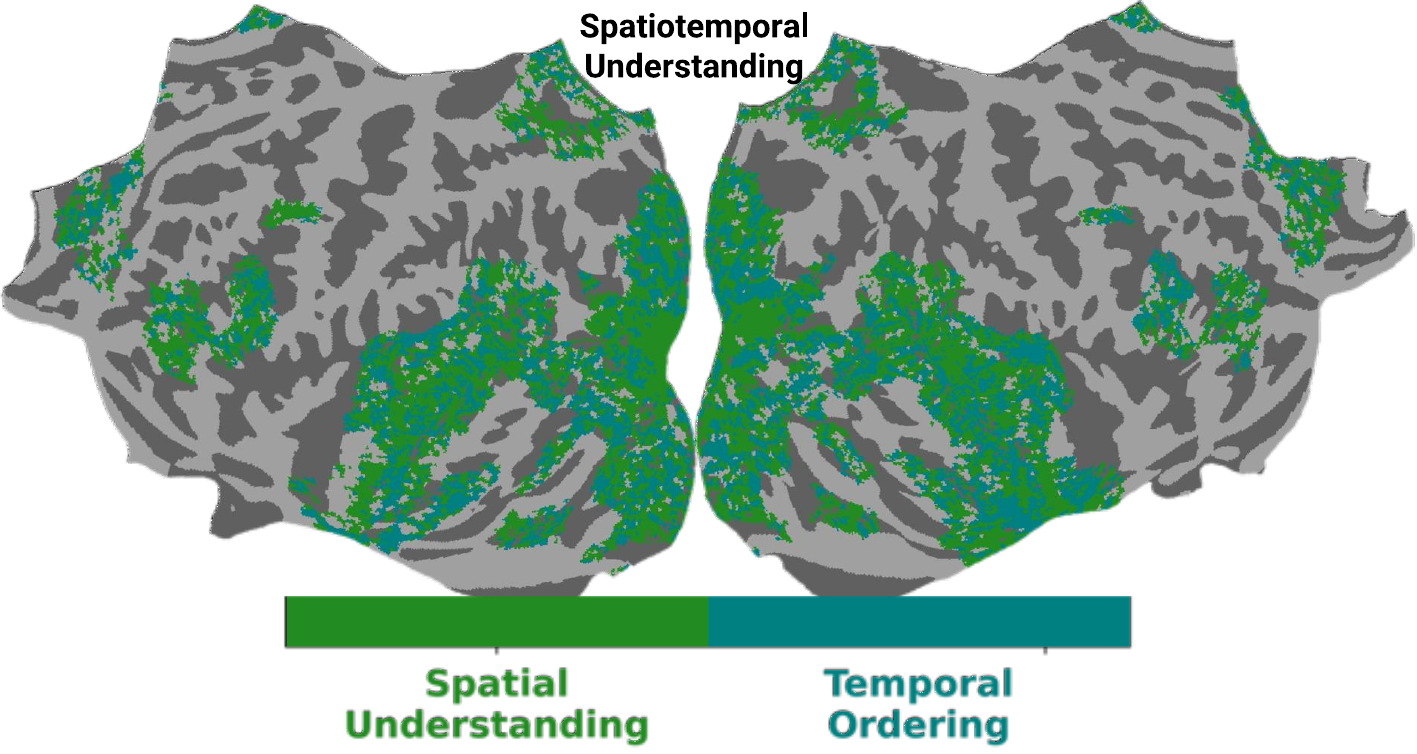}
    \includegraphics[width=0.49\linewidth]{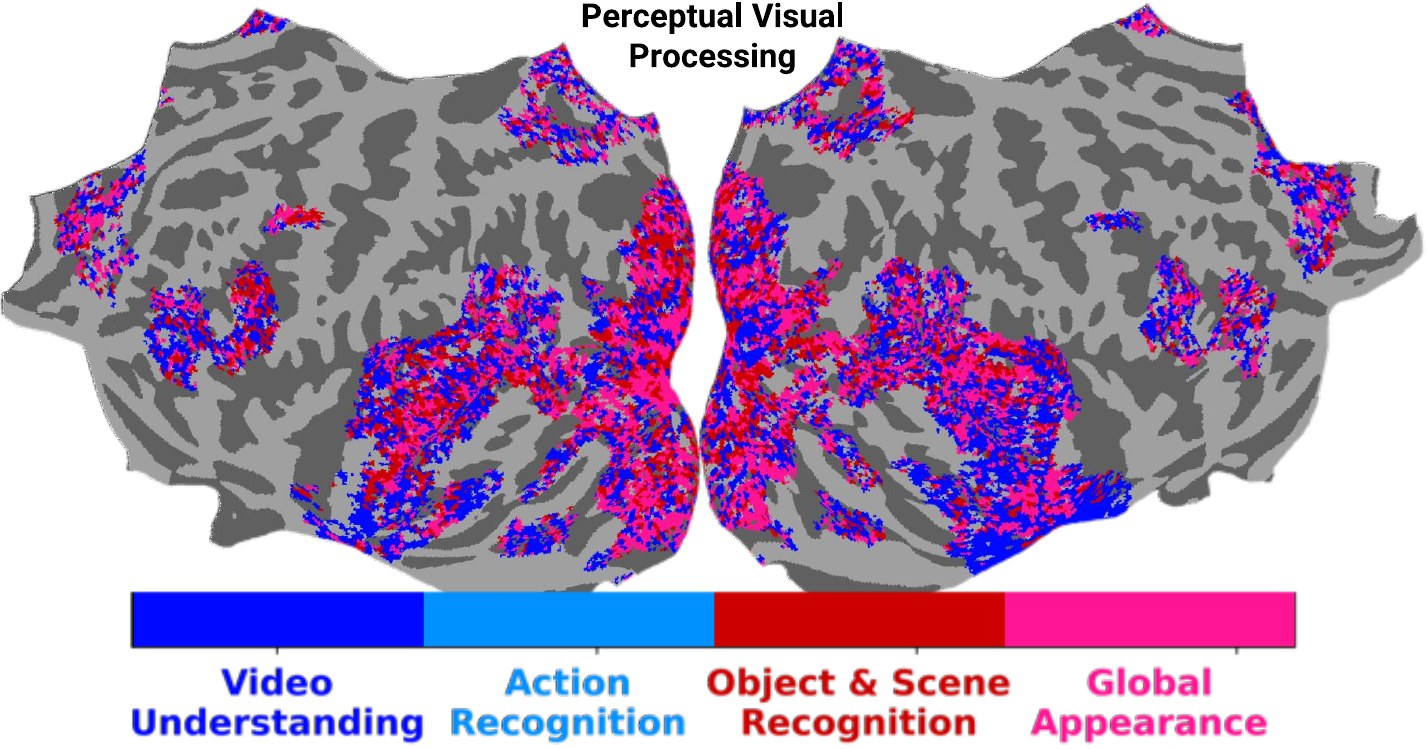}
    \includegraphics[width=0.49\linewidth]{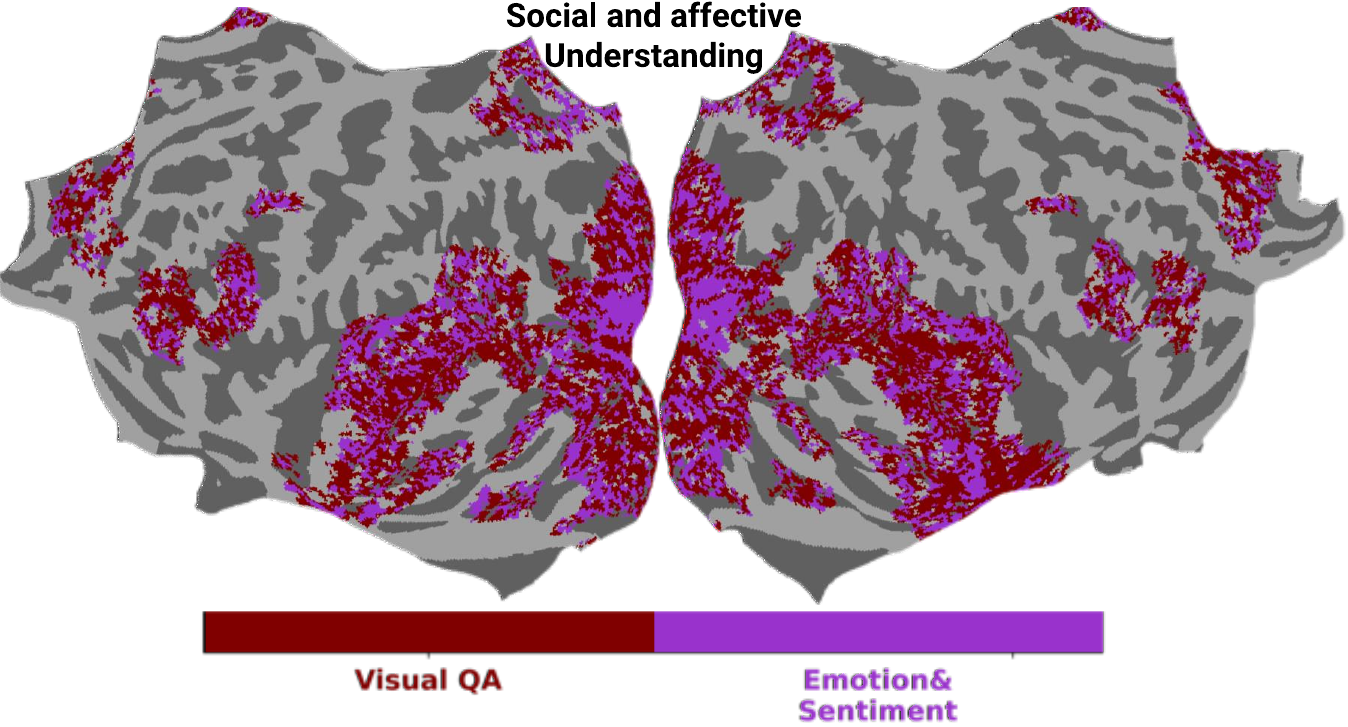}
    \caption{Semantic Task Group Analysis: Each voxel is color coded with the task instruction that led to the highest normalized brain alignment. The color bar highlights color codes for each instruction.
    % , with applied hex color codes for the 10 task instructions, 
    The voxels are projected onto the flattened cortical surface averaged across all subjects for video MLLM (Qwen-2.5-VL). While this plot shows brain maps for 2 groups, brain maps for remaining 3 task groups are in Fig.~\ref{fig:qwen_instruction_brainmap_disspciation} in Section~\ref{sec:semTaskGroupResults} in the main paper.} \label{fig:qwen_instruction_brainmap_disspciation2}
\end{figure*}

To further examine how instruction-tuned video MLLMs generate task-specific representations and reveal functional specialization in the brain, we group the 13 video tasks into five cognitively grounded categories: Perceptual visual processing, Cognitive reasoning and integration, Spatiotemporal understanding, High-level language and narrative understanding, and Social and affective understanding. 
This categorization allows us to disentangle the functional specificity of brain regions engaged by different task types. The visualizations in Fig.~\ref{fig:qwen_instruction_brainmap_disspciation} in Section~\ref{sec:semTaskGroupResults} in the main paper and Fig.~\ref{fig:qwen_instruction_brainmap_disspciation2} illustrate that this grouping captures meaningful distinctions. 

\noindent\textbf{Spatiotemporal understanding.} Temporal ordering elicits more widespread activation in the angular gyrus and posterior temporal lobe, whereas spatial understanding shows stronger engagement in the dorsal parietal cortex (part of the dorsal visual pathway) and anterior temporal lobe~\citep{zacks2007event,baldassano2017discovering}. Additionally, we observe that early visual areas are more active during the spatial understanding task, whereas early auditory cortex shows higher activity in the temporal ordering task, likely due to its role in processing sound-based events~\citep{belin2000voice}. However, the brain does not strictly separate spatial and temporal processing. These representations often co-exist, particularly in narrative and event-based cognition.
%Tasks such as spatial understanding and temporal ordering fall under the category of Spatiotemporal Understanding. As shown in Fig.~\ref{fig:qwen_instruction_brainmap_disspciation}, temporal ordering elicits more widespread activation in the angular gyrus and posterior temporal lobe, whereas spatial understanding shows stronger engagement in the dorsal parietal cortex and anterior temporal lobe~\citep{zacks2007event,baldassano2017discovering}. Additionally, we observe that early visual areas are more active during the spatial understanding task, whereas early auditory cortex shows higher activity in the temporal ordering task, likely due to its role in processing sound-based events~\citep{belin2000voice}. However, the brain does not strictly separate spatial and temporal processing. These representations often co-exist, particularly in narrative and event-based cognition.

\section{Details of explained variance partitioning}
\label{app:sharedVarianceDetails}

To further quantify how IT-MLLMs capture shared and distinct neural processes across tasks, we use a variance partitioning approach. This analysis reveals the unique and overlapping contributions of individual task-specific representations to brain responses, enhancing our understanding of how brain regions differentially process multimodal information.

\noindent\textbf{Variance partitioning.}
To disentangle task-specific instruction representations from multimodal instruction-tuned models, we used a variance partitioning approach~\citep{de2017hierarchical,lebel2021voxelwise}. This method measures the overlap in brain variance explained by different task-specific instruction representations. Specifically, variance partitioning separates the brain response variance that can be attributed to two models based on their unique and overlapping contributions~\citep{vaidya2022self,deniz2019representation}. To perform this, for every pair of instruction representations, we fit separate encoding models for each space as well as a joint encoding model, obtained by concatenating the features. Using set arithmetic, we can then derive the size of the intersection $(NBA)^{1\cap 2}_{v}$=$(NBA)^{1}_{v}$+$(NBA)^{2}_{v}$-$(NBA)^{1\cup 2}_{v}$, where NBA refers to normalized brain alignment, $v$ refers to a specific voxel, $(NBA)^{1}_{v}$ denotes alignment of model 1, $(NBA)^{2}_{v}$ denotes alignment of model 2 and $(NBA)^{1\cup 2}_{v}$ denotes alignment of the joint model. Similarly, the unique contribution of model 1's feature space is computed as $(NBA)^{1 \backslash 2}_{v}$=$(NBA)^{1}_{v}$-$(NBA)^{1\cap 2}_{v}$.

\noindent\textbf{Partitioning explained shared and unique variance between task-specific instructions}
%\vspace{-0.2cm}

While the previous analysis reveals that task-specific instructions from MLLMs modulate their representations based on distinct cognitive demands, we further examine the representations of task-specific instructions to measure the overlap in brain variance explained by MLLMs. 
%To accomplish this we use variance partitioning approach discussed in App.~\ref{app:sharedVarianceDetails}. 
%Fig.~\ref{fig:unique_shared_narrative_all} presents Venn diagrams for the whole brain, language and visual regions, depicting shared and unique variance across these regions between narrative understanding and other task instructions. Similarly, 
We show analysis for all pairs from the 13 tasks in Table~\ref{tab:variancepartitioning}. Across nearly all task pairs, the whole brain region consistently exhibits the highest shared variance. Tasks that are conceptually or functionally related exhibit high shared variance in all regions, indicating similar cognitive processing demands. Higher-level semantic and reasoning tasks (e.g., Narrative Understanding, Commonsense Reasoning, Temporal Ordering) show increased unique variance in the language network, indicating language-specific processing distinct from visual features. High visual load tasks (e.g., Action Recognition, Object and Scene Recognition, Global Appearance) contribute more uniquely in visual cortex, especially when paired with non-visual tasks.

\noindent\textbf{Shared and Unique Variance between Narrative Understanding and Remaining Task Instructions}

\begin{figure*}[!ht]
    \centering
        \begin{minipage}{0.325\textwidth}
    \centering
        \includegraphics[width=0.9\linewidth]{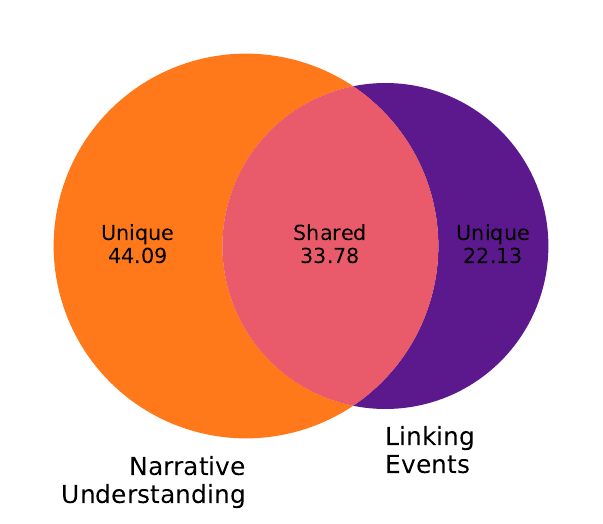}
        \\(a) Whole Brain \\
    \end{minipage}
    \hfill
    \begin{minipage}{0.325\textwidth}
    \centering
        \includegraphics[width=0.9\linewidth]{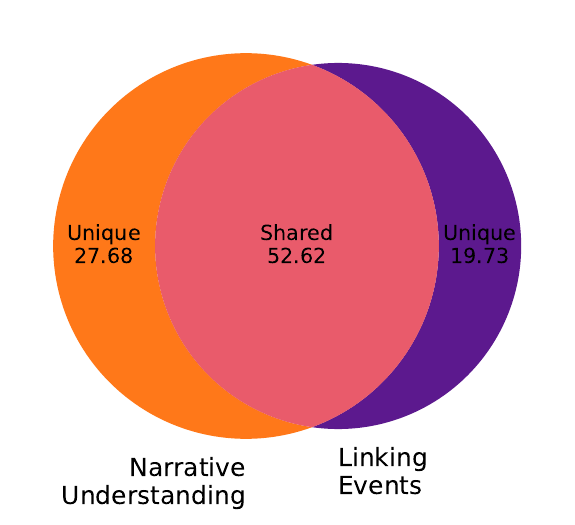}
        \\ (b) Language \\
    \end{minipage}
    \begin{minipage}{0.325\textwidth}
    \centering
        \includegraphics[width=0.9\linewidth]{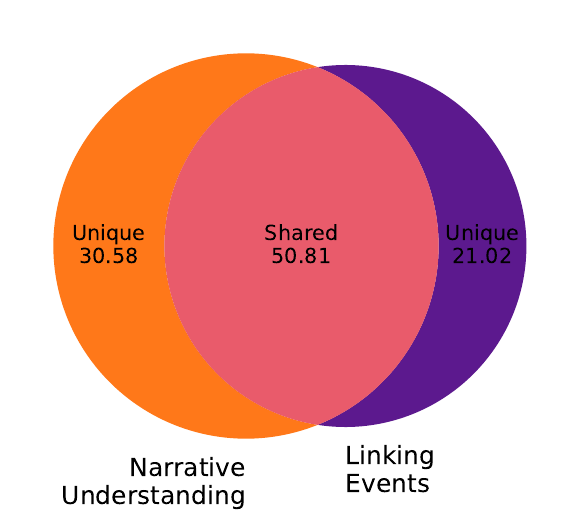}
        \\(c) Visual \\
    \end{minipage}
    
    \begin{minipage}{0.325\textwidth}
    \centering
        \includegraphics[width=0.9\linewidth]{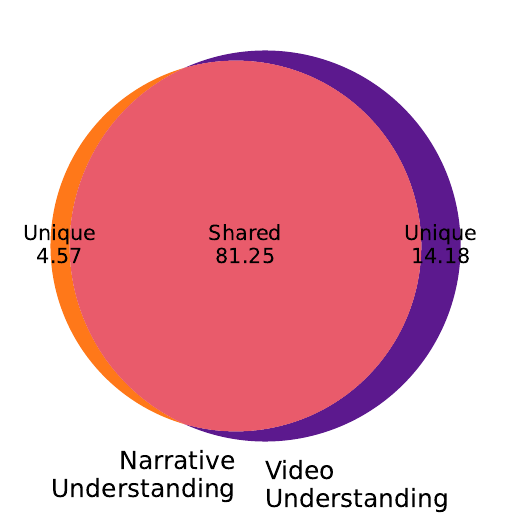}
        \\(a) Whole Brain \\
    \end{minipage}
    \hfill
    \begin{minipage}{0.325\textwidth}
    \centering
        \includegraphics[width=0.9\linewidth]{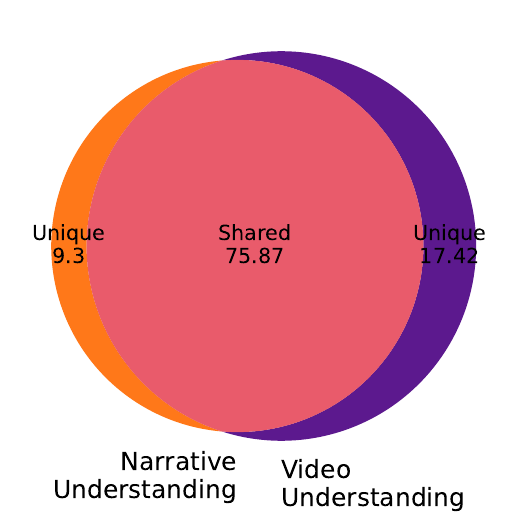}
        \\ (b) Language \\
    \end{minipage}
    \begin{minipage}{0.325\textwidth}
    \centering
        \includegraphics[width=0.9\linewidth]{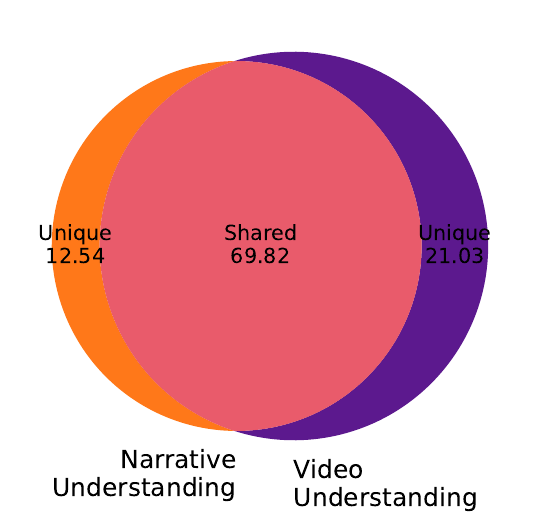}
        \\(c) Visual \\
    \end{minipage}
    
     \centering
    \begin{minipage}{0.325\textwidth}
    \centering
        \includegraphics[width=0.9\linewidth]{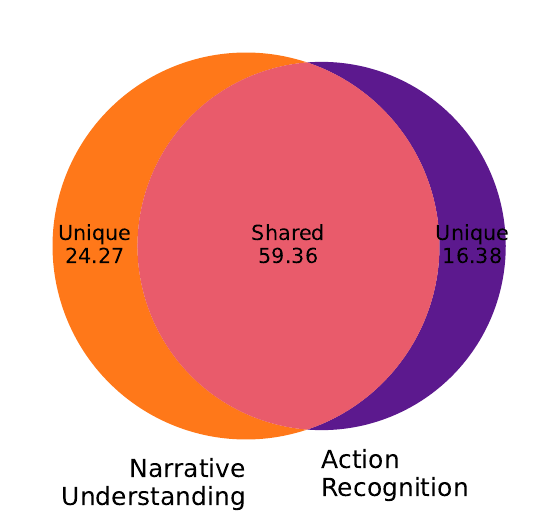}
        \\(a) Whole Brain \\
    \end{minipage}
    \hfill
    \begin{minipage}{0.325\textwidth}
    \centering
        \includegraphics[width=0.9\linewidth]{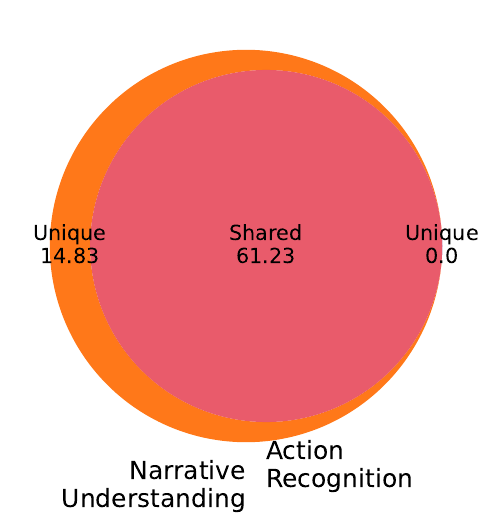}
        \\ (b) Language \\
    \end{minipage}
    \begin{minipage}{0.325\textwidth}
    \centering
        \includegraphics[width=0.9\linewidth]{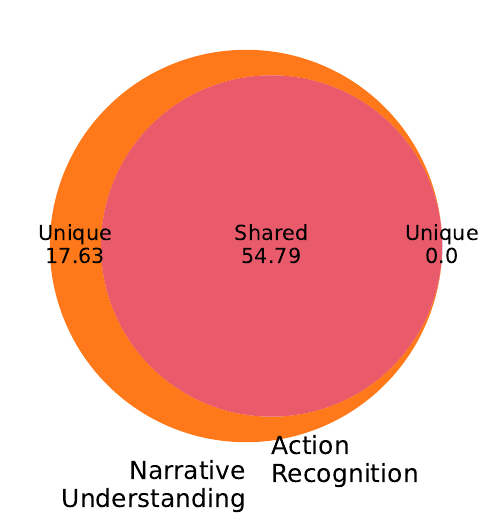}
        \\(c) Visual \\
    \end{minipage}
    \centering
    \begin{minipage}{0.325\textwidth}
    \centering
        \includegraphics[width=0.9\linewidth]{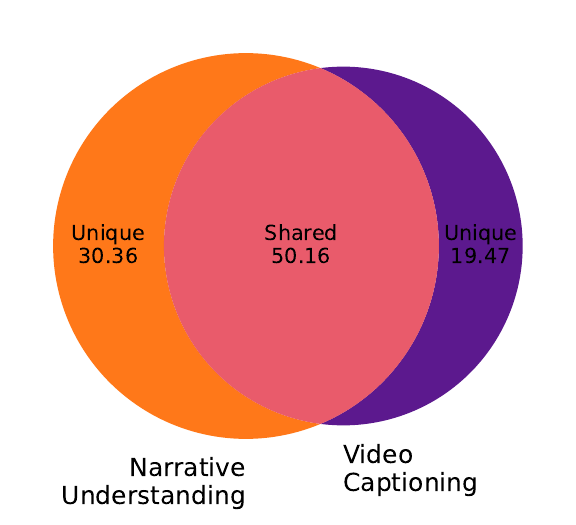}
        \\(a) Whole Brain \\
    \end{minipage}
    \hfill
    \begin{minipage}{0.325\textwidth}
    \centering
        \includegraphics[width=0.9\linewidth]{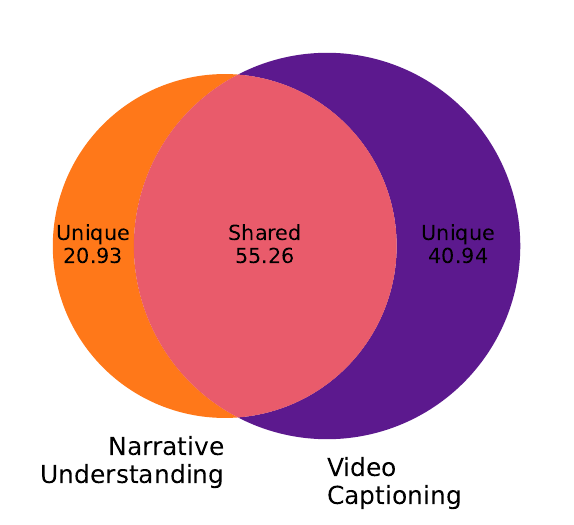}
        \\ (b) Language \\
    \end{minipage}
    \begin{minipage}{0.325\textwidth}
    \centering
        \includegraphics[width=0.9\linewidth]{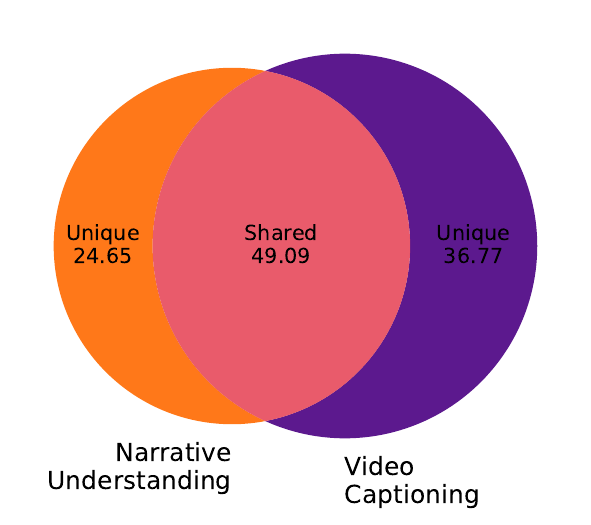}
        \\(c) Visual \\
    \end{minipage}
    \caption{Shared and Unique Variance: Narrative Understanding vs. Linking Events
Dark orange (left) shows variance unique to Narrative Understanding, indigo (right) shows variance unique to Linking Events, and the overlap indicates shared variance between both tasks.}
    \label{fig:unique_shared_narrative_all}
\end{figure*}

Fig.~\ref{fig:unique_shared_narrative_all} shows the shared variance of the Narrative Understanding task with other video tasks for Qwen-2.5-VL. 
%The voxels are projected onto the flattened cortical surface of a representative subject (S1) for the Qwen-2.5-VL video MLLM.

% \begin{figure}[t]
%     \centering
%     \includegraphics[width=0.8\linewidth]{images/shared_instructions.pdf}
%     \caption{Share variance of video tasks: 
%     % , with applied hex color codes for the 10 task instructions, 
%     The voxels are projected onto the flattened cortical surface of a representative subject (S1) for the Qwen-2.5-VL video MLLM.}
%     \label{fig:shared_unique_variance_qwen2.5-vl}
% \end{figure}

Table~\ref{tab:variancepartitioning} presents shared and unique variance explained by pairs of video tasks using brain-informed models across three neural regions: whole brain, visual cortex, and language network. The results are averaged across subjects and show how well representations from each task pair align with brain activity in specific regions.

Key Observations are as follows.
\begin{itemize}
    \item Whole Brain Shows Dominant Shared Variance: Across nearly all task pairs, the whole brain region consistently exhibits the highest shared variance (often $>$80\% in early task pairs).
For example, the pair Action Recognition and Video Understanding (1–2) shows 90.69\% shared variance, with very little unique variance from either task.
This suggests high redundancy and common processing across tasks when considering global brain activity.
\item Visual and Language Regions Yield More Balanced Partitioning: In contrast, visual and language-selective voxels exhibit lower shared variance and comparatively higher unique contributions from individual tasks. For the same task pair (1–2), shared variance in visual is 72.05\%, and in language it is 77.46\%, with higher unique components ($\sim$10-14\%).
This suggests that fine-grained processing differences are more pronounced in modality-specific regions.
\item Task Similarity Reflects in Shared Variance: Tasks that are conceptually or functionally related (e.g., Narrative Understanding-Linking Events (10-13) or Emotion and Sentiment Analysis-Linking Events (11-13)) exhibit high shared variance in all regions, indicating similar cognitive processing demands.
Conversely, task pairs with less conceptual overlap (e.g., Object Recognition-Commonsense Reasoning (5-6) or Visual QA-Object Recognition (3-5)) show lower shared variance and higher unique variance, especially in language and visual regions.
\item Language Regions Show Selectivity for High-Level Tasks: Higher-level semantic and reasoning tasks (e.g., Narrative Understanding, Commonsense Reasoning, Temporal Ordering) show increased unique variance in the language network, indicating language-specific processing distinct from visual features.
For instance, pair 6-13 (Commonsense Reasoning-Linking Events) yields 16.75\% unique variance for Linking Events in the language network.
\item Visual Cortex Captures Scene and Action Differentiation:
Tasks with high visual load (e.g., Action Recognition, Object and Scene Recognition, Global Appearance) contribute more uniquely in the visual cortex, especially when paired with non-visual tasks.
\end{itemize}

\renewcommand{\arraystretch}{0.9}
\begin{table*}[!t]
    \centering
    \scriptsize
    \begin{tabular}{|c|c|c|c|c|c|c|c|c|c|c|}
    \hline
&&\multicolumn{3}{c|}{Whole Brain}&\multicolumn{3}{c|}{Visual}&\multicolumn{3}{c|}{Language}\\\hline
Task1&Task2&Shared&Uniq1&Uniq2&Shared&Uniq1&Uniq2&Shared&Uniq1&Uniq2\\\hline
1&2&90.69&5.26&4.05&72.05&13.91&14.04&77.46&12.07&10.47\\\hline
1&3&83.53&10.05&6.42&73.67&10.28&16.05&77.05&10.72&12.23\\\hline
1&4&84.51&9.65&5.84&71.87&13.82&14.31&75.97&12.27&11.76\\\hline
1&5&79.16&13.51&7.33&66.82&14.35&18.83&73.47&13.07&13.46\\\hline
1&6&81.48&13.34&5.18&68.44&17.28&14.28&73.59&15.37&11.04\\\hline
1&7&83.07&10.44&6.49&71.99&11.88&16.13&75.20&12.30&12.50\\\hline
1&8&81.25&14.18&4.57&69.82&17.63&12.54&75.87&14.83&9.30\\\hline
1&9&86.94&7.57&5.50&73.42&10.25&16.34&78.27&9.05&12.68\\\hline
1&10&84.55&9.06&6.39&73.46&10.59&15.95&76.42&10.32&13.26\\\hline
1&11&85.44&8.51&6.05&74.92&11.12&13.96&76.56&10.96&12.48\\\hline
1&12&82.46&11.66&5.88&72.88&12.75&14.37&76.02&12.50&11.48\\\hline
1&13&91.81&4.20&3.99&74.92&11.82&13.26&80.06&10.00&9.94\\\hline
2&3&83.59&9.72&6.69&73.14&11.39&15.47&74.15&12.80&13.05\\\hline
2&4&86.25&7.40&6.36&73.32&13.52&13.16&74.41&12.14&13.45\\\hline
2&5&77.09&14.33&8.58&64.55&17.14&18.31&70.20&15.08&14.72\\\hline
2&6&79.86&13.99&6.15&69.43&17.86&12.71&73.10&14.96&11.94\\\hline
2&7&83.62&9.46&6.92&72.53&12.65&14.82&71.61&14.43&13.95\\\hline
2&8&81.30&13.10&5.60&67.98&18.96&13.05&72.05&16.07&11.88\\\hline
2&9&86.64&7.42&5.93&73.55&12.35&14.11&75.55&10.62&13.83\\\hline
2&10&85.25&7.97&6.78&72.98&12.28&14.73&73.28&12.51&14.21\\\hline
2&11&84.70&8.31&7.00&73.27&12.25&14.48&72.48&13.27&14.25\\\hline
2&12&82.97&11.16&5.88&73.06&14.41&12.54&72.99&14.99&12.02\\\hline
2&13&91.78&3.66&4.55&74.89&12.59&12.52&78.19&9.77&12.03\\\hline
3&4&68.68&13.67&17.64&68.53&18.38&13.09&71.98&14.19&13.83\\\hline
3&5&50.07&24.61&25.32&52.60&24.08&23.32&60.68&17.79&21.53\\\hline
3&6&61.39&21.67&16.94&61.59&22.97&15.44&65.21&18.68&16.12\\\hline
3&7&65.21&17.99&16.80&64.73&20.33&14.94&66.85&17.80&15.35\\\hline
3&8&66.30&20.20&13.49&61.04&23.96&15.00&62.43&21.86&15.71\\\hline
3&9&70.23&13.71&16.06&70.07&16.68&13.25&72.20&12.52&15.28\\\hline
3&10&66.99&13.00&20.01&68.60&15.97&15.42&64.43&15.79&19.78\\\hline
3&11&68.07&14.39&17.54&66.84&17.50&15.66&66.97&16.85&16.18\\\hline
3&12&61.81&19.24&18.95&65.81&19.69&14.50&67.09&17.92&14.99\\\hline
3&13&83.92&6.44&9.64&71.83&16.87&11.31&76.76&12.86&10.38\\\hline
4&5&55.03&24.36&20.61&53.05&20.94&26.00&59.06&18.82&22.13\\\hline
4&6&61.72&25.66&12.62&59.66&24.72&15.62&63.75&21.99&14.26\\\hline
4&7&69.00&17.62&13.38&66.08&17.45&16.47&67.89&17.50&14.61\\\hline
4&8&63.88&21.85&14.27&60.24&23.59&16.17&65.25&19.95&14.80\\\hline
4&9&71.16&16.55&12.28&65.51&18.15&16.34&68.66&16.14&15.19\\\hline
4&10&66.37&18.11&15.53&63.85&17.11&19.04&57.73&20.94&21.33\\\hline
4&11&72.37&13.56&14.07&70.00&13.01&16.99&70.64&13.35&16.02\\\hline
4&12&66.38&18.76&14.86&64.80&17.67&17.53&67.94&17.21&14.85\\\hline
4&13&86.69&6.09&7.23&71.23&16.28&12.49&76.56&13.87&9.57\\\hline
5&6&50.13&27.24&22.63&51.63&27.81&20.56&58.56&23.05&18.39\\\hline
5&7&49.08&24.63&26.29&53.55&25.15&21.30&55.77&24.66&19.57\\\hline
5&8&47.03&27.55&25.43&53.22&28.86&17.93&53.88&26.92&19.21\\\hline
5&9&55.06&21.61&23.34&56.84&24.75&18.42&62.62&19.24&18.15\\\hline
5&10&47.76&23.54&28.70&55.84&22.99&21.17&54.52&22.48&23.00\\\hline
5&11&52.17&22.58&25.25&57.44&22.32&20.24&57.94&22.48&19.58\\\hline
5&12&47.50&26.51&25.99&56.38&25.48&18.15&58.21&23.50&18.29\\\hline
5&13&79.36&6.98&13.67&66.31&16.96&16.74&71.80&12.91&15.29\\\hline
6&7&60.01&17.04&22.96&59.05&17.09&23.86&61.14&18.01&20.84\\\hline
6&8&54.31&21.48&24.22&57.44&21.55&21.01&62.62&18.13&19.25\\\hline
6&9&64.33&13.06&22.61&60.10&16.20&23.69&64.68&13.72&21.60\\\hline
6&10&57.84&16.91&25.25&61.41&14.59&24.00&61.01&16.15&22.84\\\hline
6&11&62.94&14.26&22.81&62.17&15.15&22.68&63.32&15.40&21.28\\\hline
6&12&55.82&19.64&24.54&60.18&17.37&22.45&60.36&18.93&20.71\\\hline
6&13&81.42&5.21&13.37&67.46&13.51&19.02&71.93&11.31&16.75\\\hline
7&8&58.19&23.15&18.65&60.58&23.47&15.95&61.00&20.86&18.13\\\hline
7&9&70.87&14.02&15.11&70.43&15.05&14.51&71.25&12.70&16.05\\\hline
7&10&68.57&12.51&18.92&67.67&13.27&19.06&63.76&14.39&21.84\\\hline
7&11&60.77&18.94&20.29&58.79&21.23&19.98&55.14&21.77&23.09\\\hline
7&12&66.57&17.86&15.57&67.97&17.05&14.98&67.18&17.38&15.44\\\hline
7&13&85.27&6.01&8.72&72.66&15.56&11.78&74.88&13.08&12.03\\\hline
8&9&62.84&15.99&21.18&63.11&15.66&21.22&68.03&13.67&18.31\\\hline
8&10&60.10&17.38&22.52&59.39&16.80&23.81&60.46&16.80&22.74\\\hline
8&11&60.31&14.63&25.07&61.67&13.24&25.09&61.38&15.64&22.98\\\hline
8&12&60.04&18.69&21.28&62.31&17.41&20.28&65.74&16.70&17.56\\\hline
8&13&81.06&5.66&13.27&68.01&14.38&17.61&74.50&11.65&13.85\\\hline
9&10&69.21&14.34&16.44&68.83&12.98&18.19&67.69&15.88&16.44\\\hline
9&11&70.80&13.15&16.05&69.96&14.08&15.96&70.82&14.04&15.15\\\hline
9&12&69.68&16.60&13.72&70.09&14.45&15.46&70.62&16.10&13.29\\\hline
9&13&87.40&5.23&7.37&72.02&15.46&12.53&77.48&12.70&9.82\\\hline
10&11&68.63&16.35&15.02&67.96&16.43&15.61&64.85&19.12&16.04\\\hline
10&12&65.06&20.66&14.27&63.79&21.85&14.36&61.84&23.65&14.50\\\hline
10&13&85.63&6.39&7.99&72.34&16.92&10.73&75.85&14.09&10.06\\\hline
11&12&61.95&22.51&15.54&65.60&19.55&14.85&63.80&21.51&14.69\\\hline
11&13&86.42&6.00&7.58&74.60&14.29&11.11&76.83&12.89&10.28\\\hline
12&13&83.82&5.77&10.41&71.56&15.38&13.06&75.37&12.20&12.43\\\hline
    \end{tabular}
    \caption{Variance partitioning for all the 13 video tasks averaged across all subjects for whole brain, visual and language regions with Qwen-2.5-VL model. Tasks are as follows: (1) Action Recognition (2) Video Understanding   (3) Visual Question Answering  (4) Video Captioning (5) Object and Scene Recognition (6) Commonsense Reasoning (7) Spatial Understanding (8) Temporal Ordering (9) Video reasoning (10) Narrative Understanding (11) Emotion and Sentiment Analysis (12) Global Appearance (13) Linking Events.}
    \label{tab:variancepartitioning}
\end{table*}

\FloatBarrier

\section{Extended Discussion}
\label{app:extendedDiscussion}
(1) ICL-MLLM representations are strongly coupled to instruction-text semantics, whereas IT-MLLM representations show weak coupling to instruction-text semantics, suggesting that IT-MLLMs produce representations organized by what the task is rather than by surface instruction wording. 

(2) Although instruction-tuned video MLLMs show strong brain alignment across the whole brain (including language, visual, and auditory regions), audio MLLMs show stronger alignment primarily in auditory and language-related areas such as the middle frontal gyrus (MFG). This suggests that instruction-tuned audio MLLMs capture features relevant to auditory processing, consistent with patterns reported in previous studies~\citep{oota2024speech,oota2024multi}. However, their performance remains comparable to non-instruction-tuned multimodal models, indicating that further improvements are needed for instruction-tuned audio MLLMs to better capture brain-relevant representations, an effort that aligns with recent work on inducing brain-relevant biases in model design~\citep{moussa2025improving,vattikonda2025brainwavlm}.

(3) Task-specific instructions are effective at predicting multimodal brain activity across different regions, indicating that both IT video and audio MLLMs produce distinct task-conditioned representations. These representations support region-differentiated brain alignment patterns, unlike prior work by~\citet{oota2025correlating}, which did not observe such differentiation when using unimodal stimuli (e.g., static images). Specifically, certain audio instructions, such as audio captioning and audio understanding, show stronger alignment with language-related regions, while tasks such as sound event detection align more with the auditory cortex and temporal lobe. These findings indicate that IT-MLLMs can serve as a useful framework for systematically varying task goals through instructions while holding the stimulus fixed, allowing researchers to examine task-conditioned brain alignment patterns.

(4) The observed correspondence between IT-MLLM layers and the brain's processing hierarchy is consistent with structured representational patterns, with shallow layers aligning more with sensory regions and deeper layers aligning more with higher-order semantic regions. This layered alignment improves interpretability and supports the use of IT-MLLMs as probes for studying how task-driven information is differentially encoded across cortical regions.

Our findings also indicate that despite the growing popularity of instruction-tuned video and audio MLLMs for handling generic task instructions, we are still far from fully understanding how language-based instructions modulate information flow through model layers and how fine-grained details are differentially represented across layers. Future work should focus on leveraging the alignment strengths of these models using more fine-grained instruction-driven prompts, similar to controlled stimulus paradigms in neuroscience, to deepen our understanding of how task-conditioned representations relate to neural responses.

% \section{LLM Usage}
% \label{app:llm_usage}
% We used OpenAI ChatGPT for grammar correction and language polishing.

\FloatBarrier

\section{Discussion on Controlling Architectural and Pretraining Differences Across MLLMs}
\label{app:architectural_differences_mllms}

MLLMs differ in pretraining corpus and design. We therefore took three complementary steps to ensure fair and scientifically valid comparisons.
\begin{itemize}
    \item (i) Controls in the evaluation protocol: Across all models, our encoding pipeline is the same as we use the same instruction templates, same video and audio datasets as input for feature extraction, use of the same cross-subject prediction scores for estimating normalized brain alignment per model.
    \item (ii) Matching scale: While all the MLLMs we tested are ~7B to 8B models across both instruction-tuned video and audio MLLMs and in-context learning pretrained MLLMs. Thus, model training sizes close to several million across all instruction-tuned video MLLMs, with varying modality mixes and task diversity.
    \item (iii) Acknowledging training-data heterogeneity and checking robustness: The instruction-tuned video MLLMs differ substantially in training sources: Some (like LLaVA-NeXT-Video and Video-LLaVA) include explicit video instruction datasets, while others (like InstructBLIPVideo and LLaVA-OneVision) primarily adapt image-based instructions to video via frame sampling.  However, across all instruction-tuned videos MLLMs show similar normalized brain alignment to our results suggesting that depth alone does not account for the observed performance differences.
\end{itemize} 

We also acknowledge that testing a larger number of models within a given class can help determine if the effect of modality is robust across various model configurations.

\noindent\textbf{Literature precedent:} Our approach follows established practice in neuro-AI, where models with differing training corpora/architectures are compared under a shared brain-encoding protocol and ceiling normalization to study representational alignment~\citep{schrimpf2021neural,toneva2019interpreting,antonello2021low,aw2022training,loong2023instruction}. Specifically, the extensive precedent in the literature, from studies comparing 43 models~\citep{schrimpf2021neural} to those examining 101 models~\citep{antonello2021low} in language models. Similarly~\citet{oota2024speech} and~\citet{antonello2024scaling} compared several text and speech models during language comprehension. These studies demonstrate that this approach is both valid and valuable for understanding the relationship between artificial and biological language processing.
It is important to observe that all the above studies utilize a number of models that are different in training architecture and training datasets, however the primary goal of all these studies is to investigate how close the semantic representations captured by each model aligns with brain-relevant semantics.
We have added pretraining corpus details across MLLMs in Table~\ref{tab:mllm_training_data}.

\begin{table*}[!ht]
\centering
\scriptsize
\caption{\textbf{Training data and scale of compared multimodal large language models (MLLMs).}}
\label{tab:mllm_training_data}
\begin{tabular}{|l|c|c|}
\hline
\textbf{Model} & \textbf{Training Data} & \textbf{Size} \\
\hline
InstructBLIP-Video &
26 public datasets: MSCOCO, TextCaps, NoCaps, VQA v2, iVQA, MSRVTT-QA, MSVD-QA &
Not specified \\
Video-LLaVA &
LAION-CC-SBU (558K), Valley (702K), LLaVA-Instruct (665K image-text + 100K video-text) &
$\sim$1.3M total \\
LLaVA-NeXT-Video &
LLaVA-Video-178K synthetic + real video QA/caption data &
$\sim$1.4M total \\
Qwen-2.5-VL &
Web-scale 4T tokens (text + multimodal), post-trained on 2M mixed samples &
4T tokens + 2M samples \\
VideoChat-R1 &
Qwen-2.5-VL backbone, 18K video instruction samples (temporal grounding, tracking, QA) &
18K samples \\
LLaVA-One-Vision &
OneVision dataset (3.2M single-image + 1.6M multi-image/video), Evo-Instruct (143K) &
$\sim$5M total \\
TVLT &
HowTo100M, YTTemporal180M (video-audio pairs) &
Large-scale (not stated) \\
VideoMAE &
Kinetics-400, Something-Something v2, Epic Kitchens 100 &
$\sim$550K videos \\
AST &
AudioSet, ESC-50, Speech Commands &
$\sim$2$+$M audio clips \\
\hline
\end{tabular}
\end{table*}

\section{Normalized brain alignment: cross-subject vs. repeat-based EV ceiling}
\label{app:normalized_alignment_crosssubject_ev}

We also compute the explainable variance (EV) using repeated test movies and perform thresholding on EV voxels. The EV is computed now based on~\citep{schoppe2016measuring}. Using EV with threshold of 0.05, the normalized brain alignment on this mask. For fair comparison, we used the same reliability threshold ($\geq$ 0.05) as in the cross-subject ceiling analysis.

\begin{figure*}[!ht] 
\centering
\begin{minipage}{\textwidth}
\centering
    \includegraphics[width=0.6\linewidth]{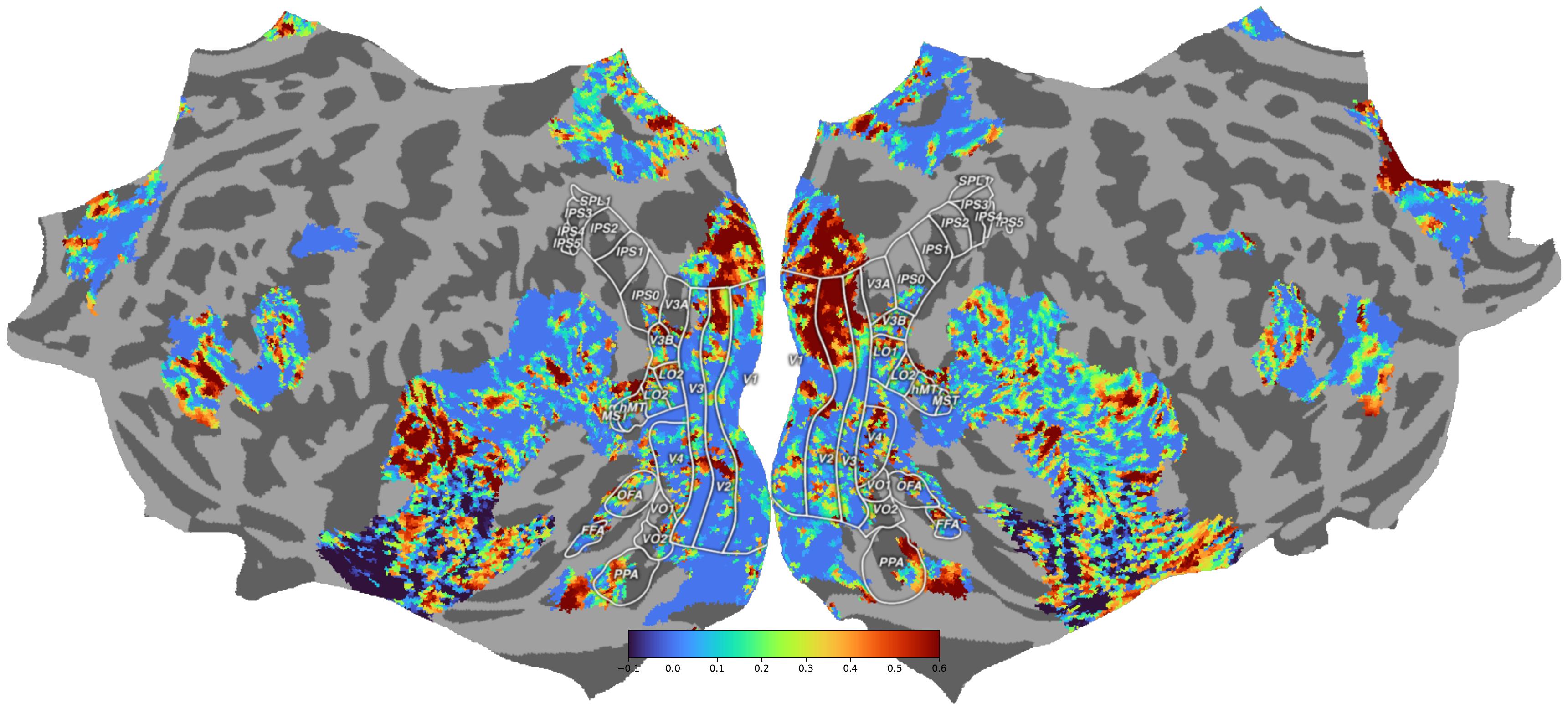}
    \\(a) Subject-01 \\
\end{minipage}
\begin{minipage}{\textwidth}
\centering
    \includegraphics[width=0.6\linewidth]{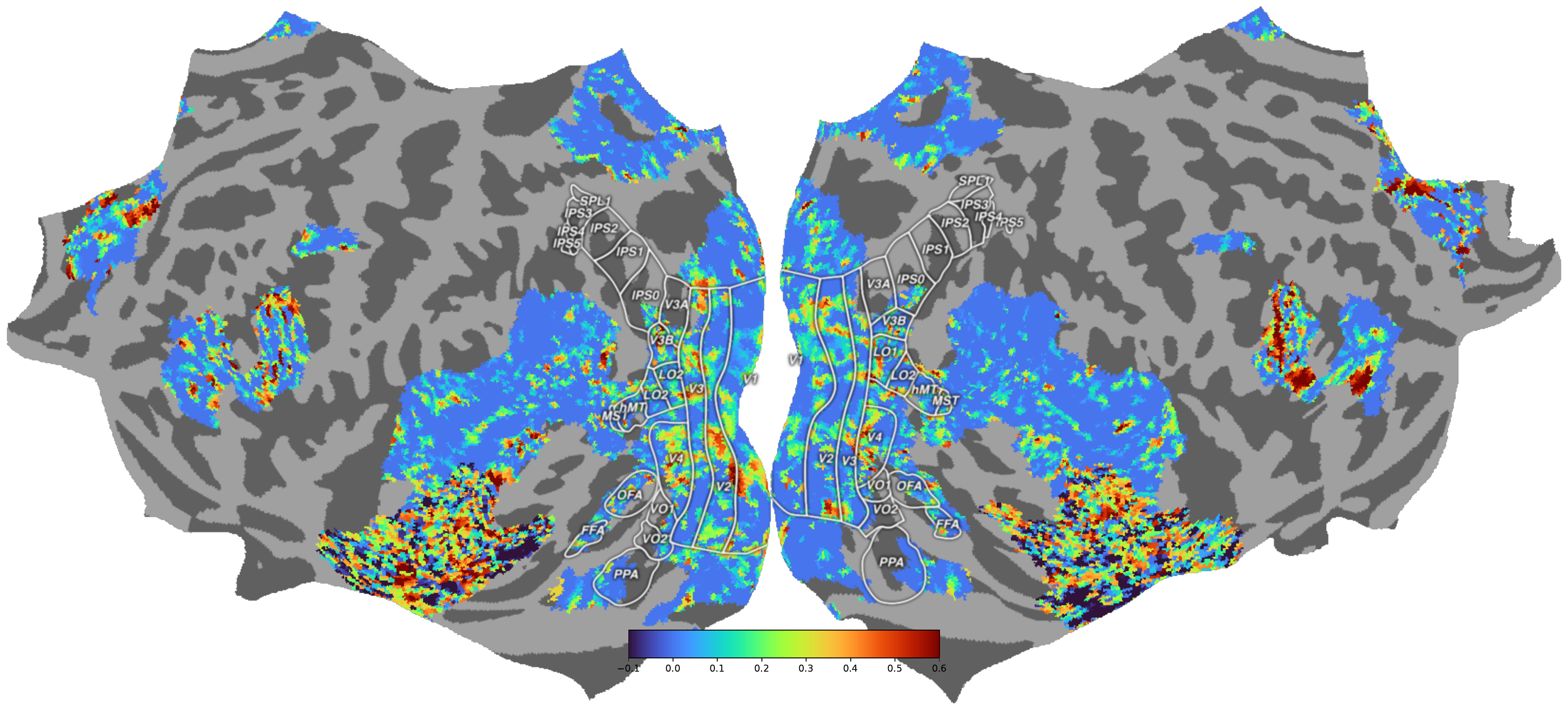}
    \\(a) Subject-02 \\
\end{minipage}
\begin{minipage}{\textwidth}
\centering
    \includegraphics[width=0.6\linewidth]{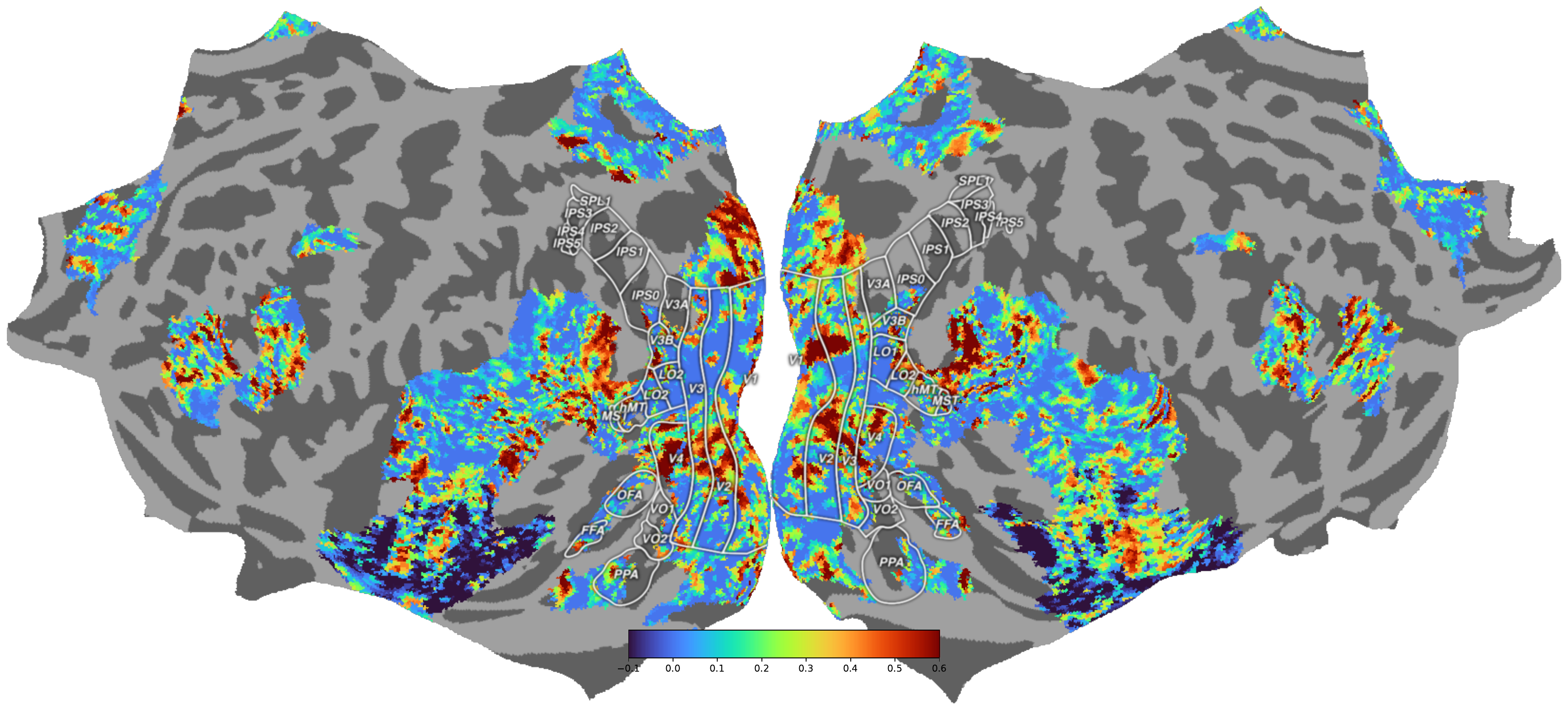}
    \\(b) Subject-03 \\
\end{minipage}
\begin{minipage}{\textwidth}
\centering
    \includegraphics[width=0.6\linewidth]{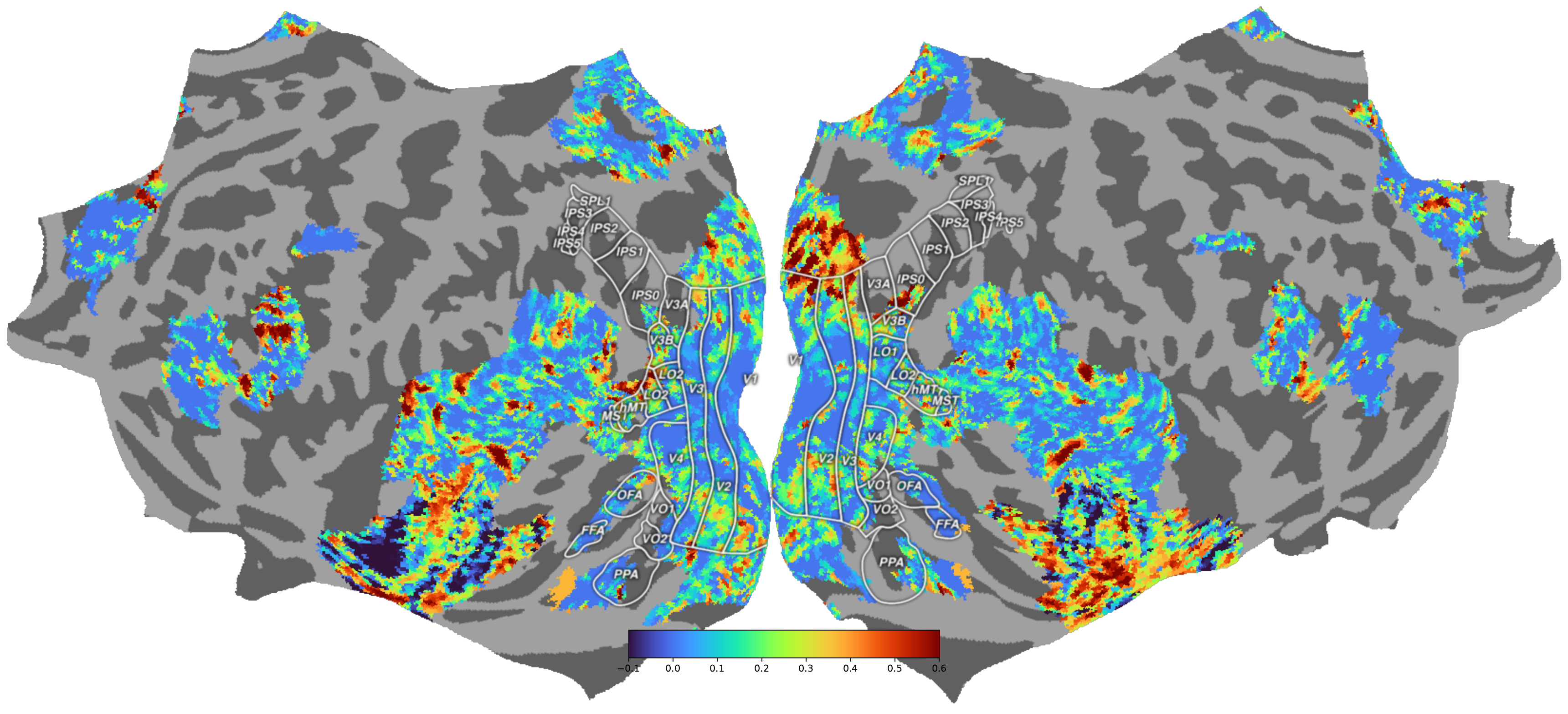}
    \\(c) Subject-05 \\
\end{minipage}
\caption{Estimated explainable variance for all four participants for the naturalistic movie watching. Explainable variance scores for each voxel in each subject are projected onto the subject’s flattened cortical surface.}
\label{fig:explainable_variance_subjects}
\end{figure*}

From the Table~\ref{tab:ceiling_comparison},  we find that the normalized brain alignment computed with the repeat-based EV ceiling is very similar to that obtained with the cross-subject ceiling; the model ranking is unchanged. We also include cortical flatmaps of repeat-based explainable variance (EV) for each participant in the Fig.~\ref{fig:explainable_variance_subjects}, showing the spatial distribution of reliable voxels (EV $\geq$ 0.05).

\begin{table}[!ht]
\centering
\scriptsize
\caption{Normalized brain alignment: cross-subject vs.\ repeat-based EV ceiling}
\label{tab:ceiling_comparison}
\begin{tabular}{|l|c|c|c|c|}
\hline
\textbf{Model} & \textbf{Cross-subject} & \textbf{Repeat EV} & $\Delta$ (Repeat $-$ Cross) & \%$\Delta$ \\
\hline
InstructBLIP Video    & 0.669 & 0.645 & $-0.024$ & $-3.64\%$ \\
Video-LLaVA           & 0.650 & 0.652 & $+0.002$ & $+0.35\%$ \\
LLaVA-NeXT-Video      & 0.678 & 0.642 & $-0.037$ & $-5.39\%$ \\
Qwen-2.5-VL-7B        & 0.746 & 0.689 & $-0.056$ & $-7.54\%$ \\
LLaVA-OneVision       & 0.666 & 0.639 & $-0.028$ & $-4.13\%$ \\
\hline
\end{tabular}
\end{table}

\section{Cross-Subject Generalization Analysis}
\label{app:cross_subject}
To address potential concerns about subject-level generalization, we conducted a held-out-subject evaluation: encoding models trained on one subject were evaluated on a held-out subject. Results across all train-test subject pairs are reported in Table~\ref{tab:cross_subject}.

\begin{table}[!ht]
\centering
\caption{Cross-subject generalization. Encoding models trained on one subject and evaluated on a different held-out subject. Cross-subject alignment is comparable to within-subject performance (Mid: 1.01×; Late: 1.03×), indicating that deeper task-conditioned representations transfer across participants.}
\label{tab:cross_subject}
\begin{tabular}{|l|c|c|c|c|c|c|c|}
\hline
Train$\rightarrow$Test & L14 & L15 & L27 & L28 & L29 & Mid Mean & Late Mean \\
\hline
S1$\rightarrow$S2 & 0.502 & 0.668 & 0.712 & 0.705 & 0.698 & 0.585 & 0.705 \\
S1$\rightarrow$S3 & 0.484 & 0.679 & 0.781 & 0.770 & 0.761 & 0.581 & 0.770 \\
S1$\rightarrow$S5 & 0.638 & 0.646 & 0.792 & 0.684 & 0.790 & 0.642 & 0.755 \\
S3$\rightarrow$S1 & 0.749 & 0.721 & 0.806 & 0.848 & 0.852 & 0.735 & 0.835 \\
S3$\rightarrow$S2 & 0.723 & 0.702 & 0.764 & 0.792 & 0.781 & 0.713 & 0.779 \\
S3$\rightarrow$S5 & 0.640 & 0.737 & 0.784 & 0.892 & 0.911 & 0.689 & 0.862 \\
S5$\rightarrow$S1 & 0.732 & 0.654 & 0.832 & 0.871 & 0.842 & 0.693 & 0.848 \\
S5$\rightarrow$S2 & 0.711 & 0.661 & 0.792 & 0.821 & 0.806 & 0.686 & 0.806 \\
S5$\rightarrow$S3 & 0.707 & 0.619 & 0.762 & 0.894 & 0.845 & 0.663 & 0.834 \\
\hline
\textbf{Average} & \textbf{0.654} & \textbf{0.676} & \textbf{0.781} & \textbf{0.808} & \textbf{0.809} & \textbf{0.669} & \textbf{0.757} \\
\hline
\end{tabular}
\end{table}

\textbf{Comparison to within-subject performance.}

\begin{center}
\begin{tabular}{|l|c|c|c|}
\hline
Setting & Mid Layers (L14-15) & Late Layers (L27-29) & Cross/Within Ratio \\
\hline
Within-subject & 0.661 & 0.734 & -- \\
Cross-subject & 0.669 & 0.757 & 1.01× / 1.03× \\
\hline
\end{tabular}
\end{center}

\textbf{Methodological note.} Because stimulus features are identical across subjects while neural responses are subject-specific, naive multi-subject concatenation would duplicate inputs with different targets, confounding interpretation. We therefore use train-on-one-subject, test-on-another-subject transfer as a cleaner test of cross-subject generalization.

\section{Limitations}
\label{app:limitations}

One possible limitation of our study lies in interpreting the differences in brain alignment between instruction-tuned video and audio MLLMs. The models we evaluate differ in several aspects, including the amount of training data and the specific objective functions used during training. To address this concern, we evaluated multiple models of each type, spanning a range of training objectives and dataset sizes, and found that our key results generalize within both video and audio MLLM categories. Still, it is possible that some of the differences in brain alignment may still be influenced by confounding factors related to model architecture, training objectives, or data scale. Future work should explore these questions using models that are more tightly controlled across these dimensions.

\end{document}